\documentclass[a4paper,fleqn,usenatbib,useAMS]{mnras}

\usepackage{graphicx}	% Including figure files
\usepackage{amsmath}	% Advanced maths commands
\usepackage{amssymb}	% Extra maths symbols
\usepackage{multicol}        % Multi-column entries in tables
\usepackage{bm}		% Bold maths symbols, including upright Greek
\usepackage{pdflscape}	% Landscape pages

\newcommand{\be}{\begin{equation}}
\newcommand{\ee}{\end{equation}}
\newcommand{\ba}{\begin{eqnarray}}
\newcommand{\ea}{\end{eqnarray}}

\newcommand{\fracb}[2]{\left(\frac{#1}{#2}\right)}

\usepackage{etoolbox}
\makeatletter
\patchcmd\@combinedblfloats{\box\@outputbox}{\unvbox\@outputbox}{}{%
   \errmessage{\noexpand\@combinedblfloats could not be patched}%
}%
 \makeatother
\usepackage[T1]{fontenc}
\usepackage{ae,aecompl}

\usepackage{newtxtext,newtxmath}
\title[Prompt GRB linear polarization]{Linear polarization in gamma-ray burst prompt emission}

\author[Gill, Granot, \& Kumar (2019)]{
Ramandeep Gill,$^{1,2,3}$\thanks{Contact e-mail: \href{mailto:rsgill.rg@gmail.com}{rsgill.rg@gmail.com}} 
Jonathan Granot,$^{1,4}$\thanks{Contact e-mail: \href{mailto:granot@openu.ac.il}{granot@openu.ac.il}} 
and Pawan Kumar$^{5}$\thanks{Contact e-mail: \href{mailto:pk@astro.as.utexas.edu}{pk@astro.as.utexas.edu}}
\\
$^{1}$Department of Natural Sciences, The Open University of Israel, 
1 University Road, PO Box 808, Raanana 4353701, Israel \\
$^{2}$Physics Department, Ben-Gurion University, P.O.B. 653, 
Beer-Sheva 84105, Israel \\
$^{3}$Institute for Theoretical Physics, Goethe University of Frankfurt, 
Max-von-Laue-Str. 1, D-60438 Frankfurt am Main, Germany \\
$^{4}$Department of Physics, The George Washington University, Washington, DC 20052, USA \\
$^{5}$Department of Astronomy, University of Texas at Austin, Austin, TX 78712, USA
}

\date{Last updated; in original form}

\pubyear{2019}

\begin{document}
\label{firstpage}
\pagerange{\pageref{firstpage}--\pageref{lastpage}}
\maketitle

%%%% ABSTRACT  %%%%%%%%%%%%%%%%%%%%%%%%%%%%%%%%%%%%%%%%%%%%%%%%%%%%%%%%%%%%%
\begin{abstract}
Despite being hard to measure, GRB prompt $\gamma$-ray emission polarization is a valuable probe of the dominant 
emission mechanism and the GRB outflow's composition and angular structure. During the prompt emission the GRB 
outflow is ultra-relativistic with Lorentz factors $\Gamma\gg1$. We describe in detail the linear polarization 
properties of various emission mechanisms: synchrotron radiation from different magnetic field structures 
(ordered: toroidal $B_{\rm tor}$ or radial $B_\parallel$, and random: normal to the radial direction $B_\perp$), 
Compton drag, and photospheric emission. We calculate the polarization for different GRB jet angular structures 
(e.g. top-hat, Gaussian, power-law) and viewing angles $\theta_{\rm obs}$. Synchrotron with $B_\perp$ can produce 
large polarizations, up to $25\%\lesssim\Pi\lesssim45\%$, for a top-hat jet but only for lines of sight just outside 
($\theta_{\rm obs}-\theta_j\sim1/\Gamma$) the jet's sharp edge at $\theta=\theta_j$. The same also holds for Compton 
drag, albeit with a slightly higher overall $\Pi$. Moreover, we demonstrate how $\Gamma$-variations during the GRB 
or smoother jet edges (on angular scales $\gtrsim 0.5/\Gamma$) would significantly reduce $\Pi$. We construct a 
semi-analytic model for non-dissipative photospheric emission from structured jets. Such emission can produce up to 
$\Pi\lesssim15\%$ with reasonably high fluences, but this requires steep gradients in $\Gamma(\theta)$.  A polarization 
of $50\%\lesssim\Pi\lesssim65\%$ can robustly be produced only by synchrotron emission from a transverse magnetic 
field ordered on angles $\gtrsim\!1/\Gamma$ around our line of sight (like a global toroidal field, $B_{\rm tor}$, for 
$1/\Gamma<\theta_{\rm obs}<\theta_j$). Therefore, such a model would be strongly favored even by a single secure 
measurement within this range. We find that such a model would also be favored if $\Pi\gtrsim20\%$ is measured in most 
GRBs within a large enough sample, by deriving the polarization distribution for our different emission and jet models.
\end{abstract}
%%%%%%%%%%%%%%%%%%%%%%%%%%%%%%%%%%%%%%%%%%%%%%%%%%%%%%%%%%%%%%%%%%%%%%%%%%%%%%%%%%%%%%%%%

% Select between one and six entries from the list of approved keywords.
% Don't make up new ones.
\begin{keywords}
Polarization -- magnetic fields -- radiation mechanisms: general -- gamma-ray bursts: general -- stars: jets
\end{keywords}

%%%%%%%%%%%%%%%%%%%%%%%%%%%%%%%%%%%%%%%%%%%%%%%%%%

%%%%%%%%%  INTRODUCTION  %%%%%%%%%%%%%%%%%%%%%%%%%%%%%%%%%%%%%%%%%%%%%%%
\section{Introduction}
The emission mechanism that produces the soft $\gamma$-ray photons during the exceptionally 
bright but brief prompt emission phase in gamma-ray bursts (GRBs) is still unclear 
\citep[see e.g.][for a review]{KZ15}. The non-thermal spectrum of the prompt emission is traditionally fit by the empirical Band-function 
\citep{Band+93}
that features two power laws that smoothly join at the photon energy $E_{\rm pk}$ where $\nu F_\nu$ peaks. A popular model for its origin is optically-thin synchrotron emission from 
relativistic electrons that are accelerated at internal shocks that form due to the collision of baryonic shells 
in a matter-dominated outflow with a variable Lorentz factor $\Gamma$  \citep[e.g.][]{Rees-Meszaros-94,PM96,SP97,DM98}. However, this model has been challenged by observations of many GRBs
for which synchrotron emission fails \citep[e.g.][]{Crider+97,Preece+98,Preece+02,Ghirlanda+03} to produce the correct 
low-energy spectral slope below $E_{\rm pk}$ \citep[however, see, e.g.][where synchrotron emission has been shown to fit the low energy spectrum with the addition of a spectral break below $E_{\rm pk}$]{Oganesyan+17,Ravasio+18}. 
This inconsistency led to the consideration of alternative models where the main radiation process is multiple inverse-Compton 
scatterings by sub-relativistic electrons below the Thomson photosphere. Such models also yield a Band-like spectrum and fall under 
a general class of dissipative photosphere models (see, e.g., \citealt{Beloborodov-Meszaros-17} for a review; and see, e.g., \citealt{Gill-Thompson-14,Thompson-Gill-14}; 
\citealt{Vurm-Beloborodov-16} for numerical treatments).

The emission mechanism and the magnetic field structure are related to the outflow composition and the dissipation mechanism.
In the standard `fireball' scenario \citep[e.g.][]{Rees-Meszaros-94} the outflow is launched radiation dominated and optically thick to 
Thomson scattering ($\tau_T>1$) due to the small number (even with a mass as small as $10^{-7}M_\odot$) of entrained baryons. 
Its initial temperature is typically around a few MeV, which results in copious production of $e^\pm$-pairs via $\gamma\gamma$-annihilation 
that further increases $\tau_T$. Adiabatic expansion of the flow under its own pressure converts 
the radiation field energy to kinetic energy of the entrained baryons. This gives rise to a 
kinetic energy or matter dominated flow, where the energy is released in internal shocks between 
multiple baryonic shells that form due to variations in $\Gamma$ within the outflow. 
On the other hand, the outflow can be launched Poynting-flux dominated 
\citep[e.g.][]{Thompson-94, LB03}, where the magnetization parameter $\sigma$ (the magnetic to particle energy flux ratio; 
see Eq.~\ref{eq:magnetization}) is initially $\sigma_0\gg1$. In this case magnetic reconnection may efficiently dissipate 
magnetic energy and accelerate particles in magnetically dominated ($\sigma>1$) regions within the outflow, which may power 
the prompt GRB emission. Such magnetic reconnection requires a flipping of the magnetic field polarity near the central 
source, which persists out to large distances, such as in a striped wind from a pulsar or magnetar, or by stochastic field 
flips during accretion onto a black hole.

There are also intermediate scenarios in which the outflow is launched Poynting flux dominated, with $\sigma_0\gg1$ near the central source, 
but then $\sigma$ gradually decreases with the distance from the source as the outflow is accelerated. Initially acceleration is tied to jet 
collimation, but in GRBs this typically saturates at $\sigma\gg1$ and the flow becomes conical. Further acceleration can proceed either through gradual 
magnetic reconnection in a striped wind over a large range of radii 
\citep[e.g.][]{Thompson-94,Lyubarsky-Kirk-01,Spruit+01,Drenkhahn-Spruit-02,Drenkhahn-02} or without magnetic dissipation in a strongly variable outflow 
\citep{GKS11}. In the latter case kinetic dominance ($\sigma<1$) may be achieved, which allows efficient energy dissipation in internal shocks, even 
though the outflow was initially magnetically dominated ($\sigma_0\gg1$).
All of these scenarios are reasonably plausible and can potentially explain the non-thermal GRB prompt emission spectrum 
\citep[see e.g.][for a review]{Granot+15}. However, the magnetic field structure in the emission region may be very different in these two scenarios, as discussed in \S\,\ref{sec:B-field-origin-config}.  

Polarization measurements of the prompt emission can shine some much needed light on the important questions 
regarding the composition of the flow, the magnetic field structure, and the dominant emission mechanism. In particular,
they can be useful for determining the dominant prompt emission mechanism, and may help distinguish between different magnetic field structures, which can both help constrain the outflow composition.
Furthermore, the degree of polarization critically depends on GRB jet's angular structure and on our viewing angle $\theta_{\rm obs}$ from its symmetry axis. Therefore, knowledge of the degree of polarization along with
the spectral properties of the burst can help distinguish between uniform jets with sharp edges (top-hat jet) 
and more smoothly varying structured jets.

In this work, we first present a comprehensive overview of the different emission mechanisms that can explain the 
typical ``Band''-like non-thermal prompt emission spectrum, and discuss their expected linear polarization signatures. 
Reviews on this topic, including theoretical modeling and/or observational results, have been 
presented, e.g., by \citet{Lazzati2006,Toma+09,Toma2013,CG16}. Here, we have endeavoured to present what we consider to be the most plausible emission mechanisms for the prompt GRB: optically-thin synchrotron radiation from 
both random and ordered magnetic fields, Compton drag, and photospheric emission. Synchrotron 
self-Compton emission has been considered in the past to explain the prompt emission spectrum, but since it is disfavored by  
the GRB energetics \citep[see e.g.][]{PSZ09} and a featureless high energy spectrum reported by \textit{Fermi}-LAT, we do not discuss 
it here. However, the expected polarization from this mechanism is discussed by \citet{Chang-Lin-14}. 

If the magnetic field coherence length is much smaller than the gyro-radius of particles, then 
synchrotron radiation, the theory for which is derived for homogeneous magnetic fields, is not the correct description of the radiative 
mechanism by which relativistic particles cool. In this case, 
the particles experience small pitch-angle scattering where their motion is deflected by magnetic field inhomogeneities by 
angles that are smaller than the beaming cone of the emitted radiation ($1/\gamma_e$). This scenario of ``jitter-radiation'' has been proposed as 
a viable alternative to synchrotron radiation \citep{Medvedev-00}, where it has been shown to yield harder spectral slopes that cannot 
be obtained in optically thin synchrotron emission. In addition, this radiation mechanism can produce much sharper spectral break at 
$E=E_{\rm pk}$, as compared to synchrotron radiation, which agrees better with observations. However, \citet{Burgess+18} 
claim that GRB spectra obtained by \textit{Fermi}-GBM are well fit by a synchrotron emission model. The small-scale magnetic fields needed in 
this scenario are produced in relativistic collisionless shocks via the Weibel instability and the expected polarization if such a field 
is completely confined to a slab that is normal to the local fluid velocity has been calculated in \citealt{Mao-Wang-13,Prosekin+16,Mao-Wang-17}. There it was 
shown that the maximum degree of polarization is obtained when the slab is viewed close to edge on. For smaller off-axis viewing angles that can 
yield measurable fluences in GRBs, jitter-radiation produces almost negligible levels of polarization. For this reason we do not consider 
this mechanism in this work.

In photospheric emission models, the jet has to be dissipative or heated as it expands from an optically thick to an optically thin state. 
Without any dissipation the radiation field that decouples from matter at the photospheric radius would have a quasi-thermal spectrum 
\citep[e.g.][]{Beloborodov-10}, where the spectrum below the peak energy $E_{\rm pk}$ would be much harder than generally observed. 
Comptonization of softer photons below the photosphere has been shown to yield a spectrum that is softer than blackbody and better agrees with observations \citep[e.g.][]{Beloborodov-10,Vurm+13,Thompson-Gill-14}. Continued heating as the jet becomes 
optically thin \citep[e.g.][]{Giannios-08,Vurm-Beloborodov-16} or even radially localized heating outside of the photosphere \citep{Gill-Thompson-14} 
can give rise to the non-thermal spectrum above the peak energy.
Since the peak and the higher energy spectrum forms through multiple Compton scattering, the polarization degree of the radiation field is 
washed away as there is no particular direction for the electric field vector. If the flow is uniform then almost negligible polarization remains 
when averaged over the entire GRB image. This symmetry can be broken in two ways. First, it has been shown, and discussed later in this work as 
well, that if the flow has a steep gradient in the LF angular profile, polarization degree of up to $\Pi\sim20\%$ can be observed \citep{Lundman+14}. 
Second, if the low energy spectrum at $E\ll E_{\rm pk}$ arises due to synchrotron emission near the photosphere \citep{Lundman+18}, then the local magnetic 
field would impart a particular direction with which the electric field vector would be aligned, resulting in polarized emission. To carry out a 
self-consistent treatment of polarized emission in a dissipative photospheric model is outside the scope of this work, and therefore only the 
non-dissipative photospheric model is discussed here.

After deriving the level of linear polarization expected from different radiative processes, outflow geometries and viewing angles, 
we perform a statistical analysis of the expected level of polarization for these different scenarios by simulating a sample of 
$10^4$ GRBs. This analysis is carried out using simple Monte Carlo (MC) simulations, where the underlying assumption is that due 
to low photon statistics a statistically significant measurement of polarization generally entails, in addition to an overall 
high fluence, integration over multiple pulses in a given emission episode. These pulses can arise from, e.g., multiple internal 
shocks between distinct shells launched intermittently by the central engine, or different magnetic reconnection sites corresponding 
to different magnetic field polarity flips at different radial locations within the outflow. In both cases $\Gamma$ is expected to 
vary between different pulses (typically by $\Delta\Gamma\sim\Gamma$), which affects the degree of polarization obtained from 
integrating over multiple pulses. A similar effect may be caused by a gradual growth in the jet half-opening angle $\theta_j$ 
throughout the course of the GRB (while $\Delta\theta_j\sim\theta_j$ may be expected, even $\Delta\theta_j\gtrsim1/\Gamma$ could have 
a large effect on the observed polarization).

Furthermore, different GRBs are observed from different viewing angles $\theta_{\rm obs}$, and a spread in $\theta_{\rm obs}$ 
will yield different levels of polarization in a given sample of GRBs. This effect is intricately linked with the geometry of 
the outflow, where the degree of polarization changes significantly between a top-hat jet and structured jet. In addition, 
$\theta_{\rm obs}$ and the jet angular structure also affect the measured fluence, which significantly drops at large off-axis 
$\theta_{\rm obs}$. This effect is much more pronounced for a top-hat jet as compared to a structured jet. The relative 
contribution of each pulse scales with its number of detected photons (or more precisely the number of Compton events that 
can be used to measure the polarization). The MC simulations conducted in this work take into account the drop in fluence for 
larger viewing angles by considering a distribution of fluence weighted viewing angles for a fixed jet half-opening (core) angle 
in the case of a top-hat (structured) jet. In addition, it accounts for the variation in $\Gamma$ when integrating over multiple pulses.

Throughout this work, we consider an axi-symmetric relativistic outflow launched by a central engine 
(a black hole or a rapidly spining magnetar) in the coasting phase, with a bulk LF $\Gamma=(1-\beta^2)^{-1/2}\gg1$ that corresponds to the dimensionless fluid velocity $\vec\beta=\vec v/c$, 
where $c$ is the speed of light.
Each pulse is assumed to originate from a single thin shell (of radial width $\Delta\ll R/\Gamma^2$) with some $\Gamma(\theta)$ distribution, where $\Gamma$ may vary between different pulses according to some probability distribution.
For simplicity we consider only radially expanding outflows, such that 
$\hat\beta=\hat r$. We consider both top-hat jets and structured jets, where in the former case, the outflow 
has an angular size with 
$10\lesssim\xi_j\equiv(\Gamma\theta_j)^2\lesssim10^3$, where $\theta_j$ is the half-opening angle of the jet. 
Angles measured with respect to the LOS are shown with a tilde, e.g. 
the polar angle measured from the LOS is $\tilde\theta$. For a top-hat jet, the emission is assumed to drop 
rapidly for $\theta>\theta_j$, effectively giving the outflow a sharp edge. When the outflow has 
an angular structure, the total energy is dominated by the core with $\xi_c\equiv(\Gamma_c\theta_c)^2$ where 
$\theta_c$ and $\Gamma_c$ are respectively the angular size and LF of the core that is surrounded by low energy 
material extending to larger polar angles $\theta$. Outside the core the LF also drops according to the given prescription, however, 
all results pertaining to the structured jet case make sure that even at large $\theta$ the LF of the material 
is $\Gamma\gtrsim10$. Therefore, all results in this work are obtained for an ultra-relativistic flow.

The outline of the paper is as follows. In \S\ref{sec:Obs}, we give a brief overview of the measurements 
of linear polarization obtained during the prompt phase as well as from early afterglow emission. 
We start by discussing the origin of polarization from synchrotron emission in \S\ref{sec:synchro}. The likely origin and 
configuration of the magnetic field in the outflow is discussed in \S\ref{sec:B-field-origin-config}. 
In \S\ref{sec:Obs-Pol}, we provide a general treatment for calculating the degree of polarization 
averaged over the entire GRB image. This formalism also applies to all other emission mechanisms discussed in 
this work. In a spherical flow, polarization arising from a random magnetic field configuration that lies 
entirely in the plane of the ejecta averages to zero. Therefore, effects due to the angular structure of the jet 
and the observer's viewing angle become important in yielding non-vanishing degree of polarization. We first 
present the general equations for the polarization treatment that apply to off-axis observers and different 
magnetic field configurations in \S\ref{sec:los-B-field-geometry}. Polarized emission from on-axis top-hat 
jets from an ordered magnetic field is treated in \S\ref{sec:on-axis} along with the temporal evolution 
of the degree of polarization over a single pulse. Off-axis top-hat jets with ordered and random magnetic fields are 
discussed in \S\ref{sec:off-axis}. A serious issue for off-axis top-hat jets is the rapid drop in fluence (\S\ref{sec:pol-fluence}) 
for viewing angles larger than the jet opening angle. This effect is important when modeling GRB polarization since 
all detectors are flux-limited and only detect emission from regions of the flow brighter than the detector threshold. The top-hat jet model, 
although simple yet instructive, is an idealization and may not be the true description of the structure of relativistic 
GRB jets. Instead, the jet may manifest angular structure and the emission may drop rather gradually outside of a 
compact core. We discuss polarization from structured jets in \S\ref{sec:struc-jets}. Alternative radiative 
mechanisms that can explain the non-thermal spectra of GRBs and also yield polarized emission are treated next. 
In \S\ref{sec:CD}, we first present the general formalism that describes the mechanism of Compton drag (\S\ref{sec:IC-general}), 
where relativistically hot electrons inverse Compton scatter ambient radiation fields. Later, we specialize to the case 
of cold electrons in a relativistic outflow (\S\ref{sec:CD-cold}) and show the degree of polarization for off-axis 
top-hat jets. In \S\ref{sec:Photospheric}, we first discuss the radiation transfer of polarized emission in a 
matter-dominated non-dissipative fireball. However, after averaging over the GRB image a spherically symmetric 
outflow would yield vanishing polarization. Analytic treatment of polarized photospheric emission, based on the 
radiation transfer solution, from a structured jet is presented for the first time in this work (\S\ref{sec:Photo-strucjet}). 
In general, the GRB prompt emission suffers from low photon statistics at high energies. This becomes an even more 
of an issue for polarization measurements. Unless the burst is exceptionally bright, one is forced to integrate over 
multiple pulses to obtain statistically significant results. We treat this topic and its effect on the net polarization 
due to varying $\Gamma$ between pulses in \S\ref{sec:multiple-pulses}. After having discussed the predictions for the 
degree of polarization arising in synchrotron emission for different viewing geometries and jet structures, 
we carry out a MC simulation of $10^4$ GRBs in \S\ref{sec:MC-Pol} to determine the most likely magnetic field configuration 
for a given measurement of linear polarization. In order to yield a robust result, we take into account the effects of 
different $\theta_{\rm obs}$ in different GRBs and integration over multiple pulses within a single GRB with fixed 
$q=\theta_{\rm obs}/\theta_j$ but varying $\Gamma$. Finally, in \S\ref{sec:diss} we discuss salient points of this work and 
present important implications of the results.

%%%% TABLE %%%%%%%%%%%%%%%%%%%%%%%%%%%%%%%%
\begin{table*}
    \centering
    \begin{tabular}{l|c|c|c|c|l}
    \hline
    GRB & $\Pi$ (\%) & PA (${}^\circ$) & $\sigma_{\rm det}~(\Pi>0\%)$ & Instrument & Ref. \\
    \hline
    021206 & $80\pm20$ & -- & $>\!5.7$ & {\it RHESSI}${}^d$ & \cite{CB03} \\
     & $0$ & & -- & & \cite{RF04} \\
     & $41^{+57}_{-44}$ & & -- & & \cite{Wigger+04} \\
    041219A & $98\pm33$ & & $\sim\!2.3$ & {\it INTEGRAL}-SPI${}^e$ & \cite{Kalemci+07} \\
     & $63^{+31a}_{-30}$ & $70^{+14}_{-11}$ & $\sim\!2$ & & \cite{McGlynn+07} \\
     & $43\pm25^b$ & $38\pm16$ & $<\!2$ & {\it INTEGRAL}-IBIS & \cite{Gotz+09} \\
     061122 & $>\!33$ ($90\%$ CL) & $160\pm20$ & -- & {\it INTEGRAL}-IBIS & \cite{Gotz+13} \\
     100826A$^c$ & $27\pm11$ & -- & $2.9$ & {\it IKAROS}-GAP & \cite{Yonetoku+11b} \\
     100826Ap1$^c$ & $25\pm15$ & $159\pm18$ & $2.0$ &  &  \\
     100826Ap2$^c$ & $31\pm21$ & $75\pm20$ & $1.6$ &  &  \\
     110301A & $70\pm22$ & $73\pm11$ & $3.7$ & {\it IKAROS}-GAP & \cite{Yonetoku+12} \\
     110721A & $84_{-28}^{+16}$ & $160\pm11$ & $3.3$ & {\it IKAROS}-GAP & \cite{Yonetoku+12} \\
     140206A & $>\!28$ ($90\%$ CL) & $80\pm15$ & -- & {\it INTEGRAL}-IBIS & \cite{Gotz+14} \\
     151006A & $<\!84$ & $-$ & $-$ & {\it AstroSat}-CZTI & \cite{Chattopadhyay+17} \\
     160106A & $69\pm24$ & $-23\pm12$ & $\gtrsim\!3$ & {\it AstroSat}-CZTI & \cite{Chattopadhyay+17} \\
     160131A & $94\pm33$ & $41\pm5$ & $\gtrsim\!3$ & {\it AstroSat}-CZTI & \cite{Chattopadhyay+17} \\
     160325A & $59\pm28$ & $11\pm17$ & $\sim\!2.2$ & {\it AstroSat}-CZTI & \cite{Chattopadhyay+17} \\
     160509A & $<\!92$ & $-$ & $-$ & {\it AstroSat}-CZTI & \cite{Chattopadhyay+17} \\
     160530A & $<\!46$ ($90\%$ CL) & -- & -- & COSI${}^g$ & \cite{Lowell+17} \\
     160607A & $<\!77$ & $-$ & $-$ & {\it AstroSat}-CZTI & \cite{Chattopadhyay+17} \\
     160623A & $<\!46$ & $-$ & $-$ & {\it AstroSat}-CZTI & \cite{Chattopadhyay+17} \\
     160703A & $<\!55$ & $-$ & $-$ & {\it AstroSat}-CZTI & \cite{Chattopadhyay+17} \\
     160802A & $85\pm30$ & $-36\pm5$ & $\gtrsim\!3$ & {\it AstroSat}-CZTI & \cite{Chattopadhyay+17,Chand+18a} \\
     160821A & $54\pm16$ & $-39\pm4$ & $\gtrsim\!3$ & {\it AstroSat}-CZTI & \cite{Chattopadhyay+17} \\
     160821A$^h$ & $66^{+26}_{-27}$ &  & $\sim\!5.3$ & {\it AstroSat}-CZTI & \cite{Sharma+19} \\
     160821Ap1$^h$ & $71^{+29}_{-41}$ & $110^{+14}_{-15}$ & $3.5$ & {\it AstroSat}-CZTI &  \\
     160821Ap2$^h$ & $58^{+29}_{-30}$ & $31^{+12}_{-10}$ & $4$ & {\it AstroSat}-CZTI &  \\
     160821Ap3$^h$ & $61^{+39}_{-46}$ & $110^{+25}_{-26}$ & $3.1$ & {\it AstroSat}-CZTI &  \\
     160910A & $94\pm32$ & $44\pm4$ & $\gtrsim\!3$ & {\it AstroSat}-CZTI & \cite{Chattopadhyay+17} \\
     %160802A & $85\pm29$ & $\sim-32$ & $\sim3$ & {\it AstroSat}-CZTI & \cite{Chand+18a} \\
     161218A & $9$ & 40 & $\sim\!1.7$ & POLAR & \cite{Zhang+19} \\
      & $<\!41$ ($99\%$ CL) & -- & -- &  &  \\
     170101A & $8$ & 164 & $\sim\!1.5$ & POLAR & \cite{Zhang+19} \\
      & $<\!30$ ($99\%$ CL) & -- & -- &  &  \\
     170114A & $4$ & 164 & $\sim\!1.5$ & POLAR & \cite{Zhang+19,Burgess+19} \\
       & $<\!28$ ($99\%$ CL) & -- & -- &  &  \\
     170114Ap1$^{f}$ & $15$ & 122 & $\sim\!1.8$ &  &  \\
     170114Ap2$^{f}$ & $41$ & 17 & $\sim\!2.8$ &  &  \\
     170127C & $11$ & 38 & $\sim\!1.9$ & POLAR & \cite{Zhang+19} \\
      & $<\!68$ ($99\%$ CL) & -- & -- &  &  \\
     170206A & $10$ & 106 & $\sim\!1.5$ & POLAR & \cite{Zhang+19} \\
     170206A & $<\!31$ ($99\%$ CL) & -- & -- &  &  \\
     171010A & $\sim\!40$ & variable  & -- & {\it AstroSat}-CZTI & \cite{Chand+18b} \\
     \hline
    \end{tabular}
    \caption{Measured degree of linear polarization and position angle in the prompt phase of GRBs. 
    The detection significance $\sigma_{\rm det}$ is the significance of measuring $\Pi>0\%$. The quoted 
    errors are at the $1\sigma$ level. 
    ${}^a$Measured for the brightest pulse of duration 66~s. ${}^b$ Measured for the second peak lasting $40\,$s. 
    ${}^c$The main prompt emission is divided into two time intervals, p1 featuring a 47~s broad flare (line 1), 
    and  53~s long p2 consisting of multiple pulses (line 3). Line 1 jointly fits p1 and p2 assuming they have the 
    same $\Pi$ but allowing and indeed finding a different PA between them. ${}^d$Reuven Ramaty High Energy Solar 
    Spectroscopic Imager. ${}^e$International Gamma-Ray Astrophysics Laboratory. $^{f}\Pi$ obtained for two equal 
    2~s time bins within a single pulse, with a significant change in PA between them. ${}^g$Compton Spectrometer 
    and Imager. ${}^h$Average polarization over the single emission episode, with a \textit{Fermi}-GBM (\textit{AstroSat}-CZTI) 
    $T_{90}=43\,$s ($42\,$s), that showed variable polarization levels and PA during three distinct time intervals 
    p1, p2, p3 within the emission episode.}
    \label{tab:pol-data}
\end{table*}
%%%%%%%%%%%%%%%%%%%%%%%%%%%%%%%%%%%%%%%%%%%%%%%

%%%% OBSERVATIONS %%%%%%%%%%%%%%%%%%%%%%%%%%%%%%%%%%%%%%%%%%%%%%%%%
\section{Observations}\label{sec:Obs}
\subsection{Measured degree of polarization of prompt emission}
To robustly measure a significantly high degree of polarization, a high signal-to-noise ratio is needed. 
Due to the dearth of photons during the prompt phase, this becomes a serious issue. Therefore, 
reports of linear polarization thus far have at best been able to establish a $\sim3\sigma$ detection significance (however, see e.g. \citealt{Sharma+19}), 
and even that only in a handful of cases. The first detection of linear polarization during the prompt phase was reported by 
\citet{CB03} for GRB 021206, where they reported a high degree of polarization (see Table \ref{tab:pol-data}). 
This result was later refuted by \cite{RF04} and \citet{Wigger+04}, who found no significant degree of polarization. 
Another controversial result was reported for GRB 041219 \citep{Kalemci+07,McGlynn+07}, but the low 
($\sim2\sigma$) statistical significance of the result did not lead to any strong conclusions. Few upper and lower 
limits, albeit only at the $90\%$ confidence level, have been reported using the INTEGRAL-IBIS and COSI data.

More robust measurements of linear polarization came from the 
``GAmma-ray bursts Polarimeter'' (GAP) on board the ``Interplanetary Kite-craft Accelerated by the 
Radiation Of the Sun'' (IKAROS) spacecraft \citep{Yonetoku+11a}. The GAP measured modest to high 
degree of polarization for three GRBs \citep{Yonetoku+11b, Yonetoku+12}. Further measurements of 
linear polarization at a detection significance of $\gtrsim 2.5\sigma$, 
with some at a lower significance, have come from the CZTI detector on board AstroSat \citep{Singh+14}. 
Upper limits on linear polarization for five GRBs with $99\%$ confidence were reported by POLAR, a dedicated 
GRB polarization detection experiment onboard China's Tiangong-2 space laboratory \citep{Zhang+19}. 
Under the assumption that all five GRBs are indeed polarized, a joint analyses revealed an average degree 
of polarization of $\langle\Pi\rangle=10\%$ with a $0.1\%$ probability that all five sources have either 
$\Pi<5\%$ or $\Pi>16\%$.

%%%%%%%% CHANGE IN POLARIZATION ANGLE %%%%%%%%%%%%%%%%%%%%%%%%%%%%%%%%%%%%%%%%%%%%%%%%%%%%%%
\subsection{Change in polarization angle}
Thus far, most measurements of linear polarization during the prompt phase have been reported with 
a fixed polarization angle (PA), and in only four cases a change in PA has been reported. In GRB 100826A, 
a change in PA was detected between two time intervals corresponding to bright emission episodes with a $3.5\sigma$ confidence level 
\citep{Yonetoku+11b}, based on a joint fit of the two intervals assuming they had the same $\Pi$ (finding $\Pi>0$ with a 
significance of $2.9\sigma$). However, when performing separate fits on these two time intervals their individual polarization detection significance is lower ($2.0\sigma$ and $1.6\sigma$; see Table~\ref{tab:pol-data}). A time-resolved analysis of GRB 170114A, which 
showed only a single pulse, revealed a large change in the PA between two $2\,$s time bins \citep{Zhang+19}, where the polarization detection significance in each time bin is moderate ($\sim\,$1.8$\sigma$ and $\sim\,$2.8$\sigma$; see Table~\ref{tab:pol-data}). 
\citet{Burgess+19} carried out a detailed spectro-polarimetric analysis of this GRB and reached similar conclusions. 
A large change in the PA was found in the time-resolved analysis of GRB 171010A over three 
time bins \citep{Chand+18b}, but with a low statistical significance. Finally, \citet{Sharma+19} found variable degree of 
polarization in a time-resolved analysis of a single emission episode from GRB 160821A, which they divided into three distinct 
time intervals. Over these intervals the burst emission gradually rises to the peak and then declines and the PA between 
the three intervals shifts by $\Delta\theta_{p,12}=81^\circ\pm13^\circ$ and $\Delta\theta_{p,23}=80^\circ\pm19^\circ$ with 
a fairly high significance of $\sim\!3.5\sigma$ and $\sim\!3.1\sigma$, respectively.

Generally, a time-resolved analysis is not possible due to small number of detected photons. This is further made 
challenging by the fact that it is actually the Compton events due to scattering in the detector that 
are used to measure polarization, and they constitute only a fraction of the total number of photons detected from the source. Therefore, to increase the sensitivity of the detection an average polarization 
as well as an average PA rather than a time-resolved one is generally obtained. However, in bright bursts with multiple pulses, 
tracking the evolution of the PA can provide critical information that can be used to further 
constrain the outflow geometry and viewing angle. As we discuss below, in the case of a top-hat 
jet if the viewing angle is very close to the edge of the jet, $\theta_{\rm obs}\approx\theta_j$, 
then change in $\Gamma$ between distinct pulses will change $\xi_j$ which can lead to a change in 
the PA by $90^\circ$. However, this only occurs in this special circumstance, and therefore, a 
change in PA between different pulses should not be so commonly observed. Alternatively, \citet{Deng+16} have 
shown, using 3D relativistic MHD simulations and a 3D multi-zone polarization-dependent radiation transfer code, 
that in the ICMART model \citep{Zhang-Yan-11} a $90^\circ$ change in the PA can arise due to magnetic reconnection where the local magnetic field 
orientation, which is orthogonal to the wave vector of the emitted photon, itself switches by $90^\circ$ as the 
field lines are destroyed and reconnected in the emission region.

On the other hand, a change in the PA by an angle $\Delta\theta_p$ that is clearly not $0^\circ$ or $90^\circ$, 
e.g. $\Delta\theta_p\sim45^\circ$, would be challenging to explain by the different emission models presented in 
this work. Any changes in the geometry or $\Gamma$ of the outflow cannot explain it, as long as the flow remains 
axi-symmetric with a symmetry axis that does not move during the GRB. The PA evolution is sensitive to changes in 
the local magnetic field direction within the visible region, and a gradual continuous change in $\theta_p$ could 
potentially arise from a similar change in the direction of the ordered magnetic field in the visible region, 
though the cause for such a change during the prompt emission is not very clear. An alternative that is worth 
mentioning is if each pulse is associated with a different ``mini-jet'' within the outflow 
\citep[e.g.][]{LB03,Narayan-Kumar-09,Kumar-Narayan-09,Lazar+09,Zhang-Yan-11}, e.g. in the context of stochastic 
magnetic reconnection events, then this would indeed produce significant deviation from axi-symmetry of the emission 
regions, and could produce different and mutually randomly oriented PA's in different pulses, leading to a total 
polarization that largely follows Eq.~(\ref{eq:patchy}). This is analogous to the suggested random afterglow polarization 
variations that may accompany variability in the afterglow lightcurve, which may be induced by a ``patchy shell'' 
model for the GRB outflow \citep{Granot-Konigl-03,NO04} or by a clumpy external medium \citep{Granot-Konigl-03}.

An alternative explanation 
for a change of $\Delta\theta_p\sim45^\circ$ in the PA that appears in \citet{Granot-Konigl-03}, in which the flow remains axi-symmetric, is a combination of an 
ordered + random field. In this case the ordered field orientation is assumed to remain fixed,\footnote{A global toroidal field still cannot work in this scenario, since some devitation from axi-symmetery is needed, and if it does not arise from the flow itself then it should be provided by the ordered field that introduces a preferred direction.} 
but the relative strength of the random (in 2D) and ordered fields changes during the GRB. In that work it was discussed 
mainly in the context of afterglows, but the physics is practically the same. One possible difference is the motivation 
for ordered and random field components. For the afterglow \citet{Granot-Konigl-03} envision an ordered field component 
to arise from shock compression of an ordered field in the external medium, while a random component may be produced at the 
shock, so that the two components are co-spatial. In the prompt emission a similar picture may arise in which an ordered upstream 
field may naturally be advected from near the central source, while the random field may either be shock-produced and co-spatial, 
or alternatively generated at a thin reconnection layer and be confined to its vicinity so that it would not occupy the same 
region as the ordered field in the bulk of the outflow.

%%%%% EARLY AFTERGLOW POLARIZATION MEASUREMENTS %%%%%%%%%%%%%%%%%%%%%%%%%%%%%%%%%%%%%%%%%
\subsection{Early afterglow polarization measurements}
Another way of probing the magnetization of the GRB outflow and the magnetic field structure is by 
obtaining polarization measurements of the early afterglow. As the relativistic ejecta slows down by 
sweeping up interstellar medium, a reverse shock propagates into it. As a result, shock heated electrons 
in the ejecta radiate synchrotron photons, the flux of which peaks in the optical at timescales of tens 
of seconds, which could give rise to the so called ``optical flash'' lasting for about 10 minutes after the 
prompt GRB. In most cases, it is not detected at all and its duration can also vary. After the reverse 
shock has fully crossed the ejecta, the shocked electrons cool adiabatically while 
the peak of their emission moves to lower frequencies, where it powers a ``radio flare'' after about 1 day.

Measurements of linear polarization up to few tens of percent have been obtained from the early optical afterglow emission 
of several GRBs. Most notable examples are: GRB 090102 with $\Pi=(10.2\pm1.3)\%$ \citep{Steele+09}; GRB 120308A with 
$\Pi=(28\pm4)\%$ with a gradual decay over the next ten minutes to $\Pi=16_{-4}^{+5}\%$ \citep{Mundell+13}. Recently, 
radio/millimeter afterglow observations of GRB 190114C, dominated by the reverse shock component at 
$t_{\rm obs}\approx2.2-5.2\,$ hrs, revealed the temporal evolution in the linear polarization from $\Pi=(0.87\pm0.13)\%$ to 
$\Pi=(0.60\pm0.19)\%$ \citep{Laskar+19}. In other cases, radio flares have only yielded low upper limits, e.g. a strict $3\sigma$ 
upper limit of $\Pi<7\%$ in GRB 991216 \citep{Granot-Taylor-05}. Both of these observations, and in particular the measurement of 
gradual rotation of the PA during the observation in GRB 190114C, challenge the model where the outflow is permeated by a large scale 
ordered toroidal magnetic field.

%%%%%%  SYNCHROTRON EMISSION  %%%%%%%%%%%%%%%%%%%%%%%%%%%%%%%%%%%%%

%%%%%%%%%%  SYNCHROTRON EMISSION   %%%%%%%%%%%%%%%%%%%%%%%%%%%%%
\section{Synchrotron Emission}\label{sec:synchro}
Relativistic electrons (or $e^\pm$-pairs) gyrating in a magnetic field cool by emitting synchrotron photons. 
In general, synchrotron emission is partially linearly polarized, where the degree of polarization depends critically 
on the structure of the magnetic field and the observer's LOS. It is simpler to first examine the 
polarization arising in the comoving frame from an infinitesimally small region (a fluid element) of the outflow. 
This will allow us to prescribe a 
particular magnetic field configuration to that region and calculate the local polarization vector from a given fluid element. 
The same can then be obtained in the observer's frame, i.e. on the plane of the sky, through the appropriate Lorentz transformation. 
Since at high energies (e.g. X-rays, $\gamma$-rays) both the prompt and the afterglow emission 
regions remain unresolved, to obtain the total degree of polarization one must sum or integrate over the entire GRB image, 
which receives flux from all of the different fluid elements in the outflow. Before we provide a general 
prescription for calculating the degree of polarization arising in synchrotron emission, we first give a brief 
overview of the different magnetic field geometries that have been considered in GRB outflows.

%%%%%%  MAGNETIC FIELD CONFIGURATION IN THE OUTFLOW  %%%%%%%%%%%%%%%%%%%%%%%%%%%%%%
\subsection{Likely origin and configuration of the magnetic field}\label{sec:B-field-origin-config}
The origin of the magnetic field in relativistic outflows that power GRBs is still a matter of active research and debate. 
Polarization measurements can help to elucidate its structure, however, so far they have not yielded any conclusive results due to 
the low statistical significance of the measurements (however, see e.g. \citealt{Sharma+19}). The magnetic field configuration 
within the outflow is expected to be affected by its degree of magnetization (the magnetic to particle energy flux ratio),
\begin{equation}\label{eq:magnetization}
    \sigma\equiv\frac{w_B'}{w_m'}=\frac{B'^2}{4\pi[\rho'c^2+\hat\gamma(\hat\gamma-1)^{-1}P']}
    \xrightarrow[{\rm cold}]{}\frac{B'^2}{4\pi\rho' c^2}~,
\end{equation}
where $w_B'$ and $w_m'$ are the comoving\footnote{All quantities measured in the outflow comoving (fluid-) frame are primed.} 
magnetic field and matter enthalpy densities, respectively, $B'$ is the comoving magnetic field strength, $\rho'$ is 
the matter rest mass density, $P'$ is its pressure, and $\hat\gamma$ is the adiabatic index. If the flow is cold, then the matter 
enthalpy density is simply its rest mass energy density with no pressure term.

The fireball model does not have a clear prediction for the magnetic field structure in the emission region. During the acceleration 
phase ($R_0<R<R_s=\eta R_0$ where $\eta$ is the energy per unit rest energy and hence the coasting Lorentz factor, and $\Gamma(R_0)\approx1$) 
$\sigma\approx\sigma_0<1$ remains unchanged.\footnote{This arises since each fluid element expands isotropically in all direction 
($\propto R$) and hence the magnetic and thermal (radiation) pressures have the same adiabatic index (4/3), so that their corresponding 
proper enthalpy densities have the same scaling ($\propto R^{-4}$) and their ratio ($\sigma$) remains unchanged.} The same also holds 
during the coasting phase until the shells, of initial radial width $\Delta_0\approx ct_\varv$ where $t_\varv$ is the source variability time, 
start to significantly spread radially at $R_\Delta\sim\Gamma^2(R_\Delta)\Delta_0\sim\eta^2\Delta_0$. However, $R_\Delta$ is also the 
radius where internal shocks are expected to occur, so in this scenario $\sigma\sim\sigma_0<1$ also in the emission region (if it is 
indeed produced by internal shocks). During the coasting phase the lateral linear size of each fluid element scales as $R$ while its 
radial size remains constant, so that flux freezing implies $B_r\propto R^{-2}$ while 
$B_{\theta,\phi}\propto R^{-1}$ so that $B_r/B_{\theta,\phi}$ decreases by a factor of $R_\Delta/R_s=\eta ct_\varv/R_0\gg1$ and the transverse 
field components strongly dominate over the radial component. For $10^{-3}\lesssim\sigma\sim\sigma_0<1$
the upstream magnetic field is large enough to form the shock transition without the need for significant magnetic field amplification beyond 
the usual shock compression \citep[e.g.][]{SS11}, so that an ordered upstream field advected from the central source is expected to dominate 
in the downstream emission region, though in this regime it appears to be difficult to accelerate electrons to a non-thermal energy distribution. 
For $\sigma<10^{-3}$ shock generated fields via the Weibel instability (which are random and lie predominantly in the plane transverse to the 
shock normal) dominate over the shock compressed upstream field just behind the shock, and non-thermal electron acceleration becomes efficient.

For outflows that are initially Poynting flux dominated the magnetic field is expected to be ordered on large scales as it is dynamically 
dominant, and tangled field features within causally connected regions would tend to either straighten out or at least partly reconnect, 
both leading to much more ordered field configurations. However, magnetic reconnection can tangle the field near the reconnection layer, 
so that the electrons that are accelerated there may radiate some or even most of their energy in a rather random field before reaching the 
ordered field in the bulk of the outflow. If kinetic energy dominance ($\sigma<1$) is reached leading to efficient dissipation in internal 
shocks, this reverts to the discussion above with the addition that in this case the upstream field is expected to be both transverse and 
ordered on large scales (angles $\gtrsim1/\Gamma$).

When $\sigma<1$, magnetic fields are dynamically subdominant and plasma motions largely dictate the magnetic field structure. 
As a result, the magnetic field can be tangled on small scales ($\theta_B\ll\theta_j$) in the plane normal to the radial 
direction. In hydrodynamic flows, energy radiated during the prompt emission is expected to be dissipated mainly in internal shocks, where in the 
emission region near-equipartition magnetic fields are typically assumed to originate via the relativistic two-stream instability 
\citep{ML99}. The fields are generated at the relativistic ion-skin depth scales $c\bar\gamma_p^{1/2}/\omega_{p,i}'\sim 10^3$~cm, 
where $\omega_{p,i}'$ is the fluid-frame ion plasma frequency and $\bar\gamma_p$ is the mean thermal energy per unit rest mass
energy of protons. The configuration of the field is random within the plane of the shock, and the field strength quickly grows 
with an e-folding time of $\sim 10^{-7}$~s to near-equipartition level. Still, the field coherence length remains much smaller than 
the outflow's angular transverse size as well as its transverse causally connected size, such that $\theta_B\ll1/\Gamma\lesssim\theta_j$.

Alternatively, if the flow is launched Poynting-flux dominated, for which $\sigma \gg 1$, the magnetic field is dynamically 
dominant. In this case, an ordered magnetic field with a large coherence length $1/\Gamma\lesssim\theta_B\lesssim\theta_j$ can be 
expected within the relativistic outflow \citep{LB03}. For an axially symmetric field configuration, the poloidal component 
of the magnetic field ($B_p\propto r^{-2}$) drops rapidly with radius. Therefore, the toroidal component 
($B_\phi\propto r^{-1}$) remains dominant at large distances from the central source.

In the following, we consider three magnetic field configurations: (i) a locally ordered field ($B_{\rm ord}$) 
that is coherent on angular scales $1\lesssim\Gamma\theta\lesssim10$ and lies entirely in the direction transverse to 
the local fluid velocity $\vec\beta=\vec \varv/c$, the direction of which is identified with the local shock normal 
and radial unit vector with $\hat\beta = \hat{r}= \hat{x}\sin\theta\cos\varphi+ \hat{y}\sin\theta\sin\varphi + \hat{z}\cos\theta$. We parameterize its direction $\hat B_{\rm ord}$ such that its projection onto the $x$-$y$ plane (normal to the jet's symmetry axis) is $\hat{B}_0=\hat{x}\cos\varphi_B+\hat{y}\sin\varphi_B$.
\footnote{This implies $\hat{B}_{\rm ord}=[\hat{\theta}\cos\theta(\cos\varphi_B\cos\varphi+\sin\varphi_B\sin\varphi)+\hat{\varphi}(\cos\varphi\sin\varphi_B-\sin\varphi\cos\varphi_B)]/[\cos^2\theta(\cos\varphi_B\cos\varphi+\sin\varphi_B\sin\varphi)^2+(\cos\varphi\sin\varphi_B-\sin\varphi\cos\varphi_B)^2]^{1/2}$ where $\hat{\theta}=\hat{x}\cos\theta\cos\varphi+\hat{y}\cos\theta\sin\varphi-\hat{z}\sin\theta$ and $\hat{\varphi}=-\hat{x}\sin\varphi+\hat{y}\cos\varphi$. The relevant region that significantly contributes to the observed prompt GRB emission and polarization is typically restricted to $\theta\ll1$, for which $\hat{B}_{\rm ord}\approx\hat{B}_0$.}
(ii) a tangled magnetic field with components both parallel ($B_\parallel$) and perpendicular ($B_\perp$) to 
$\vec\beta$. In this case, it is convenient to parameterize the field anisotropy by taking the ratio of the average 
energy density of the two field components, such that 
\begin{equation}
 b\equiv \frac{2\langle B_\parallel^2\rangle}{\langle B_\perp^2\rangle}~.
\end{equation}
When $b=0$, the configuration of the magnetic field is that of a completely tangled or random magnetic field ($B_\perp$) 
in the plane normal to the local fluid velocity which is in the radial direction here. 
On the other hand, when $b\to\infty$ the configuration of the field is that 
of an ordered field ($B_\parallel$) entirely confined in the direction parallel to the local fluid velocity; and finally 
(iii) a toroidal field ($B_{\rm tor}$) that is ordered in the transverse direction and is axisymmetric with respect to 
the jet symmetry axis, such that $\hat B_{\rm tor} = \hat{\varphi} = -\hat{x}\sin\varphi +\hat{y}\cos\varphi$.

Afterglow polarization measurements after about several hours to a few days typically give fairly low level polarization detections 
or upper limits of $\Pi\lesssim1\%-3\%$ \citep[e.g.][]{Covino03}. This is typically near the jet break time in the afterglow lightcurve,  
while GRB jet models with a shock generated field can produce $\Pi\sim10\%-20\%$ near the jet break time. This apparent discrepancy 
already tentatively suggest that $b$ may not be very far from unity, $0.5\lesssim b\lesssim 2$, in order to suppress the afterglow polarization. 

However, the recent short GRB170817A associated with the NS-NS merger gravitational wave event GW170817 provides stricter and more robust 
constraints on the value of $b$. Detailed theoretical modeling \citep{Gill-Granot-18} together with the very elaborate afterglow observations 
from this event, and in particular the detection of super-luminal motion of the radio flux centroid with an apparent velocity of 
$\beta_{\rm app}=4.1\pm0.5$ \citep{Mooley+18}, clearly imply that the late time afterglow emission arises primarily from near the energetic 
narrow core of a relativistic jet viewed from well outside of its core. The jet structure and viewing angle implied by these observations 
result in clear predictions for the afterglow linear polarization \citep{Gill-Granot-18}. A later upper limit on the radio ($2.8\;$GHz) 
linear polarization of $\Pi<12\%$ (with 99\% confitence) at $t=244\;$days \citep{Corsi+18} is very constraining for the value of $b$, and 
we find that it robustly implies $0.66\lesssim b\lesssim 1.49$ \citep{Gill-Granot-19}. It is important to keep in mind that this applies to 
the effective value of 
$b$ in the afterglow shock. However, the latter comes from all of the shocked external medium behind the afterglow shock, which experiences 
significant shear in the radial direction \citep[e.g.][]{GPS99,GPS99b}, i.e. each fluid element is stretched more in the radial direction 
than in the two transverse directions, as it is advected further downstream from the shock. Therefore, the shock produced magnetic field 
could perhaps be predominantly in the plane of the shock ($b\ll 1$) just behind the shock transition, but become more isotropic ($b\sim 1$) 
in the bulk of the emitting region due to this significant radial shear (which causes $b$ to increase with the distance behind the shock). 
This effect and its possible implications are explored in more detail in \citet{Gill-Granot-19}. Such a strong radial shear is not expected 
in internal shocks, so that there the effective value of $b$ may potentially be different (and likely lower, $b<1$) than during the afterglow.

%%%%%%  OBSERVED POLARIZATION - GENERAL TREATMENT  %%%%%%%%%%%%%%%%%%%%%%%%%%%%%%%%%
\subsection{Observed polarization - general treatment}\label{sec:Obs-Pol}
The degree of polarization for the three magnetic field configurations considered in this work has been calculated in detail in many works 
(e.g. \citet{Ghisellini-Lazzati-99,Sari99,Gruzinov99,Granot-Konigl-03,Granot03,LPB03,Granot05,Granot-Taylor-05}; see \citet{NNP16} for 
circular polarization). In the following we summarize the important results \citep[see][for a review]{Toma+09,Toma2013}.

The state of polarization of a radiation field that emanates from a given fluid element is most conveniently expressed in 
terms of the Stokes parameters $I$, $Q$, $U$, $V$. We are interested here in linear polarization for which $V=0$. Here $I$ 
is the total intensity and the \textit{local} degree of linear polarization is given by
\begin{equation}
 \Pi' = \frac{\sqrt{Q^2+U^2}}{I}~,
\end{equation}
where 
\begin{equation}
 \frac{U}{I} = \Pi'\sin 2\theta_p~,\quad \frac{Q}{I} = \Pi'\cos 2\theta_p~, 
 \quad \theta_p = \frac{1}{2}\arctan\fracb{U}{Q}~,
\end{equation}
with $\theta_p$ as the polarization position angle (PA). The Stokes parameters and PA undergo 
a Lorentz transformation from the comoving to the observer's frame, whereas the local degree of polarization is a 
Lorentz invariant (being the ratio of Stokes parameters that undergo the same Lorentz transformation). In what follows, 
we distinguish between the local degree of polarization $\Pi'$ and the \textit{global} polarization 
$\Pi$, which is obtained after integrating over the whole GRB image on the plane of the sky as described below. 

\begin{figure}
    \centering
    \includegraphics[width=0.3\textwidth]{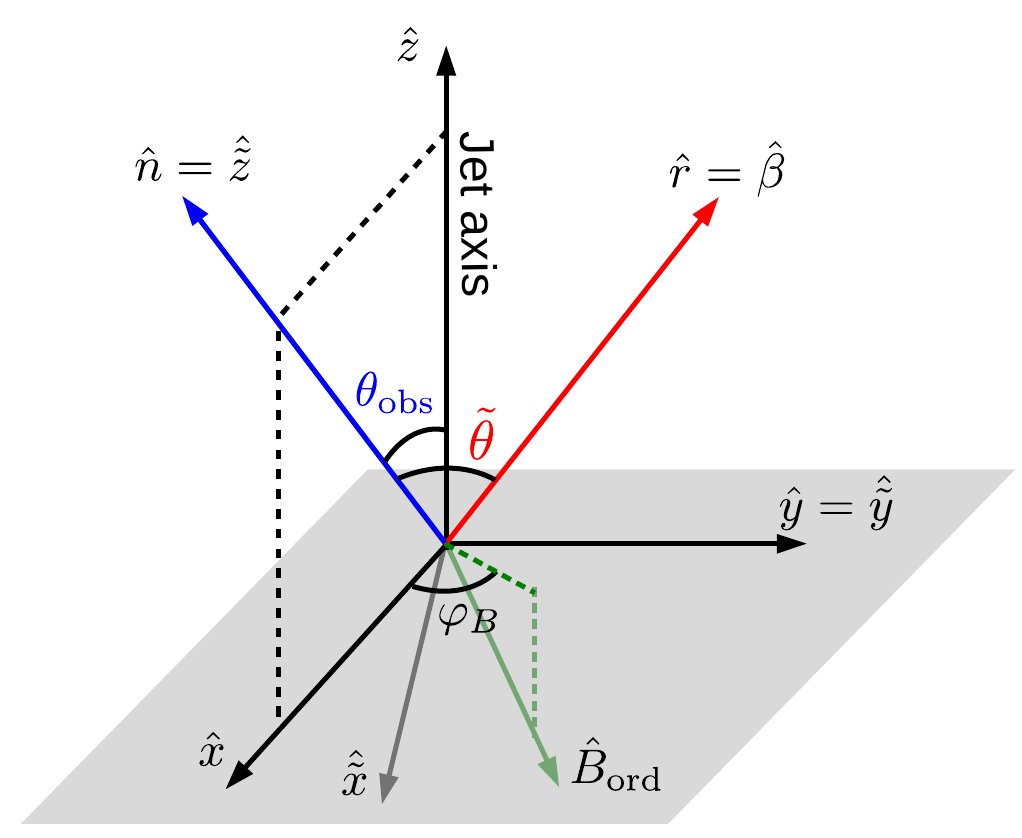}
    \includegraphics[width=0.2\textwidth]{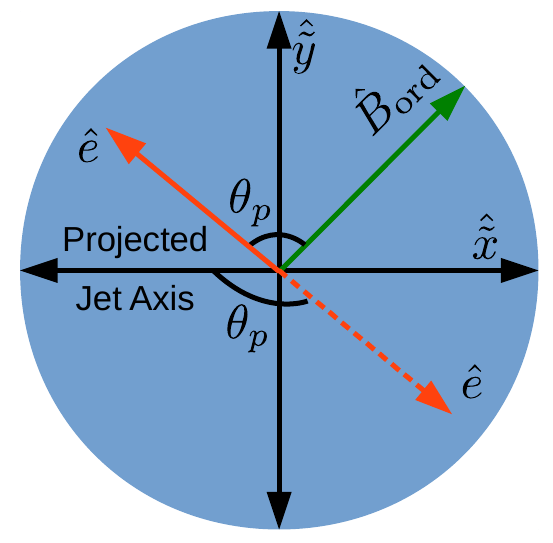}
    \caption{{\it Top}: Illustration of the coordinate system in which the polarization vector 
    associated to synchrotron emission is calculated. Here the direction of the local bulk velocity 
    is $\hat\beta=\hat r$ and the direction of the uniform magnetic field is transverse to that 
    with azimuthal angle $\varphi_B$. The polar angle $\tilde\theta$ in the lab-frame is between the directions of the 
    local bulk velocity and observed photon, with $\tilde\mu\equiv\cos\tilde\theta=\hat n\cdot\hat\beta$. 
    {\it Bottom}: The observer sees the projection of the ordered magnetic field (green arrow) 
    and polarization vector (red arrow) on the plane of the sky (shaded blue region; orthogonal to 
    the direction of the wave vector $\hat n$ of the observed photon, which points out of the page). 
    For an ordered magnetic field the polarization position angle $\theta_p$ is measured from the 
    direction of the ordered field (solid arrow), otherwise $\theta_p$ is measured from the projection 
    of the jet symmetry axis (dashed arrow).}
    \label{fig:sych_geom}
\end{figure}

At any given observer time $t_{\rm obs}$, the observer sees 
radiation emitted at different lab-frame times $t$ from different fluid elements with lab-frame coordinates ($r$, $\theta$, $\varphi$), where 
$r$ is the radial distance measured from the central engine, $\theta$ is the polar angle measured from the jet-axis, and $\varphi$ is the 
azimuthal angle. Here and what follows we use two different coordinate systems, as shown in Fig.~\ref{fig:sych_geom}. The first coordinate system 
$(x,y,z)$ is aligned with the jet's symmetry axis $(z)$, while the second, twidle-coordinate system $(\tilde x, \tilde y, \tilde z)$, is aligned 
with the direction to the observer $(\hat n = \hat z)$, and is rotated w.r.t. the first coordinate system by an angle of $\theta_{\rm obs}$ along 
the $y=\tilde y$ direction. The plane of the sky is the $\tilde x$-$\tilde y$ plane, in which we sometimes use 2D polar coordinates $(\tilde\rho,\tilde\varphi)$.

%%%%%%  FIGURE   %%%%%%%%%%%%%%%%%%%%%
%%%%%%%%%%%%%%%%%%%%%%%%%%%%%%%%%%%%%%

The measured Stokes parameters are a sum\footnote{For incoherent emission arising from distinct fluid elements, the Stokes 
parameters are additive.} over the flux $dF_\nu$ contributed by individual fluid elements, which yields \citep[e.g.][]{Granot03}
\begin{equation}\label{eq:stokes-general}
    \left\{\begin{array}{c}
        U/I  \\
        Q/I 
    \end{array}\right\}
    = \left(\int dF_\nu\right)^{-1}\int dF_\nu
    \left\{\begin{array}{c}
        \Pi'\sin 2\theta_p  \\
        \Pi'\cos 2\theta_p 
    \end{array}\right\}~,
\end{equation}
where
\begin{equation}\label{eq:dFnu}
    dF_\nu(t_{\rm obs},\hat n,r,t) = \frac{(1+z)}{d_L^2}\delta_D^2j'_{\nu'}\delta(t-t_{\rm obs}-\hat n\cdot\vec{r}/c)dtdV
\end{equation}
is the flux received from a source at a redshift $z$ with luminosity distance $d_L(z)$ emitting towards the observer in 
the direction of the unit vector $\hat n$. Here $j'_{\nu'}$ is the fluid-frame spectral emissivity, $dV$ is the lab-frame 
volume of the fluid element, and 
\begin{equation}
 \delta_D(r) = [\Gamma(1-\vec\beta\cdot{\hat n})]^{-1} = [\Gamma(1-\beta\tilde\mu)]^{-1}
\end{equation}
is the Doppler factor, where $\hat n\cdot\hat\beta=\cos\tilde\theta\equiv\tilde\mu$ 
and $\tilde\theta$ is the polar angle measured from the LOS. The delta-function term 
$\delta(t-t_{\rm obs}-\hat n\cdot\vec r/c)$ imposes the condition that for a given $t_{\rm obs}$ 
emission is received from an equal arrival time surface or volume depending on whether the emission is 
from a thin shell or a finite volume \citep[e.g.][]{GPS99,GCD08}. 

For simplicity, we ignore the radial structure of the outflow, and assume that the emission originates from an infinitely
``thin-shell.'' This approximation is valid if the timescale over which particles cool and contribute to 
the observed radiation is much smaller than the dynamical time. This implies that the emission region is a 
thin cooling layer of width (in the lab-frame) $\Delta \ll R/2\Gamma^2$. In this approximation, the flux density 
from each fluid element can be expressed as \citep{Granot05}
\begin{equation}\label{eq:flux-thin-shell}
    dF_\nu(t_{\rm obs},\hat n,r) = \frac{(1+z)}{16\pi^2d_L^2}\delta_D^3L'_{\nu'}(r)d\tilde\Omega~,
\end{equation}
where $L'_{\nu'}(r)$ is the fluid-frame spectral luminosity and $d\tilde\Omega=d\tilde\mu~d\tilde\varphi$ 
is the solid angle subtended by the fluid element w.r.t. the central source (i.e. the origin of the two coordinate systems). 

%%% FIGURE  %%%%%%%%%%%%%%%
\begin{figure*}
    \centering
    \includegraphics[width=0.3\textwidth]{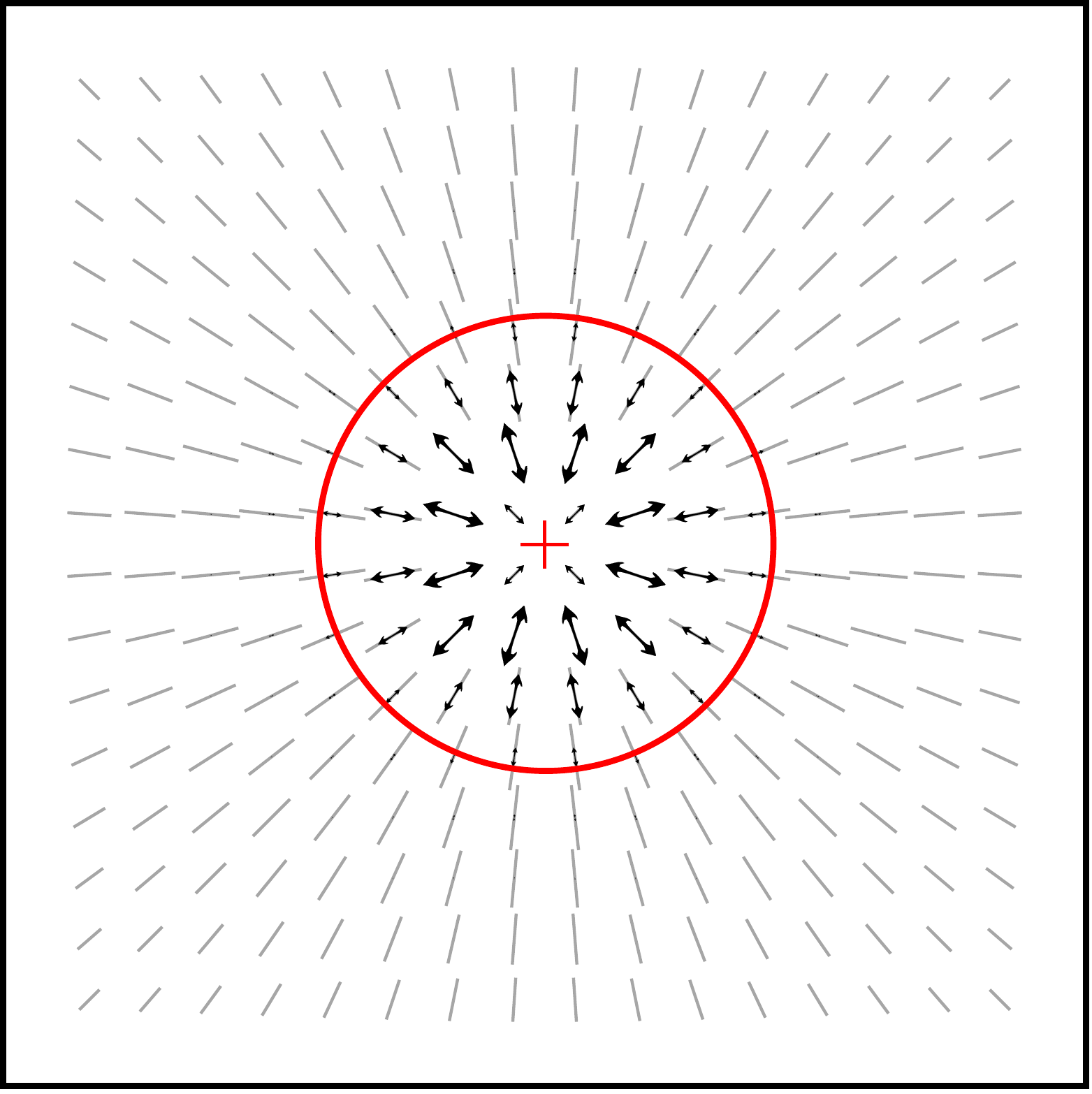}
    \includegraphics[width=0.3\textwidth]{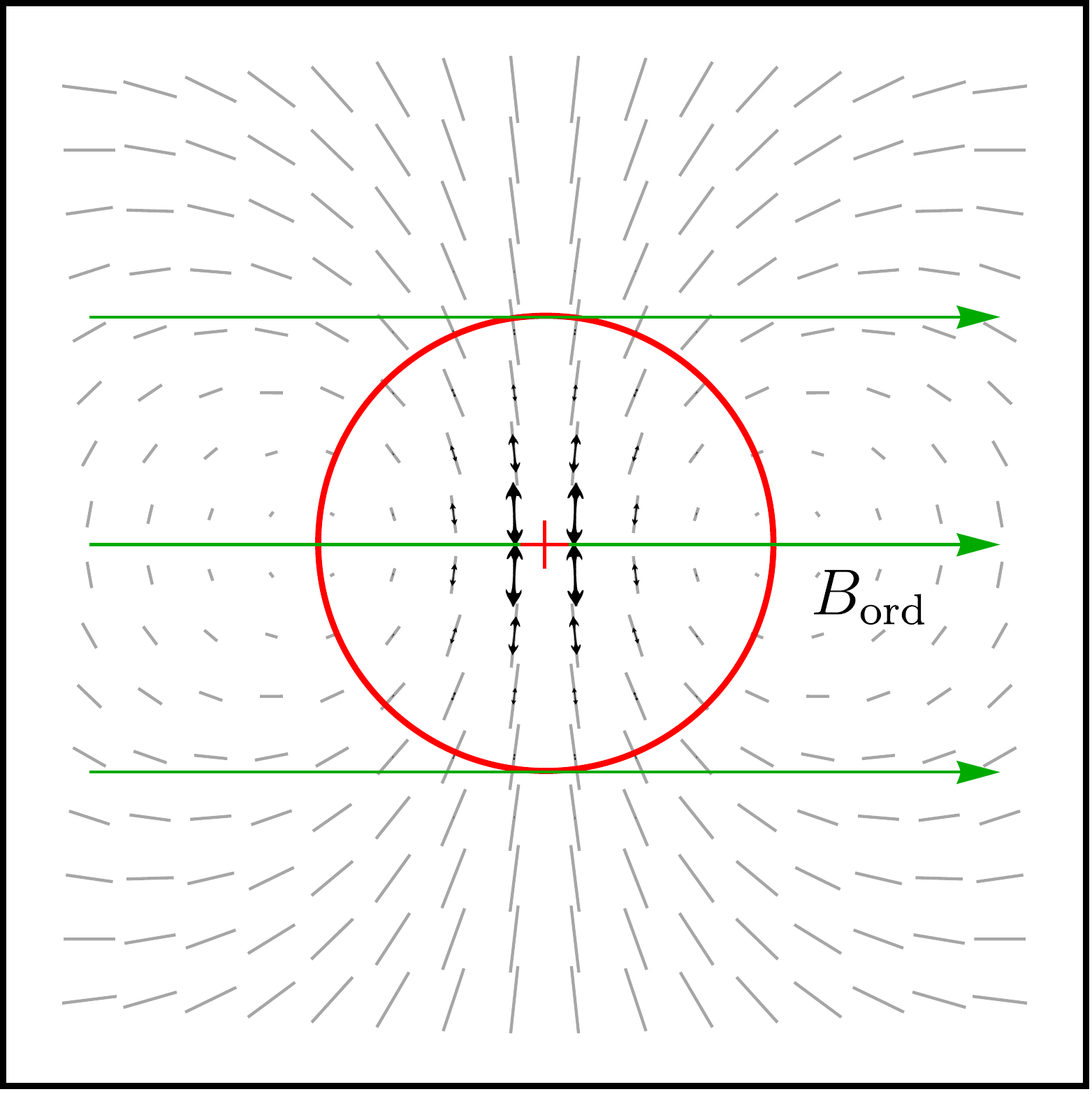}
    \includegraphics[width=0.3\textwidth]{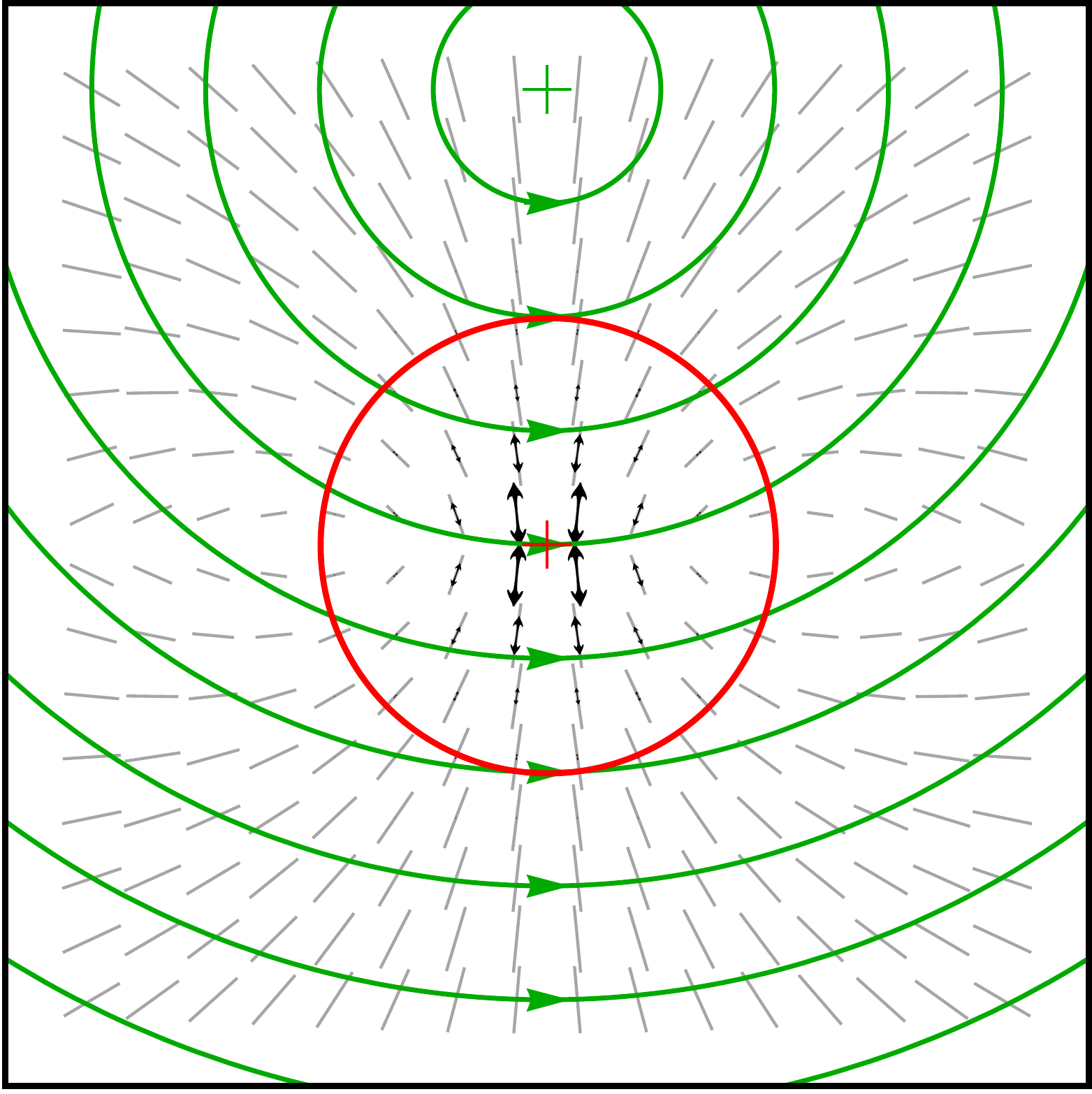}
    \caption{Local polarization map (shown for $\epsilon=1+\alpha=2$) for different magnetic field configurations 
    (shown in green for the two locally ordered field cases): ({\it Left}) Random field $B_\perp$ in the plane of the ejecta (normal to the radial direction), 
    ({\it Middle}) ordered field $B_{\rm ord}$, shown by green horizontal arrows, and ({\it Right}) toroidal field 
    ($q\sqrt\xi_j=2$ with $\sqrt\xi_j\gtrsim4.5$; $q\equiv\theta_{\rm obs}/\theta_j$ and $\xi_j\equiv(\Gamma\theta_j)^2$), 
    where the jet symmetry axis is marked with a green `+' sign. 
    The red circle shows the boundary ($\tilde{\xi}^{1/2}=\Gamma\tilde\theta=1$) of the region in the jet whose beaming cone includes our line of sight ($\tilde{\xi}=0$, marked with red `+' sign), projected on the plane of the sky. The magnitude of the black arrows reflects the polarized 
    intensity and the gray line segments show the same but normalized by $\delta_D^{(3+\alpha)}$. See also \citet{GR11}. }
    \label{fig:pol-map}
\end{figure*}
%%%%%%%%%%%%%%%%%%%%%%%%%%%

The anisotropic synchrotron spectral luminosity is expressed as \citep[e.g.][]{Rybicki-Lightman-79}
\begin{equation}\label{eq:Lnu-synchro}
 L'_{\nu'}(r)\propto (\nu')^{-\alpha}(\sin\chi')^\epsilon r^m
\propto(\nu')^{-\alpha}[1-(\hat n'\cdot\hat B')^2]^{\epsilon/2} r^m
\end{equation}
where we assume a power law spectrum and power law dependence of the emissivity on $r$. Here $\chi'$ is the angle between 
the direction of the local magnetic field and emitted photon. Since synchrotron emission 
from relativistic electrons is highly beamed in the direction of motion, $\chi'$ is also the pitch angle between the 
electron's velocity vector and the magnetic field. The power law index $\epsilon$ depends on the electron energy distribution, 
and if the latter is independent of the pitch angles then $\epsilon = 1+\alpha$. In the rest of this work, we only consider 
a constant emissivity with radius ($m=0$).

The degree to which the synchrotron emission is polarized depends on the underlying distribution of the emitting electrons, both 
in energy and pitch angle $\chi'$. We consider an isotropic electron velocity and a power law distribution in energy, 
with the number density of electrons scaling as $n_e(\gamma_e) \propto \gamma_e^{-p}$. In this case, the maximum degree of 
linear polarization from a fluid element with an ordered field is 
\begin{equation}\label{eq:Pi-max}
   % \Pi_{\rm max} = \frac{\alpha+1}{\alpha+5/3}\xrightarrow[{\rm Slow~Cooling}]{{\rm PLS~G}}\frac{p+1}{p+7/3}~,
    \Pi_{\rm max} = \frac{\alpha+1}{\alpha+5/3}=\frac{p_{\rm eff}+1}{p_{\rm eff}+7/3}~,
\end{equation}
where $\alpha=(p_{\rm eff}-1)/2$, and for optically-thin synchrotron emission $\alpha\geq-1/3$ which yields $\Pi_{\rm max}\geq1/2$. 
The value of $p_{\rm eff}$ changes depending on the different power law segments \citep[e.g.][]{GS02} of the synchrotron 
flux density, such that $p_{\rm eff}=\{2,p,p+1\}$ corresponding to $\alpha=\{1/2,(p-1)/2,p/2\}$ and PLSs \{F, G, H\}. For 
PLSs D and E, for which $\alpha=-1/3$, $\Pi_{\rm max}=1/2$ as the emission here arises from all electrons below their synchrotron frequency and therefore these 
PLSs have the lowest (optically-thin) level of polarization.

For a tangled or random field, the local degree of polarization from a given point on the emitting thin shell, after averaging over 
all directions of the random magnetic field, and under the simplifying assumption that $\epsilon=2$, is given by 
\citep[][]{Sari99,Gruzinov99,Granot-Konigl-03}
\begin{eqnarray}\label{eq:Pi_rnd}
    \frac{\Pi_{\rm rnd}'(\tilde\theta')}{\Pi_{\rm max}} && 
    = \frac{(b-1)\sin^2\tilde\theta'}{2+(b-1)\sin^2\tilde\theta'}\quad\quad(\epsilon=2) \\
    && = \left\{\begin{array}{cc}
        \displaystyle\frac{-\sin^2\tilde\theta'}{1+\cos^2\tilde\theta'} & (b=0~, B\to B_\perp) \\
        \nonumber \\
        1 & (b\to\infty~, B\to B_\parallel)
    \end{array}\right. \nonumber
\end{eqnarray}
The above result can be expressed in terms of the lab-frame angles through the aberration of light, such that
\begin{equation}\label{eq:aberration}
    \cos\tilde\theta' \equiv \tilde\mu' = \frac{\tilde\mu-\beta}{1-\beta\tilde\mu}~.
\end{equation}

To obtain the direction of the polarization vector on the plane of the sky, we start by defining the unit-vector $\hat n$ 
in the direction of the emitted photon in the lab frame. It is expressed using a coordinate system with 
$\hat z$ along the jet symmetry axis (as shown in Fig.~\ref{fig:sych_geom}), such that 
$\hat n=\sin\theta_{\rm obs}~\hat x + \cos\theta_{\rm obs}~\hat z$, where $\varphi_B$ is the azimuthal angle of the 
ordered magnetic field that is transverse to the radial vector. For synchrotron radiation, the 
polarization unit-vector in the fluid-frame $\hat e' = \hat B'\times\hat n'/|B'\times\hat n'|$ is orthogonal to both the direction of the 
local magnetic field and that of the emitted photon, both expressed in the frame of the radiating element moving with velocity 
$\vec\beta c$. In the lab-frame, the orientation of the polarization vector is obtained by the following Lorentz transformation 
\citep[see, e.g.][]{LPB03}
\begin{equation}
    \hat e = \frac{\displaystyle\hat e'+\Gamma\vec\beta\left[\frac{\Gamma}{\Gamma+1}(\hat e'\cdot\vec\beta)+1\right]}
    {\displaystyle\Gamma(1+\hat e'\cdot\vec\beta)}~.
\end{equation}
The direction of polarization naturally lies on the plane of the sky (i.e. $\hat{e}\cdot\hat{n}=0$), with 
$\hat e= (\hat e\cdot\hat{\tilde x})\hat{\tilde x} + (\hat e\cdot\hat{\tilde y})\hat{\tilde y}$,
where 
$\hat{\tilde x} = \cos\theta_{\rm obs}\hat x-\sin\theta_{\rm obs}\hat z$, $\hat{\tilde y} = \hat y$, and $\hat{\tilde z} = \hat n$.

When the magnetic field is completely tangled, for $b > 1$ ($b < 1$) the local polarization 
is $\Pi_{\rm rnd}' > 0$ ($\Pi_{\rm rnd}' < 0$) and the direction of the polarization vector is along (normal to) 
the direction of $\hat n'\times\hat r$.

%%%%%%%%%%%%%%%%%%%%%%%%%%%%%%%%%%%%%%%%%%%%%%%%%%%%%%%%%%%%%%%%%%%%%%%%%%%
\subsection{Effects of LOS and magnetic field configuration}\label{sec:los-B-field-geometry}
%%%%%%%%%%%%%%%%%%%%%%%%%%%%%%%%%%%%%%%%%%%%%%%%%%%%%%%%%%%%%%%%%%%%%%%%%%%

First we present general expressions that are valid for both on and off-axis observers. Then, in the subsequent sections 
we discuss the expected degree of polarization measured by an on-axis observer (\S\ref{sec:on-axis}) for different 
magnetic field configurations, and by off-axis observers (\S\ref{sec:off-axis}).

In the ultra-relativistic limit ($\Gamma\gg1$), approximate expressions accurate to $\mathcal{O}(\Gamma^{-2})$ 
may be used. In this limit, the Doppler factor is given by
\begin{equation}\label{eq:ultra-rel-doppler}
    \delta_D \approx \frac{2\Gamma}{(1+\tilde\xi)}\quad\quad{\rm where}\quad\quad
    \tilde\xi\equiv(\Gamma\tilde\theta)^2~,
\end{equation}
using the approximations $\tilde\mu\equiv\cos\tilde\theta\approx1-\tilde\theta^2/2$, and 
$\beta\approx1-1/(2\Gamma^2)$. From the definition of the unit-vector $\hat n$, 
and using the aberration of light, the factor related to the pitch angle in Eq.~(\ref{eq:Lnu-synchro}),
\be\label{eq:Lambda}
\Lambda\equiv\langle[1-(\hat n'\cdot\hat B')^2]^{\epsilon/2}\rangle~,
\ee
where the averaging is over the local probability distribution of $\hat B'$, can be expressed as follows for different field orientations,
\begin{eqnarray}\label{eq:sin-chi}
({\rm i})\quad & \Lambda_{\rm ord} & \approx \left[\fracb{1-\tilde\xi}{1+\tilde\xi}^2\cos^2\varphi_B + \sin^2\varphi_B\right]^{\epsilon\over2} \nonumber\\
({\rm ii})\quad & \Lambda_\perp & =
\langle\Lambda_{\rm ord}(\tilde\xi,\varphi_B)\rangle_{\varphi_B}
\nonumber\\
({\rm iii})\quad & \Lambda_\parallel & \approx \left[\frac{\sqrt{4\tilde\xi}}{1+\tilde\xi}\right]^\epsilon \\
({\rm iv})\quad & \Lambda_{\rm tor} & \approx \left[\fracb{1-\tilde\xi}{1+\tilde\xi}^2 +
\frac{4\tilde\xi}{(1+\tilde\xi)^2}\frac{(a+\cos\tilde\varphi)^2}{(1+a^2+2a\cos\tilde\varphi)}\right]^{\epsilon\over2}\,,
\nonumber
\end{eqnarray}
for (i) $B_{\rm ord}$ that is in the plane of the ejecta, (ii) for the $B_\perp$ case we average $\Lambda_{\rm ord}$ over the uniform distribution of $\varphi_B$ within the plane of the ejecta 
(see Eq.~(\ref{eq:Eq1-Sari99}) and the discussion in \S\ref{sec:random-B-field-pol}); (iii) $B_\parallel$, and (iv) $B_{\rm tor}$, for which 
$a\equiv\tilde\theta/\theta_{\rm obs}$. In the above, the angle $\varphi_B$ is measured from 
some reference direction and $\tilde\varphi$ is measured from the projection 
of the jet symmetry axis on the plane of the sky (see Fig.~\ref{fig:sych_geom} for reference). 

The polarization angle in the limit $\Gamma \gg 1$ is given by \citet[][]{Granot-Konigl-03,Granot03,Granot-Taylor-05}
\begin{eqnarray}
    ({\rm i})\quad\theta_{p} &=& \varphi_B + \arctan\left[\left(\frac{1-\tilde\xi}{1+\tilde\xi}\right)\cot\varphi_B\right] \\
    ({\rm ii})\quad\theta_p &=& \tilde\varphi \\
    ({\rm iii})\quad\theta_p &=& \left\{
    \begin{array}{cc}
        0~,\quad & \Pi' > 0 \\
        \pi/2~,\quad & \Pi' < 0 
    \end{array}\right. \\
    ({\rm iv})\quad\theta_p &=& \tilde\varphi - \arctan\left[\left(\frac{1-\tilde\xi}{1+\tilde\xi}\right)\frac{\sin\tilde\varphi}{a+\cos\tilde\varphi}\right]\ ,
\end{eqnarray}
where for the ordered field (case (i)) $\theta_p$ is measured from the local direction of the magnetic field, 
otherwise it is measured from the projection of the jet symmetry axis on the plane of the sky. For the direction of 
the PA when the magnetic field is tangled in the plane of the ejecta ($B_\perp$), see the discussion in \S\ref{sec:random-B-field-pol}.

%%%%%%%%%%%%%%%%%%%%%%%%%%%%%%%%%%%%%%%%%%%%%%%%%%%%%%%
\subsection{On-Axis Observer}\label{sec:on-axis}
%%%%%%%%%%%%%%%%%%%%%%%%%%%%%%%%%%%%%%%%%%%%%%%%%%%%%%%

%%%%%%%%%%%%%%%%%%%%%%%%%%%%%%%%%%%%%%%%%%%%%%%%%%%%%%%
\subsubsection{Top-hat jet viewed on-axis}
%%%%%%%%%%%%%%%%%%%%%%%%%%%%%%%%%%%%%%%%%%%%%%%%%%%%%%%

When the jet is ultra-relativistic ($\Gamma \gg 1$) the observer mainly 
receives photons from within a cone of semi-aperture (or beaming angle) $\tilde\theta=\Gamma^{-1}$ around the 
LOS due to relativistic beaming. Generally, $\Gamma\theta_j\gtrsim10$ and therefore the edge of the jet is not 
yet visible to an on-axis observer ($\theta_{\rm obs}=0$). In this case, the emission from the jet can be approximated 
as arising from an expanding thin spherical shell. The edge only becomes visible when the ejecta has slowed down significantly 
to $\Gamma\sim\theta_j^{-1}$, which happens around the time of the jet break. 

\begin{figure}
    \centering
    \includegraphics[width=0.47\textwidth]{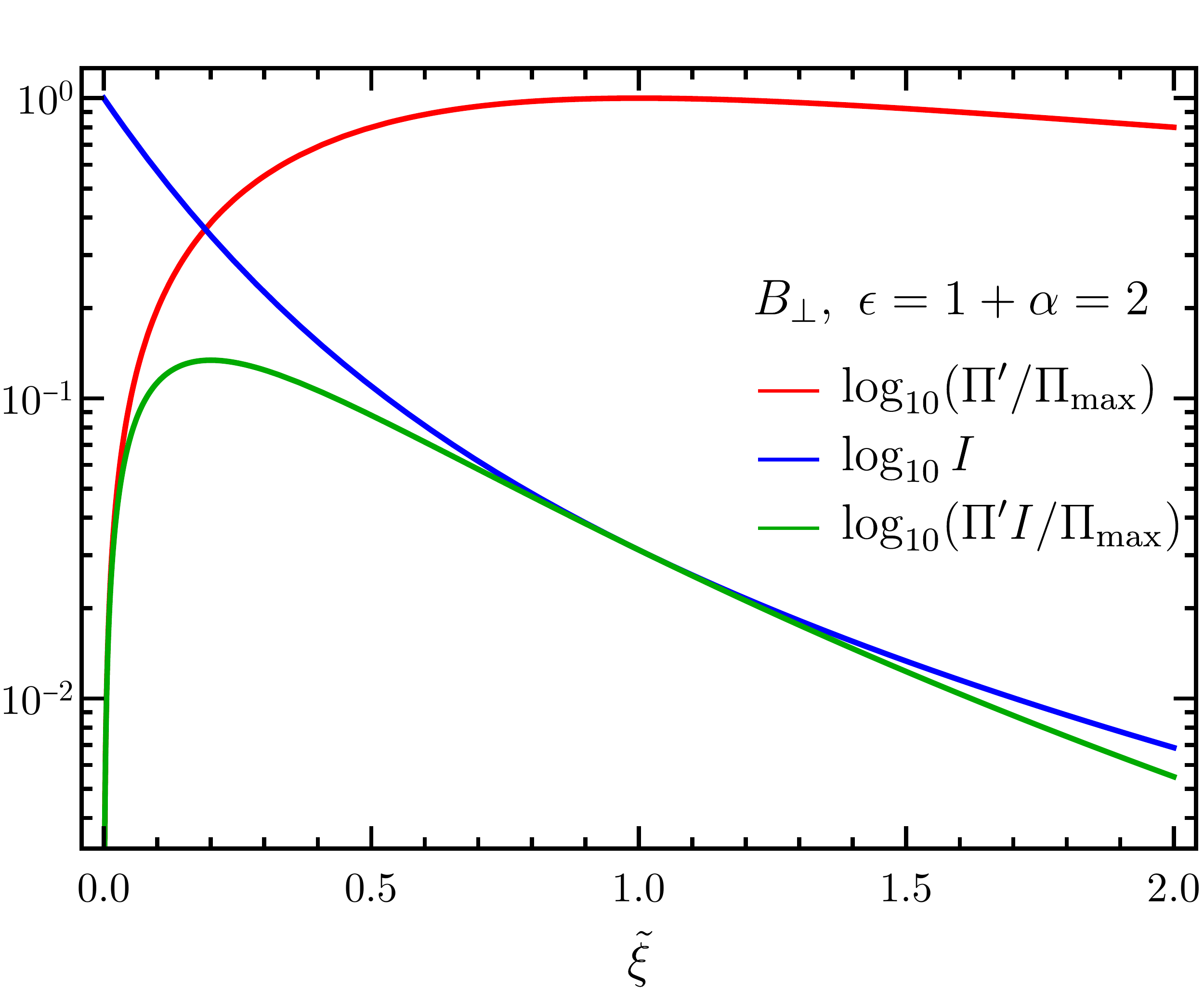}
    \caption{Comparison (for a spherical relativistic uniform emitting shell) of the  local degree of polarization, $\Pi'=\Pi_{\rm rnd}'$ 
    from Eq.~(\ref{eq:Pi_rnd}) for $b=0$, with the polarized intensity $\Pi'I$, where $I$ 
    is the intensity (obtained by averaging over the random magnetic field directions; see Eq.~(\ref{eq:Eq1-Sari99})) 
    normalized by its value along the line of sight [$\tilde\xi=(\Gamma\tilde\theta)^2=0$].}
    \label{fig:Pol-local-global}
\end{figure}

In the left and middle panels of Fig.~\ref{fig:pol-map} we show the polarization map for an on-axis observer. 
Here the length of the double-arrowed vectors shown in black represent the polarized intensity and the line segments 
in gray show the same but normalized by the Doppler factor term $\delta_D^{(3+\alpha)}$ that rapidly suppresses the intensity. 
This behaviour is more clearly shown in Fig.~\ref{fig:Pol-local-global} along with the local degree of polarization and polarized intensity 
as a function of $\tilde\xi$ for $B_\perp$ magnetic field configuration.

%%%%  FIGURE  %%%%%%%%%%%%%%%%%%%%%%%%%%%%%%%%%%%%%%%
%%%%%%%%%%%%%%%%%%%%%%%%%%%%%%%%%%%%%%%%%%%%%%%%%%%%%%

%%%%%%% FIGURE  %%%%%%%%%%%%%%%%%%%%%%%%%%%%%%%%%%%%%%
\begin{figure}
    \centering
    \includegraphics[width=0.48\textwidth]{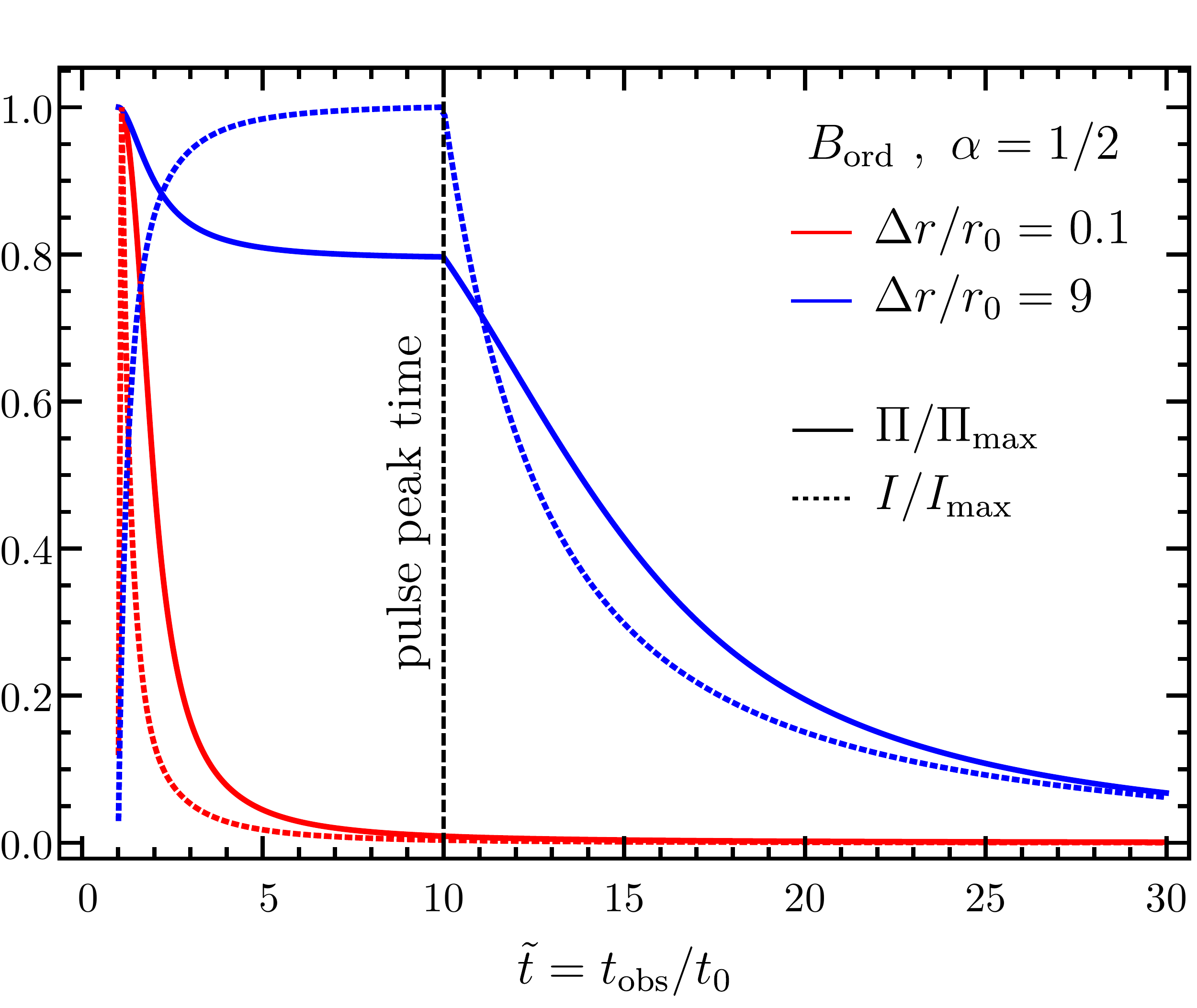}
    \caption{Temporal evolution of the degree of polarization and intensity over a single pulse, shown here 
    for an ordered magnetic field in the plane of the ejecta, for a spherical relativistic uniform emitting shell 
    \citep[after][]{NPW03}. However, when integrated over the entire pulse, both cases yield the same polarization.}
    \label{fig:pol-t-evol}
\end{figure}
%%%%%%%%%%%%%%%%%%%%%%%%%%%%%%%%%%%%%%%%%%%%%%%%%%%%%%%

%%%%%%%%%%%%%%%%%%%%%%%%%%%%%%%%%%%%%%%%%%%%%%%%%%%%%%%%%%%%%%%%%%%%%%%%%%%%%%%%%%%%%%%
\subsubsection{Temporal evolution over a single pulse}\label{sec:temp-evol}
%%%%%%%%%%%%%%%%%%%%%%%%%%%%%%%%%%%%%%%%%%%%%%%%%%%%%%%%%%%%%%%%%%%%%%%%%%%%%%%%%%%%%%%

The degree of polarization varies over the duration of a single pulse as emission from 
different radii and polar angles away from the LOS contribute to the flux at a given 
observer time $t_{\rm obs}$. In order to account for this effect, an integration over 
the equal arrival time surface (EATS) must be carried out \citep[e.g.][]{GPS99,GCD08}. 
In general, the emissivity and the spectrum can also vary over the single pulse, which 
would affect the level of polarization. Here, however, we explicitly assume, for simplicity, 
a constant emissivity and no spectral changes. More complex evolution of both and their effect 
on the time-resolved degree of polarization will be explored in a future work.

In the thin-shell approximation, after a lab-frame time $t$ the shell has moved a radial 
distance $r=\beta ct\approx ct$. In this case the EATS condition dictates that
\begin{equation}\label{eq:eats}
    t_{{\rm obs},z}\equiv\frac{t_{\rm obs}}{1+z} = t-\frac{r\tilde\mu}{c} 
    = \frac{(1-\beta\tilde\mu)}{\beta}\frac{r}{c}\approx\frac{(1+\tilde\xi)}{2\Gamma^2}\frac{r}{c}~,
\end{equation}
where the last expression is only valid in the ultra-relativistic limit. We further assume 
that the thin-shell starts radiating at radius $r=r_0$ and has a constant luminosity until 
the radius $r=r_0+\Delta r$, beyond which the emission stops. From the EATS equation, it is 
simple to deduce that for a given $t_{{\rm obs},z}$, only radii $r_{\rm min} \leq r \leq r_{\rm max}$, 
corresponding to $-1\leq\mu\leq1$, can contribute to the observed flux, where
\begin{eqnarray}
    r_{\rm min} &=& \max\left(r_0~,~\frac{\beta ct_{{\rm obs},z}}{1+\beta}\right)
    \approx\max\left(r_0~,~\frac{ct_{{\rm obs},z}}{2}\right) \\
    r_{\rm max} &=& \min\left(r_0+\Delta r~,~\frac{\beta ct_{{\rm obs},z}}{1-\beta}\right)
    \approx\min\left(r_0+\Delta r~,~2\Gamma^2ct_{{\rm obs},z}\right)\quad\quad
\end{eqnarray}
Plugging these conditions into Eq.~(\ref{eq:eats}), we find that $\tilde\xi_{\min}\leq\tilde\xi\leq\tilde\xi_{\max}$, 
where  
\begin{equation}
    \tilde\xi_{\min} = \max\left[0~,~\left(1+\frac{\Delta r}{r_0}\right)^{-1}\tilde t-1\right]
    \quad{\rm and}\quad\tilde\xi_{\max} = \tilde t-1~,
\end{equation}
with $\tilde t \equiv t_{\rm obs}/t_0$. Here $t_0 \equiv (1+z)r_0/(2\Gamma^2c)$ is the time of reception of the first photon, 
which is also equivalent to the angular time $t_{\rm obs,\theta}$ at $r_0$ within which photons from an area with angular size 
$\tilde\theta=1/\Gamma$ are received after the reception of the first photon. Then, integration over the EATS yields \citep[e.g.][]{NPW03} 
the general equation for the Stokes parameters, 
%where we define $\tilde t \equiv t_{\rm obs}/t_0$,
\begin{equation}\label{eq:Stokes-Inst}
    \left\{\begin{array}{c}
        \displaystyle\frac{U(\tilde t)}{I(\tilde t)} \\
        \\
        \displaystyle\frac{Q(\tilde t)}{I(\tilde t)} 
    \end{array}\right\}
    =\frac{\displaystyle
    \int_{\tilde\xi_{\min}(\tilde t)}^{\tilde\xi_{\max}(\tilde t)}\frac{d\tilde\xi}{(1+\tilde\xi)^{3+\alpha}}
    \int d\tilde\varphi \Lambda(\tilde\xi,\tilde\varphi)
    \left\{\begin{array}{c}
        \Pi'\sin 2\theta_{p} \\
        \Pi'\cos 2\theta_{p} 
    \end{array}\right\}
    }{\displaystyle
    \int_{\tilde\xi_{\min}(\tilde t)}^{\tilde\xi_{\max}(\tilde t)}\frac{d\tilde\xi}{(1+\tilde\xi)^{3+\alpha}}
    \int d\tilde\varphi \Lambda(\tilde\xi,\tilde\varphi)}~.
\end{equation}
In Fig.~\ref{fig:pol-t-evol}, we show the temporal evolution of the degree of polarization as well as intensity over 
a single pulse. We show two cases where $\Delta r/r_0 = 0.1$ and $\Delta r/r_0 = 9$ (corresponding to $r_f/r_0=t_f/t_0=(1+\Delta r/r_0)=10$, and explaining why the peak time is at $\tilde{t}=t_{\rm obs}/t_0=10$). In the former, 
the initial angular time $t_{\rm obs,\theta} = t_0$ dominates over the radial time $t_{{\rm obs},r} = (1+z)\Delta r/(2\Gamma^2c)$ since 
$\Delta r \ll r_0$. In the latter the radial time dominates over the initial angular time, while the final angular time at 
a radius $r_f=r_0 + \Delta r$ dominates the decaying part of the flux after it peaks. In both cases, the degree of polarization 
is maximum ($\Pi=\Pi_{\max}$) at the beginning of the pulse since only photons originating along the LOS are 
observed. However, as photons from larger angles away from the LOS are observed, the level of polarization 
declines. A sharper decline in $\Pi/\Pi_{\max}$ is seen after the peak of the pulse when high latitude 
emission dominates.

% %%%%%%% FIGURE  %%%%%%%%%%%%%%%%%%%%%%%%%%%%%%%%%%%%%%
% \begin{figure}
%     \centering
%     \includegraphics[width=0.48\textwidth]{pol-t-evol.pdf}
%     \caption{Temporal evolution of the degree of polarization and intensity over a single pulse, shown here 
%     for an ordered magnetic field in the plane of the ejecta, for a spherical relativistic uniform emitting shell 
%     \citep[after][]{NPW03}. However, when integrated over the entire pulse, both cases yield the same polarization.}
%     \label{fig:pol-t-evol}
% \end{figure}
% %%%%%%%%%%%%%%%%%%%%%%%%%%%%%%%%%%%%%%%%%%%%%%%%%%%%%%%

%%%%  FIGURE  %%%%%%%%%%%%%%%%%%%%%%%%%%%%%%%%%%%%%%%
\begin{figure}
    \centering
    \includegraphics[width=0.48\textwidth]{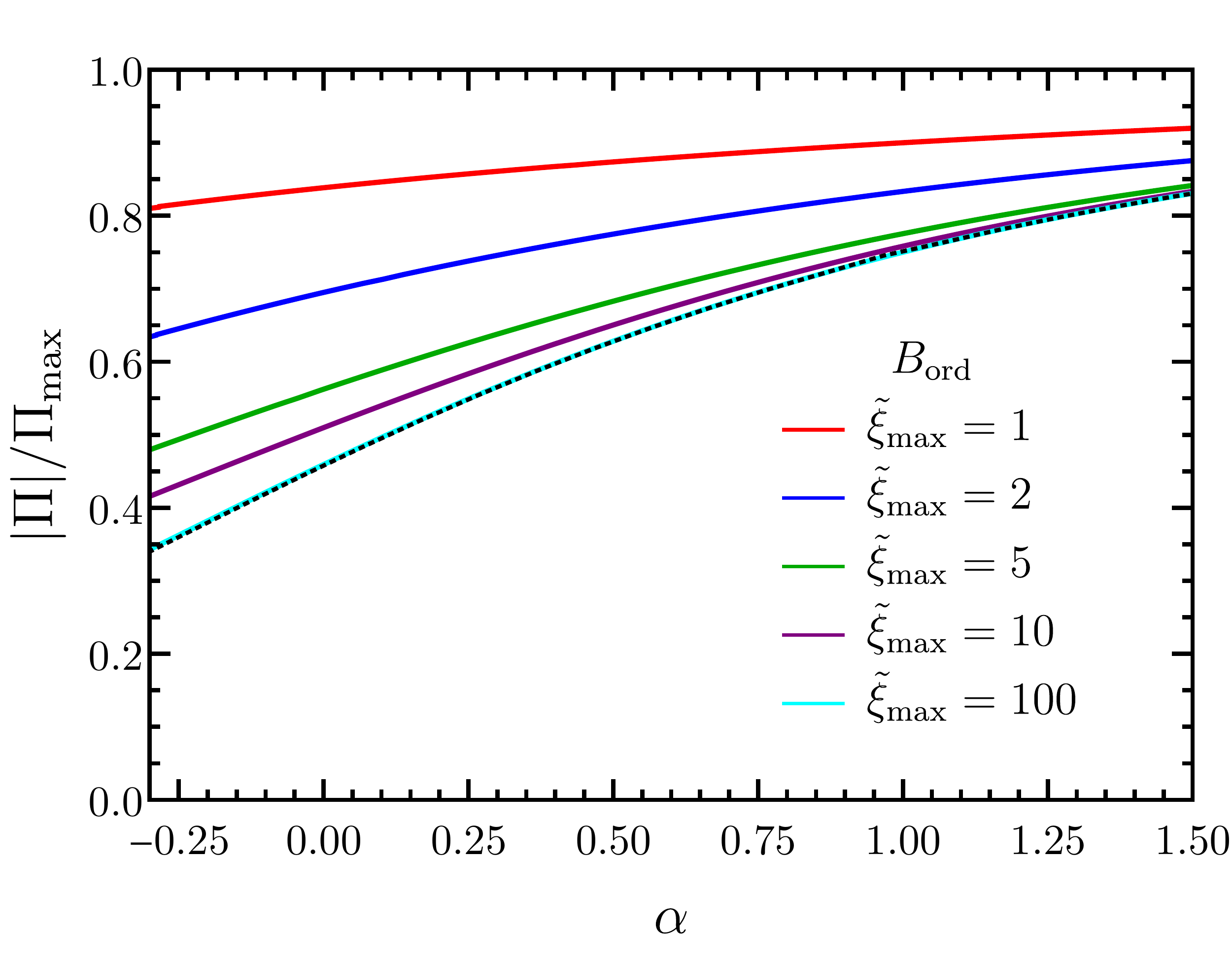}
    \caption{Degree of polarization when the magnetic field is ordered ($B_{\rm ord}$), 
    shown for different values of $\tilde{\xi}_{\rm max}=(\Gamma\theta_{\max})^2$ from Eq.~(\ref{eq:Pi-pulse-int-ximax}). 
    In the limit $\tilde{\xi}_{\rm max}\to\infty$, $\Pi$ approaches that obtained from explicit time integration 
    over a single pulse, which is shown by the black dotted line \citep[after][]{Granot03}. 
    The jet geometry is that of a spherical flow since most of the contribution arises from a region of angular size 
    $1/\Gamma$ around the LOS. Contribution from larger angles (or correspondingly larger $\tilde\xi$) 
    is suppressed by relativistic beaming. For optically-thin synchtrotron emission $-1/3\leq\alpha < p/2$ for 
    the electron energy distribution power law index $2 \lesssim p \lesssim 3$; see Eq.~(\ref{eq:Pi-max}) and the discussion that follows.}
    \label{fig:B-ord-tor-pol}
\end{figure}
%%%%%%%%%%%%%%%%%%%%%%%%%%%%%%%%%%%%%%%%%%%%%%%%%%%%%%

%%%%%%%%%%%%%%%%%%%%%%%%%%%%%%%%%%%%%%%%%%%%%%%%%%%%%%
\subsubsection{Pulse integrated polarization}
%%%%%%%%%%%%%%%%%%%%%%%%%%%%%%%%%%%%%%%%%%%%%%%%%%%%%%

In the case of prompt emission, the measured polarization is generally integrated over at least a single 
pulse, if not multiple pulses (see \S\ref{sec:multiple-pulses}). The pulse integrated Stokes 
parameters, e.g. the total intensity which is proportional to the fluence over a 
single pulse can be obtained using $dF_\nu dt_{\rm obs} \propto \Delta t'\delta_D^2L'_{\nu'}d\tilde\Omega$, 
where $\Delta t' = \delta_D dt_{\rm obs}$ is the duration of the pulse in the comoving frame 
(see Appendix~\ref{sec:app-time-integrated-pol-strucjet} for more details). This amounts 
to reducing one power of the Doppler factor in Eq.~(\ref{eq:Stokes-Inst}), and therefore the pulse integrated 
polarization can now be conveniently expressed as \citep[][]{Granot03},
\begin{equation}\label{eq:Pi-pulse-int-ximax}
    \left\{\begin{array}{c}
        U/I \\
        Q/I 
    \end{array}\right\}
    =\frac{\displaystyle
    \int\frac{d\tilde\xi}{(1+\tilde\xi)^{2+\alpha}}
    \int d\tilde\varphi \Lambda(\tilde\xi,\tilde\varphi)
    \left\{\begin{array}{c}
        \Pi'\sin 2\theta_{p} \\
        \Pi'\cos 2\theta_{p} 
    \end{array}\right\}
    }{\displaystyle
    \int\frac{d\tilde\xi}{(1+\tilde\xi)^{2+\alpha}}\int d\tilde\varphi \Lambda(\tilde\xi,\tilde\varphi)
    }~.
\end{equation}
When doing an explicit time integration in Eq.~(\ref{eq:Stokes-Inst}) another simplification can be made. Since the total 
polarization should not depend on the duration over which the radiating shell is active or equivalently $\Delta r$, a delta 
function in $r$ can be assumed by taking $\Delta r\to0$. This can also be noticed from Fig.~\ref{fig:pol-t-evol}, where 
integration over both curves yields the same polarization given a sufficiently large upper limit on $\tilde t$ when integrating 
where the polarized intensity vanishes. This effectively implies integrating over the outflow surface at a fixed radius for 
$0\leq\tilde\xi\leq\tilde\xi_{\max}$, with no dependence on $t_{\rm obs}$, and $0\leq\tilde\varphi\leq2\pi$. Therefore, 
any temporal evolution of the luminosity within a pulse does not affect the time-integrated degree of polarization when all 
else remains the same.

% %%%%  FIGURE  %%%%%%%%%%%%%%%%%%%%%%%%%%%%%%%%%%%%%%%
% \begin{figure}
%     \centering
%     \includegraphics[width=0.48\textwidth]{B-ord-pol.pdf}
%     \caption{Degree of polarization when the magnetic field is ordered ($B_{\rm ord}$), 
%     shown for different values of $\tilde{\xi}_{\rm max}=(\Gamma\theta_{\max})^2$ from Eq.~(\ref{eq:Pi-pulse-int-ximax}). 
%     In the limit $\tilde{\xi}_{\rm max}\to\infty$, $\Pi$ approaches that obtained from explicit time integration 
%     over a single pulse, which is shown by the black dotted line \citep[after][]{Granot03}. 
%     The jet geometry is that of a spherical flow since most of the contribution arises from a region of angular size 
%     $1/\Gamma$ around the LOS. Contribution from larger angles (or correspondingly larger $\tilde\xi$) 
%     is suppressed by relativistic beaming. For optically-thin synchtrotron emission $-1/3\leq\alpha < p/2$ for 
%     the electron energy distribution power law index $2 \lesssim p \lesssim 3$; see Eq.~(\ref{eq:Pi-max}) and the discussion that follows.}
%     \label{fig:B-ord-tor-pol}
% \end{figure}
% %%%%%%%%%%%%%%%%%%%%%%%%%%%%%%%%%%%%%%%%%%%%%%%%%%%%%%

%%%%  FIGURE  %%%%%%%%%%%%%%%%%%%%%%%%%%%%%%%%%%%%%%%
\begin{figure*}
    \centering
    \includegraphics[width=0.47\textwidth]{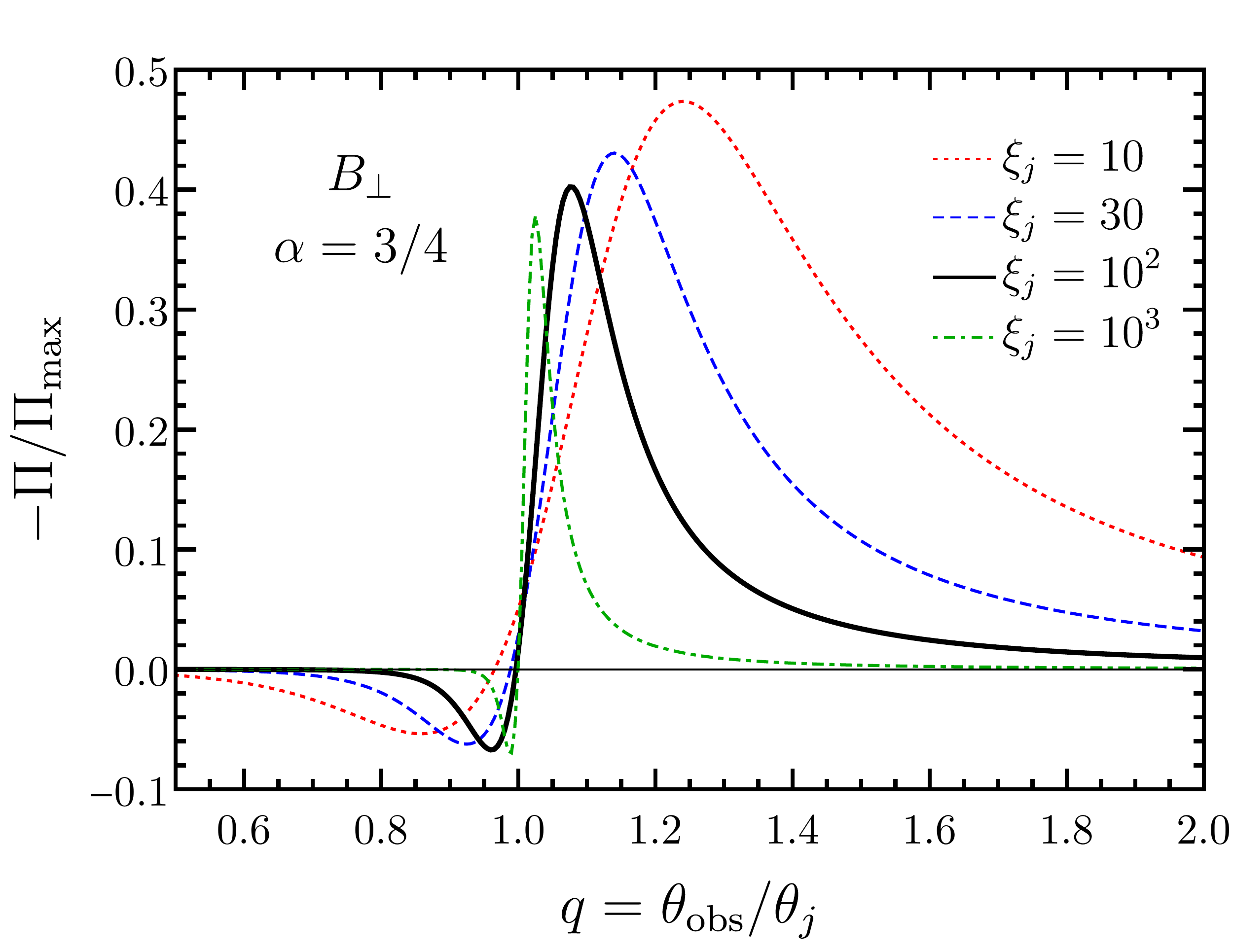}\quad\quad
    \includegraphics[width=0.47\textwidth]{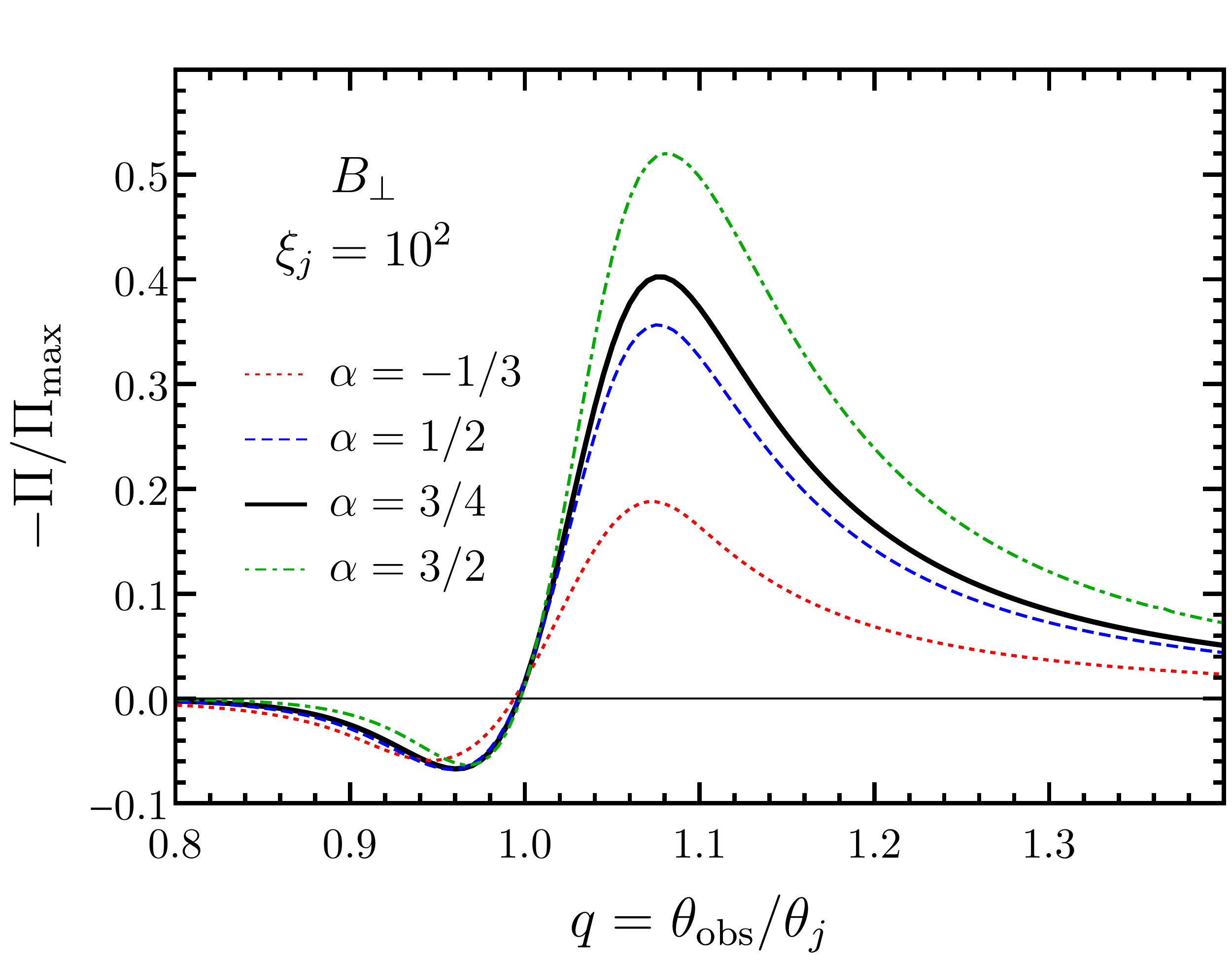}
    \includegraphics[width=0.47\textwidth]{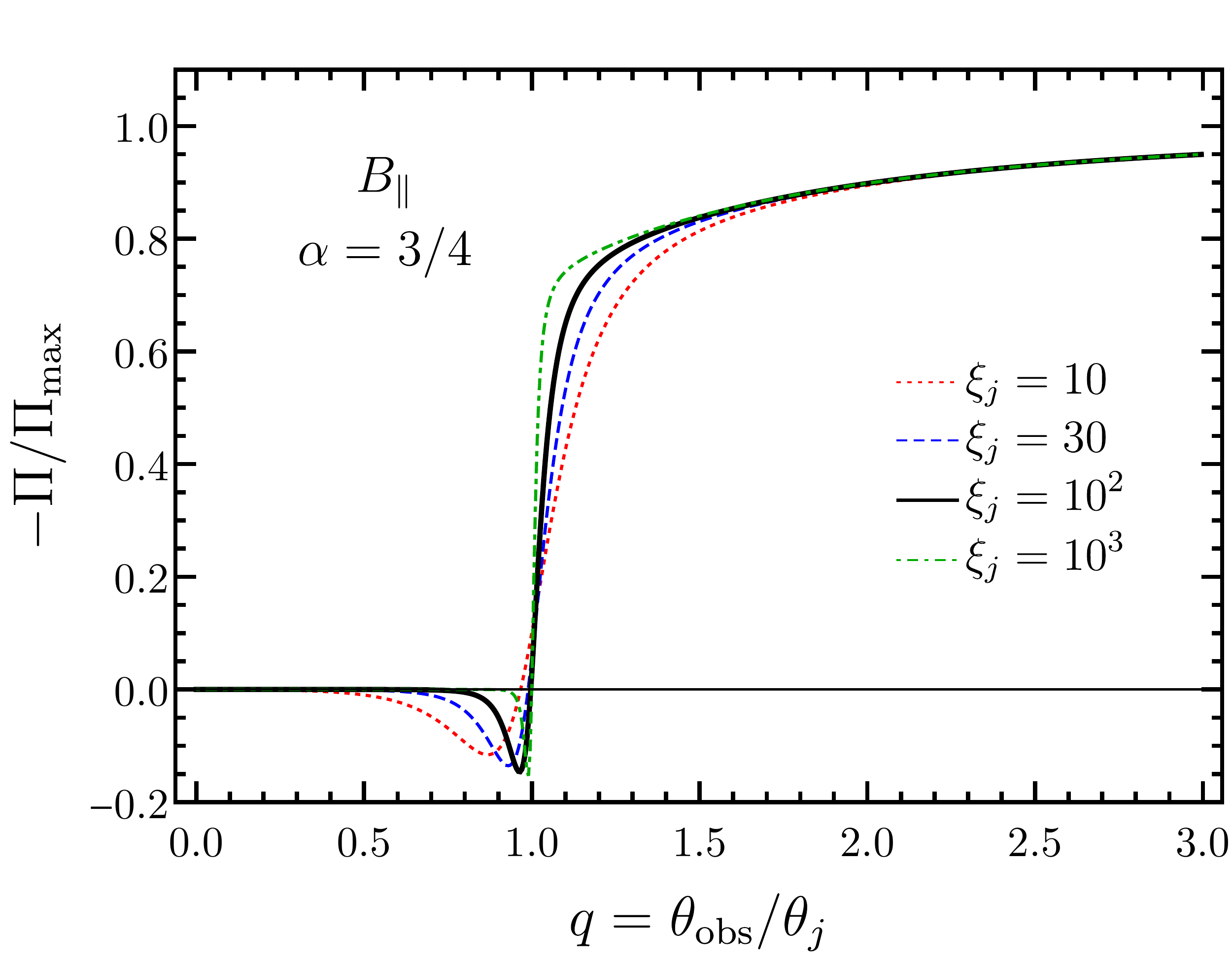}\quad\quad
    \includegraphics[width=0.47\textwidth]{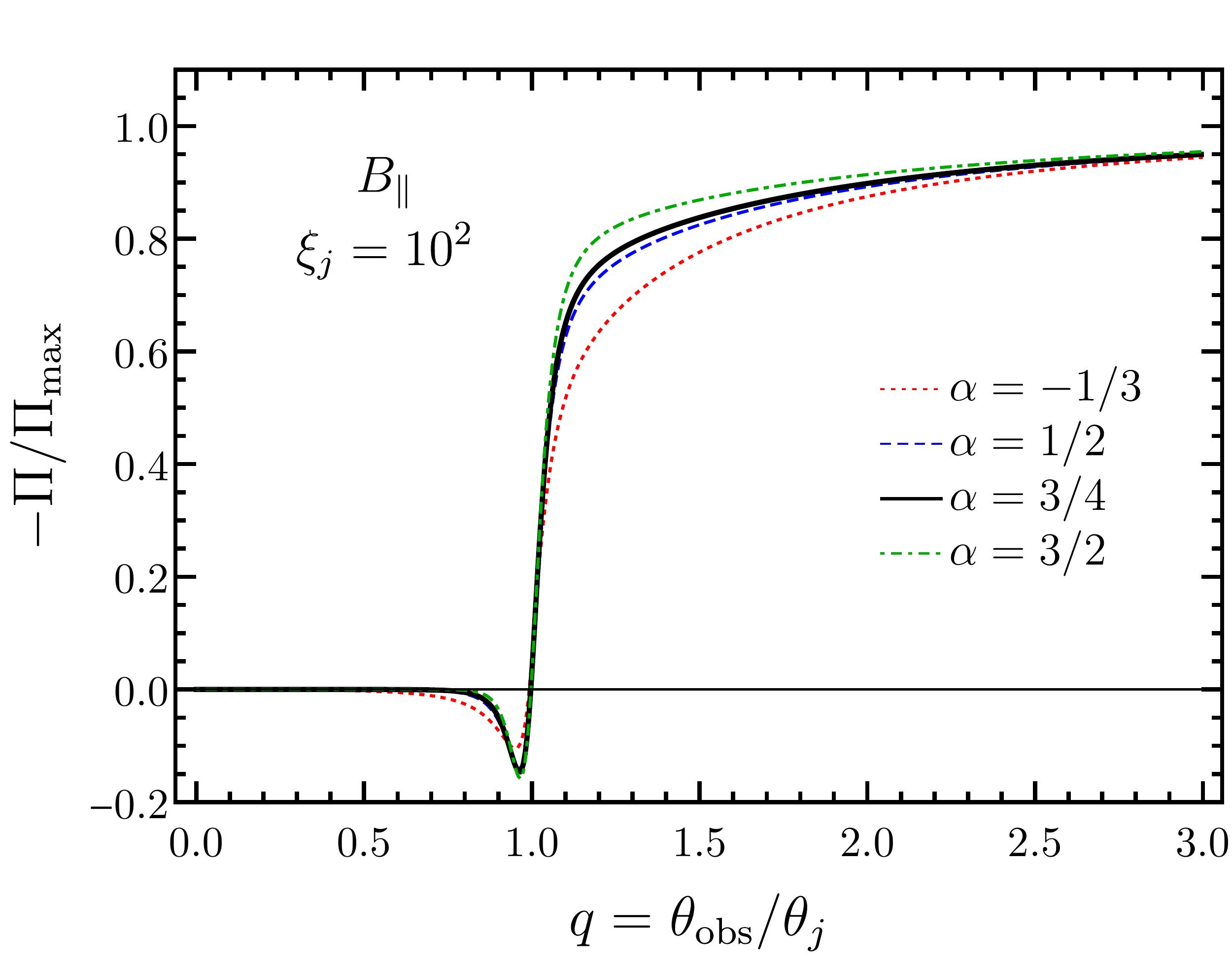}
    \includegraphics[width=0.47\textwidth]{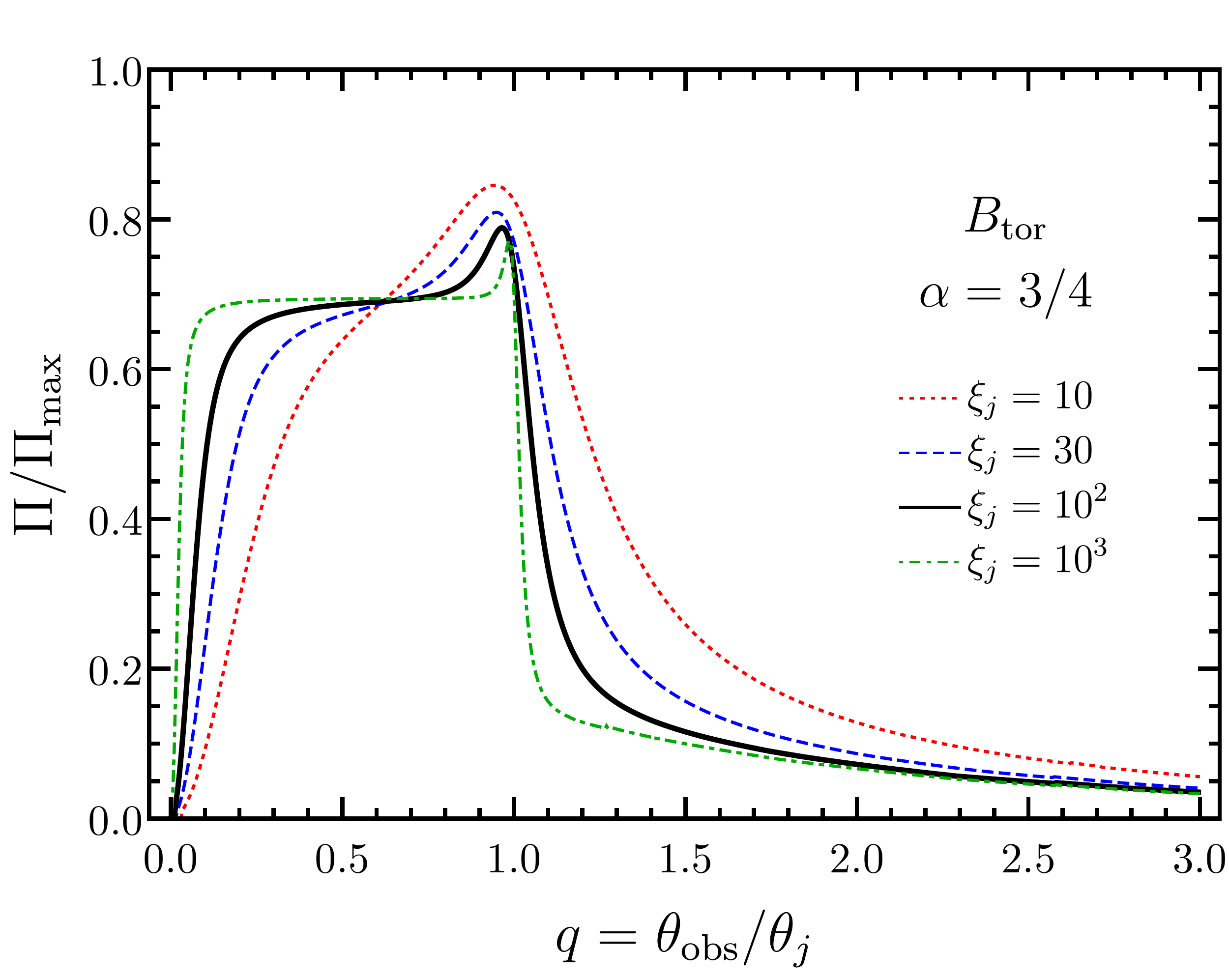}\quad\quad
    \includegraphics[width=0.47\textwidth]{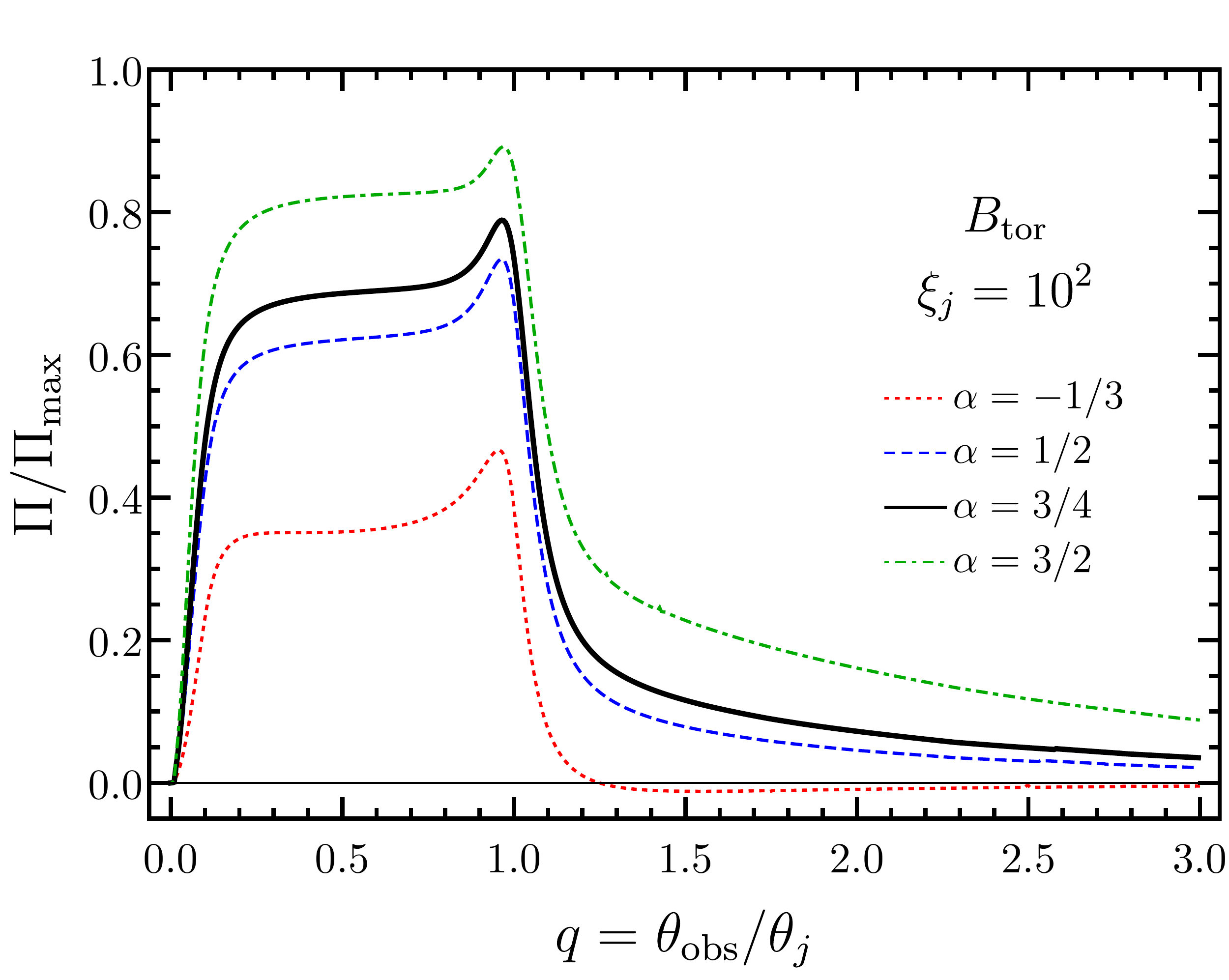}
    \caption{Pulse integrated degree of polarization arising from a top-hat jet for ({\it top}) a random magnetic field ($B_\perp$) that is normal 
    to the local velocity vector, $\vec\beta$, and lies entirely in the plane of the ejecta, ({\it middle}) a 
    locally ordered field ($B_\parallel$) with direction parallel to $\vec\beta$, and ({\it bottom}) a globally ordered toroidal field 
    ($B_{\rm tor}$). All are shown for various values of $\xi_j = (\Gamma\theta_j)^2$ (left panel) and different values of the spectral 
    index $\alpha$ (right panel) \citep[after][]{Granot03,Granot-Taylor-05}.}
    \label{fig:B-tor-rnd-pol}
\end{figure*}
%%%%%%%%%%%%%%%%%%%%%%%%%%%%%%%%%%%%%%%%%%%%%%%%%%%%%%

From symmetry considerations $U=0$ and the degree of polarization is $\Pi=\vert Q\vert/I$. 
The value of $\tilde\xi_{\rm max}=(\Gamma\tilde\theta_{\rm max})^2$ determines the maximal angle from the LOS ($\tilde{\theta}_{\rm max}$ in units of $1/\Gamma$) out to which the contribution to the observed flux is included. For a spherical shell and if the flux is integrated well into the tail of the pulse, this would correspond to $\tilde\xi_{\rm max}\gg1$. If, on the other hand, we measure the polarization of a pulse (of width $\Delta t_{\rm obs}$ and peak time $t_p$) over a time interval $t_1<t_{\rm obs}<t_2$ that contains only part of its tail (but all of its rising part), this would effectively correspond to a finite $\tilde\xi_{\rm max}\sim 1+(t_2-t_p)/\Delta{t_{\rm obs}}$. This arises since the emission at $t_{\rm obs}\sim t_p$ is dominated by the contribution from $\tilde\xi\sim1$, while during the tail it is predominantly from $\tilde\xi\sim 1+(t_{\rm obs}-t_p)/\Delta{t_{\rm obs}}$. Finally, even if the integration time extends well into the tail of the pulse, $(t_2-t_p)/\Delta{t_{\rm obs}}\gg 1$, then a line of sight close to the edge of the jet, or a rather narrow jet, can again introduce an effective $\tilde\xi_{\rm max}=(\Gamma\tilde\theta_{\rm max})^2$.

In Fig.~\ref{fig:B-ord-tor-pol}, we show the time-integrated (over the duration of a single pulse) 
degree of polarization arising from a spherical shell with an ordered magnetic field in the plane normal 
to $\vec\beta$, where for large $\tilde\xi_{\max}\sim100$ the result converges to that obtained by 
explicitly integrating over the entire pulse duration.

For an on-axis observer ($\theta_{\rm obs}=0$), if the magnetic field configuration is toroidal or random, the degree of polarization 
averaged over the GRB image vanishes due to the inherent axisymmetry of the outflow around the LOS. To break the 
symmetry, the jet must be viewed off-axis ($\theta_{\rm obs}>0$). In the case of the toroidal field, the geometry of the field 
is sufficient to break the symmetry, however, for a random field that is symmetric around the LOS the outflow must 
be sufficiently inhomogeneous in its properties as a function of $\theta$ from the jet axis, e.g. in 
(i) a top-hat jet where the jet is uniform within the initial jet half-opening angle $\theta_j$ beyond which the emissivity 
drops abruptly, effectively giving the outflow a sharp edge, or (ii) in a structured jet, where the emissivity 
$L'_{\nu'}=L'_{\nu'}(\theta)$ and/or the bulk LF $\Gamma=\Gamma(\theta)$ vary smoothly with $\theta$ outside of a compact 
core that has an angular size $\theta_c$.

%%%%%%%%%%%%%%%%%%%%%%%%%%%%%%%%%%%%%%%%%%%%%%%%%%
\subsection{Off-Axis Observer}\label{sec:off-axis}
%%%%%%%%%%%%%%%%%%%%%%%%%%%%%%%%%%%%%%%%%%%%%%%%%%

%%%%%%%%%%%%%%%%%%%%%%%%%%%%%%%%%%%%%%%%%%%%%%%%%%%%%%%%%%%%%%%%%%%%%%%%%%%%%%%
\subsubsection{Top-hat jet viewed off-axis -- Ordered magnetic field}
%%%%%%%%%%%%%%%%%%%%%%%%%%%%%%%%%%%%%%%%%%%%%%%%%%%%%%%%%%%%%%%%%%%%%%%%%%%%%%%

Here we discuss the degree of polarization obtained from ordered fields, such as a toroidal field ($B_{\rm tor}$) and 
a field ($B_\parallel$) that is parallel to the local velocity vector $\vec\beta$ which is assumed to be radial. 
In the toriodal field case, when the jet is viewed on-axis ($\theta_{\rm obs}=0$), the total polarization 
averaged over the GRB image vanishes. Therefore, the observer's LOS must be off-axis, $\theta_{\rm obs}>0$. 
The local polarization from a given point of the observed image on the plane of the sky is exactly the same 
as that from an ordered field that is entirely in the plane of the ejecta, however, the global structure of the 
magnetic field adds more complexity (see right panel of Fig.~\ref{fig:pol-map}). Therefore, after integrating over 
the solid angle subtended by the source, we find \citep[][]{Granot-Taylor-05} a time-integrated polarization
\begin{eqnarray}\label{eq:Pi_tor}
    \frac{\Pi}{\Pi_{\rm max}} = 
    &&\displaystyle\left[H(1-q)\int_0^{\xi_-}\frac{d\tilde\xi}{(1+\tilde\xi)^{2+\alpha}}\int_0^{2\pi}
    d\tilde\varphi\Lambda_{\rm tor}(\tilde\xi,\tilde\varphi,a)\cos 2\theta_p \right. \nonumber\\
    &&\displaystyle + \left.\int_{\xi_-}^{\xi_+}\frac{d\tilde\xi}{(1+\tilde\xi)^{2+\alpha}}\int_\psi^{2\pi-\psi}
    d\tilde\varphi\Lambda_{\rm tor}(\tilde\xi,\tilde\varphi,a)\cos 2\theta_p\right] \\
    &&\times\displaystyle\left[H(1-q)\int_0^{\xi_-}\frac{d\tilde\xi}{(1+\tilde\xi)^{2+\alpha}}\int_0^{2\pi}
    d\tilde\varphi\Lambda_{\rm tor}(\tilde\xi,\tilde\varphi,a) \right.\nonumber\\
    &&\displaystyle + \left.\int_{\xi_-}^{\xi_+}\frac{d\tilde\xi}{(1+\tilde\xi)^{2+\alpha}}\int_\psi^{2\pi-\psi}
    d\tilde\varphi\Lambda_{\rm tor}(\tilde\xi,\tilde\varphi,a)\right]^{-1} \nonumber
\end{eqnarray}
where $H(1-q)$ is the Heaviside step-function, and
\begin{eqnarray}
 & \cos\psi(\tilde\xi) = \frac{\displaystyle(1-q^2)\xi_j-\tilde\xi}{\displaystyle2q\sqrt{\tilde\xi\xi_j}} & \\
 & q=\theta_{\rm obs}/\theta_j~,\quad \xi_j = (\Gamma\theta_j)^2~,\quad \xi_\pm=(1\pm q)^2\xi_j~. &
\end{eqnarray}
The bottom panel of Fig.~\ref{fig:B-tor-rnd-pol} shows the pulse-integrated $\Pi$ for a toroidal field. 
The degree of polarization vanishes for $q=0$ due to symmetry, but remains high for $\xi_j^{-1/2} \lesssim q \lesssim 1+ \xi_j^{-1/2}$, 
and drops sharply for $q>1$.

The calculation for the $B_\parallel$ case follows from that presented in \citet{Granot03}, where 
the total polarization for an off-axis observer is obtained from
\begin{equation}\label{eq:Pi-off-axis}
    \Pi=\frac{\displaystyle\frac{1}{2\pi}\int_{\xi_-}^{\xi_+}
    \frac{d\tilde\xi}{(1+\tilde\xi)^{2+\alpha}}\Lambda(\tilde\xi)\Pi_{\max}\sin2\psi(\tilde\xi)}
    {\displaystyle H(1-q)\int_0^{\xi_-}\frac{d\tilde\xi\Lambda(\tilde\xi)}{(1+\tilde\xi)^{2+\alpha}} +\int_{\xi_-}^{\xi_+}d\tilde\xi\frac{\pi-\psi(\tilde\xi)}
    {\pi(1+\tilde\xi)^{2+\alpha}}\Lambda(\tilde\xi)}~,
\end{equation}
where $\Lambda(\tilde\xi)=\Lambda_\parallel(\tilde\xi)$ from Eq.~(\ref{eq:sin-chi}). The result of the 
integration are presented in the middle panel of Fig.~\ref{fig:B-tor-rnd-pol}, where the left panel shows 
the variation in $\Pi$ as the jet becomes narrow or wide, and the right panel shows dependence of $\Pi$ on the 
spectral index. Softer spectra tend to be more polarized and this trend applies to synchrotron emission regardless 
of the magnetic field configuration. The degree of polarization 
remains small for $q\lesssim1-\xi_j^{-1/2}$, but sharply increases above $q=1$ and becomes large for $q\gtrsim1+\xi_j^{-1/2}$. 
However, an important point to note here is that 
for $q>1+\xi_j^{-1/2}$, the fluence rapidly drops and such high levels of polarization in off-axis jets may only be realizable 
in nearby bursts. For bursts that are truly cosmological, one can only measure high $\Pi$ 
from this type of an ordered field for a very special geometry where $q\approx1+\xi_j^{-1/2}$. The PA undergoes a change by 
$90^\circ$ around $q=1$, and the exact value of $q$ at which the polarization curve passes $\Pi=0$ 
depends on $\xi_j$, which suggests that if $\Gamma$ varies between different pulses and $q\sim1$ then the observer 
may measure a $90^\circ$ shift in the PA. A similar behavior is observed for $B_\perp$ field case which is discussed 
next.

%%%%% FIGURE %%%%%%%%%%%%%%%%%%%%%
\begin{figure}
    \centering
    \includegraphics[width=0.48\textwidth]{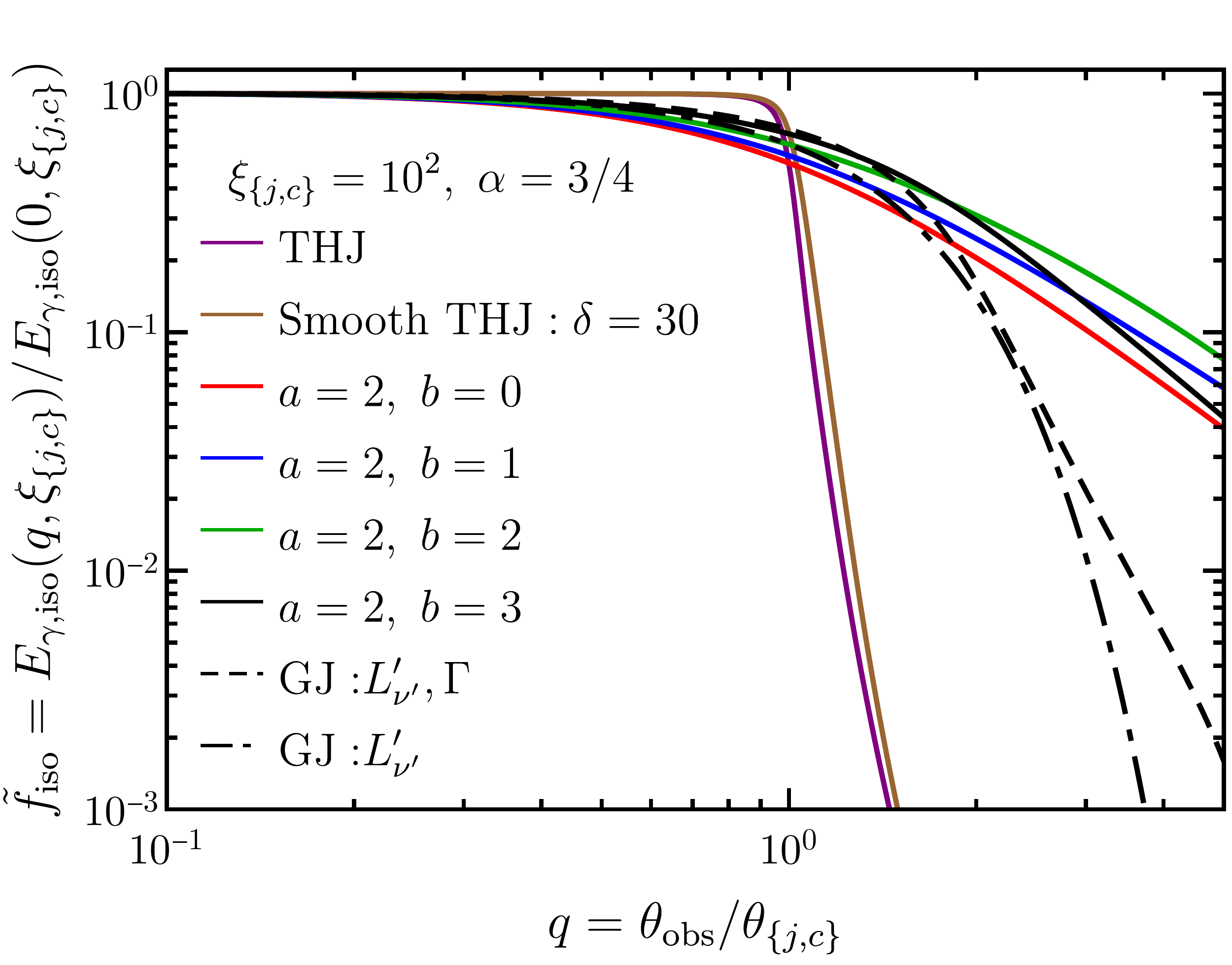}
    \caption{Ratio of off-axis to on-axis fluence, or equivalently isotropic equivalent energies, as a function of 
    the ratio of the viewing angle $\theta_{\rm obs}$ to the half jet opening or core angle $\theta_{\{j,c\}}$. 
    Shown here for different jet structures: a top-hat jet (THJ); smooth THJ ($\delta=30$; see Eq~(\ref{eq:THJ-pwrl-wings})); 
    power law structured jet (PLJ; with $a=2$ and variable $b$; see Eq.~(\ref{eq:PLJ})); and Gaussian structured Jet (GJ; with either both 
    $L'_{\nu'}$ and $\Gamma$ varying as a Gaussian or only $L'_{\nu'}$; see Eq.~(\ref{eq:GJ})).}
    \label{fig:fiso-q}
\end{figure}
%%%%%%%%%%%%%%%%%%%%%%%%%%%%%%%%%%

%%%%%%  OFF-AXIS JET -- RANDOM MAGNETIC FIELD  %%%%%%%%%%%%%%%%%%%%%%%%%%%%%%%%%%%%%%%%%%%%%%%%%%%%%%%%%%
\subsubsection{Top-hat jet viewed off-axis -- Random magnetic field}
\label{sec:random-B-field-pol}
When the magnetic field orientation is random in the plane of the ejecta, the observed 
polarization from an unresolved source vanishes upon averaging over the image on the 
plane of the sky (see left panel of Fig.~\ref{fig:pol-map}). This occurs due to the 
fact that there is no special orientation of the polarization vector and it is symmetric 
around the LOS. To break the symmetry in this case, the jet must be viewed close to its 
edge ($q\gtrsim1-\xi_j^{-1/2}$), where missing emission from $\theta>\theta_j$ results in only partial 
cancellation of the polarization when averaged over the GRB image \citep[e.g.][]{Waxman03}.

%%%%  FIGURE  %%%%%%%%%%%%%%%%%%%%%%%%%%%%%%%%%%%%%%%
\begin{figure*}
    \centering
    \includegraphics[width=0.35\textwidth]{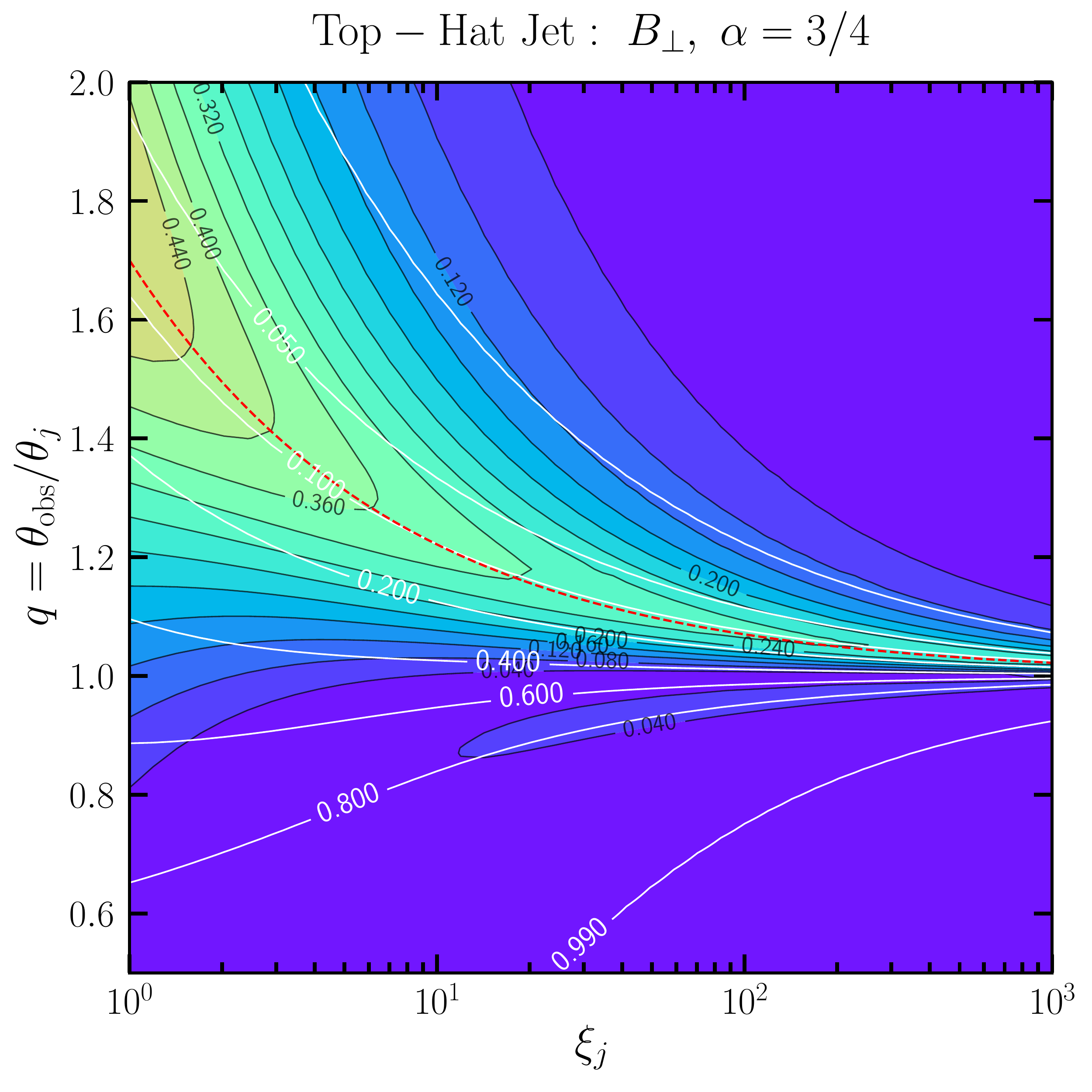}\quad\quad
    \includegraphics[width=0.35\textwidth]{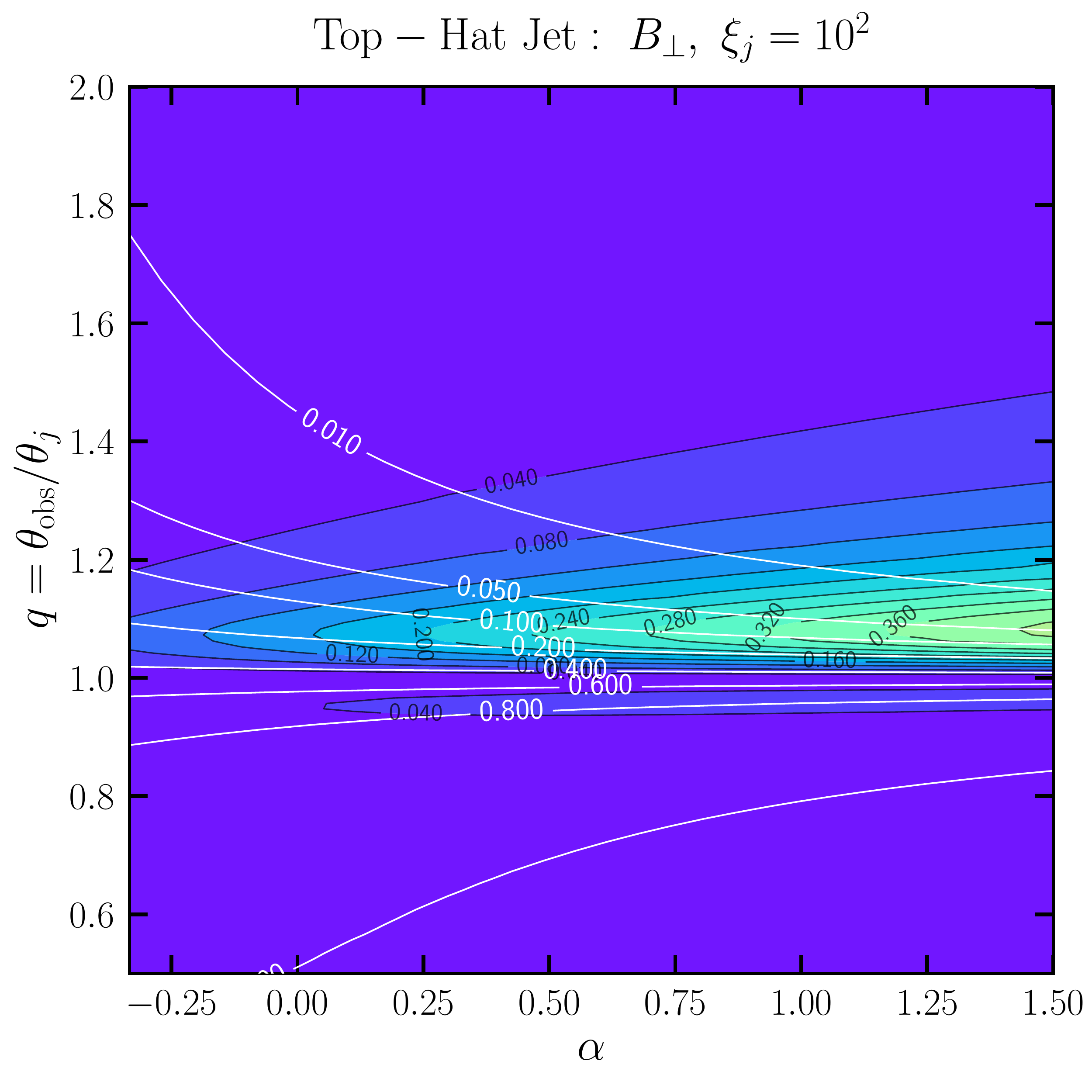}
    \includegraphics[width=0.35\textwidth]{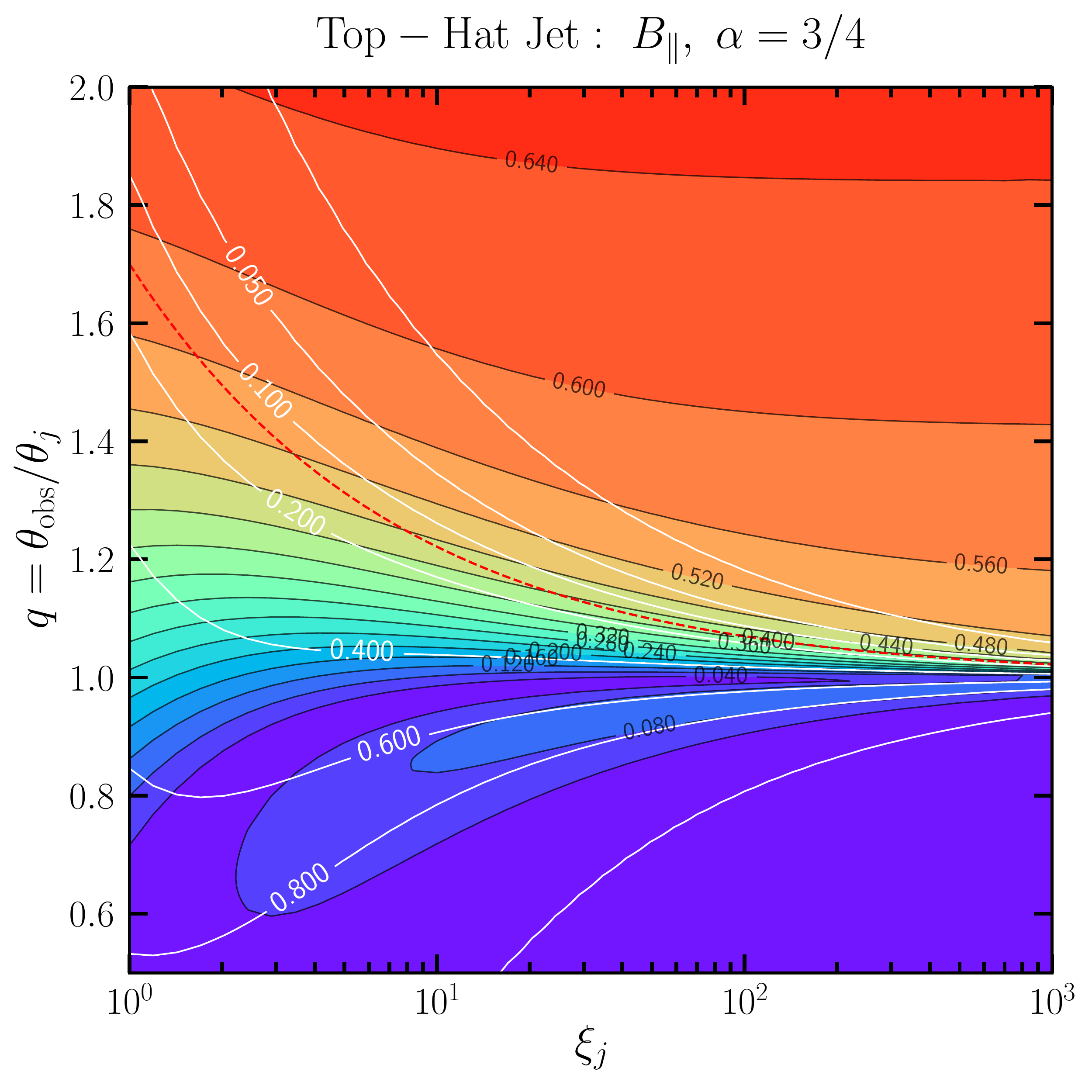}\quad\quad
    \includegraphics[width=0.35\textwidth]{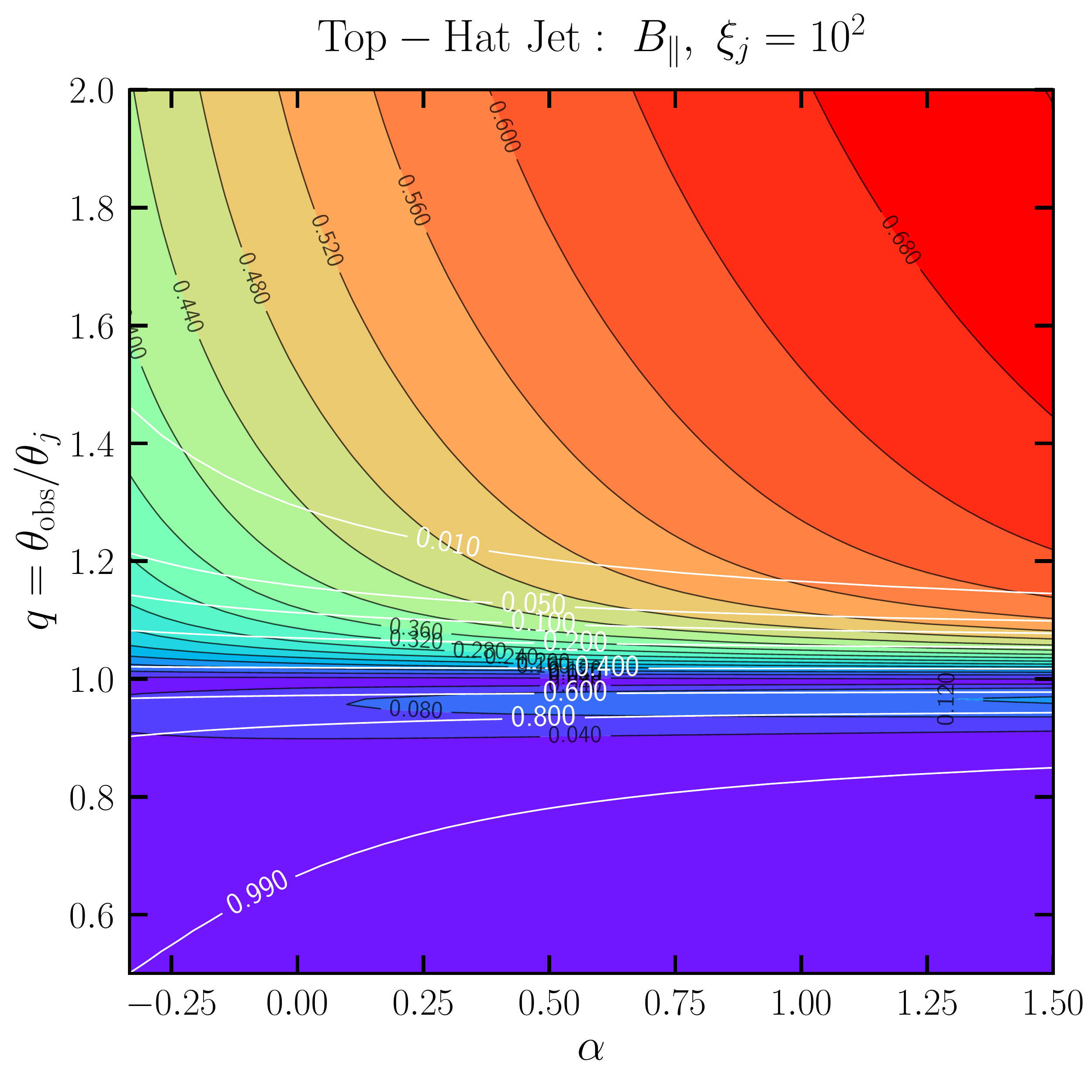}
    \includegraphics[width=0.35\textwidth]{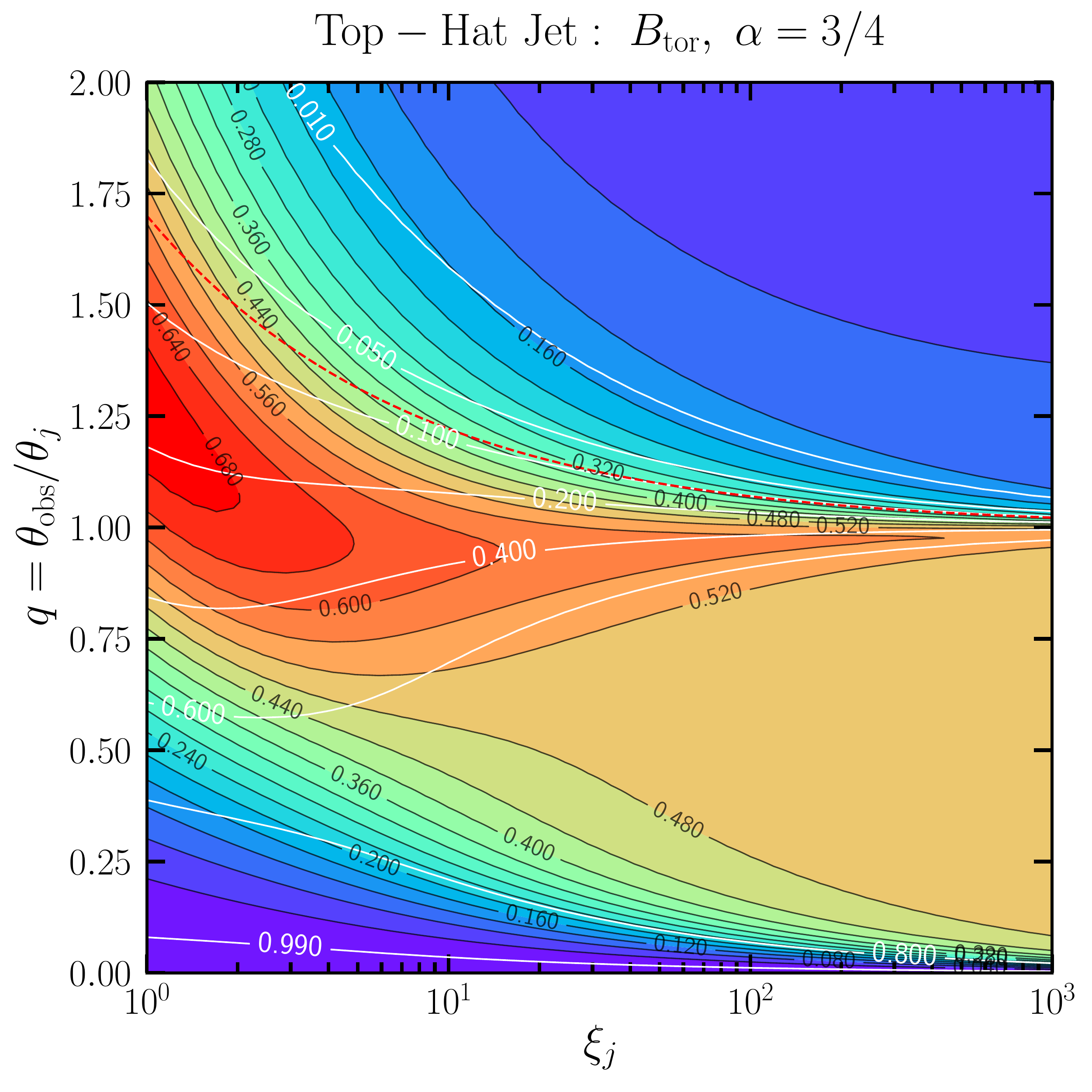}\quad\quad
    \includegraphics[width=0.35\textwidth]{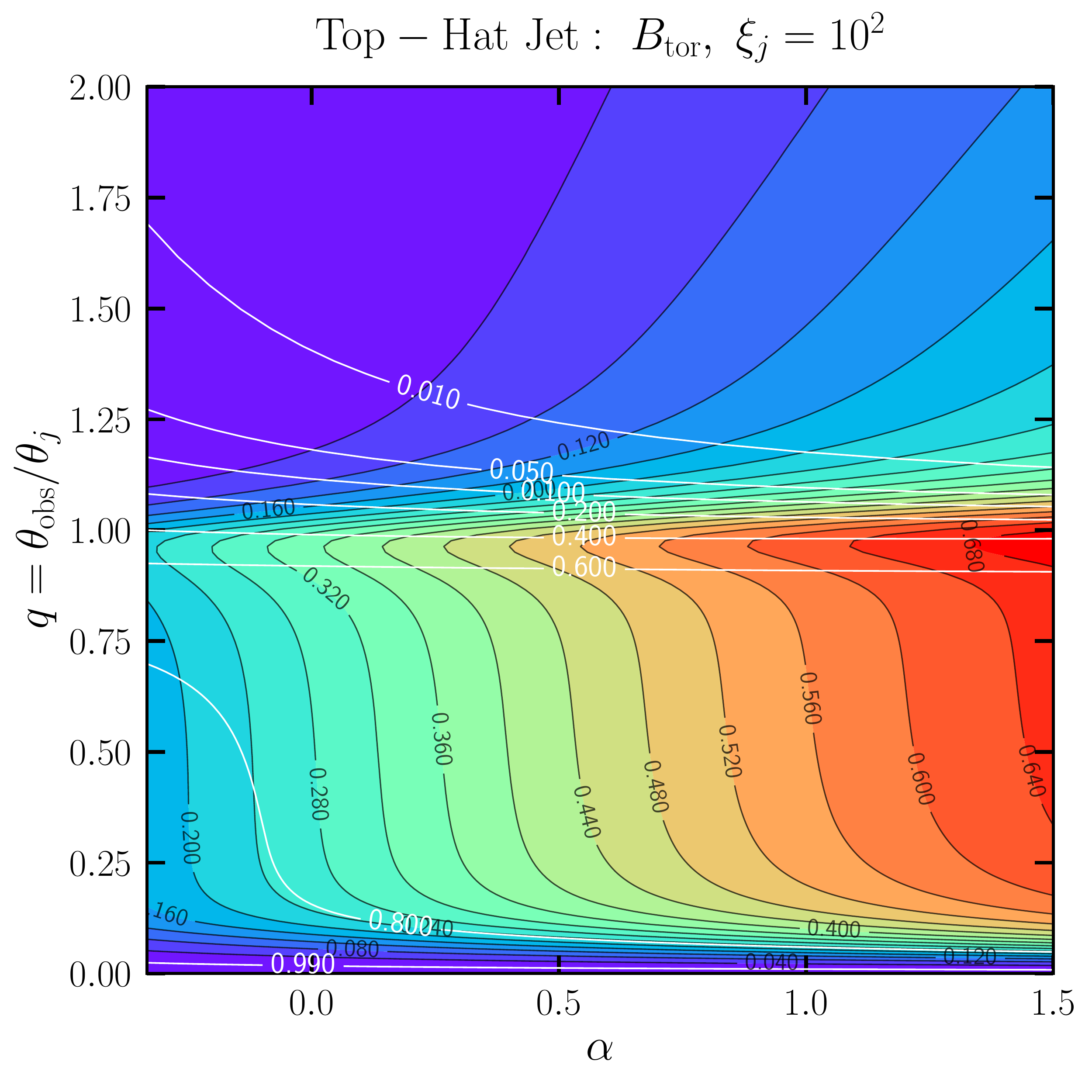}
    \caption{Contour plots of $\vert\Pi\vert$ for different magnetic field configurations: 
    (\textit{top}) random field entirely in the plane of the ejecta ($B_\perp$); (\textit{middle}) ordered 
    field parallel to the local velocity vector ($B_\parallel$); (\textit{bottom}) toroidal field ($B_{\rm tor}$). 
    The structure of the outflow is that of an ultra-relativistic top-hat jet. In the left panels $\alpha=3/4$ and 
    the red dashed line shows $q=1+0.7/\sqrt{\xi_j}$, and in the right panels 
    $\xi_j=(\Gamma\theta_j)^2=10^2$. Contours for different values of $\tilde f_{\rm iso}$ are plotted in white.}
    \label{fig:synchro-contour-plts}
\end{figure*}
%%%%%%%%%%%%%%%%%%%%%%%%%%%%%%%%%%%%%%%%%%%%%%%%%%%%%%

%%%%%%% FIGURE  %%%%%%%%%%%%%%%%%%%%%%%%%%%%%%%%%%%%%%
\begin{figure*}
    \centering
    \includegraphics[width=0.45\textwidth]{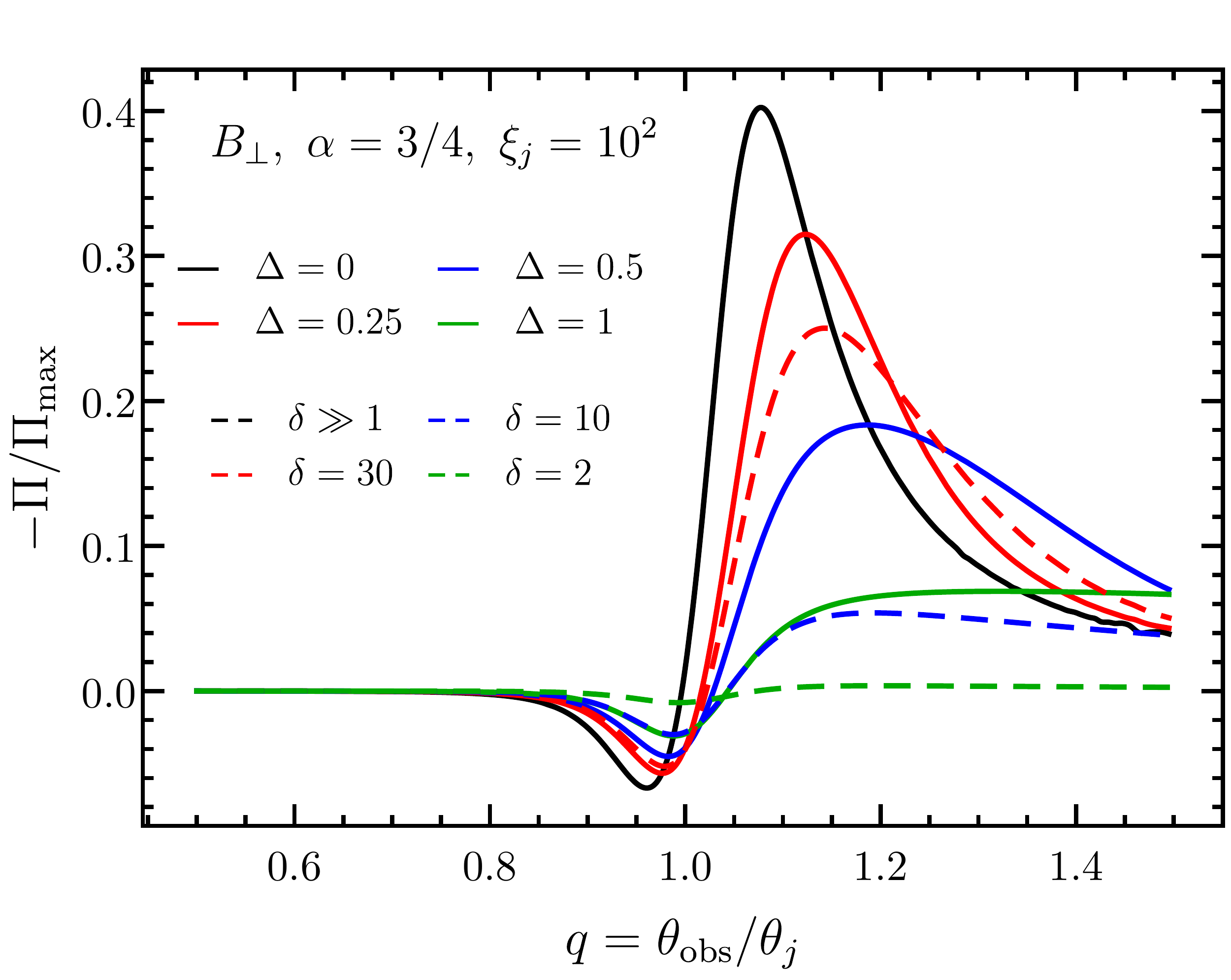}\quad\quad
    \includegraphics[width=0.45\textwidth]{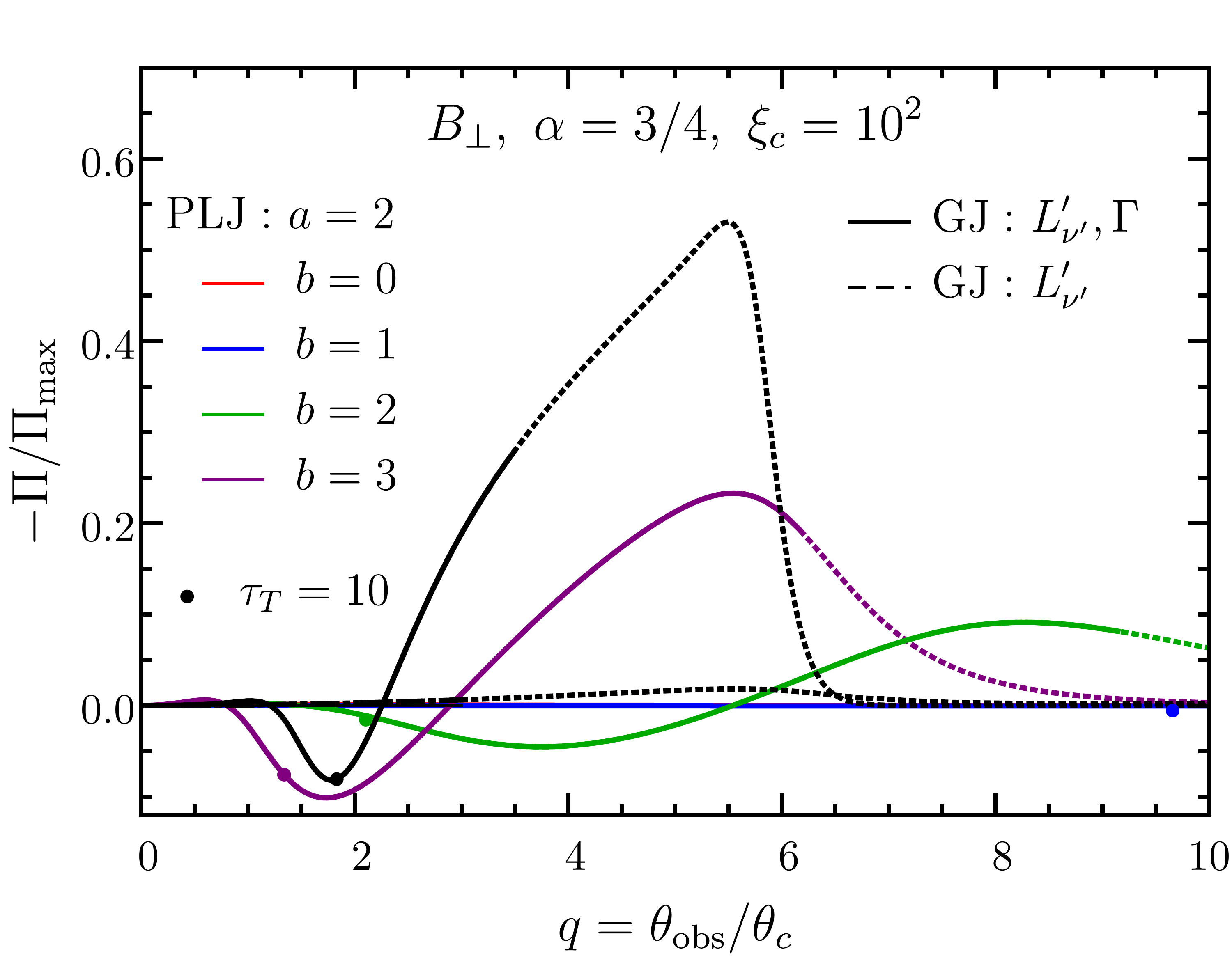}
    \includegraphics[width=0.45\textwidth]{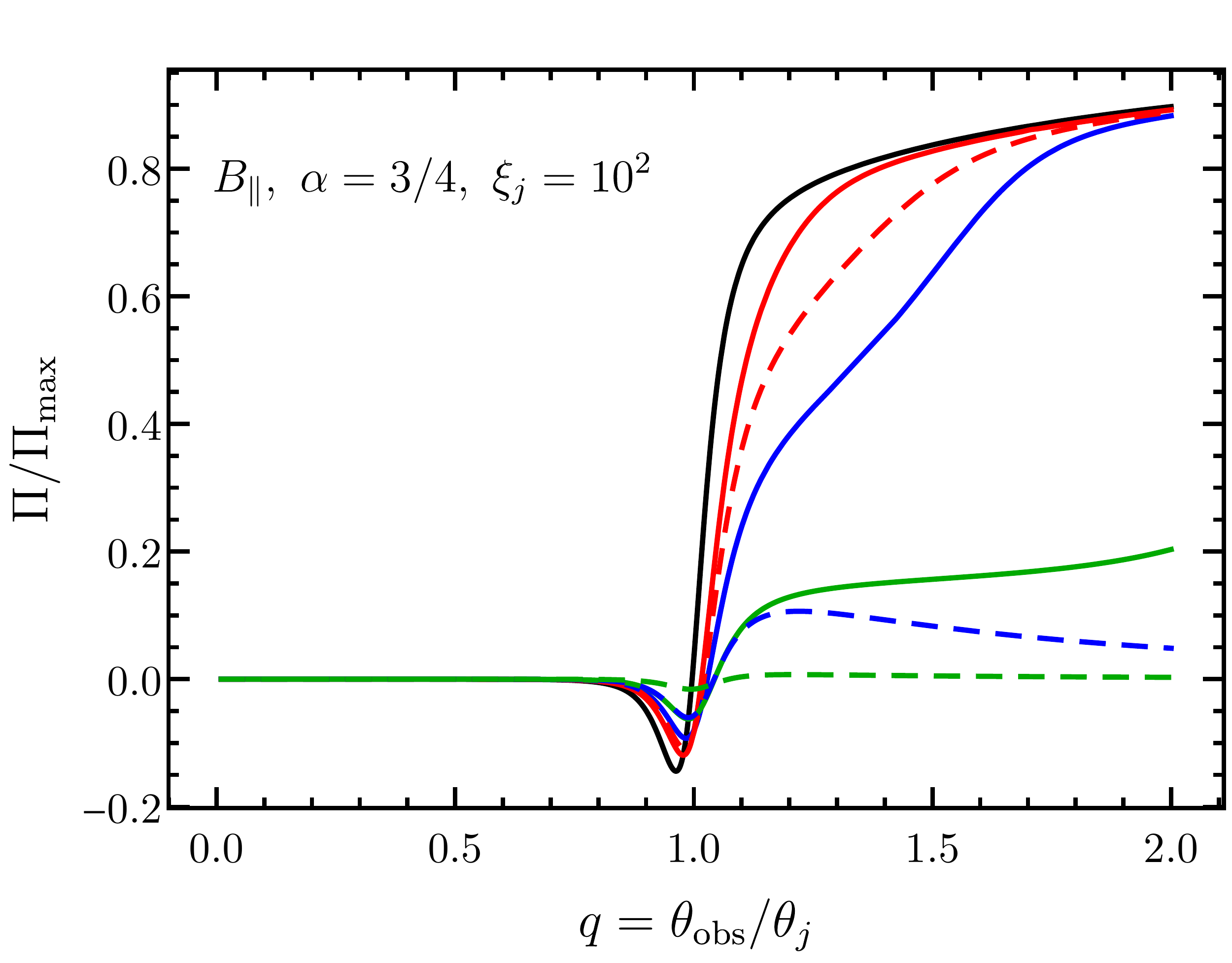}\quad\quad
    \includegraphics[width=0.45\textwidth]{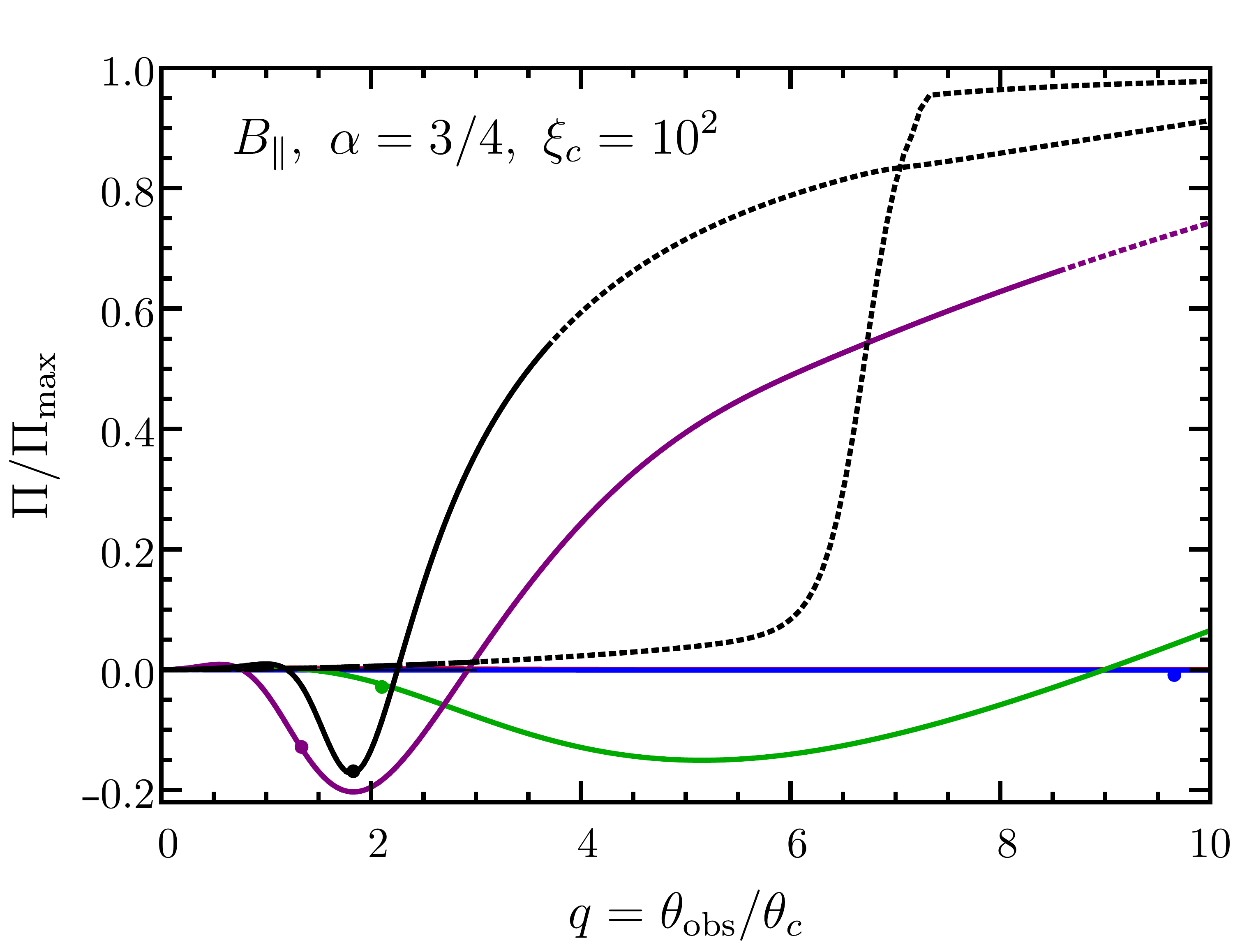}
    \includegraphics[width=0.45\textwidth]{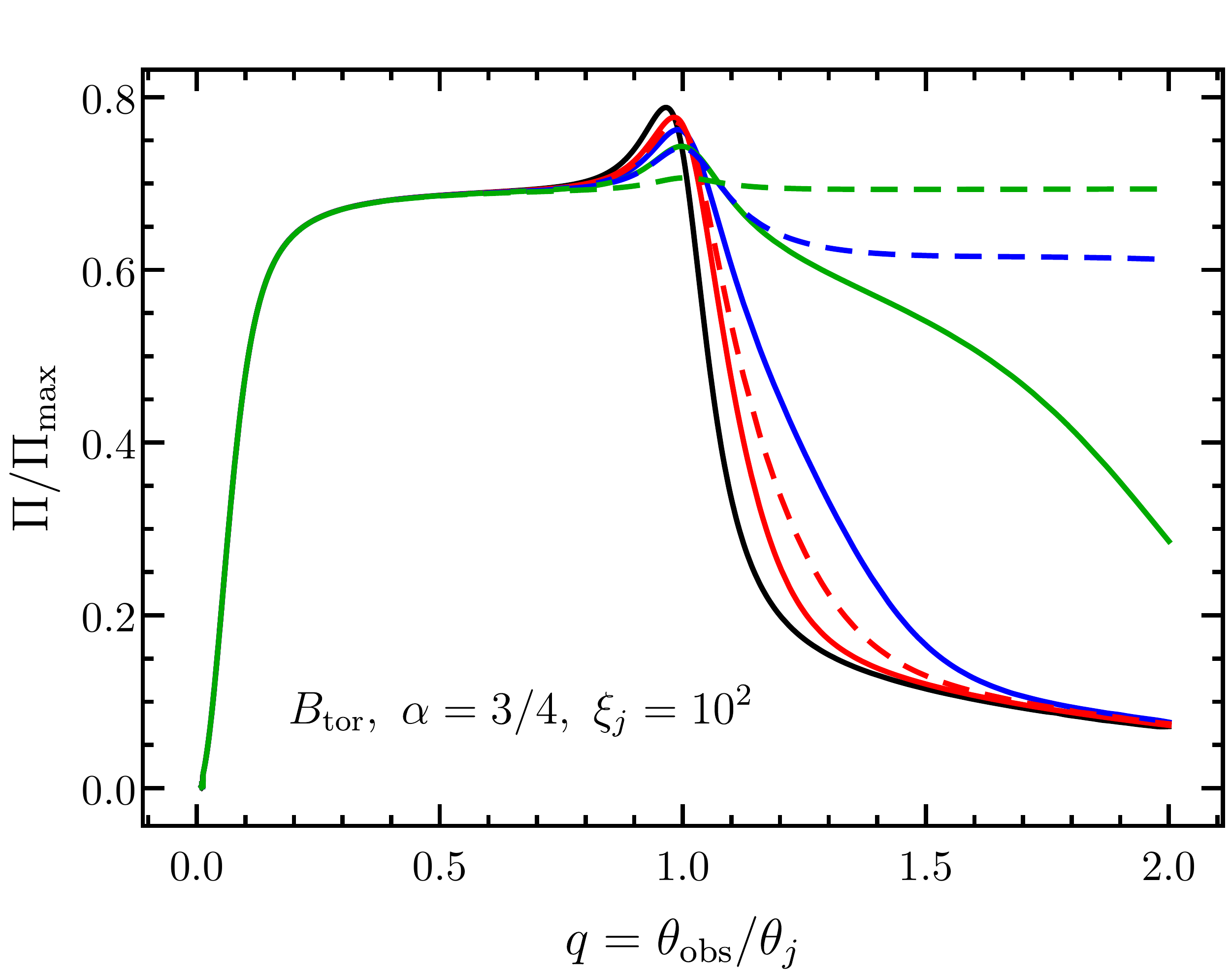}\quad\quad
    \includegraphics[width=0.45\textwidth]{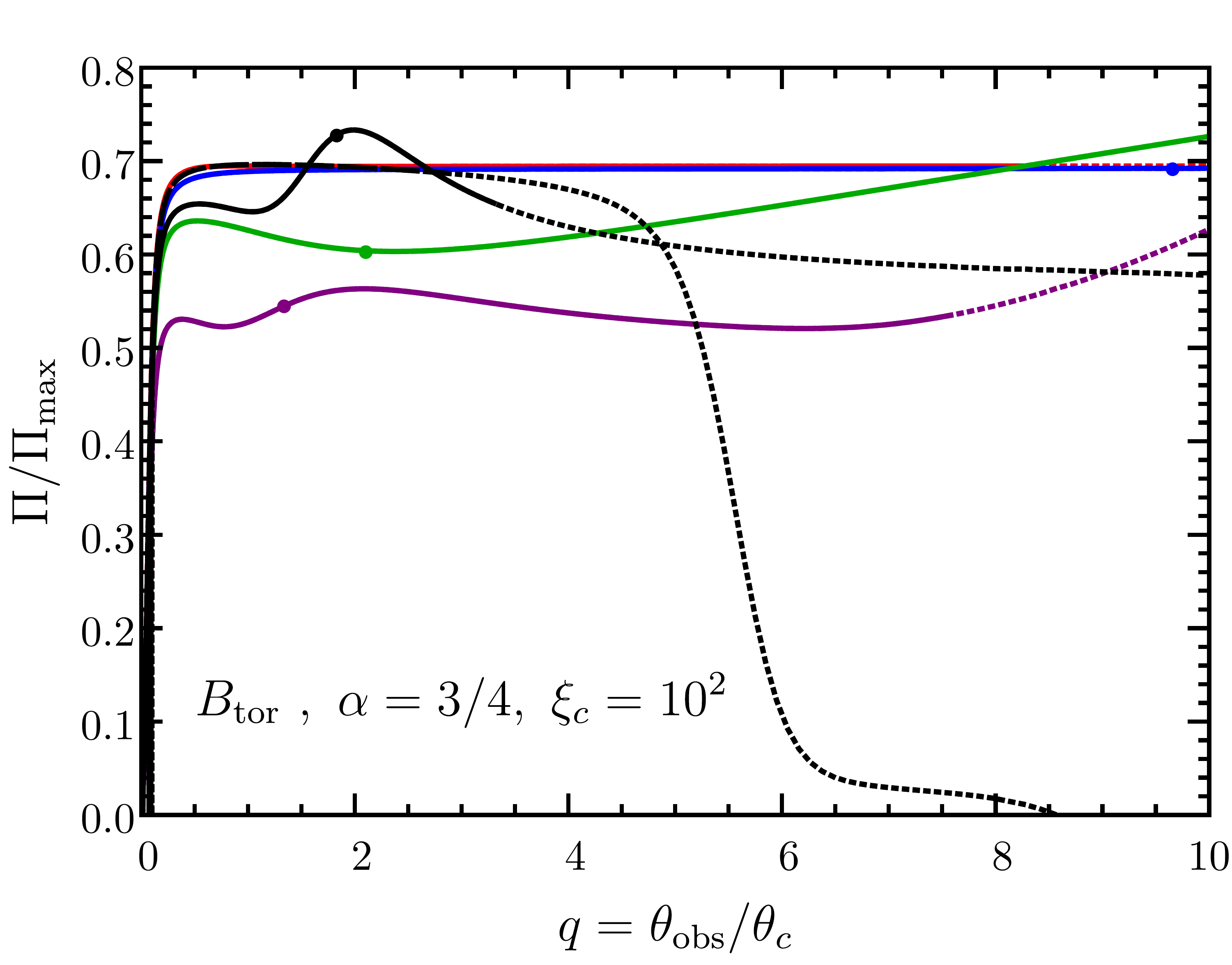}
    \caption{\textit{Left}: Pulse integrated polarization of a smooth top-hat jet with a uniform core 
    and exponential wings (solid lines) or power-law wings (dashed lines). Both are shown for different 
    magnetic field configurations and for different smoothing parameters $\Delta$ and $\delta$, 
    which control the rate at which the emissivity declines \citep[after][]{NPW03}. 
    \textit{Right}: Pulse integrated degree of polarization for a structured jet -- a power-law jet 
    (PLJ) and gaussian jet (GJ) -- shown for different field configurations. The dotted line shows the 
    trend for large $q$ values but the pulses will be dim with $\tilde f_{\rm iso}<10^{-2}$. Furthermore, 
    compactness arguments will restrict $q\lesssim2$ for sufficiently steep profiles in all emission models 
    (see \S\ref{sec:compactness} and Fig.~\ref{fig:pair-opacity}), as shown by the filled circle 
    obtained from Eq.~(\ref{eq:compactness-tauT}) for the same fiducial parameters.}
    \label{fig:pol-struc-jets}
\end{figure*}
%%%%%%%%%%%%%%%%%%%%%%%%%%%%%%%%%%%%%%%%%%%%%%%%%%%%%%%

%%%%  FIGURE  %%%%%%%%%%%%%%%%%%%%%%%%%%%%%%%%%%%%%%%
\begin{figure*}
    \centering
    \includegraphics[width=0.31\textwidth]{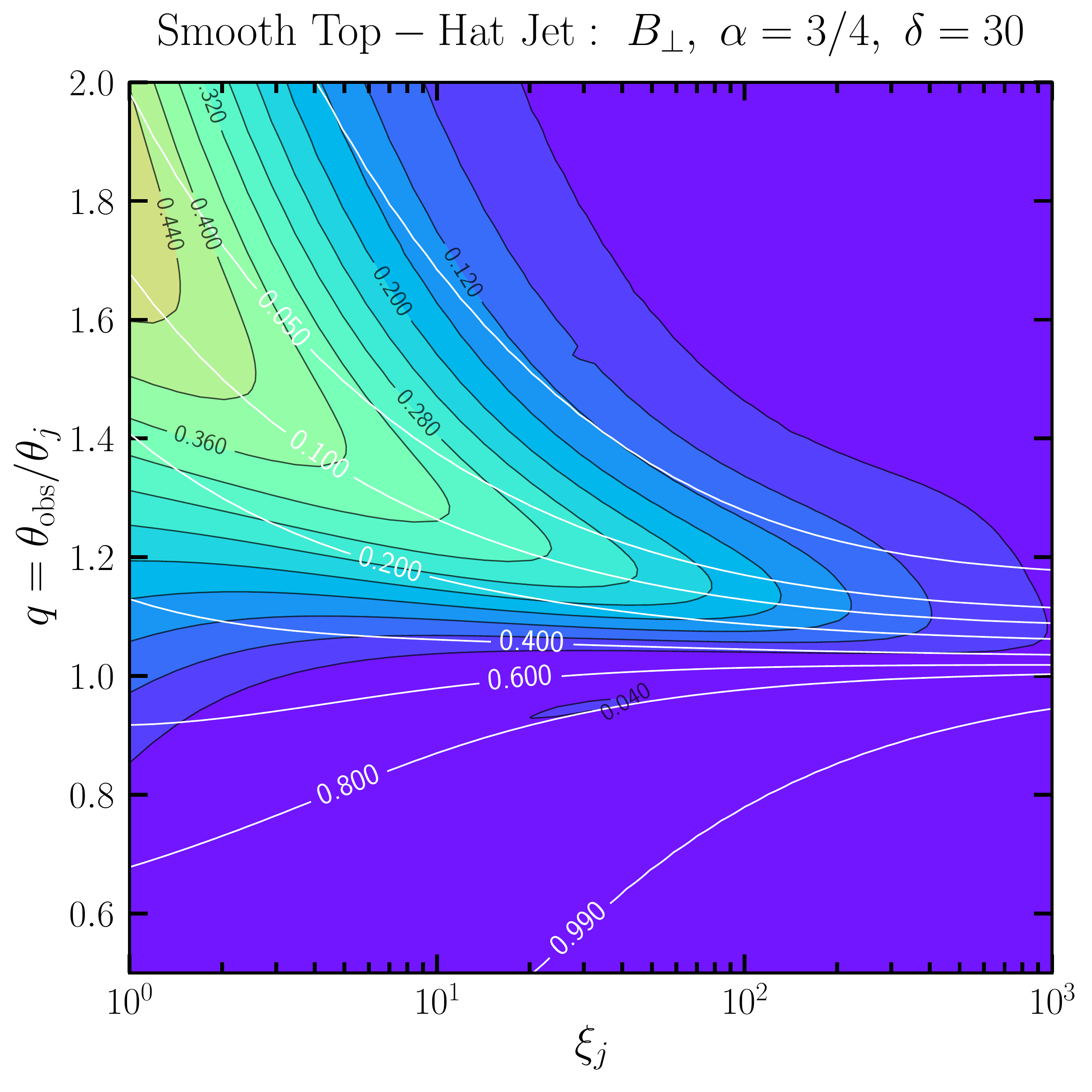}\quad\quad
    \includegraphics[width=0.31\textwidth]{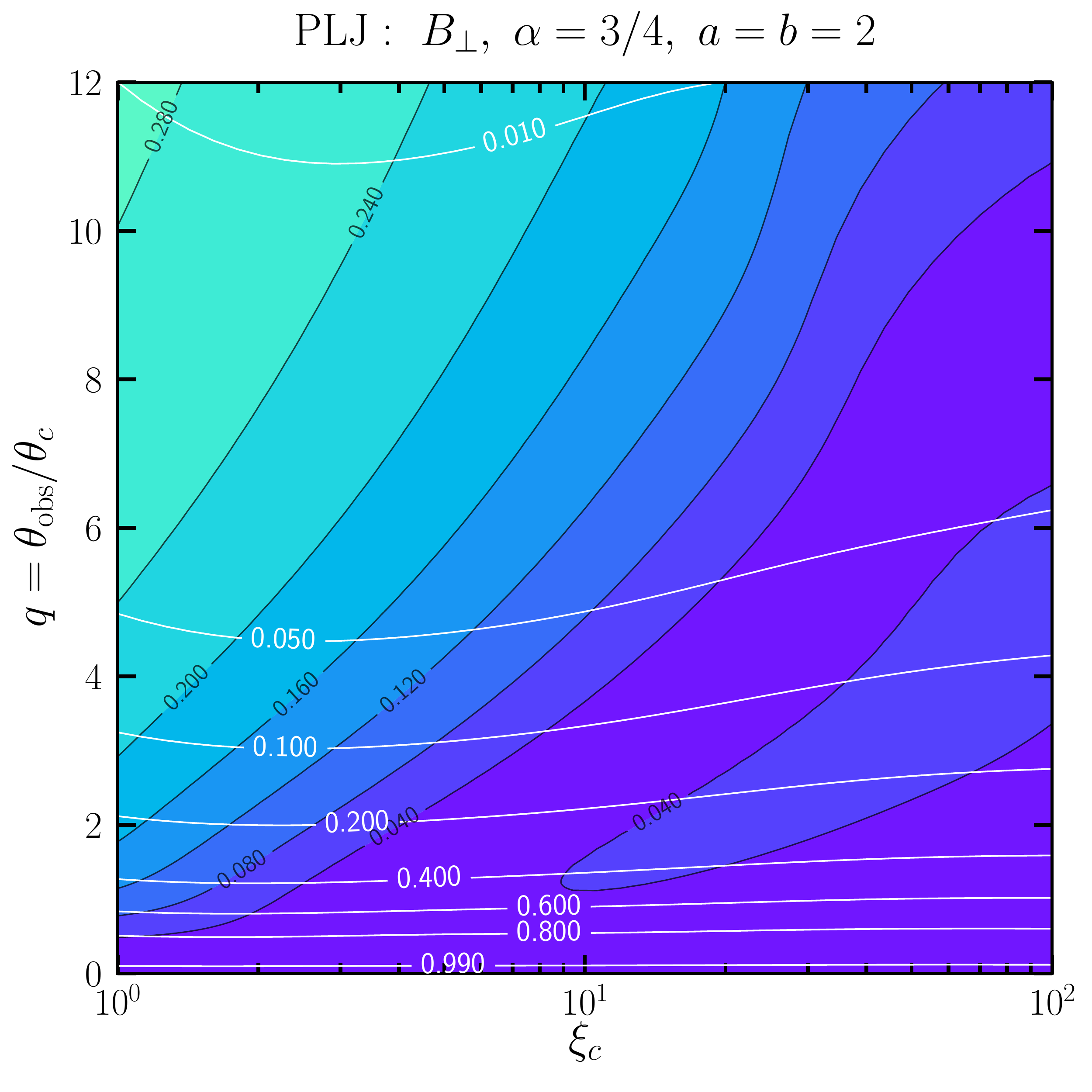}\quad\quad
    \includegraphics[width=0.31\textwidth]{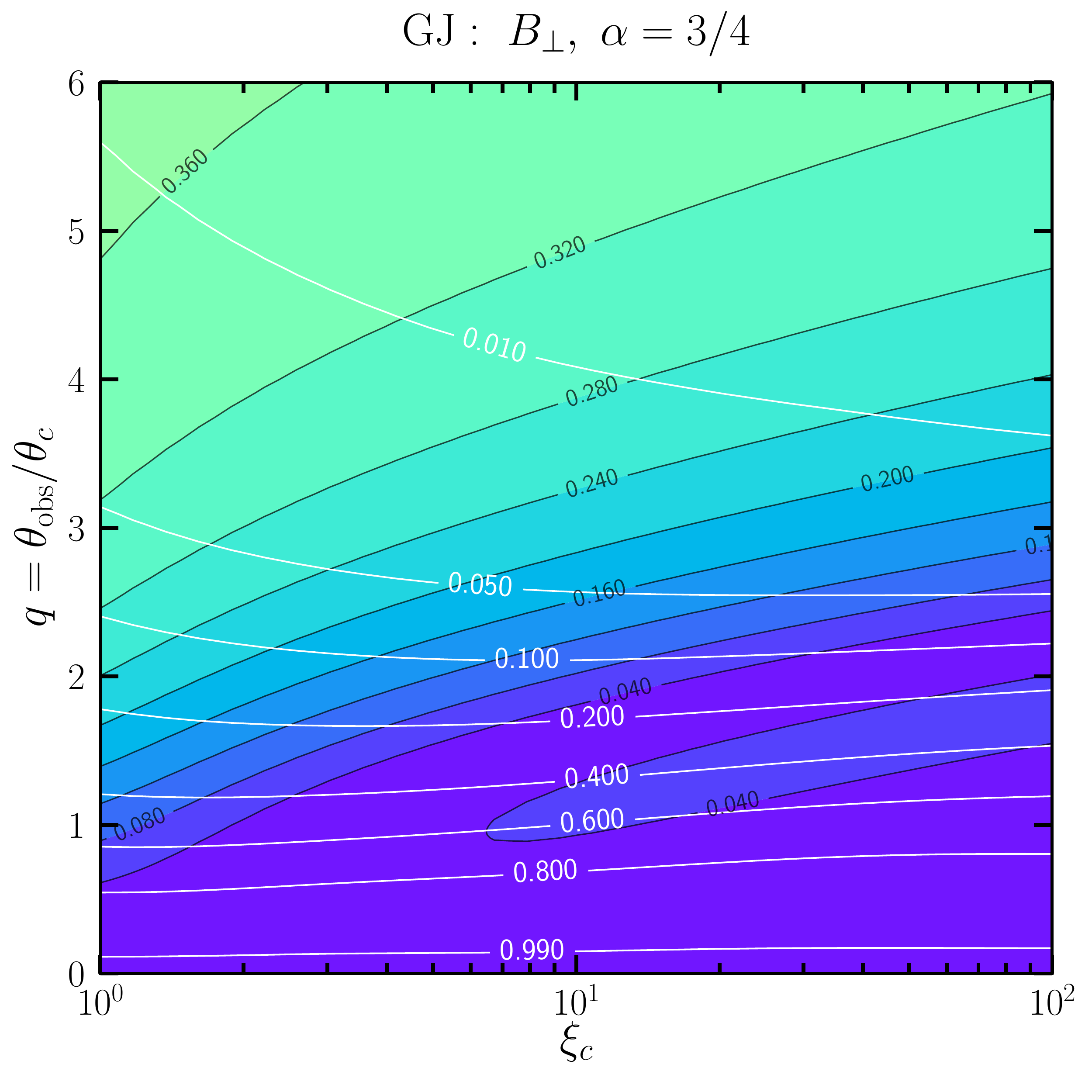}
    \includegraphics[width=0.31\textwidth]{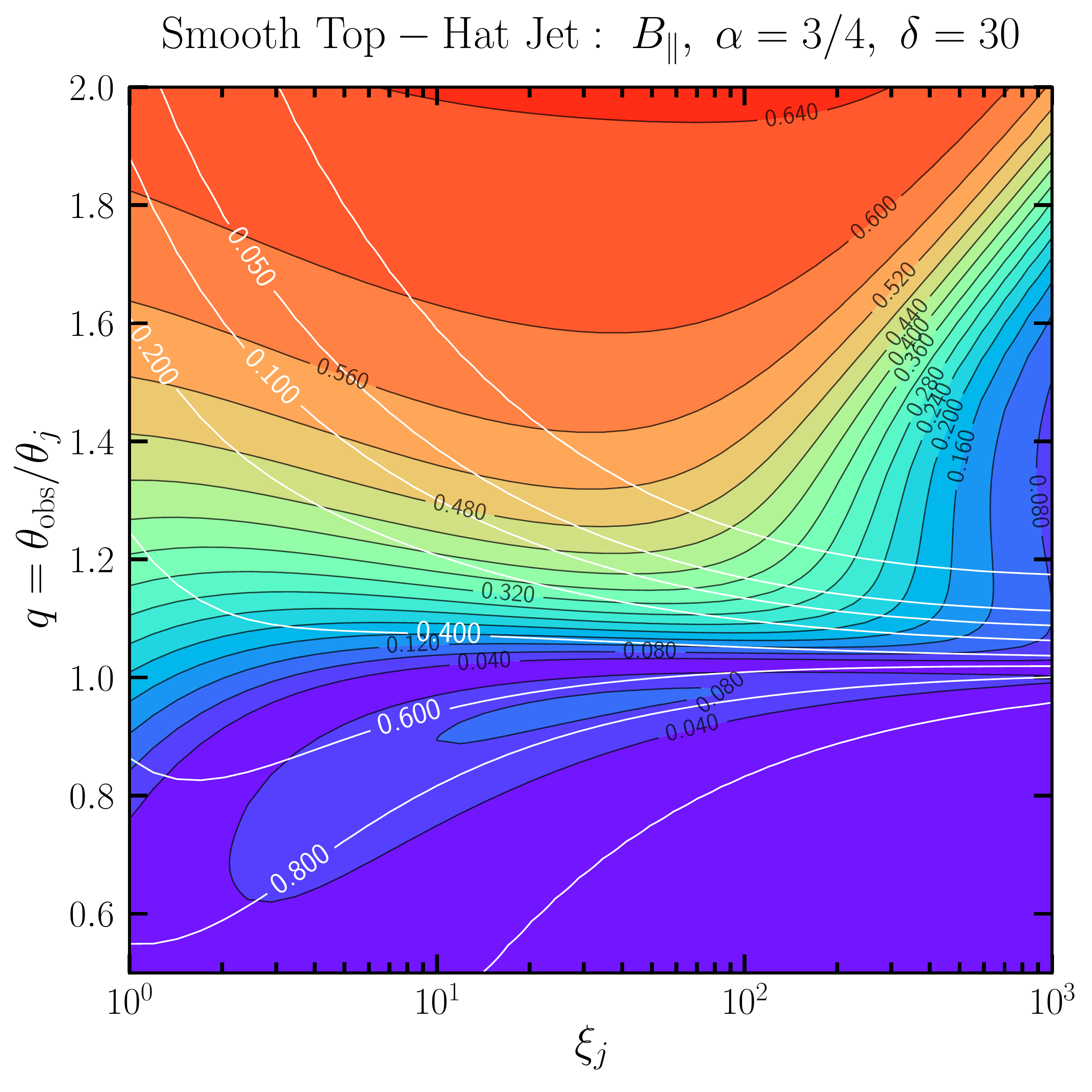}\quad\quad
    \includegraphics[width=0.31\textwidth]{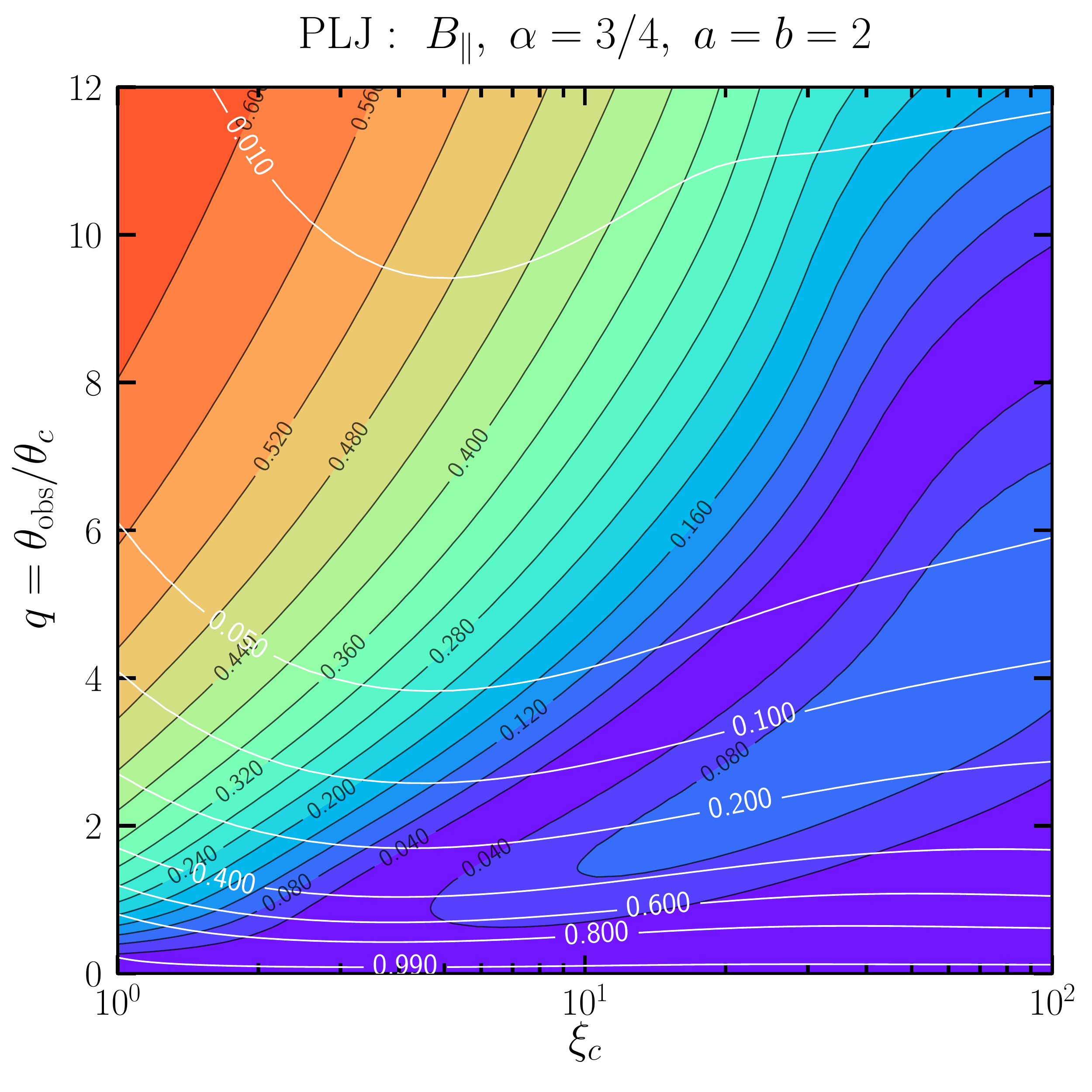}\quad\quad
    \includegraphics[width=0.31\textwidth]{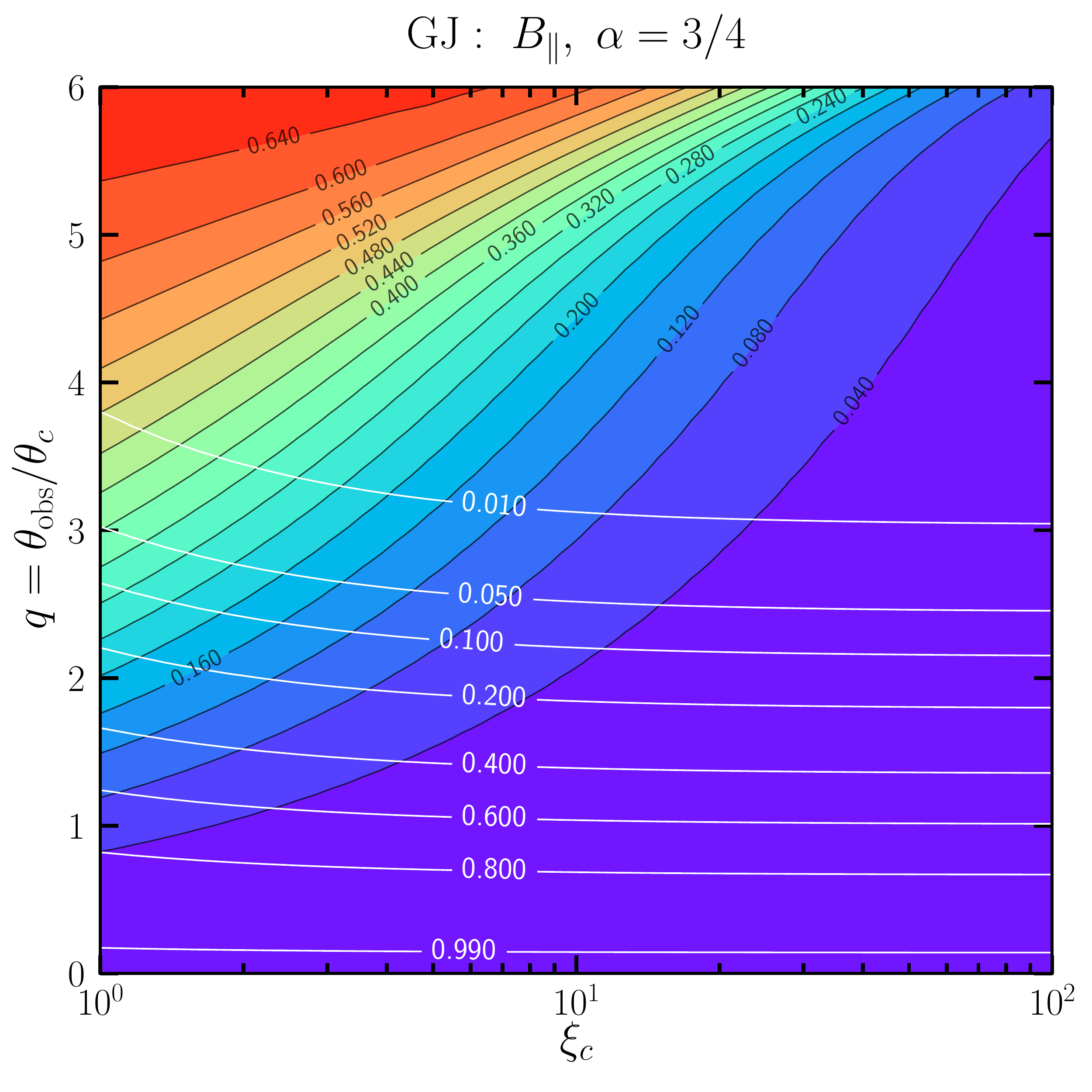}
    \includegraphics[width=0.31\textwidth]{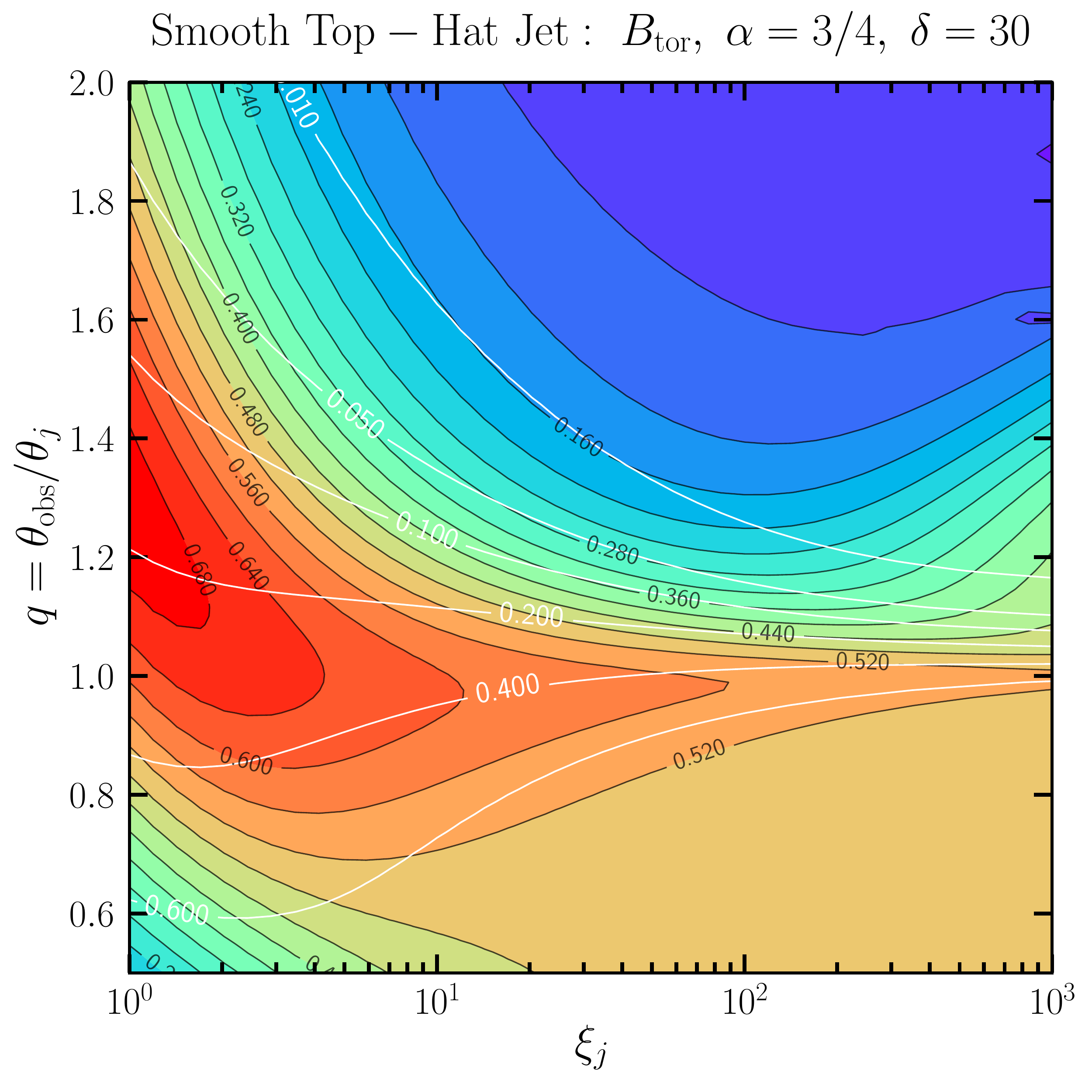}\quad\quad
    \includegraphics[width=0.31\textwidth]{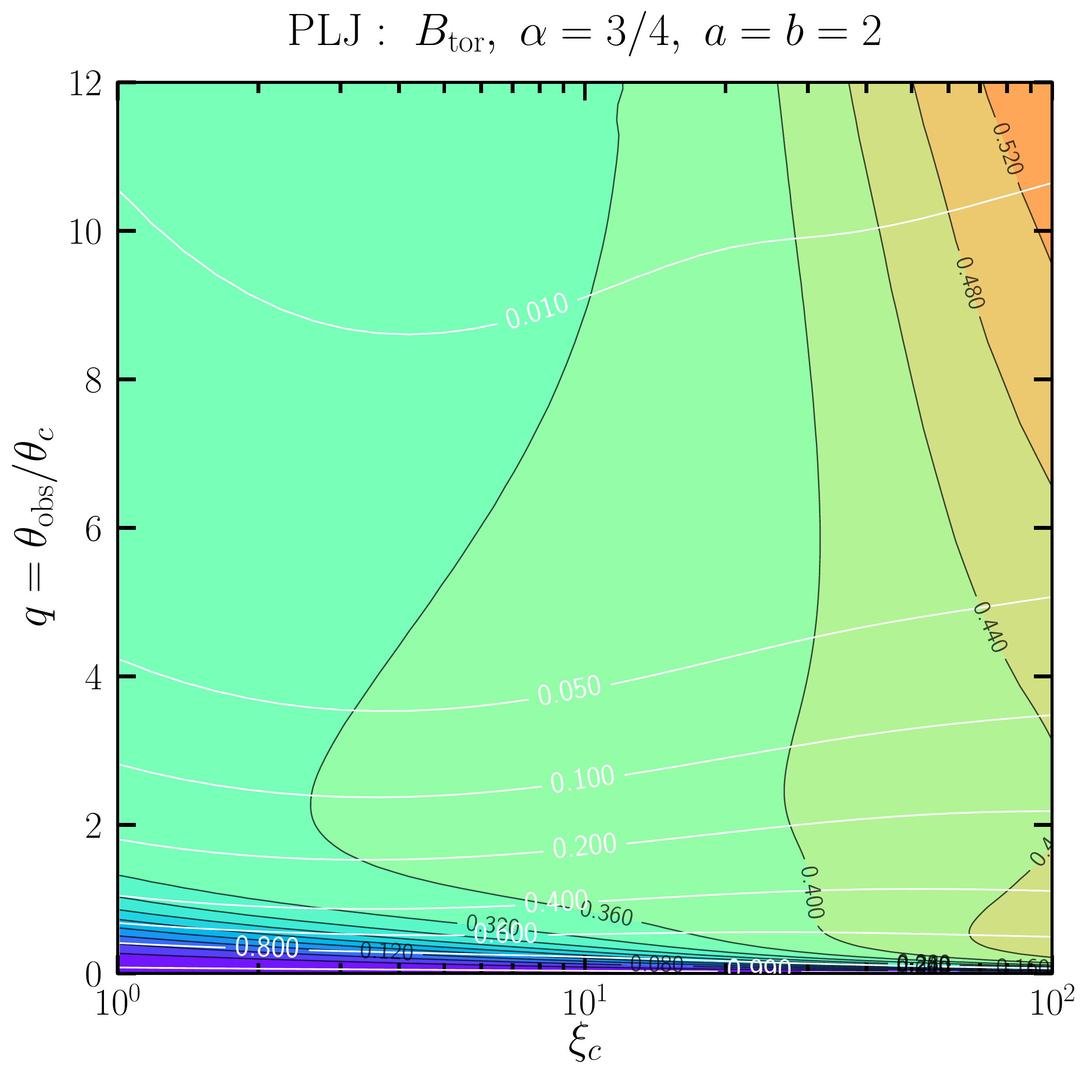}\quad\quad
    \includegraphics[width=0.31\textwidth]{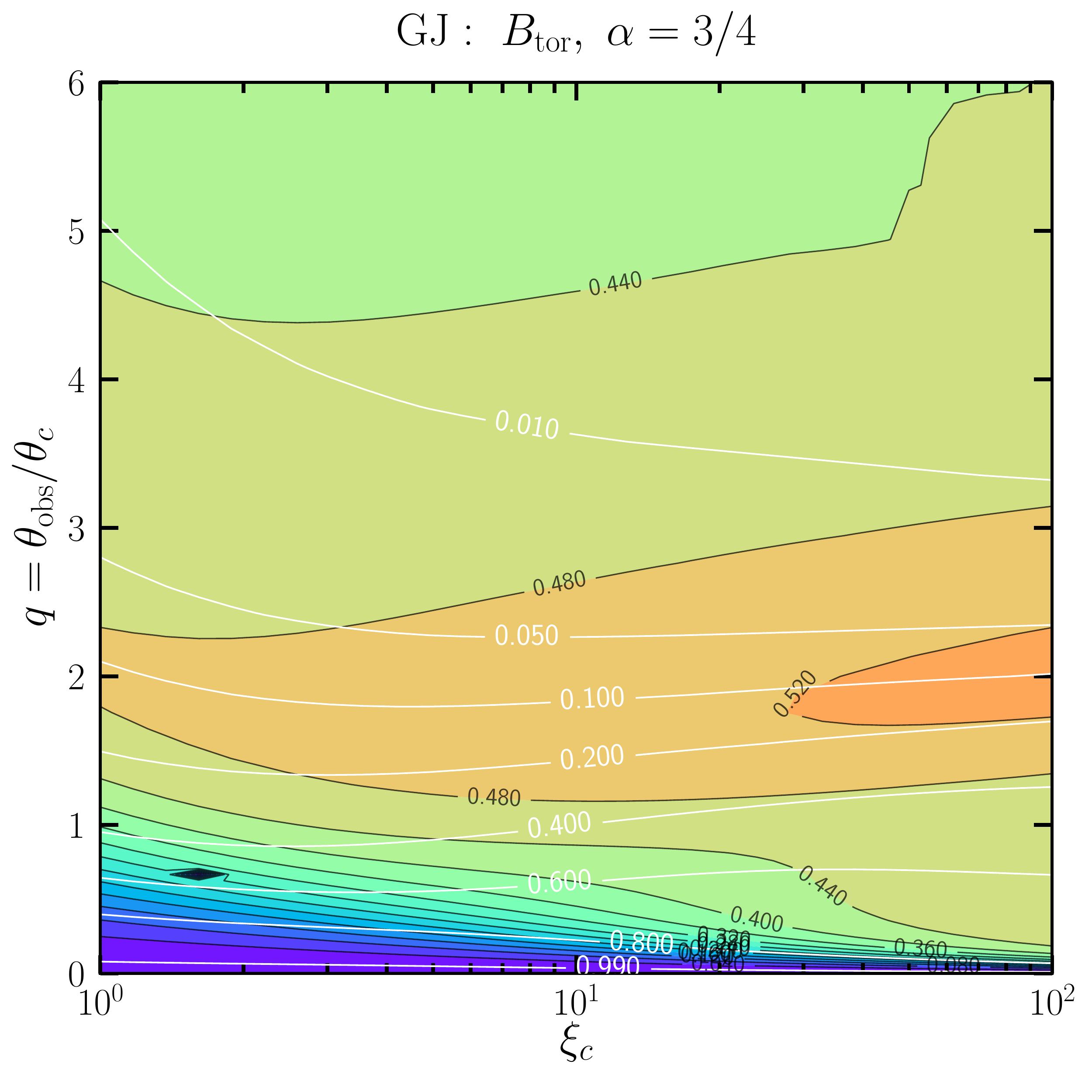}
    \caption{Contour plots of $\vert\Pi\vert$ for different magnetic field configurations and 
    different jet structures. The left column shows the degree of polarization for a smooth top-hat jet, where 
    the emissivity decays like a steep power law with smoothing parameter $\delta=30$. The center and right columns 
    correspond to structured jets with emissivity $L'_{\nu'}(\theta)$ and $\Gamma(\theta)$ both having a power 
    law ($a=b=2$; PLJ) and gaussian (GJ) profiles, respectively. The rows correspond to the different magnetic field configurations, 
    with (\textit{top}) a random field in the plane of the ejecta $B_\perp$, (\textit{middle}) an ordered field 
    in the direction parallel to the radial vector ($B_\parallel$), and ({\it bottom}) a globally ordered toroidal 
    field ($B_{\rm tor}$). Contours for different values of $\tilde f_{\rm iso}$ are plotted in white.}
    \label{fig:synchro-contour-plts-struc-jets}
\end{figure*}
%%%%%%%%%%%%%%%%%%%%%%%%%%%%%%%%%%%%%%%%%%%%%%%%%%%%%%

The degree of polarization for an off-axis observer in this case is obtained from Eq.~(\ref{eq:Pi-off-axis}), 
where $\Lambda(\tilde\xi)=\Lambda_\perp(\tilde\xi)$ from Eq.~(\ref{eq:sin-chi}). For $\epsilon=2$, we find 
from Eq.~(\ref{eq:Pi_rnd}) that the local polarization from a given magnetic field element of $B_\perp$ is 
(in the limit $b\to0$) $\Pi'(\tilde\xi)/\Pi_{\rm max}=-2\tilde\xi/(1+\tilde\xi^2)$. In the general case, when 
$\epsilon\neq2$, and for a random field that is in the plane transverse to the local velocity vector ($B_\perp$), 
the total polarization arising from a given fluid element has to 
be averaged over the various orientations of the magnetic field, which yields \citep[using Eq.~(1) of][]{Sari99}
\begin{equation}\label{eq:Eq1-Sari99}
    \frac{\Pi_\perp'(\tilde\xi)}{\Pi_{\rm max}} = 
    \frac{\displaystyle\frac{1}{\pi}\int_0^\pi\cos(2\theta_p)
    \Lambda_\perp(\tilde\xi,\varphi_B)d\varphi_B}
    {\displaystyle\frac{1}{\pi}\int_0^\pi\Lambda_\perp(\tilde\xi,\varphi_B)d\varphi_B}~,
\end{equation}
where
\begin{eqnarray}
    & \theta_p = \displaystyle\arctan\left[\left(\frac{1-\tilde\xi}{1+\tilde\xi}\right)\cot\varphi_B\right]~, & \\
    & \cos(2\theta_p) = 
    \displaystyle\left[\sin^2\varphi_B-\fracb{1-\tilde\xi}{1+\tilde\xi}^2\cos^2\varphi_B\right]
    \left[\displaystyle 1-\frac{4\tilde\xi\cos^2\varphi_B}{(1+\tilde\xi)^2}\right]^{-1}~,\quad &
\end{eqnarray}
and $\varphi_B$ is measured from some reference direction to carry out the averaging. 
Plugging in the expression for $\cos(2\theta_p)$ into eq.~(\ref{eq:Eq1-Sari99}) finally 
yields
% \footnote{Here we corrected a typo in Eq.~(3) of \cite{Granot03} 
% that is missing a power of $(\epsilon-2)/2$ in the numerator, however the final results are 
% correct.} 
\citep{Granot03}
% \begin{equation}
%     \frac{-\Pi_\perp(\tilde\xi)}{\Pi_{\rm max}} = \frac{\displaystyle\int_0^\pi d\tilde\varphi_B\left[1-\frac{4\tilde\xi\cos^2\tilde\varphi_B}{(1+\tilde\xi)^2}\right]^{(\epsilon-2)/2}
%     \left[\fracb{1-\tilde\xi}{1+\tilde\xi}^2\cos^2\tilde\varphi_B-\sin^2\tilde\varphi_B\right]}
% {\displaystyle\int_0^\pi d\tilde\varphi_B\left[1-\frac{4\tilde\xi\cos^2\tilde\varphi_B}{(1+\tilde\xi)^2}\right]^{\epsilon/2}}~.
% \end{equation}
\begin{eqnarray}
    &&\displaystyle\frac{\Pi_\perp'(\tilde\xi)}{\Pi_{\rm max}} 
    = \left\{\int_0^\pi d\varphi_B\left[1-\frac{4\tilde\xi\cos^2\varphi_B}{(1+\tilde\xi)^2}\right]^{\epsilon/2}
    \right\}^{-1} \\
    &&\times \displaystyle\int_0^\pi
    d\varphi_B\left[1-\frac{4\tilde\xi\cos^2\varphi_B}{(1+\tilde\xi)^2}\right]^{(\epsilon-2)\over2}
    \left[\sin^2\varphi_B-\fracb{1-\tilde\xi}{1+\tilde\xi}^2\cos^2\varphi_B\right]~. \nonumber
\end{eqnarray}

In the top panel of Fig.~\ref{fig:B-tor-rnd-pol}, we show the pulse-integrated degree of polarization for the 
random magnetic field scenario where the field lies entirely in the plane of the ejecta ($B_\perp$) 
for a top-hat jet. Similar to the $B_\parallel$ case, the PA changes direction by $90^\circ$ around $q=1$. Also, 
$\Pi$ now shows two distinct peaks at $q\sim1\pm\xi_j^{-1/2}$. If $\Pi<0$ ($\Pi>0$), 
then the polarization vector will lie along (normal to) the line connecting the LOS to the jet axis.

%%%%% DEGREE OF POLARIZATION Vs FLUENCE %%%%%%%%%%%%%%%%%%%%%%%%%%%%%%%%%%%%%%%%%%%%%%%%%%%%%%%%%%
\subsection{Degree of polarization Vs fluence}\label{sec:pol-fluence}
As mentioned earlier, in the case of a top-hat jet the fluence drops very rapidly for viewing angles outside of 
the sharp edges for which $q\equiv\theta_{\rm obs}/\theta_j>1$. This introduces a bias against distant off-axis 
GRBs due to the flux limitations of the detector; all high redshift GRBs that are observed during the prompt phase 
are observed within the jet aperture ($\theta_{\rm obs}\lesssim\theta_j+1/\Gamma\leftrightarrow q\lesssim1+\xi_j^{-1/2}$). 
Such a limitation also introduces a bias against measuring high degrees of polarization in the 
prompt phase from distant off-axis GRBs for a given magnetic field configuration. For example, both $B_\parallel$ and 
$B_\perp$ field configurations suffer from this bias since $\Pi$ rises significantly when $q>1$ as compared to 
its value when $q<1$.

Consider a pulse or emission episode that originated from an emission region with LF $\Gamma$ or equivalently with 
$\xi_j$ for a fixed $\theta_j$, and observed at a viewing angle $\theta_{\rm obs}$ or equivalently at some $q$. The 
fluence $S$ of the pulse can be straightforwardly obtained from the flux density defined in Eq.~(\ref{eq:dFnu}), 
where $S_\gamma = \int dt_{\rm obs}\int_{\nu_1}^{\nu_2}d\nu F_\nu(t_{\rm obs})$. This can be further used to write the 
isotropic equivalent energy $E_{\gamma,\rm iso} = 4\pi d_L^2(1+z)^{-1}S_\gamma$. Here for simplicity we assume a power 
law spectrum within the whole observed spectral range. A useful parameter to gauge the suppression in fluence 
for an off-axis observer is the ratio of the off-axis to on-axis fluence or equivalently the ratio of the off-axis to 
on-axis isotropic equivalent energies,
\begin{equation}
    \tilde f_{\rm iso} \equiv \frac{E_{\gamma,\rm iso}(q,\xi_j)}{E_{\gamma,\rm iso}(0,\xi_j)} 
    = \frac{\int_0^{\tilde\xi_{\max}}\int_0^{2\pi}d\tilde\varphi
    \delta_D^{2+\alpha}\Gamma(\theta)^{-1}\Lambda(\tilde\xi,\tilde\varphi)\mathcal L(\theta)}
    {\left[\int_0^{\tilde\xi_{\max}}\int_0^{2\pi}d\tilde\varphi
    \delta_D^{2+\alpha}\Gamma(\theta)^{-1}\Lambda(\tilde\xi,\tilde\varphi)\mathcal L(\theta)\right]_{q=0}}
\end{equation}
where the expression on the r.h.s is general and applies to any jet structure 
\citep{Granot+02,Yamazaki+03,Eichler-Levinson-04,GR11,Salafia+15,Beniamini-Nakar-18}, including a top-hat jet, and 
synchrotron emission with any magnetic field configuration as well as Compton drag. The structure of 
the jet is encoded in the dependence of the LF $\Gamma(\theta)$ and the emissivity, through $\mathcal L(\theta)=L'_{\nu'}/L'_{\nu',0}$, 
on $\theta=\theta(q,\theta_{\{j,c\}}\tilde\theta,\tilde\varphi)$ (see below and Appendix A). 
In Fig.~\ref{fig:fiso-q} we show the dependence of $\tilde f_{\rm iso}$ on $q$ for a given $\xi_{\{j,c\}}$ and 
for different jet structures, 
such as a top-hat jet, smoothed top-hat jet, and structured jets -- power law and gaussian jets -- that are discussed 
below in \S\ref{sec:struc-jets}. 
For a top-hat jet $\tilde f_{\rm iso}$ drops very sharply for $q\gtrsim1$, while in the case of a structured 
jet it decays more gradually, since the fluence is dominated by contribution from along the LOS rather than that 
from within the jet's core which is strongly suppressed at large viewing angles. Fig.~\ref{fig:synchro-contour-plts} 
shows contour plots of the degree of polarization arising in synchrotron 
emission for the different magnetic field configurations. In the left panel, we show contours of $\vert\Pi\vert$ and 
$\tilde f_{\rm iso}$ (shown with white contours) over the $q$ and $\xi_j$ parameter space with fixed $\alpha$. In the 
right panel, the same is shown over the $q$ and $\alpha$ parameter space while keeping $\xi_j$ fixed.

%%%%% POLARIZATION FROM STRUCTURED JETS %%%%%%%%%%%%%%%%%%%%%%%%%%%%%%%%%%%%%%%%%%%%%%%%%%%%%%%%%%%%%%%%%%%%%%
\subsection{Polarization from structured jets}\label{sec:struc-jets}
\subsubsection{Top-hat jet with smooth edges}
The notion that relativistic jets have sharp edges, e.g. the top-hat jet model, is highly idealized. It is conceivable 
that the emissivity does not fall sharply beyond some uniformly emitting core with angular size $\theta_j$, 
but instead it declines more gradually. Here we follow the discussion of \citet{NPW03} 
and present two models of a \textit{smooth top-hat jet}, that has a uniformly bright core with smoothly decaying wings:
\begin{enumerate}
\item \textit{Exponential wings} - the emission falls off exponentially outside of the uniform core, such that
\begin{eqnarray}
    \frac{L'_{\nu'}}{L'_{\nu',0}} = \left\{\begin{array}{cc}
        1 & \xi \leq \xi_j~, \\
        & \\
        \exp[(\sqrt{\xi_j}-\sqrt{\xi})/\Delta] & \xi > \xi_j~,
    \end{array}\right.
\end{eqnarray}
where $L'_{\nu',0}$ is the uniform spectral luminosity. 
\item \textit{Power-law wings} - the emission declines as 
a power law outside of the uniform core, such that
\begin{eqnarray}\label{eq:THJ-pwrl-wings}
    \frac{L'_{\nu'}}{L'_{\nu',0}} = \left\{\begin{array}{cc}
        1 & \xi \leq \xi_j~, \\
        & \\
        \displaystyle\fracb{\xi}{\xi_j}^{-\delta/2} & \xi > \xi_j~.
    \end{array}\right.
\end{eqnarray}
\end{enumerate}
In both cases, only the spectral luminosity is allowed to vary with $\theta$, but the dynamics remain angle independent, such that 
$\Gamma(\theta)=\Gamma_0$.

In the left panel of Fig.~\ref{fig:pol-struc-jets}, we show the degree of polarization for different magnetic field 
configurations and for the two models with exponential and power-law wings. In both cases, it is clear 
that a sharp drop in the emissivity outside of the uniformly bright core is needed to obtain a high level of 
polarization for the $B_\perp$ and $B_\parallel$ magnetic field scenarios \citep{NPW03}. However, an opposite 
trend is seen for the $B_{\rm tor}$ magnetic field case, where jets with a shallow gradients show high levels of 
polarization when $q>1$.

%%%%%%%%%  STRUCTURED JETS  %%%%%%%%%%%%%%%%%%%%%
\subsubsection{Structured jets}
In a truly structured jet the bulk LF of the emitting region must also vary with $\theta$ away from the jet 
symmetry axis. Here we consider two popular models \citep{ZM02,KG03,Granot-Kumar-03,Rossi+02,Rossi+04}:
\begin{enumerate}
    \item \textit{Gaussian Jet (GJ)}: Both the spectral luminosity and the kinetic energy of the emitting material per unit rest mass, 
    $\Gamma-1$, have a gaussian profile with a characteristic core angle $\theta_c$:
    \begin{equation}\label{eq:GJ}
        \frac{L'_{\nu'}}{L'_{\nu',0}}=\frac{\Gamma(\theta)-1}{\Gamma_c-1}
        =\max\left[\exp\left(-\frac{\theta^2}{2\theta_c^2}\right)~,~\exp\left(-\frac{\theta_*^2}{2\theta_c^2}\right)\right]~,
    \end{equation}
    where $\Gamma_c$ is the LF of the core and $\theta_*$ implies a floor, which corresponds to some finite $\beta_{\rm min}$, that is both 
    physically motivated and numerically convenient, and is chosen to be sufficiently small so that it does not affect any of the results. 
    \item \textit{Power-law Jet (PLJ)}: The spectral luminosity and the kinetic energy per unit rest mass of the emitting material decay 
    as a power law outside of the core:
    \begin{equation}\label{eq:PLJ}
        \frac{L'_{\nu'}}{L'_{\nu',0}} = \Theta^{-a}~,\quad\frac{\Gamma(\theta)-1}{\Gamma_c-1} = \Theta^{-b}~,
        \quad\quad\Theta \equiv \sqrt{1+\displaystyle\fracb{\theta}{\theta_c}^2}
    \end{equation}
\end{enumerate}

We calculate the degree of polarization for a structured jet by numerically integrating the general 
expressions that are presented in Appendix~\ref{sec:app-time-integrated-pol-strucjet}. In doing so we 
make the explicit assumption that the comoving spectral luminosity as well as the spectrum remain constant 
with shell radius $r$ as it expands. In addition, we assume that the spectrum does not depend on 
the polar angle $\theta$. The results of 
the integration are shown in the right-panels of Fig.~\ref{fig:pol-struc-jets}. To obtain high levels 
of polarization when the magnetic field configuration is that of $B_\perp$ or $B_\parallel$, sharp gradients 
in $\Gamma$ outside of an approximately uniform core are needed. However, the toroidal field case again 
shows an opposite trend where sharp gradients yield slightly lower levels of polarization. For a top-hat 
jet the fluence drops very rapidly outside of the uniform core, however, in a structured jet the observer 
has access to angular regions that are well outside the core with $q\gtrsim2$. This is demonstrated in 
the right-panels of Fig.~\ref{fig:pol-struc-jets} with the use of a dotted line for which $\tilde f_{\rm iso}<10^{-2}$.
In Fig.~\ref{fig:synchro-contour-plts-struc-jets} we show contours of $\vert\Pi\vert$ and $\tilde f_{\rm iso}$ (shown 
in white) as a function of $q$ and $\xi_j$ or $\xi_c$ for synchrotron emission and for different magnetic 
field configurations and jet structures.

%%%%%% FIGURE  %%%%%%%%%%%%%%%%%%
\begin{figure}
    \centering
    \includegraphics[width=0.48\textwidth]{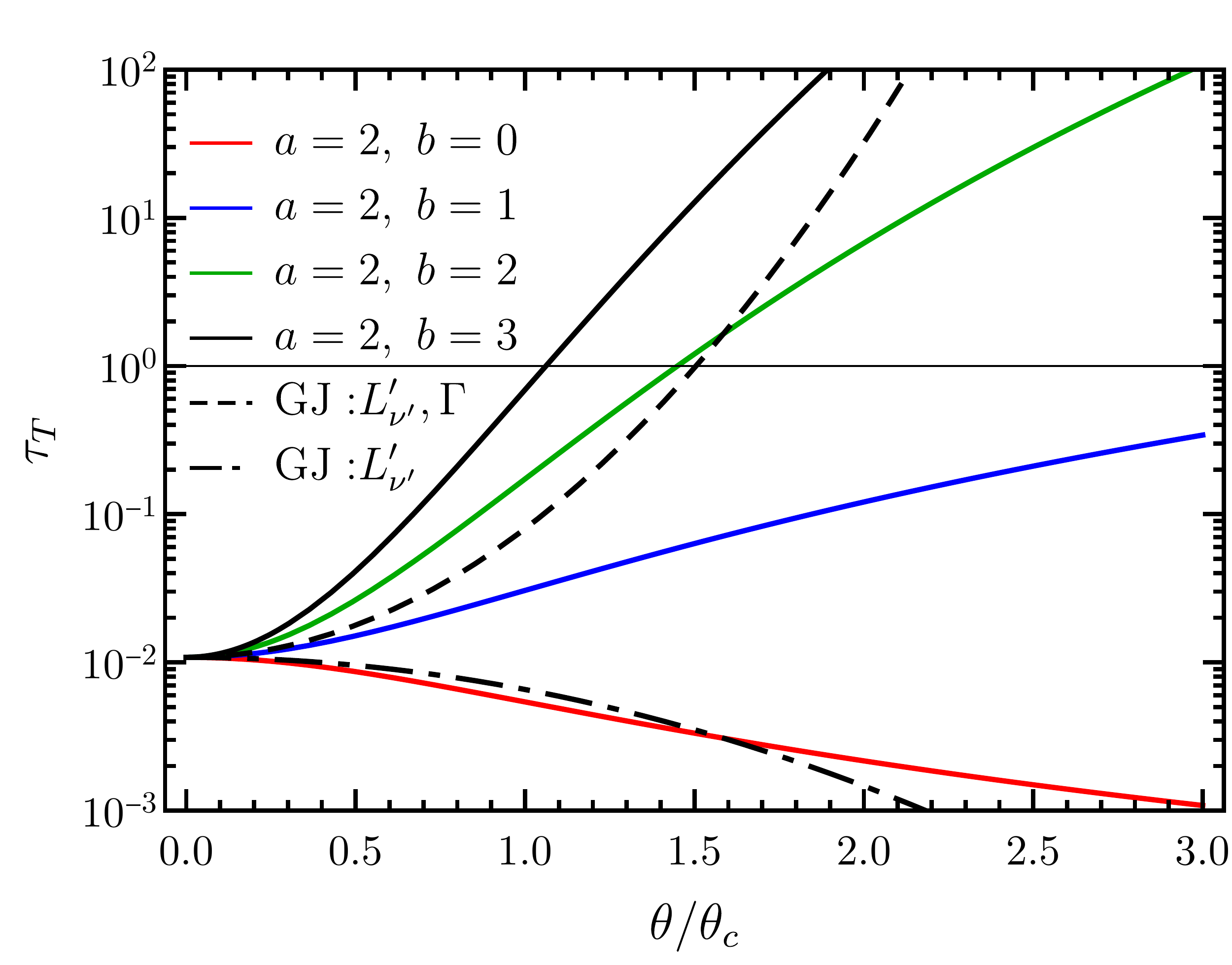}
    \caption{Thomson optical of $e^-e^+$-pairs produced (ignoring pair-annihilation) due to 
    $\gamma\gamma$-annihilation of $\gamma$-ray photons when $\Gamma$ declines with polar angle 
    from the jet symmetry axis. For a sufficiently steep angular profile for $\Gamma$, 
    the prompt emission will be highly suppressed at $q=\theta_{\rm obs}/\theta_c\gtrsim2$. 
    See caption of Fig.~\ref{fig:fiso-q} for legend labels.}
    \label{fig:pair-opacity}
\end{figure}
%%%%%%%%%%%%%%%%%%%%%%%%%%%%%%%%%

%%%%%%%%%%% COMPACTNESS LIMITATION ON Q IN STRUCTURED JETS %%%%%%%%%%%%%%%%%%%
\subsubsection{Compactness limitation on $q$ in structured jets}\label{sec:compactness}
In the case where the LF is not uniform and decreases away from the jet symmetry axis, the angular scale out to which the prompt 
emission can be observed is limited by compactness. For low values of $\Gamma$, the flow becomes optically thick to $\gamma\gamma$-
annihilation and results in the production of $e^-e^+$-pairs, which suppresses the emission of $\gamma$-ray photons. Here we 
consider an outflow carrying an isotropic power $L_{k,\rm iso}=4\pi (dL_k/d\Omega)=4\pi L_{k,\Omega}(\theta)$, 
where for a structured jet $L_{k,\Omega}(\theta)$ follows the angular distribution of the emissivity as discussed 
above for the two kinds of structured jets. The radiated power measured by a distant observer is related to the kinetic power by an 
efficiency factor $\epsilon_\gamma$, such that
\begin{equation}
    \epsilon_\gamma L_{k,\rm iso} = L_{\gamma,\rm iso} = \frac{16\pi}{3}r^2\Gamma^2cU_\gamma'~,
\end{equation}
where $U_\gamma'$ is the comoving energy density of the radiation field which is assumed to be isotropic in the comoving frame, and 
for which the lab-frame energy density is $U_\gamma = (4/3)\Gamma^2U'_\gamma$. The compactness of the radiation field is given by
\begin{equation}
    \ell_\gamma' = \sigma_T\frac{U_\gamma'}{m_ec^2}\frac{r}{\Gamma} = f_{\gamma\gamma}^{-1}\tau_T
\end{equation}
such that a fraction $f_{\gamma\gamma}$ of the total number of photons, that are above the minimum self-annihilation energy of 
$m_ec^2$ in the comoving frame, contribute a Thomson optical depth $\tau_T = \sigma_Tn_\gamma'r/\Gamma$. Here $\sigma_T$ is the 
Thomson cross-section and $n_\gamma'$ is the comoving photon number density. We further make the assumption that the dissipation 
radius is given by $r=2\Gamma^2ct_{v,z}$, where $t_{v,z}$ is the variability timescale of the burst in the cosmological rest-frame 
of the source, which finally yields
\begin{eqnarray}\label{eq:compactness-tauT}
    \tau_T&&\approx\epsilon_\gamma f_{\gamma\gamma}\frac{3\sigma_T}{8m_ec^4}\frac{L_k(\theta)}{\Gamma^5(\theta)t_{v,z}} \\
    &&\approx 10^{-2}\left(\frac{\epsilon_\gamma f_{\gamma\gamma}}{10^{-1}}\right)\kappa(\theta)L_{k,c,51}\Gamma_{2.7}^{-5}
    t_{v,z,-1}^{-1}~,
\end{eqnarray}
where $\kappa(\theta) = [L_k(\theta)/L_{k,c}][\Gamma(\theta)/\Gamma_c]^{-5}$ includes the angular dependence of $\tau_{\gamma\gamma}$, 
and $L_{k,c}$ and $\Gamma_c$ are the values of the respective distributions in the core ($\theta=0$). 
In Fig.~\ref{fig:pair-opacity}, we show the Thomson optical depth due to $\gamma\gamma$-annihilation as the emission region becomes 
more compact when $\Gamma$ declines away from the jet symmetry axis. For a sufficiently steep angular profile for $\Gamma$, prompt 
emission is only observed from regions with $q=\theta_{\rm obs}/\theta_c\lesssim2$ 
\citep[also see, e.g.][]{Beniamini-Nakar-18,Matsumoto+19}.

For LOSs that are significantly outside of the core, at $q\gtrsim2$, the compactness of the emitting region 
becomes a concern and it ultimately restricts observable emission to regions that are not too far outside of 
the bright core. This is demonstrated in the right column of Fig.~\ref{fig:pol-struc-jets}, where a filled circle 
is plotted on top of the polarization curves at which $q$ value $\tau_T=10$. Here we have assumed the 
same fiducial values for the parameters as in Eq.~(\ref{eq:compactness-tauT}).

The compactness estimate does not account for $e^+e^-$-pair annihilation which will relax the pair opacity constraint by 
reducing the Thomson optical depth by factors of a few for $\Gamma\gtrsim200$ and much more severely for more compact regions 
with $\Gamma\lesssim200$ \citep[see, e.g., the top panel of Fig. 3 in][]{Gill-Granot-18}. 
In addition, it makes the simplifying assumption of an isotropic comoving
radiation field and further adopts the ``one-zone'' approximation. Both of these assumptions may not be strictly valid and 
effects due to the spatial, temporal, and angular dependence of the radiation field can be important. A proper treatment of these 
effects can lead to a reduction by a factor $\sim2$ in the minimum $\Gamma$, below which the emission region has $\tau_{\gamma\gamma}>1$ 
\citep[see, e.g.,][]{GCD08,Hascoet+12}, permitting slightly larger $q$ values.

%%%%%% FIGURE  %%%%%%%%%%%%%%%%%%
\begin{figure}
    \centering
    \includegraphics[width=0.25\textwidth]{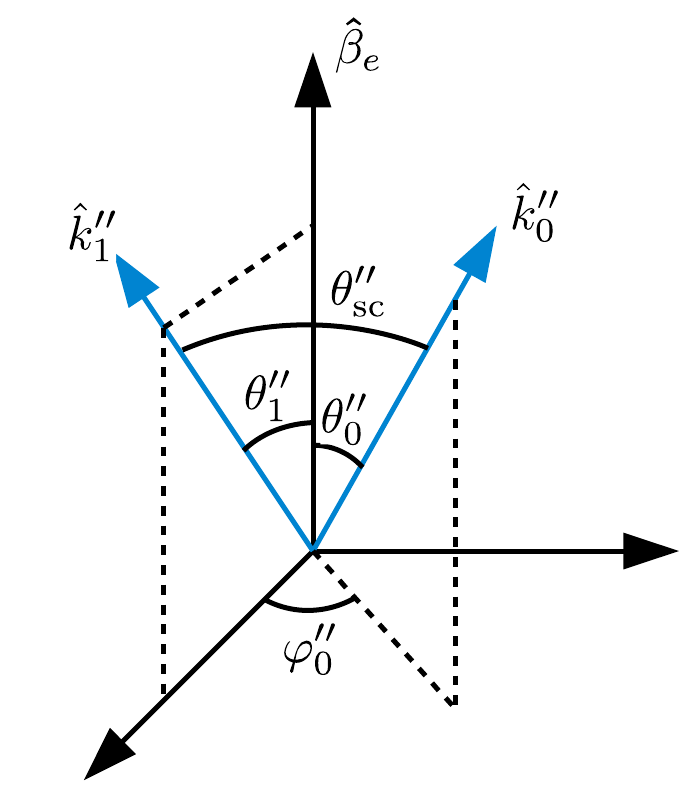}\hspace{2cm}
    \caption{Illustration of the geometry in the Compton drag model showing the directions of incoming ($\hat k_0''$) 
    and scattered ($\hat k_1''$) photons in the electron's rest frame (ERF), which is moving with velocity $\vec\beta_e'c$ 
    in the comoving frame of the outflow.}
    \label{fig:CD-geom}
\end{figure}
%%%%%%%%%%%%%%%%%%%%%%%%%%%%%%%%%

%%%%% COMPTON DRAG  %%%%%%%%%%%%%%%%%%%%%%%%%%%%%%%%%%%%%
\section{Compton Drag}\label{sec:CD}
Another radiative mechanism that can yield a high degree of linear polarization is 
inverse-Compton scattering (ICS) of softer photons by relativistic electrons. 
In this model, the electrons are assumed to be cold and the bulk LF of the outflow relative 
to the external radiation field, that is (at least roughly) isotropic in the lab-frame, is what 
causes the upscattering. This mechanism 
has been invoked not only to explain the high level of polarization ($\Pi = 80\%\pm20\%$) 
that was observed in GRB 021206 \citep{CB03}, but also to explain the non-thermal 
spectrum of GRBs in general \citep[e.g.][]{GC99,Lazzati+00,Giannios06,LB06}. 
Earlier works have discussed the potential of observing 
polarized emission via ICS in the context of electrons in the relativistic jet 
upscattering circumburst radiation fields emanating from e.g. the accretion disk 
\citep{SD95}, and in the context of a relativistic baryon-pure jet that is enveloped 
by slowly moving baryon-rich material. In the latter case, the shocked transition layer between the 
two media scatters photospheric photons and yields high levels of polarization 
under certain conditions \citep{EL03}. A proper treatment where the degree of polarization 
from Compton drag is obtained by averaging over the GRB image on the plane of the sky, 
which is different from the point source approximation adopted by earlier works, 
was presented by \citet{Lazzati+04}.

\subsection{Polarized emission due to inverse-Compton scattering: General treatment}
\label{sec:IC-general}
Relativistic electrons with energies $\gamma_em_ec^2$ propagating through a radiation field are slowed down by Compton 
scattering the soft seed photons \citep[see for e.g.][for a detailed exposition in the context of AGN jets]{BS87}. In the process, the energy of 
the incoming seed photon (in units of $m_ec^2$) $\varepsilon_0'=E_\gamma'/m_ec^2$ is increased on average to $\varepsilon_1'=(4/3)\gamma_e^2\varepsilon_0'$ 
after scattering. In the rest frame of the electron (all quantities in this frame are double-primed), the incoming photon 
has energy $\varepsilon_0''\sim\gamma_e\varepsilon_0'$, and if $\varepsilon_0'' \ll 1$ then the scattering is referred to as
\textit{coherent} or \textit{elastic} and the scattering cross-section is given by the Thomson cross-section $\sigma_T$. 
In this case, $\varepsilon_1''=\varepsilon_0''$ and the scattered radiation is polarized where the degree of polarization 
depends on the scattering angle $\theta_{\rm sc}'' = \arccos(\hat k_0''\cdot\hat k_1'')$, where $\hat k_0''$ and $\hat k_1''$ 
are the unit wave vectors of the incoming and scattered photons, respectively (see Fig.~\ref{fig:CD-geom}). In this case, 
the local degree of polarization imparted to the outgoing photon is \citep[][]{Rybicki-Lightman-79}
\begin{equation}\label{eq:Pi-head-on}
    \Pi'(\theta_{\rm sc}'') = \frac{1-\cos^2\theta_{\rm sc}''}{1+\cos^2\theta_{\rm sc}''}~.
\end{equation}
In general, $\Pi'$ is sensitive to the angle ($\theta_0''$) between the direction of the incoming photon and velocity vector 
of the electron, and the direction of the scattered photon. If the plasma is relativistically hot then the degree of 
polarization is obtained by integrating over all $\theta_0''$. For simplicity, we consider an isotropic radiation field with 
specific intensity $I'_{\nu'}(\nu')$ through which the electron with velocity $\vec\beta_e'c$ is propagating. In its rest 
frame, the electron sees an almost unidirectional radiation field with intensity 
\begin{equation}
    I''_{\nu''}(\nu'') = \delta_{D,e}^3I'_{\nu'}(\nu')\quad{\rm with}\quad\delta_{D,e} = [\gamma_e(1+\beta_e'\mu_0'')]^{-1}
\end{equation}
where $\delta_{D,e}$ is the Doppler factor associated to the electron's motion, 
$\mu_0'' \equiv \cos\theta_0''$, and $\nu'' = \delta_{D,e}\nu'$. The Stokes parameters can be expressed in the same way as 
before, such that
\begin{equation}
    \left\{\begin{array}{c}
        U/I \\
        Q/I 
    \end{array}\right\}
    =\frac{\displaystyle
    \int d\Omega_0''\delta_{D,e}^3I'_{\nu'}(\nu''/\delta_{D,e})\Pi'(\theta_{\rm sc}'')
    \left\{\begin{array}{c}
        \sin 2\theta_p'' \\
        \cos 2\theta_p'' 
    \end{array}\right\}
    }{\displaystyle
    \int d\Omega_0''\delta_{D,e}^3I'_{\nu'}(\nu''/\delta_{D,e})
    }
\end{equation}
where the solid-angle $d\Omega_0'' = d\mu_0''d\varphi_0''$. The polarization angle $\theta_p''$ in the electron rest frame 
(ERF; see Fig.~\ref{fig:CD-geom}) is obtained by first projecting the vectors $\vec\beta_e''$ and $\hat k_0''$ 
on the plane orthogonal to $\hat k_1''$ and then calculating the angle between the two. The scattering and 
polarization angles can be expressed in terms of the direction of the incoming photon ($\theta_0'',\varphi_0''$) 
and the angle ($\theta_1''$) between the scattered photon and electron's velocity vector
\begin{eqnarray}
    \mu_{\rm sc}'' &&= \mu_0''\mu_1''+\sqrt{(1-{\mu_0''}^2)(1-{\mu_1''}^2)}\cos\varphi_0'' \\
    \cos\theta_p'' &&= \frac{\mu_0''-\mu_1''\mu_{\rm sc}''}{\sqrt{(1-{\mu_1''}^2)(1-{\mu_{\rm sc}''}^2)}}~,
\end{eqnarray}
where $\mu_{\rm sc}'' \equiv \cos\theta_{\rm sc}''$ and $\mu_1'' \equiv \cos\theta_1''$. 
The polarization vector is in the direction of $\hat{e}''=(\hat k_0''\times \hat k_1'')/|\hat k_0''\times \hat k_1''|$, i.e. normal to the two wave vectors.
The Stokes parameters calculated in the comoving frame 
of the outflow heretofore apply to a single electron with Lorentz factor $\gamma_e$. To obtain the degree of polarization in the observer frame, 
the Stokes parameters have to be averaged over the velocity distribution of all electrons in the emission region. When the electron velocity is 
ultra-relativistic ($\gamma_e\gg 1,~\beta_e\simeq 1$), the radiation in the electron's rest frame is almost perfectly unidirectional and the 
``head-on'' approximation ($\mu_0''=-1$) applies \citep[][]{BS87}. In this case, the degree of polarization is simply given by Eq.~(\ref{eq:Pi-head-on}) with 
$\theta_{\rm sc}''\to\pi-\theta_1''$.

%%%%%%% FIGURE  %%%%%%%%%%%%%%%%%%%%%%%%%%%
\begin{figure}
    \centering
    \includegraphics[width=0.48\textwidth]{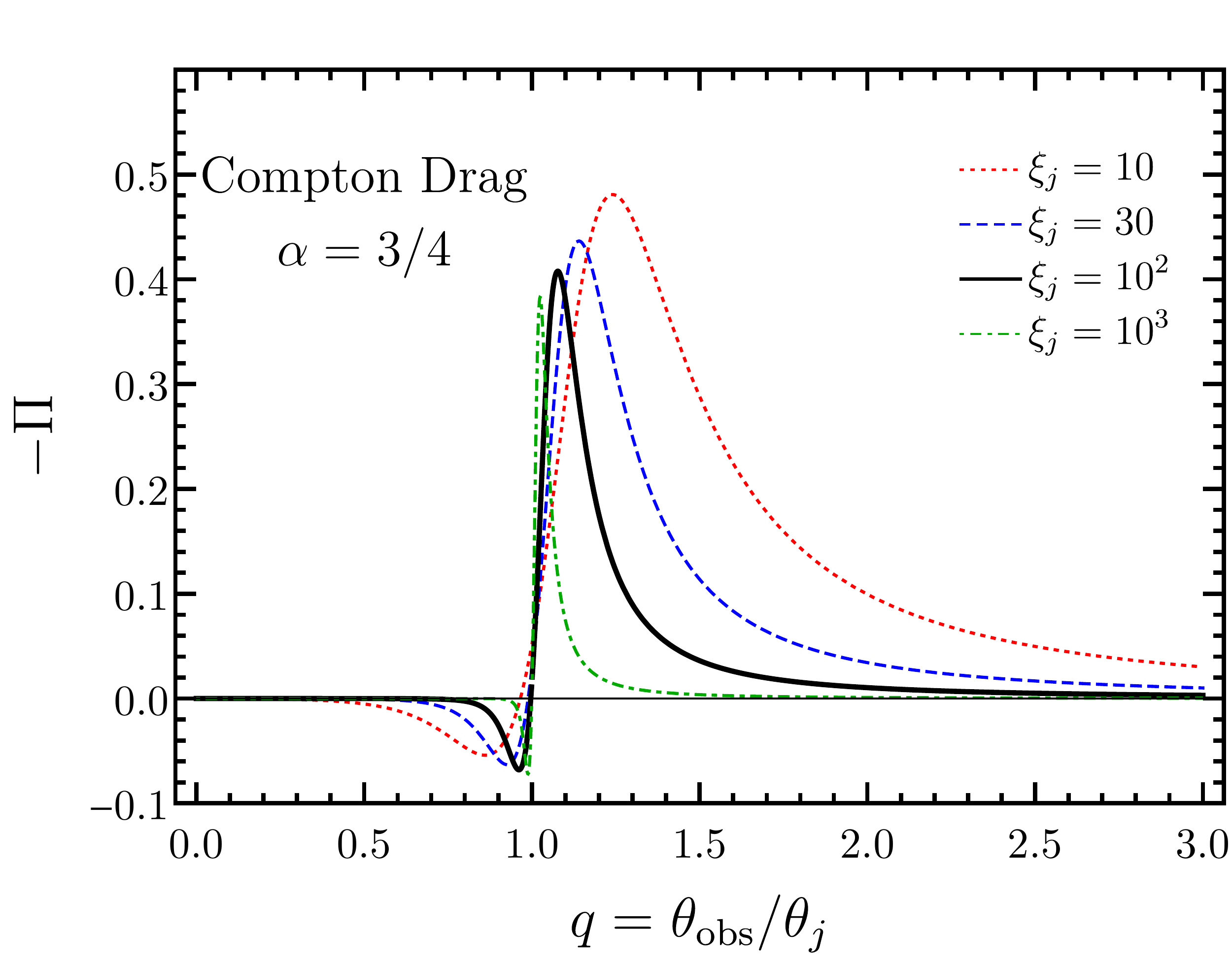}
    \includegraphics[width=0.48\textwidth]{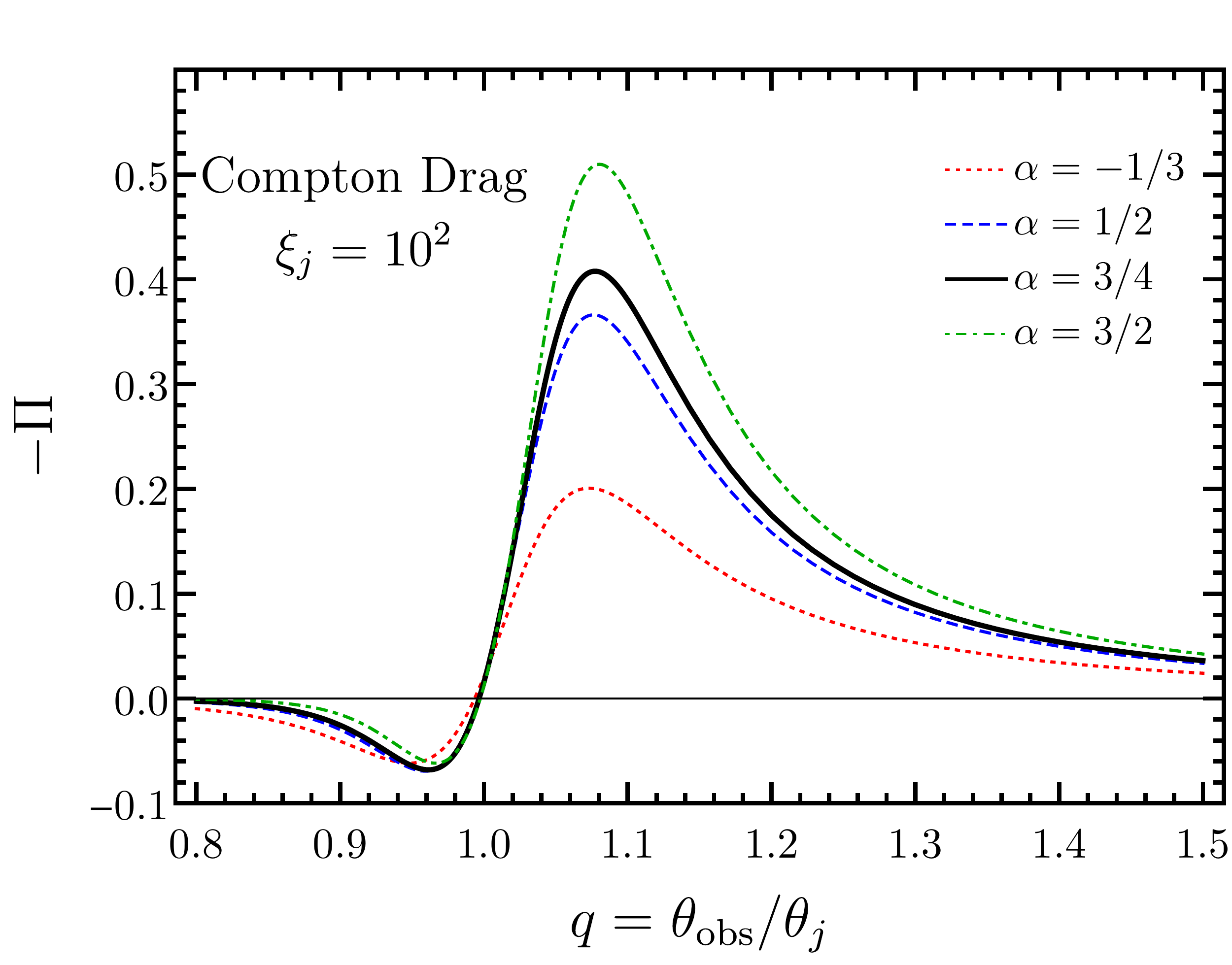}
    \includegraphics[width=0.48\textwidth]{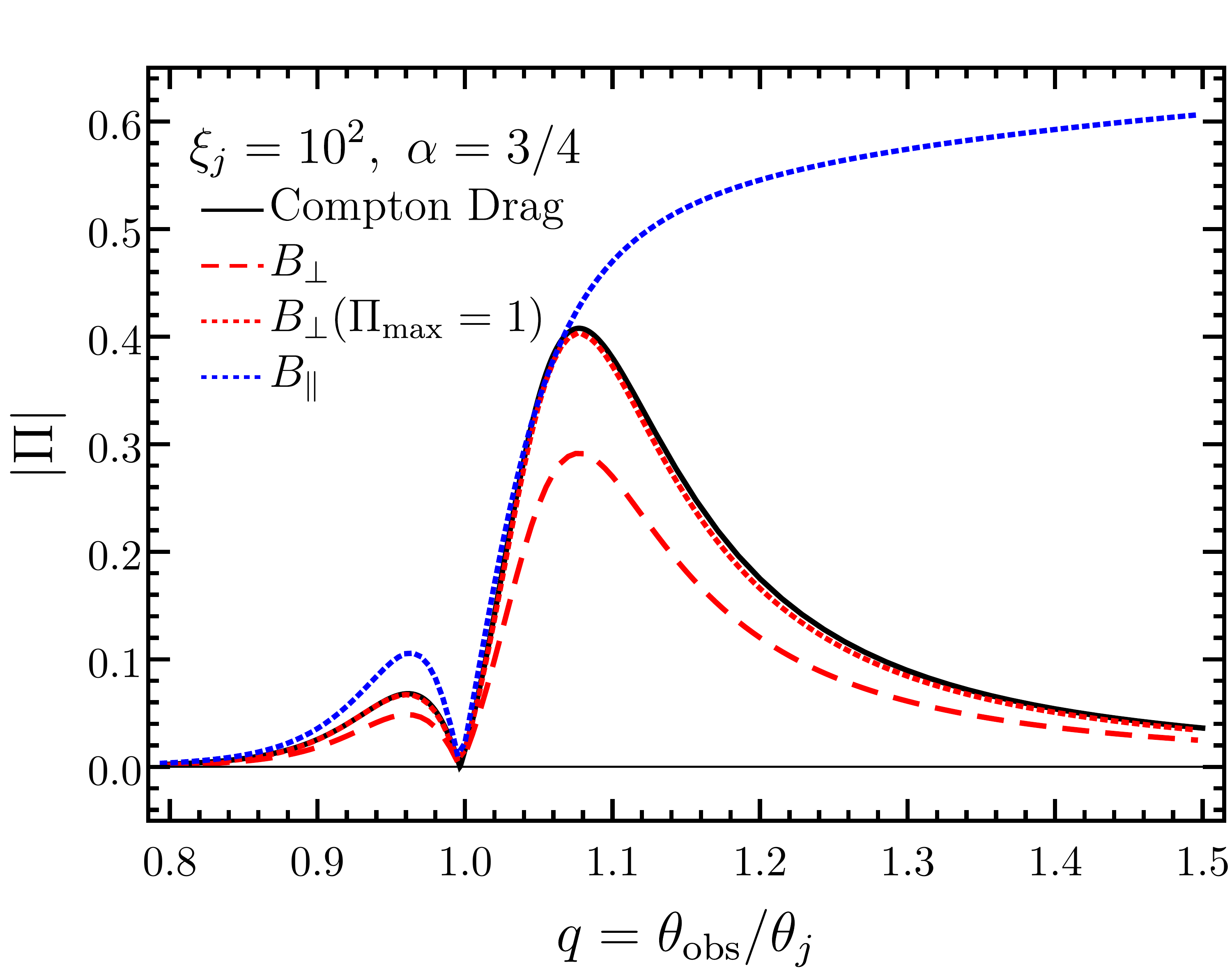}
    \caption{Degree of polarization for the Compton drag model shown as a function of viewing angle for a fixed 
    spectral index $\alpha$ and different $\xi_j = (\Gamma\theta_j)^2$ ({\it top}), and for different spectral 
    indexes and fixed $\xi_j$ ({\it middle}). In this model, cold electrons in the comoving frame of an ultra-relativistic 
    flow (top-hat jet with sharp edges) Thomson scatter unpolarized radiation \citep[also see e.g.][]{Lazzati+04,Toma+09}. 
    Also shown here ({\it bottom}) is the comparison of the degree of polarization expected in the Compton drag model to 
    that from synchrotron radiation with $B_\perp$ and $B_\parallel$ magnetic field structures.}
    \label{fig:CD-Pol}
\end{figure}
%%%%%%%%%%%%%%%%%%%%%%%%%%%%%%%%%%%%%%%%%%%%

%%%%%%  ULTRA-RELATIVISTIC JET WITH COLD ELECTRONS  %%%%%%%%%%%%%%%%%%%%%%%%%%%%%%%
\subsection{Polarized emission due to Compton-drag: Ultra-relativistic top-hat jet with cold electrons}\label{sec:CD-cold}
If the electron distribution is cold then the electrons are moving at the bulk velocity $\vec\beta_e=\vec\beta\simeq 1$ in 
the lab-frame. In this limit, the local degree of polarization is simply given by 
\begin{equation}\label{eq:Pi-local-CD}
 \Pi'(\tilde\mu') = \frac{1-\tilde\mu^{\prime 2}}{1+\tilde\mu^{\prime 2}}\approx\frac{2\tilde\xi}{1+\tilde\xi^2}~,
\end{equation}
where $\tilde\mu' = \cos\tilde\theta'$ and $\tilde\theta'$ is the polar angle of the observed photon in the comoving frame, and 
the last approximate expression is obtained for $\Gamma\gg1$, using Eq.~(\ref{eq:aberration}) for the aberration of light. To obtain the 
polarization in the observer frame to which multiple fluid elements contribute, we again perform an integration 
over the jet geometry. Due to symmetry reasons $U=0$ and $\Pi = \vert Q\vert/I$, where 
\begin{equation}\label{eq:QI-CD}
    \frac{Q}{I} = \frac{\displaystyle\int d\tilde\Omega~\delta_D^3I'_{\nu'}(\nu/\delta_D)\Pi'(\tilde\mu')\cos 2\theta_p}
    {\displaystyle\int d\tilde\Omega~\delta_D^3I'_{\nu'}(\nu/\delta_D)}~,
\end{equation}
and $\theta_p$ is always perpendicular to the plane containing the incoming and scattered photons, which means that 
$\theta_p = \tilde\varphi+\pi/2$ where both $\theta_p$ and $\tilde\varphi$ are measured from the projection of the jet 
axis on the plane of the sky. As a result, if the jet is uniform averaging the polarization over the entire image will 
yield no net polarization. Therefore, the jet must be viewed off-axis to detect any polarization. We employ the same 
methodology here to calculate the observed degree of polarization as was used for the case of synchrotron emission due 
to random magnetic fields and where the jet was viewed off-axis. This can be calculated using 
Eq.~(\ref{eq:Pi-off-axis}, \ref{eq:Pi-head-on}, \ref{eq:QI-CD}) with $\Lambda(\tilde\xi)=\Lambda_C(\tilde\xi)$. When the 
incoming radiation is completely unpolarized, the intensity of the scattered radiation varies with $\tilde\xi$ 
\citep[e.g.][]{Rybicki-Lightman-79}, such that
\begin{equation}
    \Lambda_C = \frac{1}{2}(1+\tilde\mu'^2)\approx\frac{1+\tilde\xi^2}{(1+\tilde\xi)^2}~.
\end{equation}
In the following, we assume that the incoming radiation field is unpolarized, which yields 
\begin{equation}
    \Pi=\frac{\displaystyle
    (2\pi)^{-1}
    \int_{\xi_-}^{\xi_+}d\tilde\xi\frac{\Lambda_C(\tilde\xi)}{(1+\tilde\xi)^{2+\alpha}}\frac{2\tilde\xi}{1+\tilde\xi^2}\sin2\psi(\tilde\xi)}
    {\displaystyle H(1-q)\int_0^{\xi_-}d\tilde\xi\frac{\Lambda_C(\tilde\xi)}{(1+\tilde\xi)^{2+\alpha}} + 
    \int_{\xi_-}^{\xi_+}d\tilde\xi\frac{[\pi-\psi(\tilde\xi)]\Lambda_C(\tilde\xi)}{\pi(1+\tilde\xi)^{2+\alpha}}}~.
\end{equation}

In the top two panels of Fig.~\ref{fig:CD-Pol}, we show the degree of polarization for the Compton drag model for 
different viewing angles while assuming a top-hat jet. It is very similar to the corresponding polarization curves 
for synchrotron emission from $B_\perp$, with a somewhat higher normalization, corresponding $\Pi_{\rm max}\to100\%$ 
for the synchrotron-$B_\perp$ model. This is nicely demonstrated by the dotted red line in the bottom panel of 
Fig.~\ref{fig:CD-Pol}, which is almost on top of the curve for Compton drag (solid black line). Therefore, the 
degree of polarization of the synchrotron-$B_\perp$ model is lower than that for Compton drag by a factor of 
$\approx\Pi_{\rm max}=(\alpha+1)/(\alpha+5/3)\sim 0.5-0.75$. We expect the same behavior to persist also for structured 
jets. In particular, we expect the Compton drag polarization from a structured jet to closely follow that for the 
synchrotron-$B_\perp$ model, which is shown in the top-right panel of Fig.~\ref{fig:pol-struc-jets} \citep[see also][]{Lazzati+04}, 
with a somewhat higher normalization, as described above.

%%%%% POLARIZATION OF PHOTOSPHERIC EMISSION  %%%%%%%%%%%%%%%%%%%%
\section{Photospheric Emission}\label{sec:Photospheric}
Photospheric emission from a hot and relativistically expanding fireball was first considered by \citet{Goodman1986} and 
\citet{Paczynski1986} while suggesting that GRBs are cosmological sources. 
% at a luminosity distance $d_L\sim10^{28}$~cm. 
% In that case, the typically measured flux of $F\sim10^{-6}~{\rm erg~cm}^{-2}~{\rm s}^{-1}$ would 
% require the outflow to carry an isotropic-equivalent luminosity 
% $L_{\gamma,{\rm iso}}=4\pi d_L^2F\sim10^{51}~{\rm erg~s}^{-1}$. With such a large luminosity, mostly 
% in the form of radiation, injected by the central engine into a compact region of size $r_0\lesssim10^7$~cm, 
% the outflow quickly becomes optically thick to pair-production ($\gamma\gamma\to e^-e^+$). Therefore, 
The flow starts as optically thick to scattering due to copious production of $e^\pm$-pairs and expands 
adiabatically under its own pressure. 
% at which point the LF of the fireball grows linearly with radius, 
% $\Gamma(r)\propto r$. Depending on the amount of baryons carried by the flow $\Gamma(r)$ saturates at 
% $r = r_s = \eta r_0$, where $\eta=L/\dot M_bc^2$ is the total energy per unit rest energy, $\dot M_b$ is 
% the mass flux of baryons and $L$ is the total jet power. At $r>r_s$ the fireball coasts at a constant 
% $\Gamma=\eta$ and becomes optically thin at the photospheric radius $r=r_{\rm ph}$ where the radiation 
% field decouples from matter. 
Initially, the LF of the expanding fireball grows linearly with radius, $\Gamma(r)\propto r$, until all of 
the initial energy is transferred to the kinetic energy of the entrained baryons. Beyond this point, the 
fireball coasts at a constant $\Gamma$ and becomes optically thin at the photospheric radius $r=r_{\rm ph}$, 
where the radiation field decouples from matter. A passively 
expanding fireball with no energy dissipation would only give rise to a quasi-thermal spectrum \citep{Beloborodov-10}, 
which does not agree with the typical non-thermal spectrum of the prompt GRB emission. Therefore, some form of dissipation is needed 
in the flow, both below the photosphere and above it. Photospheric emission in dissipative jets has been considered as another 
alternative to synchrotron radiation in many works for the underlying mechanism of the prompt emission 
\citep[e.g.][]{Thompson-94,EL00,MR00,RM05,Lazzati+09,PR11,Begue+13,Thompson-Gill-14,Gill-Thompson-14,Vurm-Beloborodov-16}.

It has been shown by \citet[][B11 hereafter]{Beloborodov2011} that prior to decoupling, the radiation field 
becomes significantly anisotropic in the comoving frame when the flow is matter dominated, such that $\rho'c^2\gg U_\gamma'$, 
where $\rho'$ and $U_\gamma'$ are the baryon rest mass density and the radiation field energy density, respectively, in the fluid's comoving rest frame. Because of the large anisotropy, the scattered radiation becomes linearly polarized at the photosphere, in a qualitatively similar manner as in Compton drag that was discussed in the previous section. 
On the other hand, if the flow is radiation dominated, the angular distribution of the 
radiation field is preserved as the flow becomes optically thin. 
Since the radiation field must be isotropic in the optically thick regions, it remains so after last scattering 
which produces no polarization.

Here we consider a matter-dominated outflow in the form of a spherical shell expanding relativistically with 
bulk LF $\Gamma\gg1$. For simplicity, we only discuss a passively expanding (non-dissipative) outflow that is 
carrying cold electrons (or $e^\pm$-pairs). We follow the treatment of B11 in writing down the spherically 
symmetric and frequency integrated equations of radiation transfer for the Stokes parameters in the comoving frame
\begin{eqnarray}
    \centering
    \label{eq:Iprime}
    &&\frac{\partial I'}{\partial\ln r} 
    = -(1-\mu'^2)g\frac{\partial I'}{\partial\mu'}-4(1-\mu'g)I'+\tau_T\frac{(S'-I')}{1+\mu'}~, \\
    &&\frac{\partial Q'}{\partial\ln r} 
    = -(1-\mu'^2)g\frac{\partial Q'}{\partial\mu'}-4(1-\mu'g)Q'+\tau_T\frac{(R'-Q')}{1+\mu'}~,
    \label{eq:Qprime}
    \quad\quad
\end{eqnarray}
where the degree of polarization is given by $\Pi=\vert Q'\vert/I'$. In the above equation, $S'$ and $R'$ are the source functions 
\citep[][]{Chandrasekhar1960,Sobolev1963}
\begin{eqnarray}
    && S'(\mu',r) = I'_0+\frac{3}{8}(3\mu'^2-1)\left(I_2'-\frac{I'_0}{3}+Q'_0-Q'_2\right) \\
    && R'(\mu',r) = \frac{9}{8}(1-\mu'^2)\left(I_2'-\frac{I_0'}{3}+Q_0'-Q_2'\right)~,
\end{eqnarray}
where 
\begin{equation}
    \{I_m'(r)~,~Q_m'(r)\} = \frac{1}{2}\int_{-1}^{1}\{I'(\mu',r)~,~Q'(\mu',r)\}\mu'^md\mu'    
\end{equation}
are the moments of total and polarized intensities. The quantity 
\begin{equation}
    g(r) = 1-\frac{d\ln\Gamma(r)}{d\ln r}
\end{equation}
expresses the acceleration profile of the flow; for a coasting flow $g=1$. In this case, the Thomson optical depth 
of a relativistically expanding outflow along a radial trajectory is \citep{Abramowicz+91}
\begin{equation}\label{eq:tauT-r}
    \tau_T(r) = \int_r^\infty n_e'(r)\sigma_T\Gamma(1-\beta)dr = 
    \frac{n_e'(r)\sigma_Tr}{2\Gamma}.
\end{equation}
The comoving volume of the outflow scales as $V'=4\pi r^2\Delta'\propto r^2$, 
where it has comoving width $\Delta'$. As a result, the number density of electrons scales as 
$n_e'\propto r^{-2}$ and therefore $\tau_T(r)\propto r^{-1}$. At the photospheric radius 
$\tau_T(r_{\rm ph}) \equiv 1$ and the Thomson optical depth can simply be expressed as
\begin{equation}
    \tau_T(r) = \frac{r_{\rm ph}}{r}~.
\end{equation}

Deeper in the flow, at $r\ll r_{\rm ph}$ where $\tau_T\gg1$, matter and radiation are tightly coupled 
via Compton scattering that causes the radiation field to be isotropic. The flow expands adiabatically under its 
own pressure where the radiation field loses energy to $PdV$ work, such that the comoving intensity declines over radius as
\begin{equation}\label{eq:Inorm}
    I'(r) = I'_{\rm ph}\fracb{r}{r_{\rm ph}}^{-8/3}~,
\end{equation}
where $I_{\rm ph}'$ is the normalization of the intensity at the photosphere. The above equation is 
strictly valid at $\tau_T\gtrsim10$ and it begins to break down near the photosphere where the radiation field becomes highly anisotropic (B11). However, the difference is of order unity, and therefore we will 
assume that the adiabatic cooling of the radiation field approximately applies all the way up to the photosphere.

The total isotropic equivalent power carried by the outflow is 
$L_{\rm tot,iso}=L_{\gamma,{\rm iso}}+L_{\rm k,iso}$, where $L_{\rm k,iso}$ is the kinetic power of the 
baryons and $L_{\gamma,{\rm iso}}$ is the luminosity of the radiation field which is given by
\begin{equation}\label{eq:L-gamma-iso}
    L_{\gamma,{\rm iso}}(r) = 4\pi r^2F = 4\pi r^2\int I(\mu,r)\mu d\Omega
    =16\pi^2r^2I_1~.
\end{equation}
The first moment of the lab-frame intensity $I_1$ can be expressed in terms of the comoving-frame 
quantities via Lorentz transformation, which gives (B11)
\begin{equation}\label{eq:I1}
    I_1 = \Gamma^2[\beta(I_0'+I_2')+(1+\beta^2)I_1'] \approx \frac{4}{3}\Gamma^2\beta I'~.
\end{equation}

Depending on the amount of baryons carried by the flow $\Gamma(r)$ saturates at 
$r = r_s = \eta r_0$, where $\eta=L/\dot M_bc^2$ is the total energy per unit rest energy, $\dot M_b$ is 
the mass flux of baryons, $L$ is the total jet power, and $r_0$ is the radius at which the flow was launched. 
For $r>r_s$ the radiation field provides no acceleration and the flow simply coasts at a constant $\Gamma=\eta$. At this point, 
the enthalpy density of the radiation field equals that of matter, $4e_\gamma'/3=e_m'+p_m'$. Here, $e_\gamma'$ 
and $e_m'$ are the comoving energy densities of the radiation field and matter (including its rest mass energy), 
respectively, and $p'_m$ is the thermal pressure of the matter component. This also implies that 
$L_{\gamma,{\rm iso}}(r_s)=L_{\rm k,iso}(r_s)=(1/2)L_{\rm tot,iso}$, which by combining 
Eq.(\ref{eq:Inorm},\ref{eq:L-gamma-iso},\ref{eq:I1}) yields the powers measured by an observer 
at $r=\infty$ for the two components,
\begin{eqnarray}\label{eq:L-gamma-infty}
    L_{\gamma,{\rm iso},\infty} &\approx& \frac{64\pi^2}{3}r_{\rm ph}^2\Gamma^2\beta I'_{\rm ph} \\
    L_{\rm k,iso,\infty} &\approx& \fracb{r_{\rm ph}}{r_s}^{2/3}L_{\gamma,{\rm iso},\infty}
    \approx\epsilon_\gamma^{-1} L_{\gamma,{\rm iso},\infty}~,
    \label{eq:L-k-infty}
\end{eqnarray}
where the adiabatic factor is defined as 
\begin{equation}
    \epsilon_\gamma\equiv \frac{L_{\gamma,{\rm iso},\infty}}{L_{\gamma,{\rm iso}}+L_{\rm k,iso}}
    \approx\frac{L_{\gamma,{\rm iso},\infty}}{L_{{\rm k,iso},\infty}}
    \approx \fracb{r_{\rm ph}}{r_s}^{-2/3}
\end{equation}

The isotropic power carried by a passively expanding cold flow, for which $e_m' = n_e'm_pc^2$ and $p_m'=0$, 
is given by
\begin{equation}
    L_{k,\rm iso}(r) = 4\pi r^2\Gamma^2\beta n_e'm_pc^3\approx 4\pi r^2\Gamma^2n_e'm_pc^3~,
\end{equation}
where $m_p$ is the proton mass and $n_e'$ is the density of the baryonic electrons. From the 
expression for the Thomson optical depth in Eq.~(\ref{eq:tauT-r}), we find that the photospheric 
radius for this outflow is
%assuming that the observed isotropic equivalent luminosity is 
%$L_{\rm \gamma,iso}=\epsilon_\gamma (L_{k,\rm iso}+$L_{\rm \gamma,iso})\approx\epsilon_\gamma L_{k,\rm iso}$,
\begin{equation}\label{eq:rph}
    r_{\rm ph} = \frac{\sigma_T L_{\rm k,iso}}{8\pi\Gamma^3c^3m_p}
    \approx 5.5\times10^{12}\Gamma_2^{-3}L_{\rm k,iso,52}~{\rm cm}~.
\end{equation}

%%%%%%%  FIGURE  %%%%%%%%%%%%%%%%%%%%%%%%%%%%%%%%%
\begin{figure}
    \centering
    \includegraphics[width=0.41\textwidth]{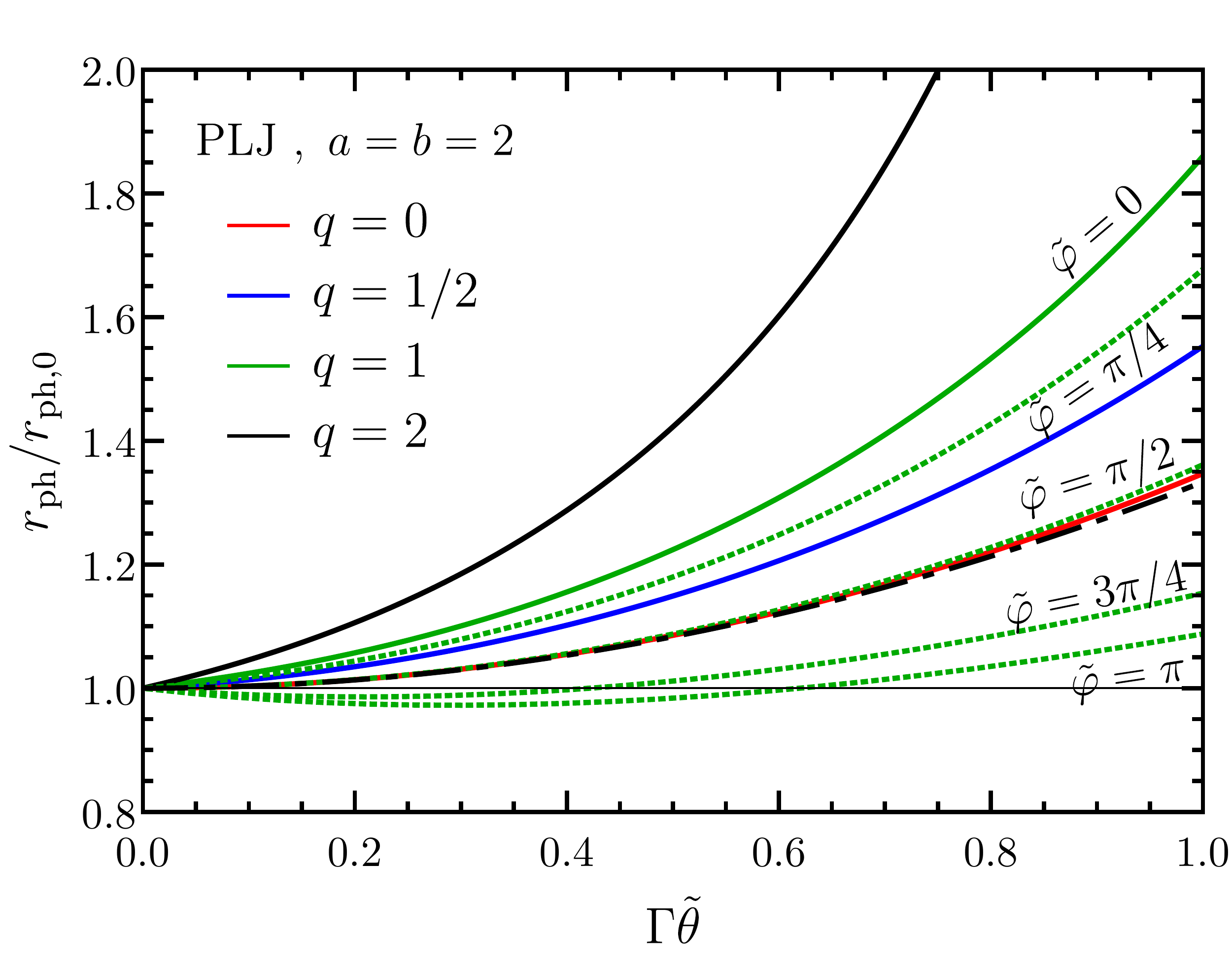}\quad
    \includegraphics[width=0.41\textwidth]{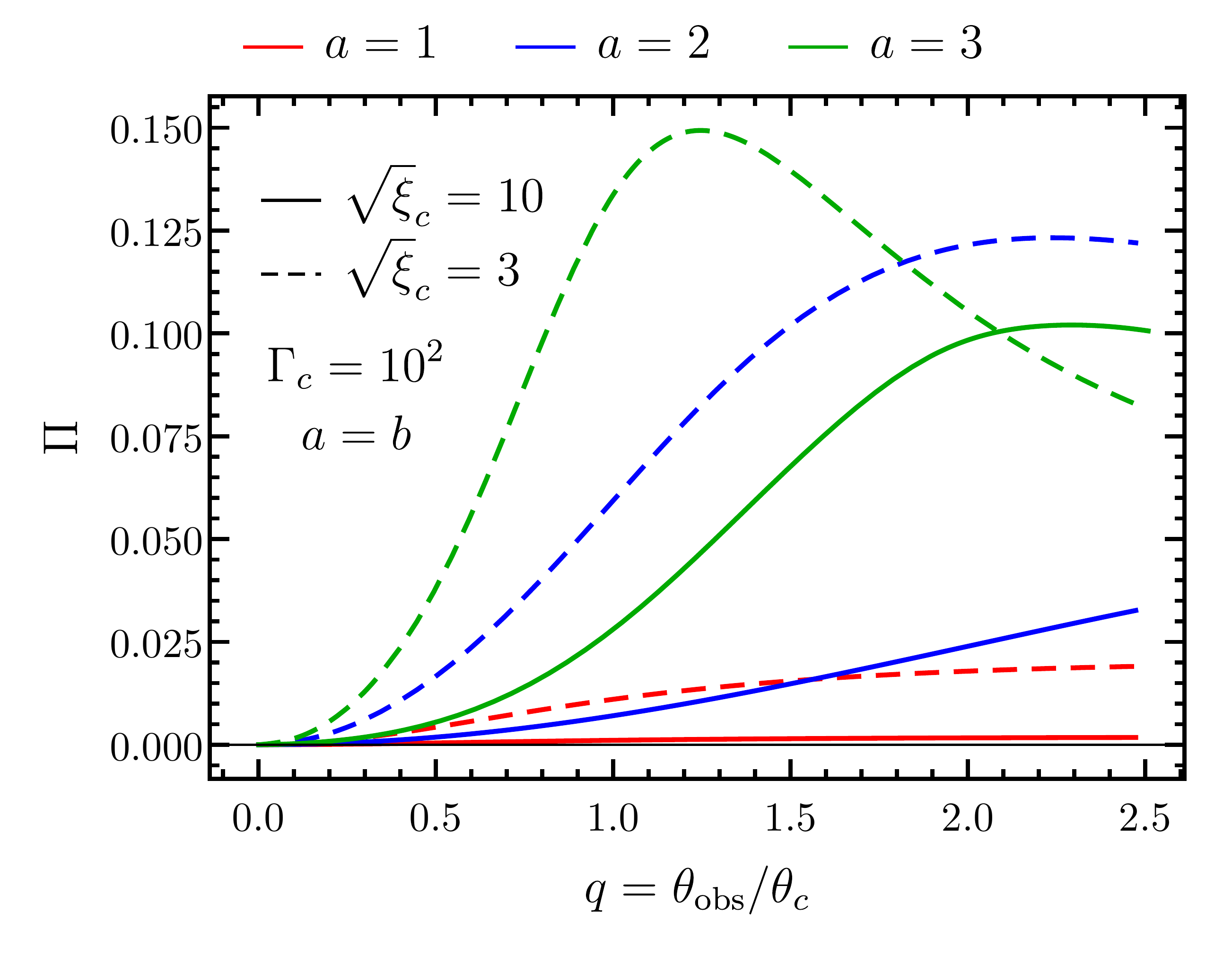}\quad
    \includegraphics[width=0.41\textwidth]{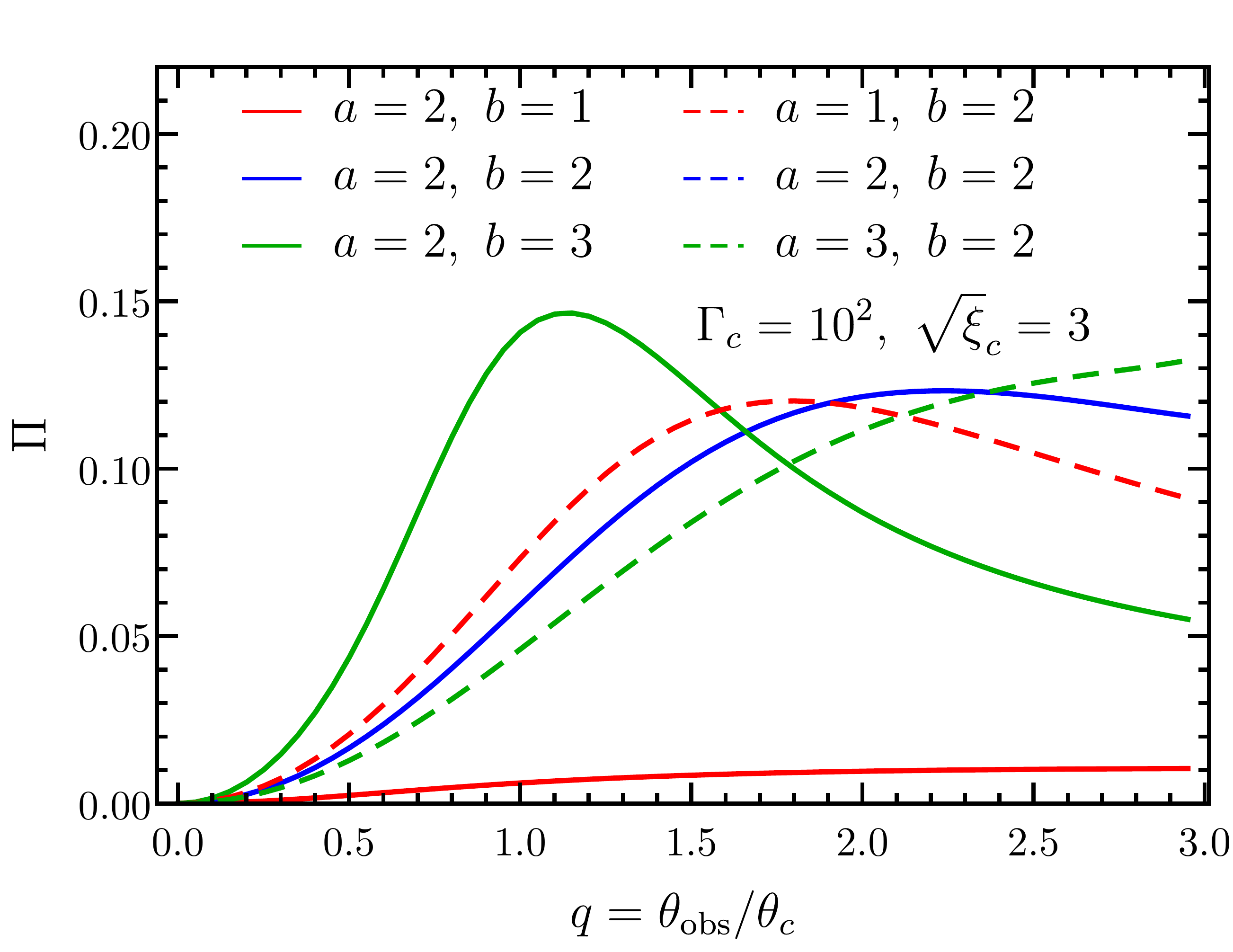}
    \caption{\textit{Top}: Deviation of the photospheric radius within the beaming cone with respect to that obtained 
    along the LOS, shown here for different values of $q\equiv\theta_{\rm obs}/\theta_c$. 
    The change in $r_{\rm ph}$ with azimuthal angle is shown here for $q=1$ and different values of $\tilde\varphi$ as 
    green dotted lines. The black dash-dotted line shows the result for a spherical flow with no angular structure. 
    Here we assumed $\Gamma_c=10^3$ and $\xi_c\equiv(\Gamma_c\theta_c)^2=10^2$. 
    \textit{Middle}: Degree of polarization arising from photospheric emission in a power law jet (PLJ), shown 
    for a narrow ($\sqrt\xi_c=3$) and wide ($\sqrt\xi_c=10$) jet. \textit{Bottom}: The polarization trend for 
    different values of the power law indices ($a,b$) is shown.
    }
    \label{fig:rph}
\end{figure}
\subsection{$\Pi$ from a structured jet viewed off-axis}\label{sec:Photo-strucjet}
To obtain the observed polarization, integration over the GRB image must be performed. An important consequence of this 
integration is that radiation emerging from within the beaming cone (of angular size $\Gamma^{-1}$) experiences different 
Thomson optical depths, such that $\tau_T = \tau_T(\tilde\theta,\tilde\varphi)$. Therefore, the matter-radiation decoupling 
radius also varies with angle around the LOS, $r_{\rm ph} = r_{\rm ph}(\tilde\theta,\tilde\varphi)$, which leads to 
variations in $\Pi=\Pi(\tilde\theta,\tilde\varphi)$ around the LOS. If the properties of the flow are symmetric around 
the LOS, the observed polarization vanishes (similarly to Compton drag or synchrotron for $B_\perp$ or $B_\parallel$, in which there is symmetry around the local radial direction). Therefore, the outflow must either be structured or the intensity must be 
inhomogeneous.

In general, due to the statistical nature of last scattering, the photospheric radius is a random variable (B11). As a 
consequence, the matter-radiation decoupling doesn't occur at a sharp boundary, but instead it is radially extended where 
roughly $2/3$ of the photons undergo last scattering at $r_{\rm ph}/3<r<3r_{\rm ph}$. This leads to the notion of a 
``fuzzy'' photosphere (B11). For simplicity, here we adopt the sharp photosphere.

In the following we consider a power-law structured jet that was as discussed earlier. The Thomson optical depth measured 
in the direction of the observer ($\hat n$) around the LOS along some photon trajectory $\mathcal{S}$, with length 
$s=r\cos\tilde\theta=\tilde z$, is
\begin{equation}
    \tau_T = \int n_e'(r,\theta)\sigma_Tds' 
    = \int_{\tilde z}^\infty n_e'(r,\theta)\sigma_T\Gamma(\theta)[1-\beta(\theta)\tilde\mu]ds~,
\end{equation}
where we made use of the fact that $ds=\delta_Dds'$. For $\tilde\mu=1$, one recovers the expression in 
Eq.~(\ref{eq:tauT-r}). The transverse distance from the LOS to the path $\mathcal{S}$ is a constant, such that 
$r\sin\tilde\theta=r_{\rm ph}\sin\tilde\theta_{\rm ph}$, which results from the fact that light travels in a straight 
path. This can be used to write the integral over the more useful quantity $\tilde\theta$ instead of $s$ through the 
Jacobian of transformation
\begin{equation}
    ds = \frac{ds}{d\tilde\theta}d\tilde\theta = -\frac{r^2}{r_{\rm ph}\sin\tilde\theta_{\rm ph}}d\tilde\theta~.
\end{equation}
Finally, by noticing that $\tau_T=1$ at the photospheric radius, we find
\begin{equation}
    r_{\rm ph}(\tilde\theta,\tilde\varphi) = \frac{\sigma_T}{\sin\tilde\theta}
    \int_0^{\tilde\theta}n'_e(r,\theta)r^2\Gamma(\theta)[1-\beta(\theta)\tilde\mu^{\prime\prime}]d\tilde\theta''~,
\end{equation}
where $\tilde\theta''$ is a dummy variable, and $\theta=\arccos\mu$ can be expressed in terms of the LOS coordinates ($\tilde\theta,\tilde\varphi$) using
\begin{equation}
    \mu = \tilde\mu\tilde\mu_{\rm obs}-\cos\tilde\varphi\sqrt{(1-\tilde\mu^{2})(1-\tilde\mu_{\rm obs}^2)}~.
\end{equation}

For an ultra-relativistic structured jet, the isotropic equivalent kinetic power is
\begin{equation}
     L_{k,{\rm iso}}(\theta) = 4\pi r^2\Gamma^2(\theta)n'_e(r,\theta)m_pc^3~.
\end{equation}
From here it is easy to see that 
\begin{equation}
    r^2 n'_e(r,\theta)\equiv\hat n_e'(\theta)
    \propto \frac{L_{k,{\rm iso}}(\theta)}{\Gamma^2(\theta)}
    \approx \frac{L_{\rm tot,iso}(\theta)}{\Gamma^2(\theta)}
\end{equation}
is a completely $r$ independent quantity and it only varies with polar angle $\theta$. 
Then, along the LOS, for which $\theta=\theta_{\rm obs}$ and $\tilde\theta=0$, the photospheric radius lies at 
\begin{equation}
    r_{\rm ph,0}=\frac{\sigma_T\hat n_e'(\theta_{\rm obs})}{2\Gamma(\theta_{\rm obs})}~.   
\end{equation}
The deviation of the photospheric radius along photon trajectories that originate at different $\tilde\theta$ and 
$\tilde\varphi$ around the LOS is shown in the top panel of Fig.~\ref{fig:rph}.

The comoving intensity will also be modified due to the angular structure of the outflow. Its angular dependence 
can be obtained by expressing the normalization $I'_{\rm ph}$ in terms of $L_{\rm k,iso,\infty}(\theta)$ and 
$\Gamma(\theta)$ from Eq.(\ref{eq:L-k-infty} \& \ref{eq:rph}), such that
\begin{equation}
    I'_{\rm ph}(\theta)\propto\frac{\Gamma(\theta)^{20/3}}{L_{\rm k,iso}^{5/3}(\theta)}\equiv\kappa(\theta)~.
\end{equation}
The flux measured by a distant observer is given by
\begin{equation}
    F = \frac{1}{d_L^2}\int\delta_D^4I'(r)dS_\perp
    = \frac{1}{d_L^2}\int\delta_D^4I'(r)\tilde\rho d\tilde\rho d\tilde\varphi~,
\end{equation}
where $dS_\perp$ is the differential area on the plane of the sky and 
$\tilde\rho=r_{\rm ph}\sin\tilde\theta_{\rm ph}$ is the transverse distance from the LOS. 
For convenience, the above integral can be performed over the polar angle $\tilde\theta$ via a 
simple transformation,
\begin{equation}
    d\tilde\rho = \frac{d\tilde\rho}{d\tilde\theta}d\tilde\theta
    = r_{\rm ph}\tilde\mu\left(1+\frac{\sqrt{1-\tilde\mu^2}}{\tilde\mu}
    \frac{d\ln r_{\rm ph}}{d\tilde\theta}\right)d\tilde\theta~,
\end{equation}
which finally yields,
\begin{equation}
    dS_\perp = r_{\rm ph}^2\left(\tilde\mu+\sqrt{1-\tilde\mu^2}
    \frac{d\ln r_{\rm ph}}{d\tilde\theta}\right)d\tilde\mu d\tilde\varphi~.
\end{equation}
Now, the degree of polarization measured by a distant observer can be expressed as
\begin{equation}
    \Pi=\frac{Q}{I} = \frac{\displaystyle\int\delta_D^4Q'(r_{\rm ph})\kappa(\theta)\cos(2\tilde\varphi)dS_\perp}
    {\displaystyle\int\delta_D^4I'(r_{\rm ph})\kappa(\theta)dS_\perp}~,
\end{equation}
where $I'(r_{\rm ph})$ and $Q'(r_{\rm ph})$ are obtained from the radiative transfer 
equations for a spherically symmetric flow, with the angular structure embedded in $\kappa(\tilde\theta,\tilde\varphi)$ and $r_{\rm ph}(\tilde\theta,\tilde\varphi)$.

To be able to use the radiative transfer solutions from Eq.~(\ref{eq:Iprime} \& \ref{eq:Qprime}) that assume a spherical outflow, to calculate the 
degree of polarization when the outflow has an angular structure, an important consideration is choosing the correct 
angular scale $\Delta\theta$ over which the properties of the outflow don't change significantly. 
The properties of the flow change significantly over angular scales $\delta\theta_\epsilon$, where 
the fractional change in the energy per unit solid angle of the outflow is of order unity, such that $\Delta\epsilon/\epsilon\sim1$ (similar considerations also apply for the angular dependence of $\Gamma$). 
Therefore, the spherically symmetric solution to the radiative transfer equations is approximately 
valid on angular scales
\begin{equation}
    \Delta\theta \ll \delta\theta_\epsilon 
    \equiv \theta\left\vert\frac{d\ln\epsilon}{d\ln\theta}\right\vert^{-1}~.
\end{equation}
For a structured jet with a uniform core and power-law wings, the energy per unit solid angle is 
$\epsilon(\theta)\propto\theta^{-a}$ outside of the core, which yields $\delta\theta_\epsilon=\max[\theta_c,\theta/a]$. 

Next, we compare the angular scale $\Delta\theta$ with the typical angular scale over which photons are 
scattered $\tilde\theta_{\rm sc}$ while they try to 
diffuse from deep within the flow outwards. In the comoving frame, their diffusion length can be expressed as
$\ell'_{\rm diff}\sim\sqrt{N_{\rm sc}}\lambda'$, where $N_{\rm sc}$ is the mean number of scatterings they 
undergo and $\lambda'$ is their mean free path. In a relativistically expanding flow, $N_{\rm sc}\sim\tau_T$ 
rather than $\tau_T^2$, where the Thomson optical depth of the flow is 
$\tau_T=r/(\Gamma\lambda')$. This finally yields the diffusion length of photons $\ell'_{\rm diff}\sim r/(\Gamma\sqrt{\tau_T})$, 
which suggests that deeper in the flow, where $\tau_T\gg1$, photons only diffuse a very short distance and are 
instead advected with the flow. If the photons diffuse a mean transverse distance 
$r\sin\tilde\theta_{\rm sc}\sim r\tilde\theta_{\rm sc}$ of the order of the diffusion distance, such that
$r\tilde\theta_{\rm sc}\sim\ell'_{\rm diff}$, then the mean scattering angle is 
$\tilde\theta_{\rm sc}\sim(\Gamma\sqrt{\tau_T})^{-1}$. Finally, letting $\Delta\theta=\tilde\theta_{\rm sc}$, 
yields the constraint
\begin{equation}
    \Gamma\theta\gg\frac{a}{\sqrt{\tau_T}}~.
\end{equation}
In a structured jet, outside of the uniform core that has angular size $\theta_c$, the LF decays with angle $\theta$ 
away from the jet symmetry axis (see Eq.(\ref{eq:PLJ})). For $\theta>\theta_c$, 
$\Gamma\theta\sim\Gamma_c\theta_c(\theta/\theta_c)^{1-b}$, where $\Gamma_c\theta_c\sim3-10$ and $\Gamma_c\gg1$. 
In this case, deeper in the flow the above condition is almost always satisfied, however, close to the photosphere, where $\tau_T\sim1$, it breaks down for $b>1$ beyond some critical angle. This is one caveat 
of the approximation made here. However, $\Gamma\theta<1$ implies lateral causal contact, for which the flow dynamics naturally tend to wash out lateral gradiants.

In the middle and bottom panels of Fig.~\ref{fig:rph}, we show the degree of polarization of photospheric emission 
when the outflow is structured with having a power law jet (PLJ; Eq.~\ref{eq:GJ}) profile. In this case, 
when the bulk LF falls sharply outside of the narrow core for a given profile of the kinetic power, we 
find a moderate level of polarization with $\Pi\lesssim15\%$ for narrow jets ($\sqrt\xi_c=3$) and $\Pi\lesssim10\%$ 
for wider jets ($\sqrt\xi_c=10$). This result is broadly consistent with that found 
from Monte Carlo simulations of photospheric emission from structured jets \citep{Ito+14,Lundman+14}, where it was 
found that $\Pi\sim20\% - 30\%$ when $\xi_c\sim10$ and $\Pi\lesssim20\%$ when $\xi_c\sim10^2$. 
The exact result depends on the angular structure assumed in such simulations. Looking at the trend of $\Pi$ as 
the power law indices of the $\Gamma(\theta)$ and $L_{k,\rm iso}(\theta)$ profiles are changed, it is clear 
from the bottom panel of Fig.~\ref{fig:rph} that a steeper $\Gamma(\theta)$ profile yields higher degree of polarization. 
On the other hand, steeper kinetic power profiles only translate the polarization curve to larger 
$q$ values while approximately maintaining the same maximum level of $\Pi$.

\section{Integration over multiple pulses}\label{sec:multiple-pulses}
Unless the source is nearby or particularly bright, observations of the prompt phase in GRBs are 
typically photon starved. To increase photon statistics observers generally have to 
average over multiple pulses, which washes out any temporal dependence. This is especially true for 
polarization measurements. On the one hand, this may be the only way to derive a statistically significant 
measurement of the level of polarization, while on the other hand this operation guarantees the loss of crucial 
information such as the temporal dependence of the polarization angle. More importantly, since the 
properties of the outflow, e.g. $\Gamma$ or equivalently $\xi_j$ for a fixed $\theta_j$, can also change from 
pulse to pulse, this can affect the level of polarization when integrating over multiple pulses.

%%%%%%%  FIGURE  %%%%%%%%%%%%%%%%%%%%%%%%%%%%%%
\begin{figure}
    \centering
    \includegraphics[width=0.39\textwidth]{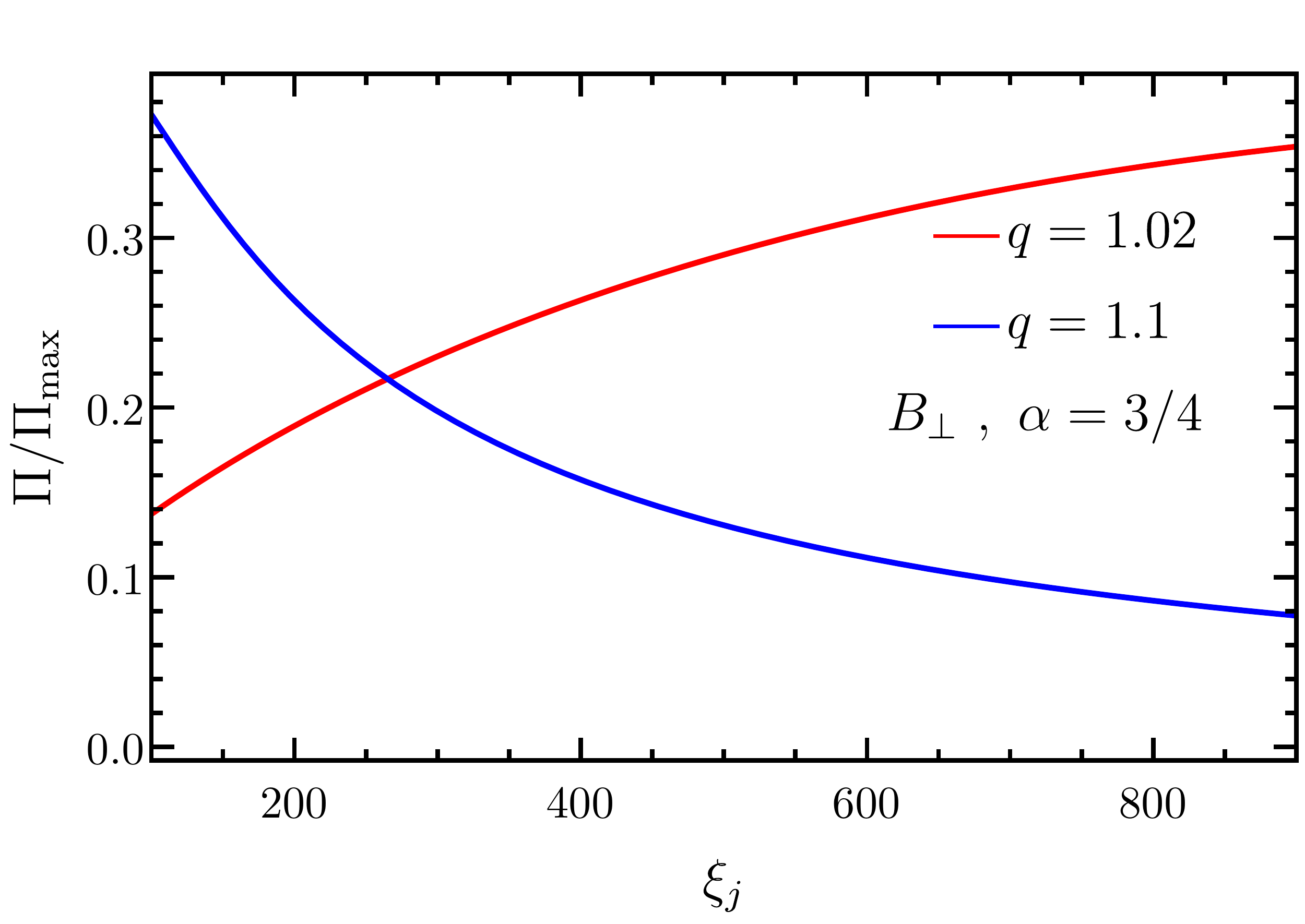}
    \includegraphics[width=0.39\textwidth]{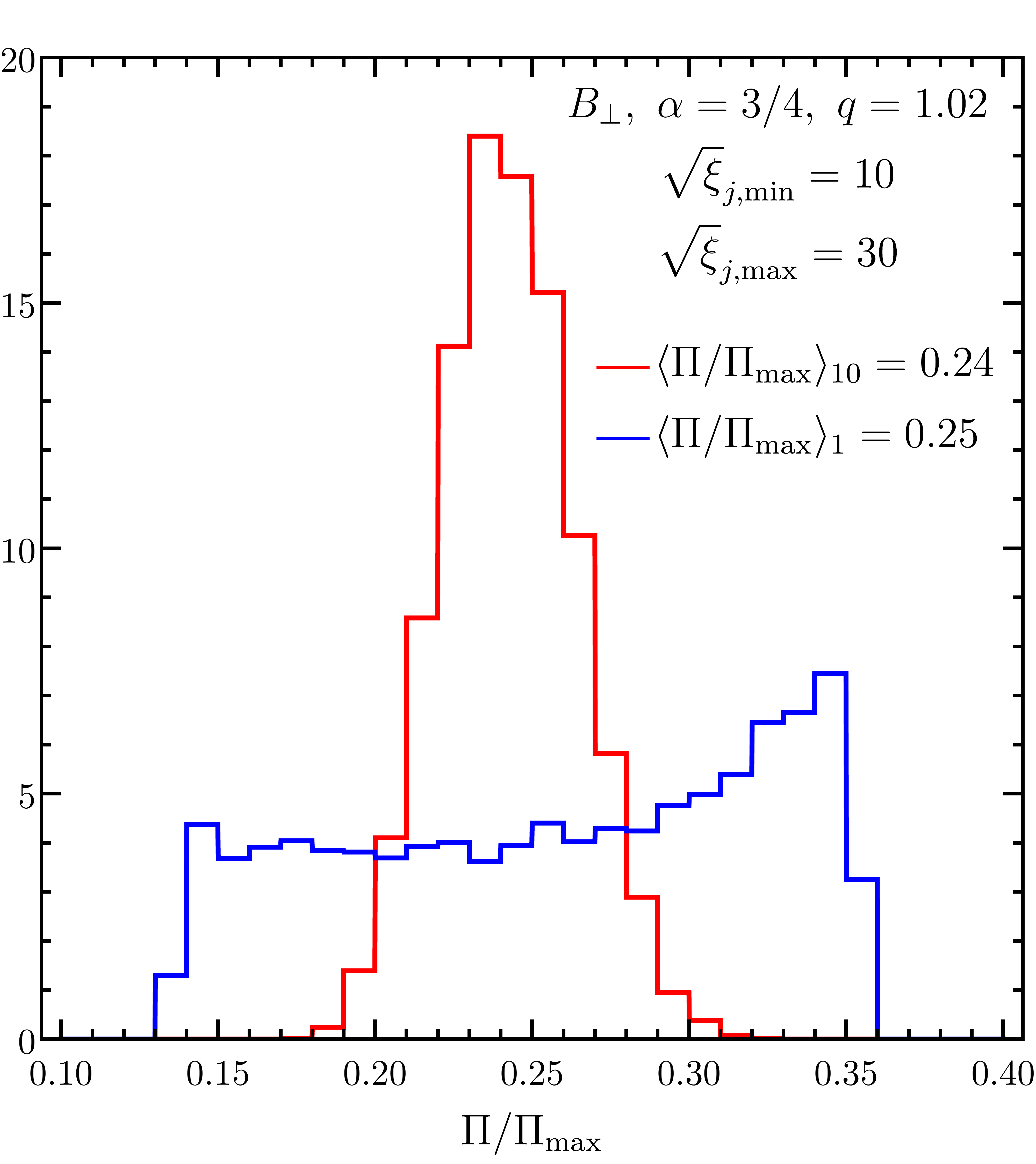}
    \includegraphics[width=0.39\textwidth]{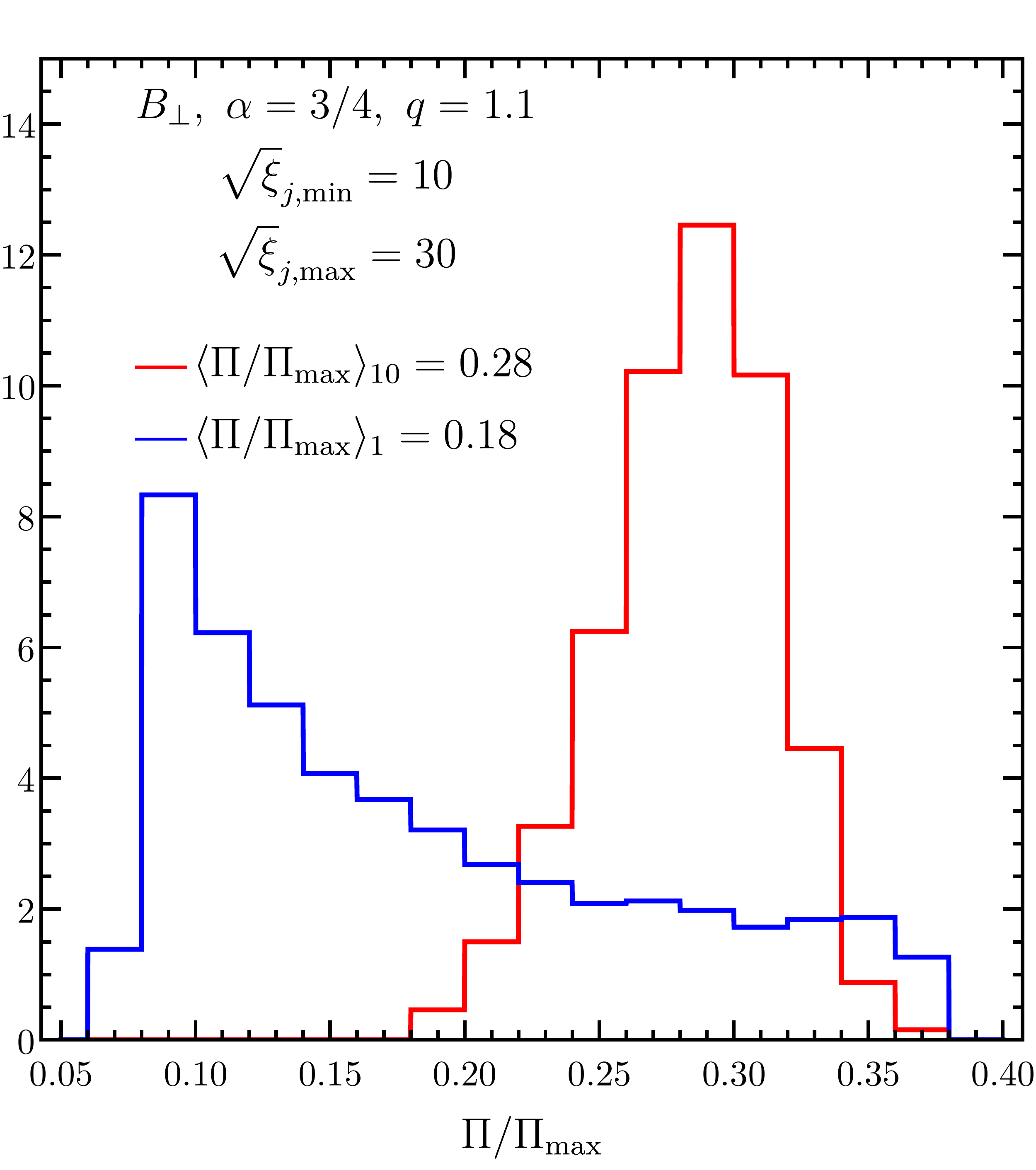}
    \caption{\textit{Top}: The trend of $\Pi$, either monotonically decaying 
    ($q=\theta_{\rm obs}/\theta_j=1.1$) or rising ($q=1.02$), as $\xi_j=(\Gamma\theta_j)^2$ of the individual pulses is varied. 
    Shown here for the case of a random magnetic field completely in the plane of the ejecta. 
    The chosen values of $q$ are not special, but yield high levels of polarization in this 
    particular case. \textit{Middle \& Bottom}: Distribution of $\Pi$ when 
    obtained from a single pulse (blue) or after having integrated over multiple 
    pulses (red; $N_p=10$) in an emission episode. Shown here for two different values 
    of $q$ for which the trend of $\Pi$ is opposite when $\xi_j^{1/2}$ is varied between 
    pulses. Here $\xi_{j,\min}^{1/2}\leq\xi_j^{1/2}\leq\xi_{j,\max}^{1/2}$ is distributed uniformly.}
    \label{fig:q-gamma-pol-dist-Brnd-perp}
\end{figure}
%%%%%%%%%%%%%%%%%%%%%%%%%%%%%%%%%%%%%%%%%%%%%%%

The total polarization of an emission episode, which is a sum of $N_p>1$ pulses, is obtained from summing up their respective Stokes parameters,
\begin{equation}
    \Pi = \frac{Q}{I}=\frac{\sum_{i=1}^{N_p}Q_i}{\sum_{i=1}^{N_p}I_i}~.
\end{equation}
In the case of an ordered magnetic field with coherence 
length as large as the size of the emission region that produces a single pulse, multiple pulses arising from 
such mutually incoherent patches will yield a lower degree of polarization. This occurs due to the fact that the PAs of emission 
from different patches are randomly oriented which leads to cancellations and leave a lower level of net polarization. Adding up the polarization from $N_p$ pulses is essentially a random walk for $Q$ while $I$ adds up coherently. Therefore, the total polarization for $N_p$ pulses can be deduced from the above equation to obtain \citep{GW99}
\begin{equation}\label{eq:patchy}
    \Pi\sim\frac{\Pi_{\max}}{\sqrt{N_p}}~.
\end{equation}
For other magnetic field configurations, cancellation of polarization between different pulses due to the change of 
sign of $\Pi_i=Q_i/I_i$ (i.e. of $Q_i$, since $I_i>0$) may not occur.

%%%%%%% DISTRIBUTION OF PI IN A SINGLE BURST %%%%%%%%%%%%%%%%%%%%%%%%%%%%%%%%%%%%%%%%%%%%%%%%%%%%%%
\subsection{Distribution of $\Pi$ in a single burst}
Here we consider a single burst and an emission episode with an agglomeration of multiple pulses 
that may be produced by emission regions with different $\Gamma$. The distribution of $\Gamma$ in the different pulses that correspond to different emission regions is not known, and assumed to be drawn from some probability distribution over a finite range  
$\Gamma_{\min}\leq\Gamma\leq\Gamma_{\max}$. For simplicity we assume here 
that $\theta_j$, and therefore also $q=\theta_{\rm obs}/\theta_j$, remains fixed over the entire GRB. 
Therefore, a distribution of $\Gamma$ is equivalent to that of 
$\xi_{j,\min}^{1/2}\leq\xi_j^{1/2}\leq\xi_{j,\max}^{1/2}$. In what follows, we consider a uniform distribution 
of $\xi_j^{1/2}$, such that $P\left(\xi_j^{1/2}\right)=\left(\xi_{j,\max}^{1/2}-\xi_{j,\min}^{1/2}\right)^{-1}$. The following analysis can be 
easily extended to other distributions, 
however, there's no straightforward way of discerning one from the other. 
%consider two different probability density functions $P(\Gamma)$: (i) a 
%uniform distribution with $P(\Gamma)=(\Gamma_{\max}-\Gamma_{\min})^{-1}$, and (ii) distribution of $\log\Gamma$ 
%with $P(\log\Gamma) = 1/\log(\Gamma_{\max}/\Gamma_{\min})$. 
To demonstrate the effect of averaging over multiple pulses, for simplicity, we will consider in this section a 
top-hat jet with a random magnetic field ($B_\perp$) in the plane of the ejecta. We carry out a Monte-Carlo (MC) 
simulation where we draw $10^4$ random samples, where each sample represents an emission episode with $N_p$ pulses. 

On average integration over multiple pulses can yield $\Pi$ that is higher or lower in comparison to 
a single pulse. This depends on the viewing angle, in particular $q$, and the trend of $\Pi$ as $\xi_j$ 
is varied (see top panel of Fig.~\ref{fig:q-gamma-pol-dist-Brnd-perp}). We illustrate this with two cases, 
as shown in the middle and bottom panel of Fig.~\ref{fig:q-gamma-pol-dist-Brnd-perp}, 
where the value of $q$ is chosen so that the trend of $\Pi$ is opposite. Since the jet is fairly wide with 
$\xi_j > 10^2$, there is no cancellation of the polarization as $\Pi_i$ never switches sign in this case. However, 
for narrower jets with $\xi_j<10$ multiple pulses with even smaller $\xi_j$ and $q\lesssim1$ can have opposite 
signs for the PA leading to cancellation and lower net polarization.

%%%%%%%  STATISTICAL INFERENCE OF MAGNETIC FIELD STRUCTURE FROM PI  %%%%%%%%%%%%%%%%%%%%%%%%%%%%%%%%%%
\section{Statistical inference of magnetic field structure from polarization}\label{sec:MC-Pol}
A firm detection of linear polarization can provide valuable insight into the structure of the magnetic field 
in the outflow, which can be further used to constrain the jet composition. In order to derive meaningful 
inference about the magnetic field structure from the measured degree of polarization, there are three basic 
quantities that determine the outcome: (i) $\xi_j^{1/2}=\Gamma\theta_j$, which determines how narrow the jet is and 
varies between different pulses due to variation in $\Gamma$ while $\theta_j$ is assumed here to be fixed for a given burst; 
(ii) $q=\theta_{\rm obs}/\theta_j$, which determines the viewing angle and remains fixed for the different pulses 
but varies between different bursts; and (iii) $\tilde f_{\rm iso}(q,\xi_j)\equiv E_{\gamma,{\rm iso}}(q,\xi_j)/E_{\gamma,{\rm iso}}(0,\xi_j)$ 
or equivalently the off-axis to on-axis fluence ratio, which depends on both $\xi_j$ and $q$ and varies between 
bursts as well as different pulses. The appropriate relative weight ($E_{\gamma,{\rm iso}}$) is assigned to each pulse when adding up the Stokes parameters for different pulses that are added up in order to increase the observed signal.

Additional effects that characterize the spectrum, viz. the $\nu F_\nu$-peak energy and the spectral indices above and below it, 
can also have an effect \citep[see, e.g.,][]{Toma+09}. For instance, if the spectral peak is located in a given frequency band, 
$\nu_1<\nu_{\rm pk}<\nu_2$, then its temporal evolution will be reflected in the temporal evolution of the polarization. As it was 
shown earlier, the degree of polarization depends on the spectral index in both synchrotron and Compton drag emission mechanisms, where 
softer spectra yield a larger degree of polarization. Therefore, dominance of a given spectral component is reflected in the corresponding 
level of polarization, making this spectro-polarimetric correlation a useful probe of the underlying emission mechanism. The evolution 
of the spectral properties over multiple pulses is not considered in this work to limit the degrees of freedom, and therefore to 
ensure the robustness of the results. The formalism developed in this work can be easily extended to include spectral and temporal effects.

\begin{figure}
    \centering
    \includegraphics[width=0.48\textwidth]{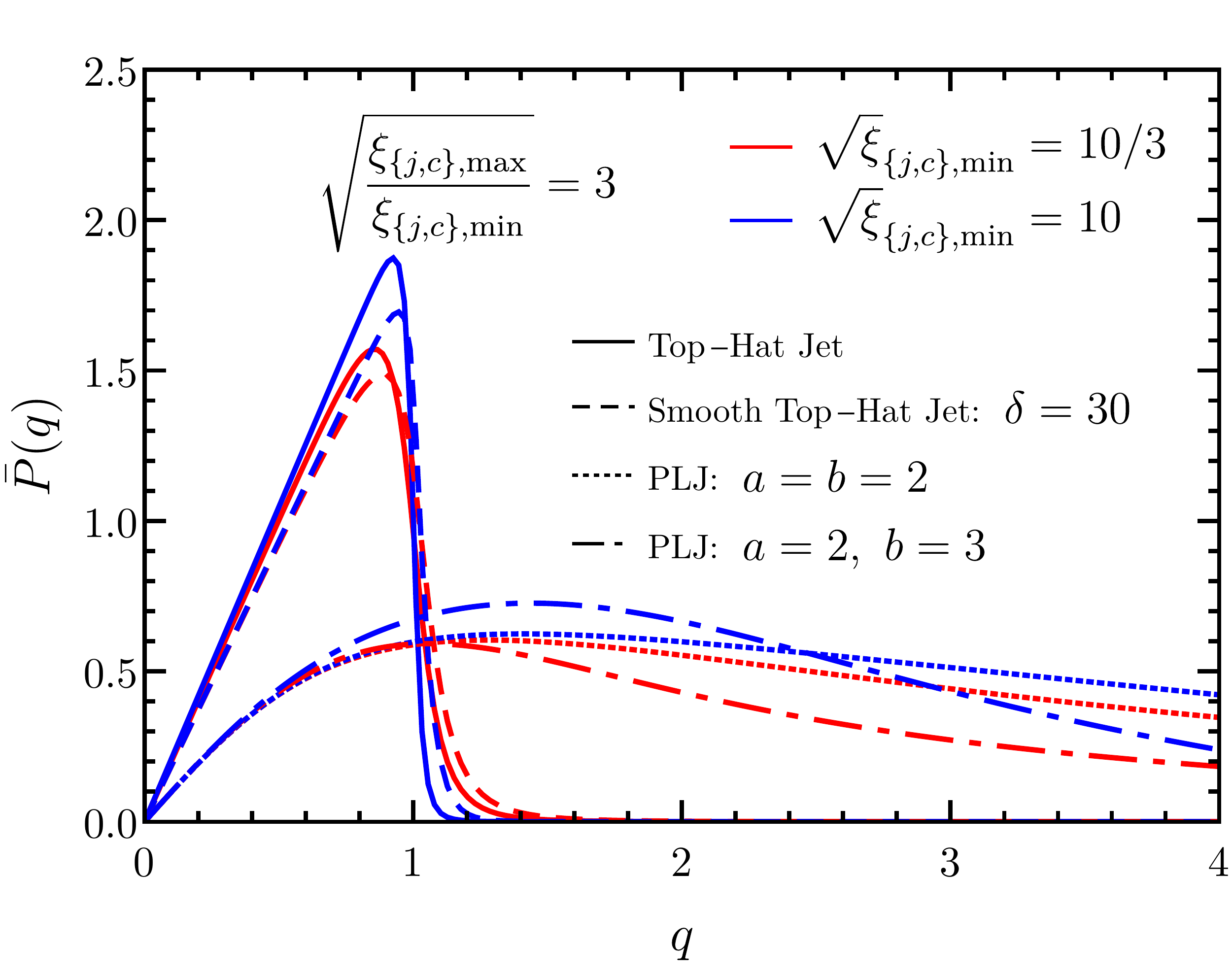}
    \caption{Fluence weighted distribution of $q=\theta_{\rm obs}/\theta_j$ that is 
    marginalized over a uniform distribution of $\sqrt\xi_{\{j,c\},\min}\leq\sqrt\xi_{\{j,c\}}\leq\sqrt\xi_{\{j,c\},\max}$. 
    Shown here for four different jet structures.}
    \label{fig:Pq-dist}
\end{figure}

For simplicity we consider a fixed initial jet opening angle $\theta_j$. This renders the distribution of the three basic parameters to 
arise due to the spread in the viewing angle $\theta_{\rm obs}$ between different GRBs and of $\Gamma$ also between different pulses within 
the same GRB. First we consider a uniform distribution of $\Gamma$ or equivalently of
$\xi_{\{j,c\},\min}^{1/2}\leq\xi_{\{j,c\}}^{1/2}\leq\xi_{\{j,c\},\max}^{1/2}$, where the subscript $j$ applies when discussing a 
top-hat jet with sharp or smooth edges, and the subscript $c$ applies when discussing a structured jet with a compact core. For 
brevity, only the subscript $j$ is used in the following discussion. The viewing angle is distributed according to the solid angle, such that 
$P(\theta_{\rm obs})d\theta_{\rm obs}=\sin\theta_{\rm obs}d\theta_{\rm obs}$. Finally, the off-axis to on-axis fluence 
ratio depends on the distribution of $\xi_j$ and $q$. These are obtained for a fixed $\theta_j$, such that 
$P(\xi_j)d\xi_j=P\left(\xi_j^{1/2}\right)d\xi_j^{1/2}$ and $P(q)dq=P(\theta_{\rm obs})d\theta_{\rm obs}$, which yields
\begin{eqnarray}
    && P(\xi_j) = \frac{P\left(\xi_j^{1/2}\right)}{2\xi_j^{1/2}} \\
    && P(q) = \theta_jP(\theta_{\rm obs}) \propto q~.
\end{eqnarray}
Since $P(q)\propto q$, it would favour larger $q>1$ values, for which the fluence will be too small. Therefore, a more meaningful distribution 
of $q$ should account for the rapid drop in fluence for $q>1$. To include this effect, we define a fluence weighted distribution 
for $q$ with $\bar P(q)=[\int\bar f_{\rm iso}(q)P(q)dq]^{-1}\bar f_{\rm iso}(q)P(q)$, where 
\begin{equation}
    \bar f_{\rm iso}(q) = \int_{\xi_{j,{\min}}}^{\xi_{j,{\max}}}\tilde f_{\rm iso}(q,\xi_j)P(\xi_j)d\xi_j
\end{equation}
is the distribution of $\tilde f_{\rm iso}(q,\xi_j)$ with $q$ but marginalized over the distribution of $\xi_j$. 
Fig.~\ref{fig:Pq-dist} shows the fluence weighted distribution of $\bar P(q)$ for two different jet opening angles and 
for a uniform distribution in $\sqrt\xi_j$. The suppression in $\bar P(q)$ for $q\gtrsim1$ is caused by the sharp 
(gradual) drop in fluence for a top-hat (structured) jet. This has important implications for the distribution of detected 
GRBs since for flux limited detectors the drop in fluence reduces the effective volume probed by the detector. For a 
top-hat jet the drop in fluence is so sharp that the total number of GRBs in a given volume can be obtained from 
$N_{\rm tot}=f_bN_{\rm obs}$, where $f_b = 4\pi/\Delta\Omega=4\pi/(1-\cos\theta_j)\approx4/\theta_j^2$ is the beaming factor for $\theta_j\ll1$. 
However, for a structured jet, this estimate must take into account the dependence on $q$.

To simulate different bursts we carried out MC simulations with $10^4$ sample bursts and $N_p=10$ multiple 
pulses for each burst. For each sample burst, a value of $q$ was randomly drawn from $\bar P(q)$ and for 
each pulse the distribution of $\xi_j^{1/2}$ was randomly sampled. To further eliminate the polarization contribution 
from pulses with low fluence, only pulses with $\tilde f_{\rm iso}(q,\xi_j)>10^{-2}$ were included. This threshold 
implies that the detectors are flux limited and can only detect bursts/pulses that are dimmer by a factor of $10^{-2}$ 
from their absolute on-axis fluence. Since dimmer pulses would fall below the detector threshold and won't 
be detected, their contribution to $\Pi$ should be removed when calculating the total polarization. Here we don't take 
into account the limitation imposed by the compactness of the flow for even modestly steep $\Gamma(\theta)$ profiles. 
Such a constraint would restrict viewing angles to even smaller values as compared to the constraint on $q$ imposed by 
$\tilde f_{\rm iso}$. In order to incorporate this effect, more detailed spectral modeling than conducted in this work would be 
needed which is outside the scope of this work.

%%%%%%%%  FIGURE  %%%%%%%%%%%%%%%%%%%
%%%%%%%%%%%%%%%%%%%%%%%%%%%%%%%%%%%%%

%%%%%% FIGURE %%%%%%%%%%%%%%%%%%%%%%%%%%%%%%%%
\begin{figure*}
    \centering
    \includegraphics[width=0.4\textwidth]{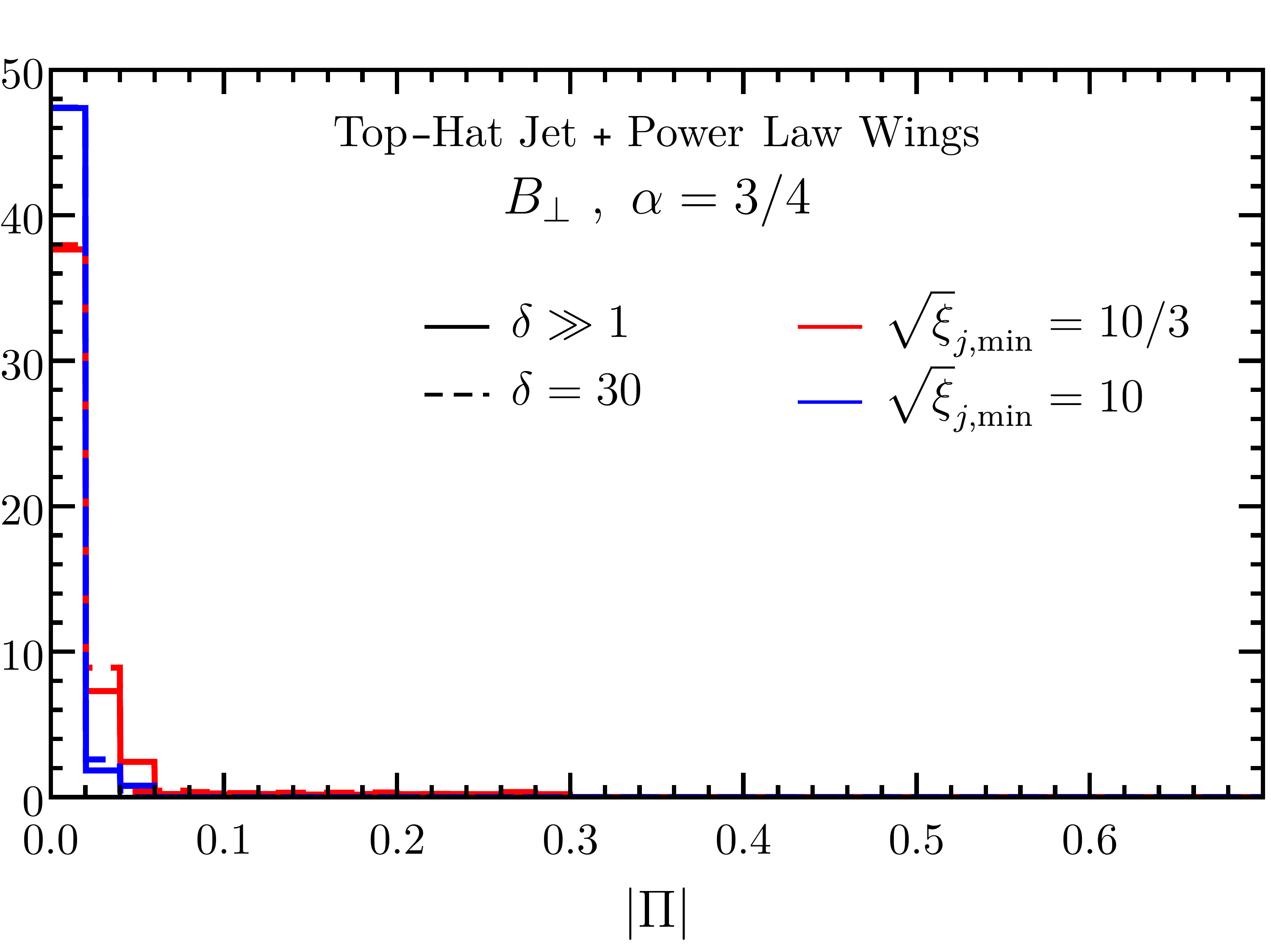}\quad\quad
    \includegraphics[width=0.4\textwidth]{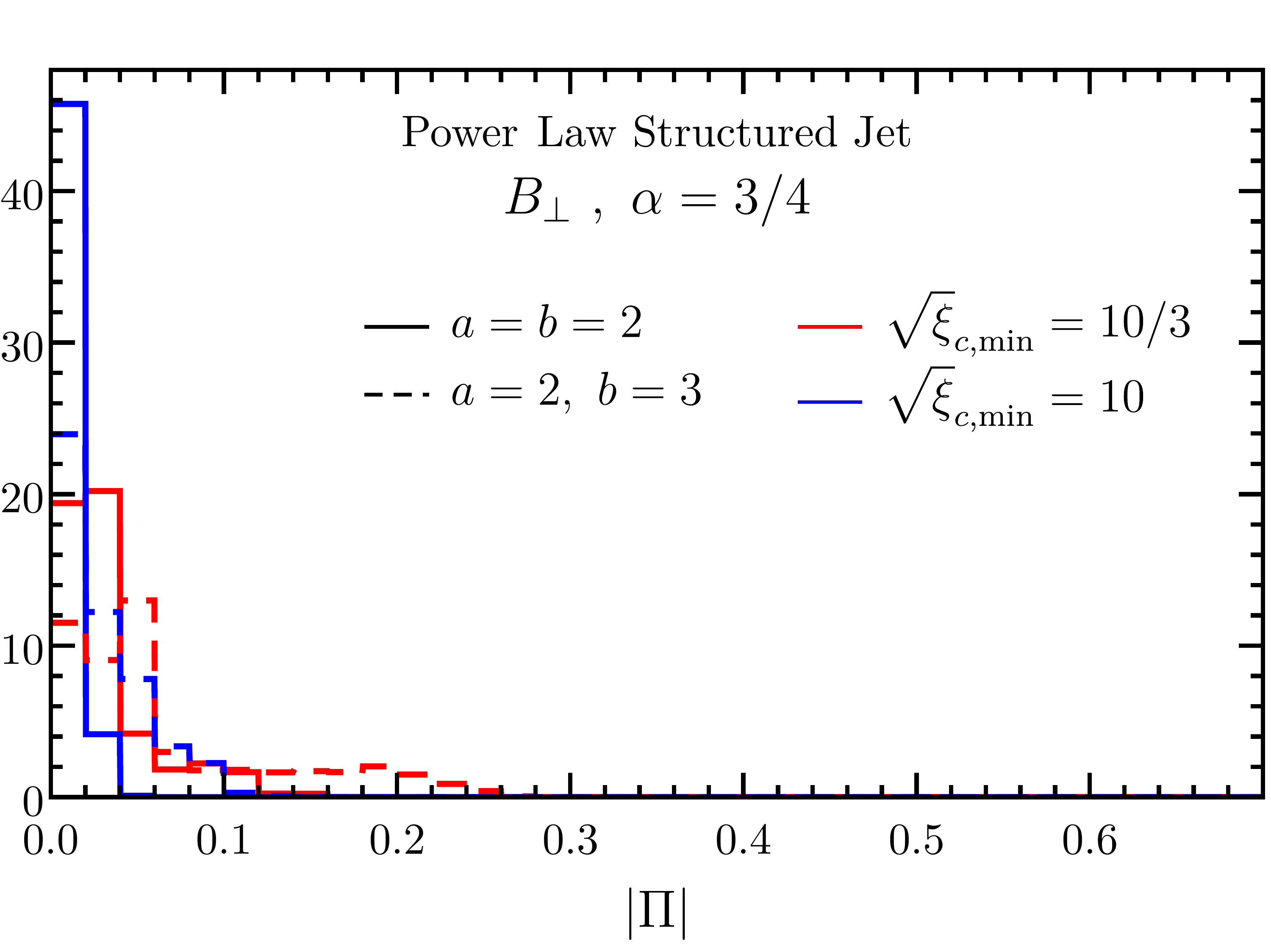}
    \includegraphics[width=0.4\textwidth]{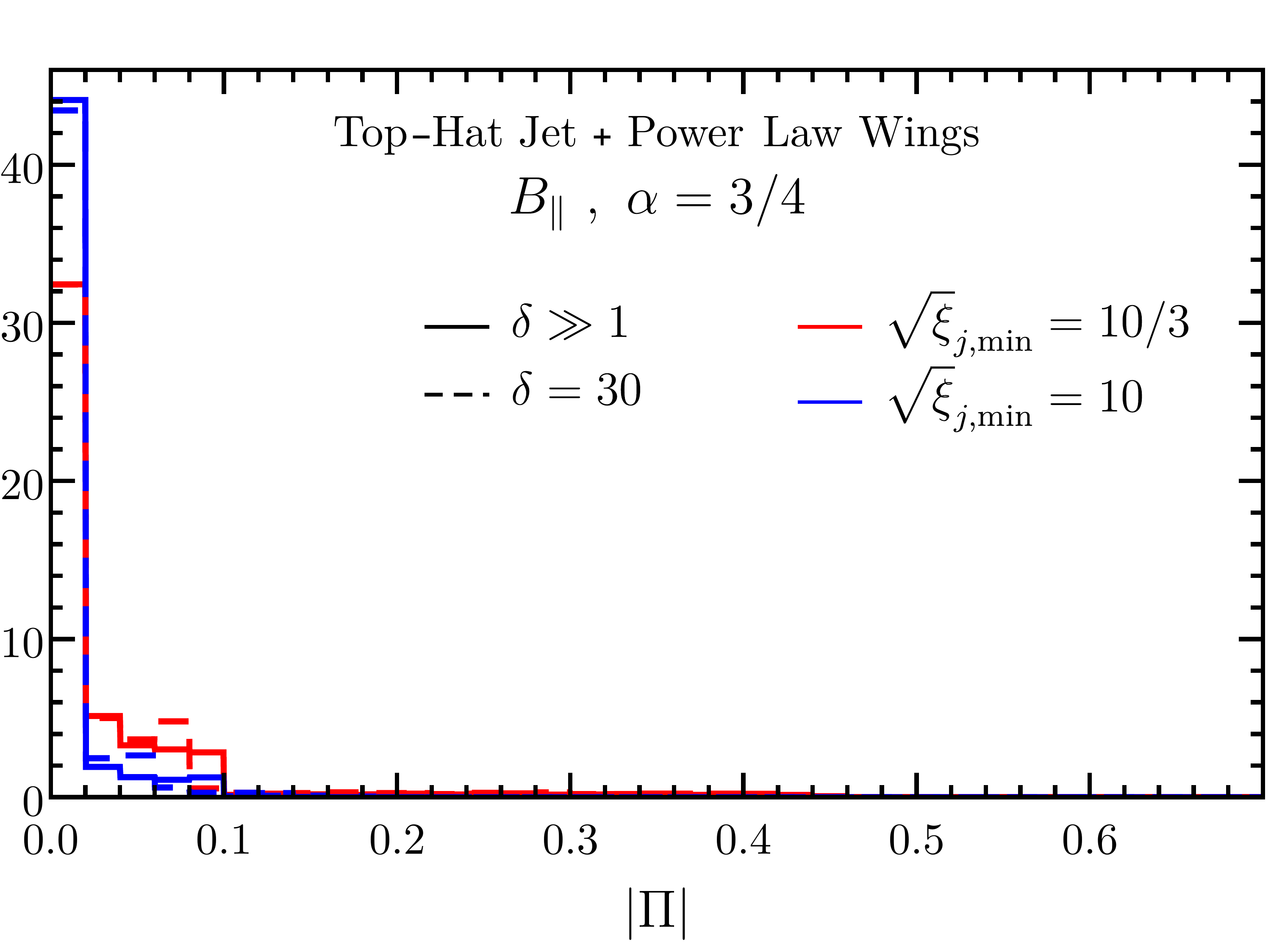}\quad\quad
    \includegraphics[width=0.4\textwidth]{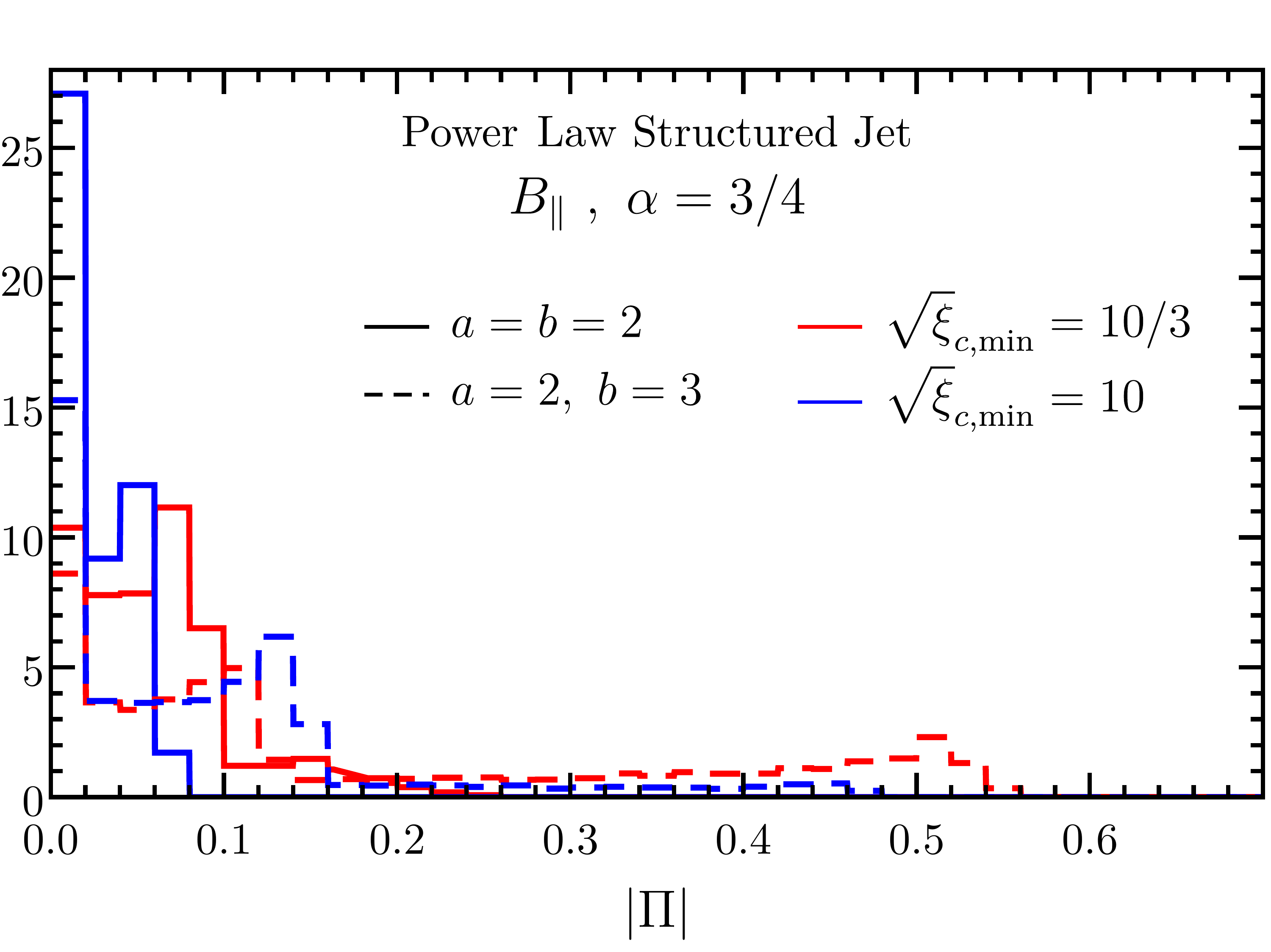}
    \includegraphics[width=0.4\textwidth]{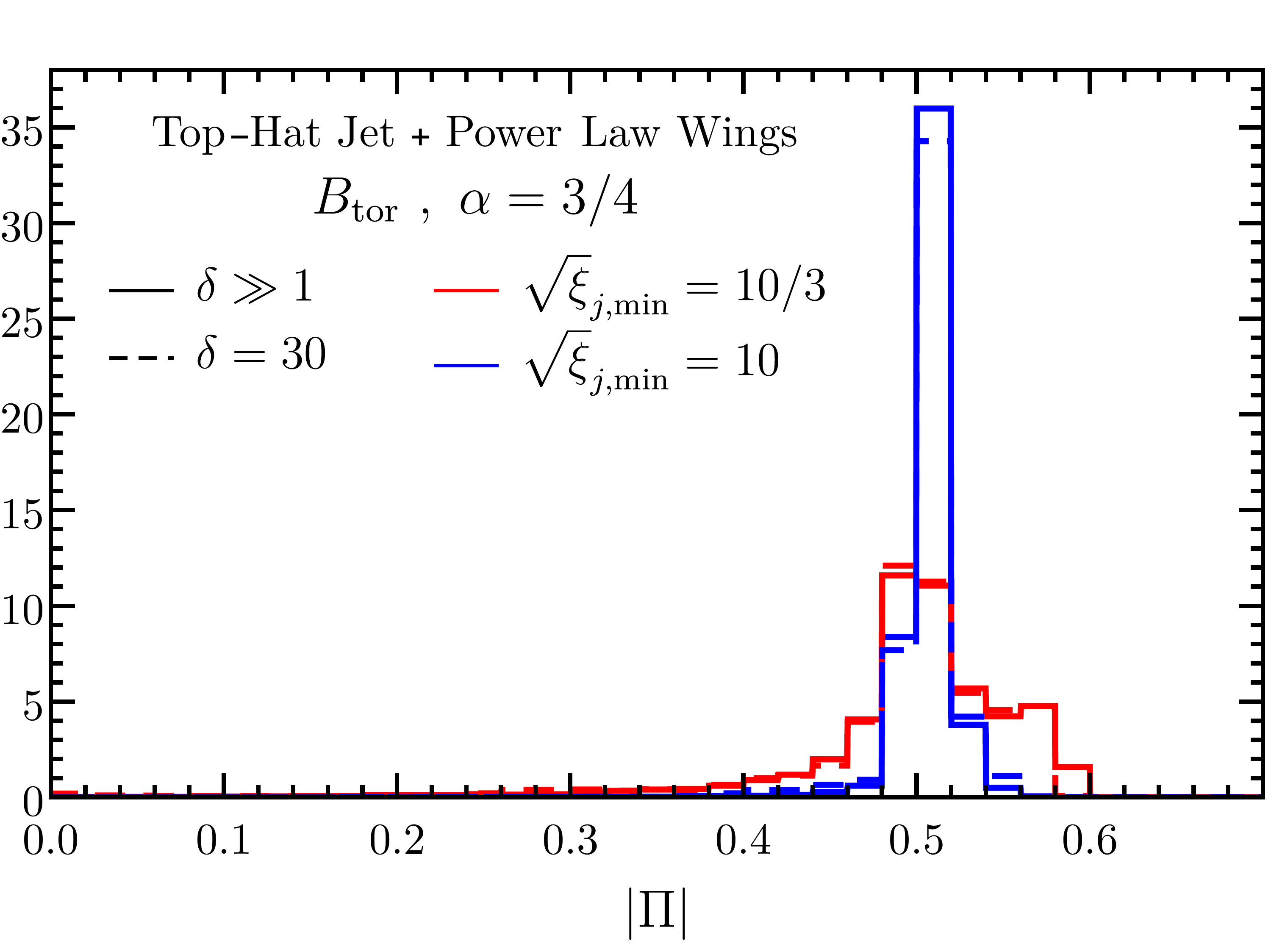}\quad\quad
    \includegraphics[width=0.4\textwidth]{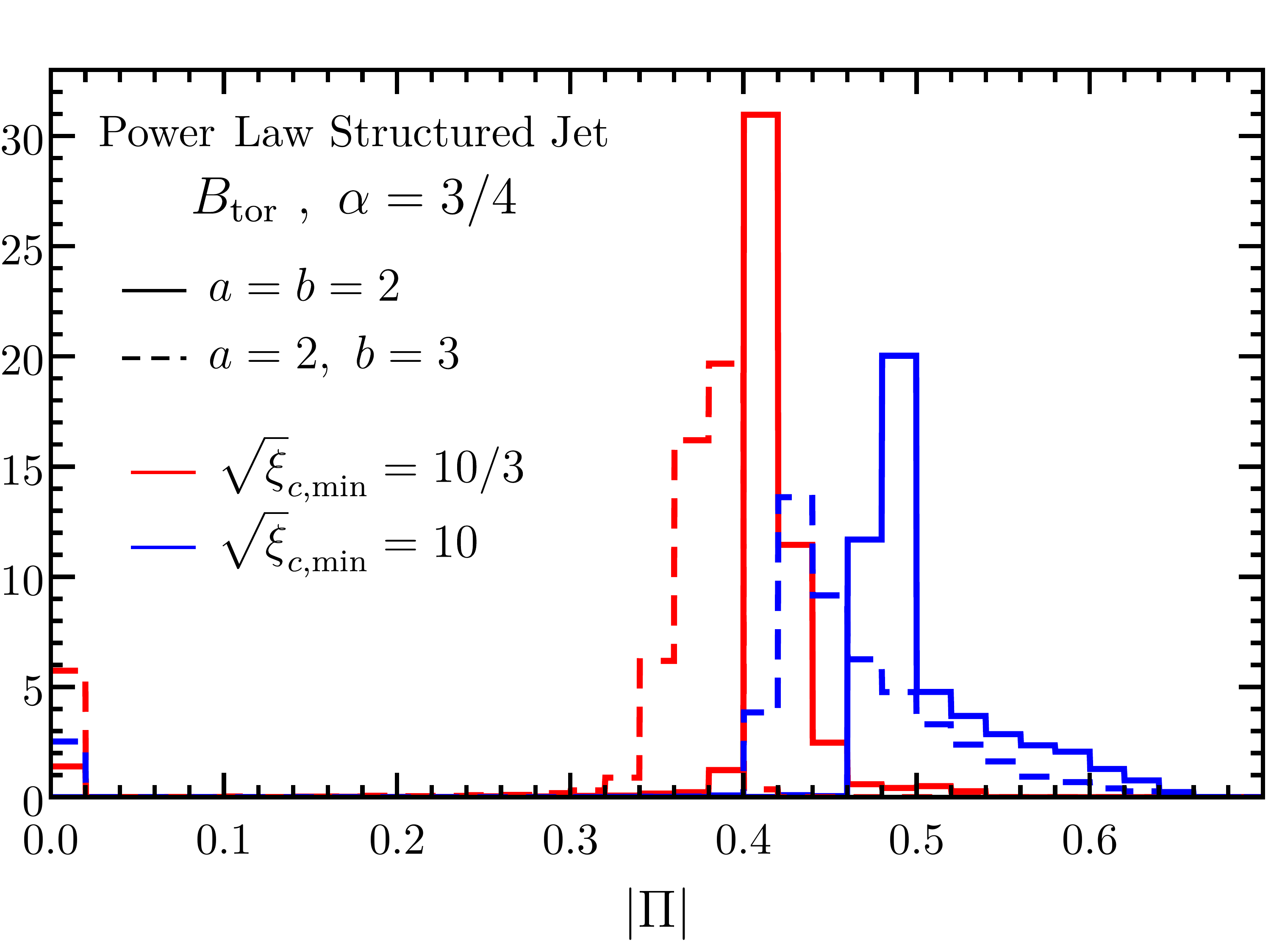}
    \includegraphics[width=0.4\textwidth]{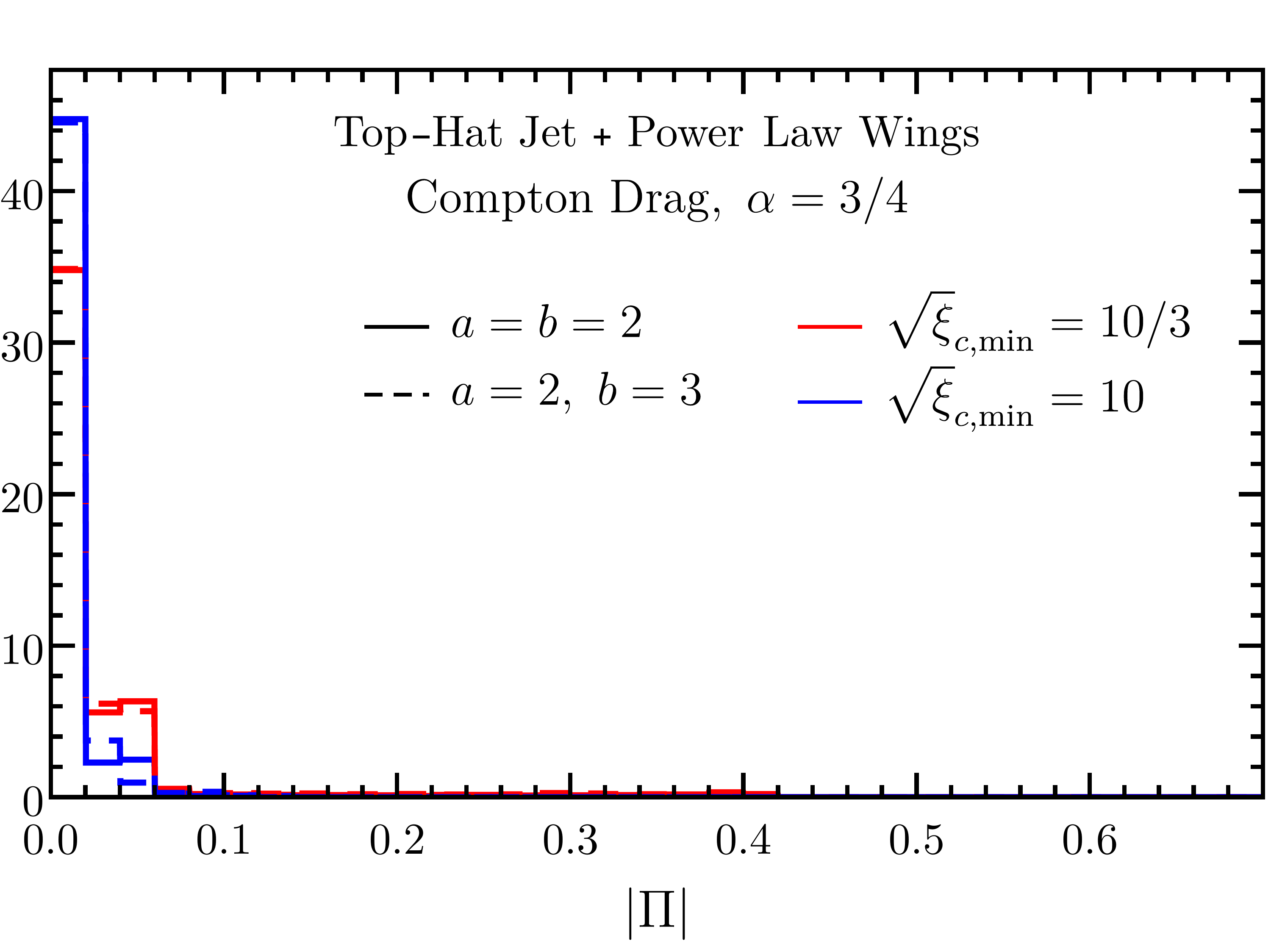}\quad\quad
    \includegraphics[width=0.4\textwidth]{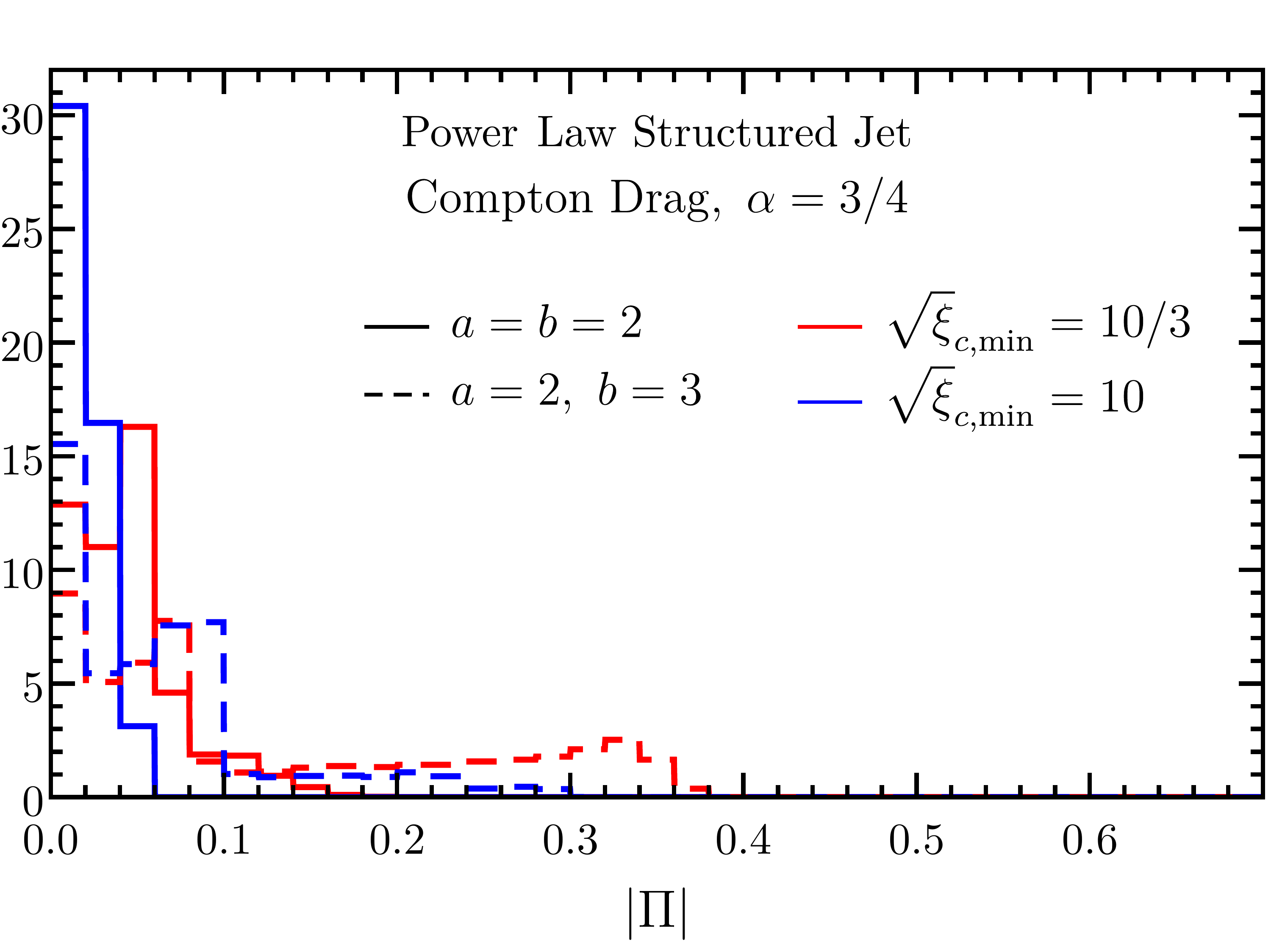}
    \caption{Distribution of $\vert\Pi\vert$ arising from synchrotron radiation and Compton drag when 
    integrated over multiple ($N_p=10$) pulses for different magnetic field configurations and jet structures, 
    with spectral index $\alpha=3/4$. The total sample consists of $10^4$ simulated 
    GRBs with fluence weighted distribution of $q=\theta_{\rm obs}/\theta_j$ or $q=\theta_{\rm obs}/\theta_c$. 
    For each burst, the multiple pulses are randomly sampled from a uniform distribution of 
    $\sqrt\xi_{\{j,c\},\min}\leq\sqrt\xi_{\{j,c\}}\leq\sqrt\xi_{\{j,c\},\max}$, with $\sqrt\xi_{\{j,c\},\max}=3\sqrt\xi_{\{j,c\},\min}$.}
    \label{fig:multiple-pulses}
\end{figure*}
%%%%%% FIGURE %%%%%%%%%%%%%%%%%%%%%%%%%%%%%%%%

In Fig.~\ref{fig:multiple-pulses} we show the distribution of $\vert\Pi\vert$ arising from Compton drag and from synchrotron 
emission for different configurations of the magnetic field and different jet geometries. It is clear that only a 
globally ordered field, such as a toroidal field, can yield high levels of polarization. Any random field component 
($B_\perp$) or a locally ordered ($B_\parallel$) field will statistically most likely produce $\Pi\lesssim5\%-10\%$ only 
if the jet is structured with moderately sharp gradients in $\Gamma$. For a top-hat jet both field configurations yield 
$\Pi\lesssim1\%$. The same is true for the case of Compton drag. Broadly similar results were obtained by \citet{Pearce+19}.

Since the true distribution of $\Gamma$ is unclear, we have tested the robustness of the results shown in 
Fig.~\ref{fig:multiple-pulses} by using two additional distributions of $\xi_j^{1/2}$: (i) a uniform distribution 
in $\ln\xi_j^{1/2}$, and (ii) a log-normal distribution, which are expressed as the following
\begin{eqnarray}
    &&(i)\quad\quad P(\ln\xi_j^{1/2}) = \left[\ln\left(\frac{\xi_{j,\max}^{1/2}}{\xi_{j,\min}^{1/2}}\right)\right]^{-1} \\
    &&(ii)\quad\quad P(\xi_j^{1/2}) = \frac{1}{\xi_j^{1/2}\sigma\sqrt{2\pi}}\exp\left[-\frac{(\ln\xi_j^{1/2}-\mu)^2}{2\sigma^2}\right]~,
\end{eqnarray}
where $\mu$ and $\sigma$ are the mean and standard deviation, respectively, of the distribution which results after taking 
the natural logarithm of the log-normally distributed $\xi_j^{1/2}$. In a population synthesis study carried out by 
\citet{Ghirlanda+13} using a large sample of {\it Swift}/BAT, {\it Fermi}/GBM and {\it CGRO}/BATSE GRBs, it was found that 
the distribution of $\Gamma$ is best represented by a log-normal distribution with $\mu_\Gamma\sim4.5$ and $\sigma_\Gamma\sim1.5$. This 
result was obtained under the assumption that both the $(\nu F_\nu)$-peak and true jet energies in the comoving 
frame are clustered around typical values in a large sample of GRBs. In addition, it was assumed that the product $\theta_j^{2.5}\Gamma=\,$const. 
Here we assume the same underlying distribution of $\Gamma$ with ($\mu_\Gamma,\sigma_\Gamma$), and also assume a fixed $\theta_j=10^{-1}$ in order to switch from $P(\Gamma)$ to $P(\xi_j^{1/2})$.

\begin{figure*}
    \centering
    \includegraphics[width=0.45\textwidth]{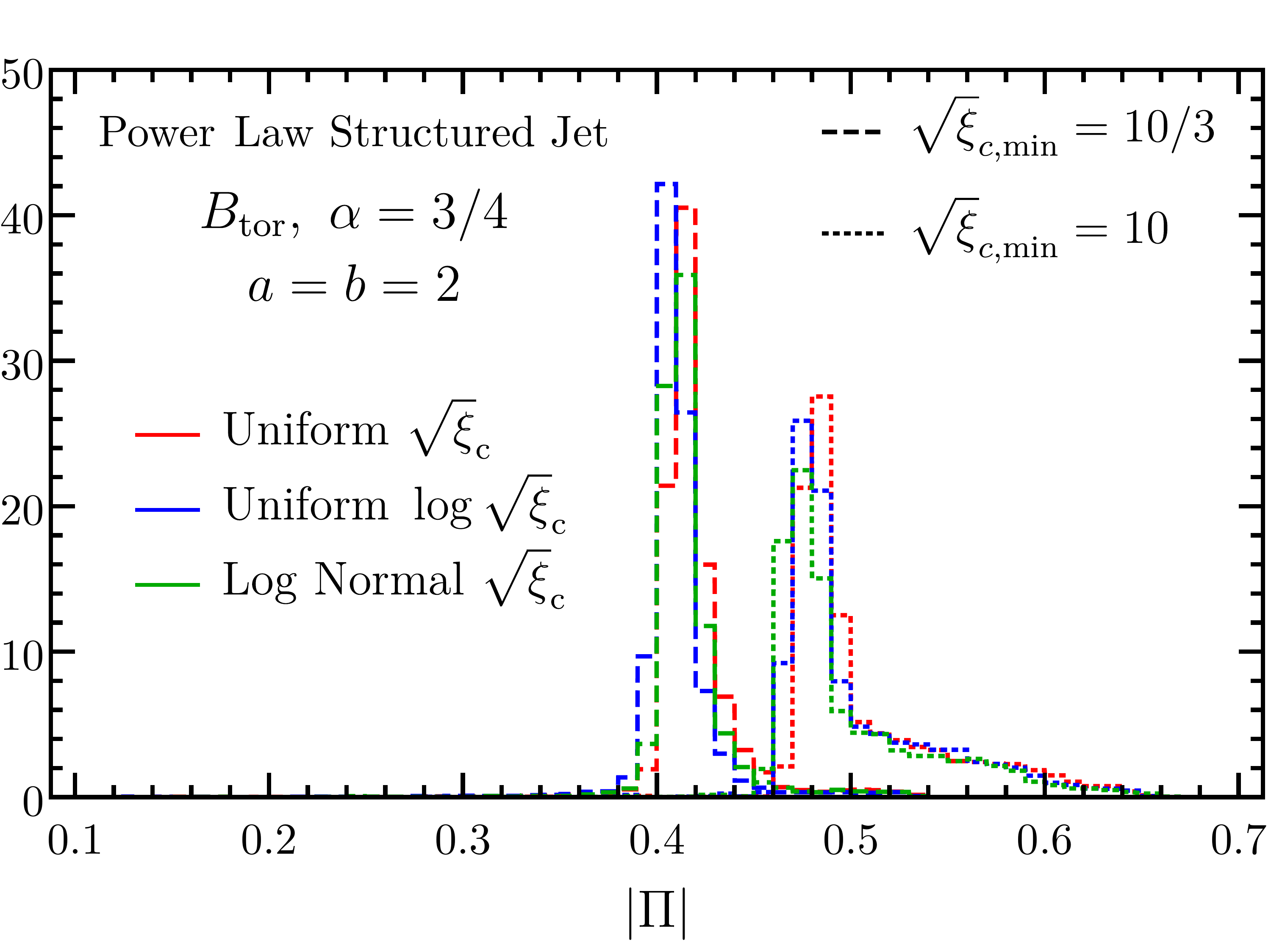}\quad\quad
    \includegraphics[width=0.45\textwidth]{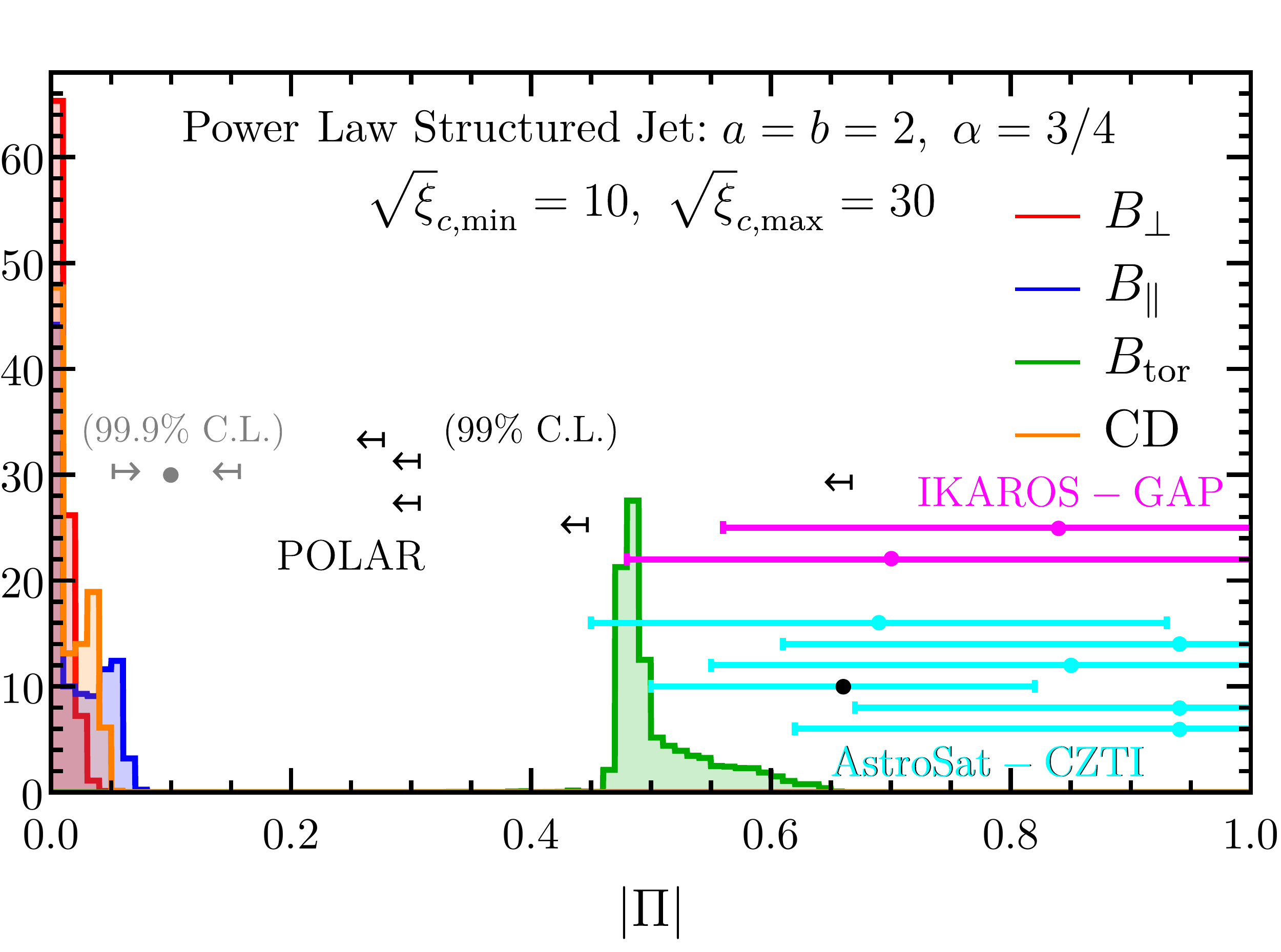}
    \caption{{\it Left}: Comparison of $\vert\Pi\vert$ obtained for different distributions of $\sqrt\xi_c$ when 
    integrating over multiple pulses ($N_p=10$). The magnetic field configuration is that of a globally ordered toroidal 
    field and the outflow has a power law angular structure. All three distributions sample $\sqrt\xi_c$ between 
    $\sqrt\xi_{c,\min}\leq\sqrt\xi_c\leq\sqrt\xi_{c,\max}$ with $\sqrt\xi_{c,\max}=3\sqrt\xi_{c,\min}$. For the log-normal 
    distribution $\mu=2.2$ and $\sigma=0.8$. {\it Right}: Comparison of $\Pi$ arising from synchrotron emission with 
    different magnetic field configurations as well as from Compton drag (CD), for a power law structured 
    jet, with measurements of GRB prompt emission polarization. Data with $\geq3\sigma$ detection significance is 
    shown with $1\sigma$ error bars. Upper limits with 99\% confidence for the POLAR detected five GRBs are shown with black arrows, 
    whereas $\Pi$ and limits derived from the joint analysis of these five GRBs are shown in gray (see Table~\ref{tab:pol-data}). 
    The \citet{Sharma+19} measurement of the average $\Pi=66^{+26}_{-27}\%$ ($\sim\!5.3\sigma$) over the emission episode obtained 
    using \textit{AstroSat}-CZTI is shown with a black dot with cyan error bars. For the models, the spectral index $\alpha=3/4$ 
    where a different value might introduce a small variation, and $\sqrt\xi_c$ of the $N_p=10$ pulses is distributed uniformly.
    }
    \label{fig:Btor-MP-gdist}
\end{figure*}

In the left panel of Fig.~\ref{fig:Btor-MP-gdist}, we compare the results of the three distributions when the magnetic field 
configuration is given by $B_{\rm tor}$ and the outflow has a power law angular structure. We find that all three distributions 
of $\xi_c^{1/2}$ produce very similar predictions for $\vert\Pi\vert$ with a small spread ($<10\%$) which shows that the 
results are quite robust. In the right panel, we compare the predictions of the synchrotron model to measurements of polarization in the prompt 
emission of GRBs that have at least $3\sigma$ detection significance. Apart from a small variation introduced by different 
spectral indices $\alpha$ in the given bursts in the model distributions, the measured high degree of polarization 
appear to favour a globally ordered toroidal field configuration of the outflow magnetic field.

In the right panel of Fig.~\ref{fig:Btor-MP-gdist}, we compare the degree of polarization expected from a power law structured jet, 
when the prompt $\gamma$-ray emission mechanism is either synchrotron or Compton drag, to statistically significant measurements 
and upper limits of $\Pi$. Results from different magnetic field configurations for the synchrotron case are shown. The model distributions 
take into account the limitation on the observability of emission observed at higher $q$ values due to the drop in fluence, as discussed 
earlier. In addition, they factor in the effect of integrating over multiple pulses ($N_p=10$) sampled from a uniform distribution in 
$\sqrt\xi_{c,\rm min}\leq\sqrt\xi_c\leq\sqrt\xi_{c,\rm max}$. The high statistical significance ($\gtrsim3\sigma$) measurements from 
IKAROS-GAP and AstroSat-CZTI are consistent with each other and both show that the prompt $\gamma$-ray emission is highly polarized with 
$50\%\lesssim\Pi\lesssim95\%$ (though with fairly high uncertainties). On the other hand, although the upper limits obtained by POLAR are marginally consistent with the results 
of IKAROS-GAP and AstroSat-CZTI, a joint analysis of five GRBs detected by POLAR shows only a modest level of polarization with a mean 
polarization of $\langle\Pi\rangle\sim10\%$. This result is in tension with those from earlier measurements that showed $\Pi\gtrsim50\%$. 
However, the current sample size is still small and the uncertainties on each measurement are fairly large, which together prevent us from reaching any firm conclusions regarding the dominant emission mechanism of GRB prompt $\gamma$-ray emission.

%%%%%% FIGURE %%%%%%%%%%%%%%%%
%%%%%%%%%%%%%%%%%%%%%%%%%%%%%%

%%%%% DICUSSION & CONCLUSIONS %%%%%%%%%%%%%%%%%%%%%%%%%%%%%%%%%%%%%%%%%%%%%%%
\section{Discussion \& Conclusions}\label{sec:diss}
The measurement of linear polarization in the prompt emission of GRBs is of great interest as 
it offers very useful insights into the composition of the outflow and the structure of its
magnetic field. This can further be used to pin down the exact radiation mechanism that gives 
rise to the prompt GRB gamma-ray emission. In this work, we discuss relevant radiation mechanisms that 
have been proposed to explain the prompt emission and that can also yield different levels of 
linear polarization. Furthermore, we have used the predictions for the polarization from these 
mechanisms (which depend on the jet geometry, viewing angle, magnetic field structure, and the 
spectral parameters), to ask the question what is the most likely explanation for a given polarization 
measurement. We have shown that either a single secure measurement of $50\%\lesssim\Pi\lesssim65\%$ 
or measuring $\Pi\gtrsim20\%$ in most GRBs within a large enough sample (using MC simulations), would 
strongly favor synchrotron emission from a transverse magnetic field ordered on angles 
$\gtrsim\!1/\Gamma$ around our line of sight (like a global toroidal field, $B_{\rm tor}$, for 
$1/\Gamma<\theta_{\rm obs}<\theta_j$).

In \S\ref{sec:off-axis}, we showed the predictions for $\Pi$ from synchrotron emission 
for three different magnetic field configurations in a top-hat jet. In the case of the random magnetic 
field that is completely in the plane of the ejecta ($B_\perp$), high levels of 
$\Pi$ are only achieved for a particular jet geometry and LOS. In this case, the jet has to 
be narrow with a uniform core and fairly sharp edges. On top of that, the observer's LOS 
must be very close to the edge of the jet with $q\sim1+\xi_j^{-1/2}$. The probability of observing close to the 
edge is $\sim(\Gamma\theta_j)^{-1}$, where typically $\Gamma\theta_j\sim10$, and so roughly 
$10\%$ of the bursts from a top-hat jet are seen slightly off-axis from near the edge of the jet. 
Majority of the bursts, especially at high redshift, must then be observed on-axis with $q<1$, 
otherwise the sharp drop in fluence for $q>1$ would render the burst too dim to be observed (let alone to be bright enough for their polarization to be measured). For 
this very reason, measurement of high levels of polarization arising for off-axis observers ($q>1$) 
when the outflow magnetic field is parallel to the local velocity vector everywhere ($B_\parallel$) 
will be challenging. In the case of a top-hat jet and for $q\lesssim1$ only an ordered transverse magnetic field, 
such as a globally ordered toroidal field, in the outflow can yield the highest degree of polarization from synchrotron emission.

On the other hand, a structured jet offers a better chance for measuring higher $\Pi$ for off axis 
observers for all magnetic field configurations. However, as shown in \S\ref{sec:struc-jets}, 
in the case of $B_\perp$ and $B_\parallel$ steep 
gradients in $\Gamma(\theta)$ are needed, otherwise it yields negligible polarization. In the case of the 
top-hat jet the necessity of having a sharp gradient in $\Gamma$ was replaced
by the jet having a sharp edge. 
The $B_{\rm tor}$ configuration yet again yields the highest levels of polarization and does not require 
steep gradients in $\Gamma$. This model overcomes the problematic requirement of 
having a special LOS to observe a high degree of polarization, which makes this configuration robust from an 
observational standpoint. It also implies that majority of GRBs should show high polarization levels with 
$\Pi\gtrsim20\%$. This can be potentially tested as the observed sample grows and measurements become better 
with upcoming more sensitive instruments.

An important consideration in the case of structured jets that are viewed off-axis is that compactness arguments 
require $\Gamma(\theta)$ to be shallow, e.g. $b\lesssim1$ for a power law jet. However, such profiles don't yield 
any detectable polarization when the magnetic field is not ordered on large scales, such as in the case of 
$B_\perp$ and $B_\parallel$; the same is also true for Compton drag. For steeper profiles, the observer can only 
see emission from close to the core and cannot be too off axis with $q\lesssim2$. This constraint would also favour 
a large scale ordered magnetic field if $\Pi>20\%$ is observed even in a single burst.

The Compton drag model (\S\ref{sec:CD}) suffers from the same difficulty as 
the synchrotron model with $B_\perp$ and will mostly yield low levels of $\Pi$ unless $q\gtrsim1$ 
and the jets are quite narrow. It was shown in \citet{Lazzati+04} that the top-hat jet must be narrow with 
$\xi_j\lesssim25$ in order to obtain $\Pi\gtrsim40\%$ while getting $\Pi\gtrsim95\%$ for 
extremely narrow jets with $\xi_j=4\times10^{-2}$. To distinguish between the synchrotron emission model, 
especially with $B_\perp$ and $B_\parallel$ field configurations, and the Compton drag scenario, one will have 
to rely on spectral modeling. In synchrotron emission the spectral index is rather limited to $-1/3\leq\alpha\lesssim3/2$ 
which also limits the local maximum degree of polarization to $50\%\leq\Pi_{\max}\lesssim75\%$. There is no such 
limitation on $\Pi_{\max}$ in the Compton drag model. Therefore, detecting spectrally harder bursts that violate 
the synchrotron line-of-death can be one way to discriminate between the two emission models.

In the photospheric emission model (\S\ref{sec:Photospheric}), with no dissipation below the photosphere, $\Pi$ is rather 
limited to $\lesssim15\%-20\%$. In order to achieve even this level of polarization the jet must be structured 
and have steep gradients in its energy per unit solid angle and $\Gamma$ with $\theta$. The angular structure 
of the jet is unclear and in the simplest scenario of a top-hat jet the photospheric model will 
yield negligible polarization for $q<1$. Spectrally, this model can be distinctly recognized as it 
produces a quasi-thermal spectrum, which has only been seen in a handful of bursts. On the other hand, 
dissipative photosphere models yield Band-like spectrum where the peak forms as a result of 
multiple Compton scatterings by heated electrons (or $e^\pm$-pairs) below the photosphere. Therefore, 
the peak itself will have negligible polarization, however, if the source of soft photons is 
synchrotron, which will be the dominant component below the peak, then the best case 
scenario can yield $\Pi\lesssim50\%$ \citep{Lundman+18}.

Finally, only an ordered magnetic field that has a coherence length comparable or larger than the size of the visible emitting region can consistently produce high levels of polarization with $\Pi\sim\Pi_{\max}$. 
However, if the size of coherent patches is smaller than that of the visible region so that $N_p$ patches contribute to a single emission episode, or alternatively $N_p$ intrinsicaly coherent (single-patch) but mutually incoherent pulses are integrated over in the same GRB, the this will reduce the maximum polarization by a factoir of $\sim\sqrt{N_p}$. 
In addition, since the PA will be randomly oriented for emission from any given patch (or pulse), time-resolved (pulse-resolved) polarization analysis should reveal significant oscillations of the PA between pulses. This prediction is in contrast with other field 
configurations where a constant PA should be observed, except for a $90^\circ$ flip. \citet{Inoue+11} studied the 
creation of ordered magnetic fields via the Richtmyer-Meshkov instability (RMI) in internal shocks using special relativistic 
magneto-hydrodynamic (MHD) simulations. It was realized there that the RMI would generate a large number ($\sim10^3$) 
of incoherent patches which would lead to $\Pi\sim2\%$. Measurements of higher levels of polarization would 
necessarily violate this estimate and point either to another mechanism of producing such ordered fields or the 
outflow having a large scale globally ordered field.

%%%%%% IMPLICATION OF MEASURING PI > 20% %%%%%%%%%%%%%%%%%%%%%%%%%%%%%%%%%%%%%%%%%%%%%%%%%
\subsection{Implications of measuring $\Pi>20\%$}
High degrees of polarization have been measured now in the prompt emission of several GRBs albeit with only modest 
statistical significance. A firm detection of $\Pi>20\%$ in several GRBs would point towards a globally 
ordered transverse magnetic field configuration in the outflow, for which a good candidate is toroidal magnetic field. It will 
also strongly indicate that the underlying dominant emission mechanism for the GRB prompt emission is synchrotron. For the 
toroidal field case and for the typical value of the jet parameter $\xi_j=10^2$, the range of the observed degree of polarization for 
a single burst is $0.4\lesssim\Pi/\Pi_{\max}\lesssim0.85$ for different values of the spectral index $-1/3\leq\alpha\leq3/2$, 
which corresponds to $20\%\lesssim\Pi\lesssim68\%$, but it will never be larger than $75\%$. Also, in this case, both 
top-hat and structured jets would yield similar levels of polarization in a large sample of GRBs, with $\Pi\sim40\%-50\%$ 
for $\alpha=3/4$. This will make it hard to distinguish between the two jet geometries based on polarization alone. 

A firm detection of GRB gamma-ray polarization requires high-fluence sources, and in turn viewing angles within or very close to the jet core, $q\lesssim1$. This limit on $q$ is further substantiated by compactness arguments. 
If only a small fraction ($\sim10\%$) of GRBs show $\Pi\gtrsim20\%$ this would favor models in which there is no net polarization for a spherical flow or LOS well within a uniform jet ($q<1-\xi_j^{-1/2}$), and require instead a special line of sight, $q\sim1+\xi_j^{-1/2}$. Such models include emission from a top-hat jet and either synchrotron with $B_\perp$ or $B_\parallel$, or Compton drag. 
It would naturally also disfavor synchrotron emission from a large scale ordered magnetic field such as $B_{\rm tor}$.

Statistically significant measurements of GRB prompt emission polarization will increase with the advent of new high-energy 
polarimeters and with the observations of very bright GRBs with currently operating instruments 
\citep[see, e.g.][for a review of various instruments]{McConnell-17}. Comparison of the moderately statistically significant 
measurements ($\gtrsim3\sigma$) with the different emission models and magnetic field configurations strongly favour the 
existence of a toroidal (or other transverse and globally ordered) magnetic field in the outflow and that the underlying prompt GRB emission mechanism is synchrotron.

The models considered in this work have assumed an axisymmetric jet or outflow angular structure, which leads to a constant PA, $\theta_p$, or at most a change of $\Delta\theta_p=90^\circ$ in $\theta_p$. However, it is important to keep in mind that non-axisymmetric effects can lead to arbitrary changes in $\theta_p$. In particular, a ``mini-jet'' type of emission model, in which each spike is produced by plasma moving relativistically w.r.t the bulk outflow frame and in a random direction within that frame, could produce a random $\theta_p$ for each pulse. In such a case the polarization from different pulses would add up incoherently. This is analogous to the patchy shell model in which the outflow has a single bulk $\Gamma$ but the angular distribution of the emission brightness is highly non-uniform. However, such strong variations within the visible region of $1/\Gamma$ around the LOS, which is also in lateral causal contact, would be very hard to maintain in the flow, while a mini-jet model does not suffer from such a difficulty. 

%%%%%%%%%%%%%%%%%%%%%%%%%%%%%%%%
\section*{Acknowledgements}
%%%%%%%%%%%%%%%%%%%%%%%%%%%%%%%%
We thank the anonymous referee for useful comments. 
R.G. and J. G. are supported by the Israeli Science Foundation under grant No. 719/14. 
We thank Merlin Kole for organizing and the hospitality at the conference 
`Shedding new light on Gamma-Ray Bursts with polarization data' in Geneva where part of 
this work was completed.

%%%%% BIBLIOGRAPHY %%%%%%%%%%%%%%%%%%%%%%%%%%%%%%%%%%%%%%%%%%%%%

%%%%%% APPENDIX %%%%%%%%%%%%%%%%%%%%%%%%%%%%%%%%%%%%%
\appendix
\section{Time-integrated polarization for an ultra-relativistic structured jet}
\label{sec:app-time-integrated-pol-strucjet}
Here we present a general formalism for obtaining time-integrated polarization for a structured jet that is 
emitting synchrotron radiation. This can be easily generalized further to other 
radiation mechanisms discussed in this work. Also, general expressions valid for a uniform jet are pointed out. 
The flow is assumed to be ultra-relativistic with $\Gamma\gg1$ and the emission is 
assumed to arise from an infinitely thin shell. The instantaneous degree of polarization follows from 
Eq.~(\ref{eq:stokes-general},\ref{eq:flux-thin-shell},\ref{eq:Lnu-synchro}) and can be expressed as 
%\begin{eqnarray}
%    \Pi(t_z) = \frac{Q(t_z)}{I(t_z)} 
%    && = \frac{\int\delta_D^3L'_{\nu'}\Pi'\cos(2\theta_p)d\tilde\Omega}{\int\delta_D^3L'_{\nu'}d\tilde\Omega} \\
%    && = \frac{\int\delta_D^{3+\alpha}\Lambda\Pi'\cos(2\theta_p)d\tilde\Omega}
%    {\int\delta_D^{3+\alpha}\Lambda d\tilde\Omega}
%\end{eqnarray}
\begin{equation}
    \Pi(t_z) = \frac{Q(t_z)}{I(t_z)} 
    = \frac{\int\delta_D^3L'_{\nu'}\Pi'\cos(2\theta_p)d\tilde\Omega}{\int\delta_D^3L'_{\nu'}d\tilde\Omega}
\end{equation}
where $t_z$ is the arrival time of photons in the cosmological rest frame of the source and is expressed 
through the equal arrival time condition (Eq.~\ref{eq:eats} in the text)
\begin{equation}
    \frac{t_{\rm obs}}{(1+z)}\equiv t_z = t-\frac{r\tilde\mu}{c}~.
\end{equation}
The time-integrated polarization is obtained from
\begin{equation}
    \Pi = \frac{\int_{t_{z,0}}^{t_{z,\max}}Q(t_z)dt_z}{\int_{t_{z,0}}^{t_{z,\max}}I(t_z)dt_z}
\end{equation}
where $t_{z,0}$ is the arrival time of the first photon. To analytically integrate over all the arrival times, we can express 
$dt_z$ in terms of $dr$ from the equal arrival time condition for constant $\tilde\mu$ and $\tilde\varphi$, such that
\begin{equation}
    dt_z = \frac{(1-\beta\tilde\mu)}{\beta c}dr = \frac{\delta_D^{-1}}{\Gamma\beta c}dr\approx\frac{\delta_D^{-1}}{\Gamma c}dr
\end{equation}
For a uniform jet, the factor of $\Gamma$ would be constant with polar angle $\theta$ (and also assumed constant with $r$ here) 
and cancel in the final expression for $\Pi$, however, it won't cancel if the jet is structured since $\Gamma=\Gamma(\theta)$.

Next, the radial integral can be collapsed to a delta function in $r$ since the integrand or any other parameters are assumed 
to be independent of $r$, which simplifies the treatment and yields the time-integrated polarization
%\begin{equation}
%    \Pi = \frac{Q}{I} = \frac{\displaystyle\int\frac{\delta_D^{2+\alpha}}{\Gamma}\Lambda\Pi'\cos(2\theta_p)d\tilde\Omega}
%    {\displaystyle\int\frac{\delta_D^{2+\alpha}}{\Gamma}\Lambda d\tilde\Omega}~.
%\end{equation}
\begin{equation}\label{eq:Pi-r-independ}
    \Pi = \frac{Q}{I} = \frac{\displaystyle\int\frac{\delta_D^2}{\Gamma}L'_{\nu'}\Pi'\cos(2\theta_p)d\tilde\Omega}
    {\displaystyle\int\frac{\delta_D^2}{\Gamma}L'_{\nu'} d\tilde\Omega}~.
\end{equation}

We can express $d\tilde\Omega = d\tilde\mu d\tilde\varphi$, where for small angles 
$d\tilde\mu\approx\tilde\theta d\tilde\theta = \frac{1}{2}d(\tilde\theta^2) = \frac{1}{2\Gamma_c^2}d(\Gamma_c^2\tilde\theta^2) = \frac{1}{2\Gamma_c^2}d\tilde\xi$. 
Here we have normalized the polar angle measured from the LOS in terms of the beaming angle of the core 
emission $\Gamma_c^{-1}$ in a structured jet, which is a constant, and further defined the useful quantity 
$\tilde\xi\equiv\Gamma_c^2\tilde\theta^2$. Notice that this is the same parameterization as used 
in Eq.~(\ref{eq:ultra-rel-doppler}) since for a top-hat jet $\Gamma\to\Gamma_c$ as it doesn't vary with 
polar angle $\theta$. The Doppler factor can also be expressed using the same parameterization in the 
ultra-relativistic limit, which yields
\begin{equation}
    \delta_D = \frac{1}{\Gamma(1-\beta\tilde\mu)}\approx\frac{2\Gamma}{1+\Gamma^2\tilde\theta^2}
    = 2\Gamma_c\frac{\hat\Gamma}{1+\hat\Gamma^2\tilde\xi} = 2\Gamma_c\frac{\hat\Gamma}{1+\hat\xi}~,
\end{equation}
where $\hat\xi\equiv\Gamma^2\tilde\theta^2=\hat\Gamma^2\tilde\xi$. Again, for a uniform jet $\hat\Gamma\equiv\Gamma/\Gamma_c=1$ 
and therefore $\delta_D\propto(1+\tilde\xi)^{-1}$.

If the azimuthally symmetric jet has angular structure, its kinetic energy per unit solid angle, 
$\epsilon(\theta)\equiv dE_k(\theta)/d\Omega$, and LF, $\Gamma(\theta)$, would vary with polar angle away 
from the jet symmetry axis. The corresponding isotropic equivalent 
kinetic energy is given by $E_{k,\rm iso}(\theta)=4\pi\epsilon(\theta)$. If this energy is radiated with 
efficiency $\epsilon_\gamma$ over a lab-frame time $\Delta t_{\rm lab}=\Delta r/\beta c$, then 
the radiated power, which is a Lorentz invariant, can be expressed as
\begin{equation}\label{eq:Liso-a}
L'_{\rm iso}(\theta) = \frac{dE_{\rm rad,iso}'}{dt'} = \frac{dE_{\rm rad,iso}}{dt_{\rm lab}} 
= \frac{4\pi\epsilon_\gamma\epsilon(\theta)}{(\Delta r/\beta c)}~.
\end{equation}
Next we assume that the normalization of the fluid-frame isotropic spectral luminosity, without 
the factor $\Lambda$ (defined in Eq.~(\ref{eq:Lambda}) in the text) that is associated to a 
particular LOS, is given by an infinite power law, such that 
$L'_{\nu'} = L'_{\nu_p'}(\nu'/\nu_p')^{-\alpha}$ for $\nu'\geq\nu_p'$, 
where $\nu_p'$ is a characteristic frequency at which most of the power is radiated, 
i.e. where $\nu'L'_{\nu'}$ peaks. 
Integration over $\nu'$, while neglecting any contribution from frequencies $\nu'<\nu_p'$ (accounting for this contribution would slightly modify the factor $\chi$ below, which would generally remain of order unity), 
which is assumed negligible here, then yields the bolometric power, 
\begin{equation}\label{eq:Liso-b}
L'_{\rm iso} = \int_{\nu_p'}^{\nu_{\max}'} d\nu' L'_{\nu'} 
= \frac{\nu_p'L'_{\nu_p'}}{(\alpha-1)}\left[1-\fracb{\nu_{\max}'}{\nu_p'}^{1-\alpha}\right]
= \chi\nu_p'L'_{\nu_p'}~
\end{equation}
for $\alpha>1$. When $(\nu'_{\max}/\nu_p')\gg1$, $\chi\to(\alpha-1)^{-1}$. 

Equating Eqs.~(\ref{eq:Liso-a} \& \ref{eq:Liso-b}) yields the comoving spectral luminosity in 
terms of the energy per unit solid angle of the flow, $\epsilon = \epsilon_c\Theta^{-a}$, which e.g. is 
assumed here to vary as a power law,
\begin{equation}\label{eq:Lnu}
L'_{\nu'} = \frac{4\pi\epsilon_\gamma\epsilon(\theta)}{\chi\nu_p'(\Delta r/\beta c)}\fracb{\nu'}{\nu'_p}^{-\alpha}
= \frac{4\pi\epsilon_\gamma\beta c}{\chi\nu_p'\Delta r}\fracb{\nu'}{\nu'_p}^{-\alpha}
\epsilon_c\Theta^{-a}\ ,
\end{equation}
When assuming a similar power law dependence for $\Gamma(\theta)-1$ (the kinetic energy per unit rest energy), the following expressions
(as described by Eq.~(\ref{eq:PLJ}) in the text) are obtained, 
\begin{eqnarray}\label{eq:Lnu-Gamma}
    &\displaystyle\frac{L'_{\nu'}}{L'_{\nu',0}} = \Theta^{-a},\quad\quad\frac{\Gamma(\theta)-1}{\Gamma_c-1} = \Theta^{-b}& \\
    &\Theta = \sqrt{1+\frac{\theta^2}{\theta_c^2}} = \sqrt{1+\frac{\Gamma_c^2\theta^2}{\Gamma_c^2\theta_c^2}} = \sqrt{1+\frac{\xi}{\xi_c}}~.&
\end{eqnarray} 
In the ultra-relativistic limit, 
$(\Gamma(\theta)-1)/(\Gamma_c-1)\approx\Gamma(\theta)/\Gamma_c\equiv\hat\Gamma(\theta)$. 
Plugging in Eqs.~(\ref{eq:Lnu} \& \ref{eq:Lnu-Gamma}) and the factor $\Lambda$ in 
Eq.~(\ref{eq:Pi-r-independ}) gives
\begin{equation}
\Pi = \frac{Q}{I} 
	= \frac{\displaystyle\int\delta_D^{2+\alpha}\hat\Gamma^{-1}\Theta^{-a}\Lambda\Pi'\cos(2\theta_p)d\tilde\Omega}
    {\displaystyle\int\delta_D^{2+\alpha}\hat\Gamma^{-1}\Theta^{-a}\Lambda d\tilde\Omega}~.
\end{equation}
Here we have explicitly assumed $\nu_p'$ to be a constant, however, in general 
$\nu_p' = \nu_p'(t',\theta)$. In addition, the above expression is valid so long 
$\nu'=\delta_D^{-1}\nu > \nu_p'$ or in general, $\nu'$ falls in the same power law segment 
for which $L'_{\nu'}\propto\nu'^{-\alpha}$. 

The emissivity and $\Gamma(\theta)$ profiles, and the variable $\hat\xi$, depend on $\xi\propto\theta^2$ and the relation between these 
is obtained from the geometry of the problem. In general,
\begin{equation}
    \mu = \mu_{\rm obs}\tilde\mu-\cos\tilde\varphi\sqrt{(1-\mu_{\rm obs}^2)(1-\tilde\mu^2)}~,
\end{equation}
where $\mu\equiv\cos\theta$, $\mu_{\rm obs}\equiv\cos\theta_{\rm obs}=\cos(q\theta_c)$ with $q=\theta_{\rm obs}/\theta_c$, 
and $\tilde\mu\equiv\cos\tilde\theta$. For the ultra-relativistic case, and in the small angle limit, the above relation simplifies to
\begin{eqnarray}
    &&\theta^2 \approx \tilde\theta^2 + q^2\theta_c^2 + 2q\tilde\theta\theta_c\cos\tilde\varphi \\
    \Rightarrow&&\xi = \tilde\xi + q^2\xi_c + 2q\sqrt{\tilde\xi\xi_c}\cos\tilde\varphi~,
\end{eqnarray}
where the second equation was obtained by simply multiplying both sides by $\Gamma_c$. This defines all the relevant set of equations 
that are needed to calculate time-integrated $\Pi$, for which the final expression becomes
\begin{equation}
    \Pi = \frac{\int_0^{\tilde\xi_{\max}}d\tilde\xi\int_0^{2\pi}d\tilde\varphi\delta_D^{2+\alpha}\hat\Gamma^{-1}
    \Lambda(\hat\xi,\tilde\varphi)\Theta^{-a}\Pi'(\hat\xi,\tilde\varphi)\cos(2\tilde\varphi)}
    {\int_0^{\tilde\xi_{\max}}d\tilde\xi\int_0^{2\pi}d\tilde\varphi\delta_D^{2+\alpha}\hat\Gamma^{-1}\Lambda(\hat\xi,\tilde\varphi)\Theta^{-a}}~,
\end{equation}
where $\tilde\xi_{\max}>10^2$ is chosen appropriately which guarantees a converged result.

\label{lastpage}

\begin{thebibliography}{}

\bibitem[Abramowicz, Novikov, \& Paczy\'{n}ski(1991)]{Abramowicz+91}
Abramowicz, M. A., Novikov, I. D., \& Paczy\'{n}ski, B. 1991, \apj, 369, 175

\bibitem[Band et al.(1993)]{Band+93}
Band, D. et al. 1993, \apj, 413, 281

\bibitem[Begelman \& Sikora(1987)]{BS87} 
Begelman, M. C. \& Sikora, M. 1987, \apj, 322, 650

\bibitem[B\'{e}gu\'{e} et al.(2013)]{Begue+13}
B\'{e}gu\'{e}, D., Siutsou, I. A., \& Vereshchagin, G. V. 2013, \apj, 767, 139

\bibitem[Beloborodov(2010)]{Beloborodov-10}
Beloborodov, A. M. 2010, MNRAS, 407, 1033

\bibitem[Beloborodov(2011)]{Beloborodov2011} 
Beloborodov, A. M. 2011, \apj, 737, 68

\bibitem[Beloborodov \& M\'{e}sz\'{a}ros(2017)]{Beloborodov-Meszaros-17}
Beloborodov, A. M. \& M\'{e}sz\'{a}ros, P. 2017, Space Sci. Rev., 207, 87

\bibitem[Beniamini \& Nakar(2018)]{Beniamini-Nakar-18}
Beniamini, P. \& Nakar, E. 2018, ArXiv eprints, arXiv:1808.07493

\bibitem[Burgess et al.(2018)]{Burgess+18}
Burgess, J. M. et al. 2018, arXiv eprints: arXiv:1810.06965

\bibitem[Burgess et al.(2019)]{Burgess+19}
Burgess, J. M. et al. 2019, arXiv eprints: arXiv:1901.04719

\bibitem[Chattopadhyay et al.(2017)]{Chattopadhyay+17}
Chattopadhyay, T. et al. 2017, ArXiv eprints, arXiv:1707.06595

\bibitem[Coburn \& Boggs(2003)]{CB03}
Coburn, W. \& Boggs, S. E. 2003, Nature, 423, 415

\bibitem[Covino et al.(2003)]{Covino03}
Covino, S., Ghisellini, G., Lazzati, D., \& Malesani, D. 2003, in ASP Conf. Ser.
312, Third Rome Workshop on Gamma-Ray Bursts in the Afterglow Era, ed.
M. Feroci, F. Frontera, N. Masetti, \& L. Piro (San Francisco: ASP), 169

\bibitem[Covino \& G\"{o}tz(2016)]{CG16}
Covino, S. \& G\"{o}tz, D. 2016, A\&AT, 29, 205

\bibitem[Crider et al.(1997)]{Crider+97}
Crider, A., Liang E. P., \& Smith, I. A. 1997, \apj, 479, L39

\bibitem[Chand et al.(2018a)]{Chand+18a}
Chand, V. et al. 2018a, \apj, 862, 154

\bibitem[Chand et al.(2018b)]{Chand+18b}
Chand, V. et al. 2018b, eprint (arXiv:1807.01737)

\bibitem[Chandrasekhar(1960)]{Chandrasekhar1960} 
Chandrasekhar, S. 1960, Radiative Transfer (New York: Dover)

\bibitem[Chang \& Lin(2014)]{Chang-Lin-14}
Chang, Z. \& Lin, H.-N. 2014, \apj, 765, 36

\bibitem[Corsi et al.(2018)]{Corsi+18}
Corsi, A., et al. 2018, ApJ, 861, L10

\bibitem[Daigne \& Mochkovitch(1998)]{DM98}
Daigne, F. \& Mochkovitch, R. 1998, MNRAS, 296, 275

\bibitem[Deng et al.(2016)]{Deng+16}
Deng, W. et al. 2016, ApJ, 821, L12

\bibitem[Drenkhahn(2002)]{Drenkhahn-02}
Drenkhahn, G. 2002, A\&A, 387, 714

\bibitem[Drenkhahn \& Spruit(2002)]{Drenkhahn-Spruit-02}
Drenkhahn, G. \& Spruit, H. C. 2002, A\&A, 391, 1141 

\bibitem[Eichler \& Levinson(2000)]{EL00}
Eichler, D. \& Levinson, A. 2003, \apj, 529, 146

\bibitem[Eichler \& Levinson(2003)]{EL03}
Eichler, D. \& Levinson, A. 2003, \apj, 596, L147

\bibitem[Eichler \& Levinson(2004)]{Eichler-Levinson-04}
Eichler, D. \& Levinson, A. 2004, \apj, 614, L13

\bibitem[Giannios(2006)]{Giannios06}
Giannios, D. 2006, A\&A, 457, 763

\bibitem[Giannios(2008)]{Giannios-08}
Giannios, D. 2008, A\&A, 480, 305

\bibitem[Gill \& Granot(2019)]{Gill-Granot-19}
Gill, R. \& Granot, J. 2019, arXiv e-prints: arXiv:1910.05687

\bibitem[Gill \& Granot(2018)]{Gill-Granot-18}
Gill, R. \& Granot, J. 2018, MNRAS, 475, L1

\bibitem[Gill \& Thompson(2014)]{Gill-Thompson-14}
Gill, R. \& Thompson, C. 2014, \apj, 796, 81

\bibitem[Ghisellini \& Celotti(1999)]{GC99}
Ghisellini, G. \& Celotti, A. 1999, \apj, 511, L93

\bibitem[Ghisellini \& Lazzati(1999)]{Ghisellini-Lazzati-99}
Ghisellini, G. \& Lazzati, D. 1999, MNRAS, 309, L7

\bibitem[Ghirlanda, Celotti, Ghisellini(2003)]{Ghirlanda+03}
Ghirlanda, G., Celotti, A., \& Ghisellini, G. 2003, A\&A, 406, 897

\bibitem[Ghirlanda et al. (2013)]{Ghirlanda+13}
Ghirlanda, G. et al. 2013, MNRAS, 428, 1410

\bibitem[Ghirlanda et al. (2015)]{Ghirlanda+15}
Ghirlanda, G. et al. 2015, MNRAS, 420, 483

\bibitem[Goodman(1986)]{Goodman1986}
Goodman, J. 1986m \apj, 308, L47

\bibitem[G\"{o}tz et al.(2009)]{Gotz+09}
G\"{o}tz, D. et al. 2009, \apj, 695, L208

\bibitem[G\"{o}tz et al.(2013)]{Gotz+13}
G\"{o}tz, D. et al. 2013, MNRAS, 431, 3550

\bibitem[G\"{o}tz et al.(2014)]{Gotz+14}
G\"{o}tz, D. et al. 2014, MNRAS, 444, 2776

\bibitem[Granot(2003)]{Granot03} 
Granot, J. 2003, \apj, 596, L17

\bibitem[Granot(2005)]{Granot05} 
Granot, J. 2005, \apj, 631, 1022

\bibitem[Granot, Komissarov \& Spitkovsky(2011)]{GKS11}
Granot, J., Komissarov, S. S., \& Spitkovsky, A. 2011, MNRAS, 411, 1323

\bibitem[Granot \& Sari(2002)]{GS02}
Granot, J. \& Sari, R. 2002, \apj, 568, 820

\bibitem[Granot, Cohen-Tanugi, \& Do Couto E Silva(2008)]{GCD08}
Granot, J., Cohen-Tanugi, J., \& Do Couto E Silva, E. 2008, \apj, 677, 92

\bibitem[Granot et al.(2015)]{Granot+15}
Granot, J. et al. 2015, Space Sci. Rev., 191, 471

\bibitem[Granot \& Kumar(2003)]{Granot-Kumar-03}
Granot, J. \& Kumar, P. 2003, \apj, 591, 1086

\bibitem[Granot \& K\"{o}nigl(2003)]{Granot-Konigl-03} 
Granot, J. \& K\"{o}nigl, A. 2003, \apj, 594, L83

\bibitem[Granot et al.(2002)]{Granot+02}
Granot, J. et al. 2002, \apj, 570, L61

\bibitem[Granot, Piran, \& Sari(1999a)]{GPS99}
Granot, J., Piran, T., \& Sari, R. 1999a, \apj, 513, 679

\bibitem[Granot, Piran, \& Sari(1999b)]{GPS99b}
Granot, J., Piran, T., \& Sari, R. 1999b, \apj, 527, 236

\bibitem[Granot \& Ramirez-Ruiz(2011)]{GR11} 
Granot, J. \& Ramirez-Ruiz, E. 2011, in Kouveliotou, C., Woosley, S. E. Wijers, R. A. M. J., 
Gamma-ray Bursts, Cambridge Univ. Press, Cambridge (arXiv:1012.5101)

\bibitem[Granot \& Taylor(2005)]{Granot-Taylor-05} 
Granot, J. \& Taylor, G. 2005, \apj, 625, 263

\bibitem[Gruzinov(1999)]{Gruzinov99} 
Gruzinov, A. 1999, \apj, 525, L29

\bibitem[Gruzinov \& Waxman(1999)]{GW99}
Gruzinov, A. \& Waxman, E. 1999, \apj, 511, 852

\bibitem[Hasco\"{e}t et al.(2012)]{Hascoet+12}
Hasco\"{e}t, R., Daigne, F., Mochkovitch, R., \& Vennin, V. 2012, MNRAS, 421, 525

\bibitem[Inoue et al.(2011)]{Inoue+11}
Inoue, T. et al. 2011, \apj, 734, 77

\bibitem[Ito et al.(2014)]{Ito+14}
Ito, H. et al. 2014, \apj, 789, 159

\bibitem[Kalemci et al.(2007)]{Kalemci+07}
Kalemci, E. et al. 2007, \apjs, 169, 75

\bibitem[Korchakov \& Syrovatskii(1961)]{KS61}
Korchakov, A. A. \& Syrovatskii, S. I. 1961, Soviet Astr.--A. J., 5, 678

\bibitem[Kumar \& Granot(2003)]{KG03}
Kumar, P. \& Granot, J. 2003, \apj, 591, 1075

\bibitem[Kumar \& Narayan(2009)]{Kumar-Narayan-09}
Kumar, P. \& Narayan, R. 2009, MNRAS, 395, 472

\bibitem[Kumar \& Zhang(2015)]{KZ15}
Kumar, P. \& Zhang, B. 2015, Phys. Rep., 561, 1

\bibitem[Laskar et al.(2019)]{Laskar+19}
Laskar, T. et al. 2019, \apj, 878, L26

\bibitem[Lazar et al.(2009)]{Lazar+09}
Lazar, A., Nakar, E., \& Piran, T. 2009, ApJ, 695, L10

\bibitem[Lazzati(2006)]{Lazzati2006}
Lazzati, D. 2006, New J. Phys., 8, 131

\bibitem[Lazzati et al.(2000)]{Lazzati+00}
Lazzati, D., Ghisellini, G., Celotti, A., \& Rees, M. J. 2000, 529, L17

\bibitem[Lazzati et al.(2004)]{Lazzati+04} 
Lazzati, D. et al. 2004, MNRAS, 347, L1

\bibitem[Lazzati, Morsony, \& Begelman(2009)]{Lazzati+09}
Lazzati, D., Morsony, Morsony, B. J., \& Begelman, M. C. 2009, \apj, 700, L47

\bibitem[Lazzati \& Begelman(2006)]{LB06}
Lazzatti, D. \& Begelman, M. C. 2006, \apj, 725, 1137

\bibitem[Lowell et al.(2017)]{Lowell+17}
Lowell, A. W. et al. 2017, \apj, 848, 119

\bibitem[Lundman, Pe'er, \& Ryde(2014)]{Lundman+14}
Lundman, C., Pe'er, A. \& Ryde, F. 2014, MNRAS, 440, 3292

\bibitem[Lundman, Vurm, \& Beloborodov(2018)]{Lundman+18}
Lundman, C., Vurm, I., \& Beloborodov, A. M. 2018, \apj, 856, 145

\bibitem[Lyubarsky \& Kirk(2001)]{Lyubarsky-Kirk-01}
Lyubarsky, Y. \& Kirk, J. G. 2001, \apj, 547, 437

\bibitem[Lyutikov \& Blandford(2003)]{LB03} 
Lyutikov, M. \& Blandford, R. 2003, ArXiv eprints, arXiv:astro-ph/0312347

\bibitem[Lyutikov, Pariev, \& Blandford(2003)]{LPB03} 
Lyutikov, M., Pariev, V. I., \& Blandford, R. 2003, \apj, 597, 998

\bibitem[Matsumoto, Nakar, \& Piran(2019)]{Matsumoto+19}
Matsumoto, T., Nakar, E., \& Piran, T. 2019, MNRAS, 486, 1563

\bibitem[Mao \& Wang(2013)]{Mao-Wang-13}
Mao, J. \& Wang, J. 2013, ApJ, 776, 17

\bibitem[Mao \& Wang(2017)]{Mao-Wang-17}
Mao, J. \& Wang, J. 2017, ApJ, 838, 78

\bibitem[McConnell(2017)]{McConnell-17}
McConnell, M. L. 2017, NewAR, 76, 1

\bibitem[McGlynn et al.(2007)]{McGlynn+07}
McGlynn, S. et al. 2007, A\&A, 466, 895

\bibitem[Medvedev(2000)]{Medvedev-00}
Medvedev, M. V. 2000, \apj, 540, 704

\bibitem[Medvedev \& Loeb(1999)]{ML99}
Medvedev, M. V. \& Loeb, A. 1999, \apj, 526, 697

\bibitem[M\'{e}sz\'{a}ros \& Rees(2000)]{MR00}
M\'{e}sz\'{a}ros, P. \& Rees, M. J. 2000, \apj, 530, 292

\bibitem[Mooley et al.(2018)]{Mooley+18}
Mooley, K. P.., et al. 2018, Nature, 561, 355

\bibitem[Mundell et al.(2013)]{Mundell+13}
Mundell, C. G. et al. 2013, Nature, 504, 119

\bibitem[Nakar \& Oren(2004)]{NO04}
Nakar, E., \& Oren, Y. 2004, ApJ, 602, L97

\bibitem[Nakar, Piran, \& Waxman(2003)]{NPW03}
Nakar, E., Piran, T., \& Waxman, E. 2003, JCAP, 10, 005

\bibitem[Narayan \& Kumar(2009)]{Narayan-Kumar-09}
Narayan, R. \& Kumar, P. 2009, MNRAS, 394, L117

\bibitem[Nava, Nakar, \& Piran(2016)]{NNP16} 
Nava, L., Nakar, E., \& Piran, T. 2016, MNRAS, 455, 1594

\bibitem[Oganesyan et al.(2017)]{Oganesyan+17}
Oganesyan, G. et al. 2017, \apj, 846, 137

\bibitem[Paczy\'{n}ski(1986)]{Paczynski1986}
Paczy\'{n}ski, B. 1986, \apj, 308, L43 

\bibitem[Papathanassiou \& M\'{e}sz\'{a}ros(1996)]{PM96}
Papathanassiou, H. \& M\'{e}sz\'{a}ros, P. 1996, \apj, 471, L91

\bibitem[Pearce et al.(2019)]{Pearce+19}
Pearce, M. et al. 2019, APh, 104, 54 

\bibitem[Pe'er \& Ryde(2011)]{PR11}
Pe'er, A. \& Ryde, F. 2011, \apj, 732, 49

\bibitem[Piran, Sari, \& Zou(2009)]{PSZ09}
Piran, T., Sari, R., \& Zou, Y. C. 2009, MNRAS, 1107, 1113

\bibitem[Preece et al.(1998)]{Preece+98}
Preece, R. D., Briggs, M. S., Mallozzi, R. S. et al. 1998, \apj, 506, L23

\bibitem[Preece et al.(2002)]{Preece+02}
Preece, R. D., Briggs, M. S., Mallozzi, R. S. et al. 2002, \apj, 518, 1248

\bibitem[Prosekin et al.(2016)]{Prosekin+16}
Prosekin, A. Y., Kelner, S. R., \& Aharonian, F. A. et al. 2016, Phys. Rev. D, 94, 063010

\bibitem[Ravasio et al.(2018)]{Ravasio+18}
Ravasio, M. E. et al. 2018, A\&A, 613, A16

\bibitem[Rees \& M\'{e}sz\'{a}ros(1994)]{Rees-Meszaros-94}
Rees, M. J. \& M\'{e}sz\'{a}ros, P. 1994, \apj, 430, L93

\bibitem[Rees \& M\'{e}sz\'{a}ros(2005)]{RM05}
Rees, M. J. \& M\'{e}sz\'{a}ros, P. 2005, \apj, 628, 847

\bibitem[Rossi, Lazzati, \& Rees(2002)]{Rossi+02}
Rossi, E. M., Lazzati, D., \& Rees, M. J. 2002, MNRAS, 332, 945

\bibitem[Rossi et al.(2004)]{Rossi+04}
Rossi, E. M., Lazzati, D., Salmonson, J. D. et al. 2004, MNRAS, 354, 86 

\bibitem[Rybicki \& Lightman(1979)]{Rybicki-Lightman-79} 
Rybicki, G. B., \& Lightman, A. P. 1979, Radiative Processes in Astrophysics, Wiley \& Sons (New York)

\bibitem[Rutledge \& Fox(2004)]{RF04}
Rutledge, R. E. \& Fox, D. B. 2004, MNRAS, 350, 1288

\bibitem[Sari(1999)]{Sari99} 
Sari, R. 1999, \apj, 524, L43

\bibitem[Sari \& Piran(1997)]{SP97}
Sari, R. \& Piran, T. 1997, MNRAS, 287, 110

\bibitem[Salafia et al.(2015)]{Salafia+15}
Salafia, O. S. et al. 2015, MNRAS, 450, 3549

\bibitem[Singh et al.(2014)]{Singh+14}
Singh, K. P. et al. 2014, SPIE, 9144, 1

\bibitem[Sironi \& Spitkovsky(2011)]{SS11}
Sironi, L., \& Spitkovsky, A. 2011, ApJ, 726, 75

\bibitem[Sharma et al.(2019)]{Sharma+19}
Sharma, V., Iyyani, S., Bhattacharya, D. et al. 2019, ApJL, 882, L10

\bibitem[Shaviv \& Dar(1995)]{SD95}
Shaviv, N. J. \& Dar, A. 1995, \apj, 447, 863

\bibitem[Sobolev(1963)]{Sobolev1963} 
Sobolev, V. V. 1963, A Treatise on Radiative Transfer (Princeton, NJ: Van Nostrand-Reinhold)

\bibitem[Spruit et al.(2001)]{Spruit+01}
Spruit, H. C., Daigne, F., \& Drenkhahn, G. 2001, A\&A, 369, 694 

\bibitem[Steele et al.(2009)]{Steele+09}
Steele, I. A. et al. 2009, Nature, 462, 767

\bibitem[Thompson(1994)]{Thompson-94}
Thompson, C. 1994, MNRAS, 270, 480

\bibitem[Thompson \& Gill(2014)]{Thompson-Gill-14}
Thompson, C. \& Gill, R. 2014, \apj, 791, 30

\bibitem[Toma et al.(2009)]{Toma+09} 
Toma, K. et al. 2009, \apj, 698, 1042

\bibitem[Toma(2013)]{Toma2013} 
Toma, K. 2013, ArXiv eprints, arXiv:1308.5733

\bibitem[Vurm \& Beloborodov(2016)]{Vurm-Beloborodov-16}
Vurm, I. \& Beloborodov, A. M. 2016, \apj, 831, 175

\bibitem[Vurm, Lyubarsky, \& Piran(2013)]{Vurm+13}
Vurm, I., Lyubarsky, Y., \& Piran, T. 2013, \apj, 764, 143

\bibitem[Waxman(2003)]{Waxman03}
Waxman, E. 2003, Nature, 423, 388

\bibitem[Wigger et al.(2004)]{Wigger+04}
Wigger, C. et al. 2004, \apj, 613, 1088

\bibitem[Yamazaki et al.(2003)]{Yamazaki+03}
Yamazaki, R., Ioka, K., \& Nakamura, T. 2003, \apj, 606, L33

\bibitem[Yonetoku et al.(2011a)]{Yonetoku+11a}
Yonetoku, D. et al. 2011, PASJ, 63, 625

\bibitem[Yonetoku et al.(2011b)]{Yonetoku+11b}
Yonetoku, D. et al. 2011, \apj, 743, L30

\bibitem[Yonetoku et al.(2012)]{Yonetoku+12}
Yonetoku, D. et al. 2012, \apj, 785, L1

\bibitem[Zhang et al.(2019)]{Zhang+19}
Zhang, S.-N. et al. 2019, Nature Astronomy, doi:10.1038/s41550-018-0664-0 

\bibitem[Zhang \& M\'{e}sz\'{a}ros(2002)]{ZM02}
Zhang, B. \& M\'{e}sz\'{a}ros, P. 2002, \apj, 571, 876

\bibitem[Zhang \& Yan(2011)]{Zhang-Yan-11}
Zhang, B. \& Yan, H. 2011, ApJ, 726, 90

\end{thebibliography}
\end{document}